\documentclass[a4paper,11pt]{book}

\usepackage{bold-extra}

\usepackage{graphicx}
\usepackage{subfig}
\usepackage{pdfpages}
\graphicspath{ {./Figures/} }

\usepackage{geometry}
\usepackage{comment}

\pdfoutput=1
\usepackage{color}
\usepackage{cancel}
\usepackage{todonotes}
\usepackage{booktabs}
\usepackage{array}
\usepackage{multirow,rotating}
\usepackage{amsmath,amssymb,bm,graphicx,wasysym,mathrsfs,dsfont}
\usepackage{cite}
\usepackage{slashed,relsize}
\usepackage[colorlinks=true,linkcolor=black,anchorcolor=black,citecolor=blue,filecolor=cyan,menucolor=red,runcolor=filecolor,urlcolor=blue,bookmarks=true,bookmarksnumbered=true]{hyperref}

\usepackage[font=small,labelfont=bf]{caption}

\usepackage[normalem]{ulem}

\def\beq{\begin{equation}\displaystyle\displaystyle}
	\def\eeq{\end{equation}}
\def\bea{\begin{eqnarray}\displaystyle} 
	\def\eea{\end{eqnarray}}

\def\({\left(}
\def\){\right)}
\def\bry{\begin{array}}
	\def\ery{\end{array}}

\def\f{\frac}

\def\sign{{\rm{sign}}}
\def\Re{{\rm Re}}

\begin{document}
\includepdf[pages={1}]{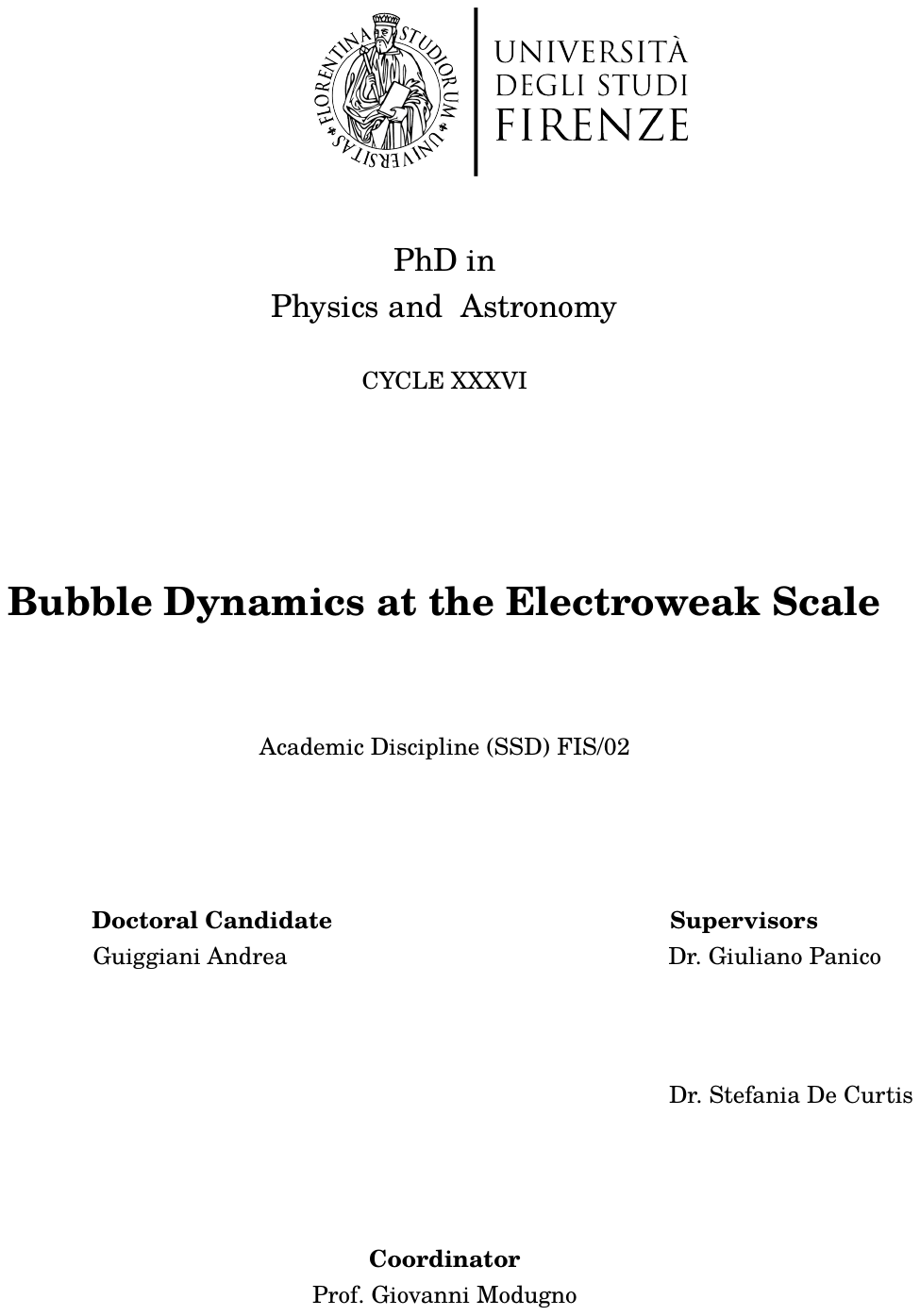}
\newpage
	\baselineskip=14pt
	\arraycolsep=2pt
 
	\begingroup
	\tableofcontents
	\endgroup 
	
	\setcounter{equation}{0}
	\setcounter{footnote}{0}
	\setcounter{page}{1}
	
	\newpage

\chapter*{Introduction}
\addcontentsline{toc}{chapter}{Introduction}

Understanding the fundamental laws of nature is the ultimate and most ambitious goal of physics. This goal can only be achieved by interpreting within a suitable theoretical framework the results of different observations and experiments. The phenomena that we observe strongly depend on the scales at which we are looking at. On subatomic scales our best description of nature is given by the Standard Model (SM) of Electroweak (EW) and strong interactions. It allows us to describe matter as fermionic fields interacting via the exchange of bosons that mediate three of the four fundamental forces of nature, namely the strong, the weak and the electromagnetic forces. At large scales, namely at cosmological scales our description of nature is vastly different. At such scales the universe appears isotropic and homogeneous and the dominant interaction that determines its dynamics is gravity. In such a case physics is well described by general relativity (GR) which instead models gravity using the language of geometry.

A great success of the SM and GR is standard cosmology which describes the evolution of the universe. This framework is strongly supported by the correct prediction of deuterium and He$_4$ abundance
originated during Big Bang Nucleosynthesis (BBN), and by the observations on the cosmic microwave background radiation (CMBR) that was generated when photons scattered for the last time with free protons and electrons after recombination. However, our knowledge still lacks a comprehension of the dynamics that occurred during earlier stages of the cosmological history of our universe. This limitation is related to our ignorance about physics at high energy scales,
in particular around the TeV and above.

Despite the great success of the SM which has been tested with a high degree of accuracy by current experiments, it is clear that it does not constitute the ultimate theory of nature. It is now widely accepted that rather to be fundamental, the SM is an effective description of nature valid up to some cutoff scale. The exact value of such cutoff scale can only be determined by future experiments probing the theory in extreme situations, but an upper bound on its value is certainly provided by the Planck mass $M_{pl}$ where quantum gravity effects become important. At such a scale it is expected that an unified theory of the SM and GR should emerge.

Additional limitations to the SM arise from cosmological observations and they involve phenomena taking place at scales far below $M_{pl}$. An example is given by the observed dark matter (DM) abundance whose existence has been widely supported by experiments. The SM spectrum, however, lacks a stable particle that could explain our current observations and the nature of DM still remains one of the main mysteries that guides the research field of high-energy physics.

Another difficulty is represented by the origin of the matter-antimatter asymmetry. Current observation shows that the universe is mostly made of matter rather than antimatter. Nevertheless such an asymmetry cannot be justified within our current models representing nature. One possible way out of this dilemma, strongly supported by cosmological observations, is the idea that this asymmetry can be explained by assuming that at some point in the evolution of the universe a baryon asymmetry was generated. However, it is not possible to assume that the baryon asymmetry was an initial condition of the universe evolution since this possibility is ruled out in natural scenarios by inflationary models~\cite{Linde:1981mu,Albrecht:1982wi}. The latter, introduced to explain the homogeneity and isotropy of CMB, predicts an evolution stage of the universe characterized by an accelerated expansion. Hence any asymmetry generated before the inflationary stage would be washed out and only an unnatural big asymmetry could be compatible with the current cosmological observation. As a consequence the asymmetry must be generated by a dynamical process which, however, is not provided by the SM.

Many theories have been proposed over the years to address the origin of the baryon asymmetry. Among them, one of the most elegant solution is Electroweak baryogenesis (EWBG)~\cite{Kuzmin:1985mm,Shaposhnikov:1986jp,Farrar:1993hn}. Differently from the leptogenesis scenario~\cite{Fukugita:1986hr} which cannot be probed by near future experiments, EWBG affects the dynamics at TeV scale and can be tested by future collider experiments. This scenario is in fact intimately related to the Higgs boson physics and, in particular, to one of the most important mechanisms in the SM: spontaneous symmetry breaking (SSB).


One remarkable aspect regarding SSB in gauge theories is that, just as in ferromagnetism, in the presence of a sufficiently high-temperature broken symmetries are restored~\cite{Kirzhnits:1972iw,Kirzhnits:1972ut}. Such an environment was present in the early Universe which was filled by an extremely hot and dense plasma. This feature, together with the universe expansion that cooled down the plasma, opens the intriguing possibility that, during its thermal history, the universe underwent a phase transition (PT) at the moment of SSB.


In the EWBG mechanism the baryon asymmetry is generated during the Electroweak phase transition (EWPT), which is triggered by a non-vanishing VEV acquired by the Higgs field when the universe cools below the EW scale ($T\sim 100$ GeV). In order for EWBG mechanism to successfully generate the observed baryon asymmetry the EWPT needs to be a first-order phase transition (FOPT), i.e.~it must be characterized by the presence of an energy barrier in the scalar potential that separates the broken and the symmetric phase. Because of this barrier, FOPT proceeds through the nucleation of bubbles of true vacuum on a background of the metastable phase. The bubbles expand and eventually collide and coalesce completing the PT.

The bubble dynamics that takes place during a first-order EWPT has many interesting aspects. The presence of expanding bubbles breaks thermal equilibrium, providing one of the three necessary conditions for baryogenesis known as Sakharov conditions~\cite{Sakharov:1967dj} (the other two being CP and baryon number violation). In addition it provides many interesting cosmological relics such as DM remnants~\cite{Konstandin:2011dr,Falkowski:2012fb,Hambye:2013dgv,Hambye:2018qjv,Bai:2018dxf,Baker:2019ndr,Azatov:2021ifm,Baldes:2020kam,Hong:2020est,Asadi:2021pwo,Baldes:2021aph}, primordial black holes~\cite{Kodama:1982sf,Hawking:1982ga,Gross:2021qgx,Baker:2021nyl,Baker:2021sno,Kawana:2021tde}, magnetic fields~\cite{Vachaspati:1991nm} and other topological defects~\cite{Kibble:1976sj,Kibble:1995aa,Borrill:1995gu} and a gravitational wave (GW) signal~\cite{Witten:1984rs,Kamionkowski:1993fg,Hogan:1986qda}. Among the many experimental signatures, the emission of GW is certainly one of the most interesting both from the experimental and the theoretical viewpoint. 

GW are sourced by three different mechanisms that take place at different stages of the bubble evolution, namely bubble collision, sound waves and turbulence. Each of these mechanisms is characterized by the release of a huge amount of energy that builds a GW signal which is potentially detectable at future space based intereferometers.

Although in the SM the EWPT is extremely weak (lattice simulations~\cite{Kajantie:1996mn,Rummukainen:1998as} indicate that the EWPT is actually a crossover), 
many BSM theories actually predict a first-order EWPT~\cite{Carena:1996wj, Delepine:1996vn,Fromme:2006cm, Dorsch:2013wja, Dorsch:2014qja, Dorsch:2016nrg, Benincasa:2022elt,Espinosa:2011ax,Cline:2012hg,Curtin:2014jma,Profumo:2014opa,Kakizaki:2015wua,Vaskonen:2016yiu,Kurup:2017dzf,DeCurtis:2019rxl}. The success of precision cosmology and the new opportunities offered by GW interferometry open new channels for testing such models. The sensitivity regions of future experiments, such as the 
European interferometer LISA~\cite{Caprini:2015zlo,Caprini:2019egz}, the Japanese project DECIGO~\cite{Kawamura:2006up,Kawamura:2011zz} and the Chinese Taiji~\cite{Hu:2017mde,Ruan:2018tsw} and TianQin~\cite{TianQin:2015yph} proposals, will indeed probe a range of the expected peak frequencies of EWPT ultimately leading to a synergy with collider experiments for the quest of probing the high-energy scales, in particular for theories potentially affecting the dynamics of the EW symmetry breaking.


An accurate understanding of the PT dynamics is crucial for a quantitative determination of the phenomenological signatures of a first-order EWPT. In particular, this is essential to get a precise characterization of the emitted GW signal and to study the possibility to achieve a successful EWBG.
Four main quantities control the PT dynamics: the temperature at which bubbles are nucleated $T_n$, the strength of the phase transition, namely the amount of energy released in the process $\alpha$, the inverse of time duration of the phase transition $\beta$ and the terminal velocity $v_w$ of the bubble wall at the moment of collision. While we have a solid framework for the determination of the first three quantities, the terminal velocity is the observable we have less theoretical control on. Its computation, deeply studied in the literature~\cite{Moore:1995ua,Moore:1995si,John:2000zq,Moore:2000wx,Konstandin:2014zta,Kozaczuk:2015owa,Bodeker:2017cim,Cline:2020jre,Laurent:2020gpg,BarrosoMancha:2020fay,Hoche:2020ysm,Azatov:2020ufh,Balaji:2020yrx,Cai:2020djd,Wang:2020zlf,Friedlander:2020tnq,Cline:2021iff,Cline:2021dkf,Bigazzi:2021ucw,Ai:2021kak,Lewicki:2021pgr,Gouttenoire:2021kjv,Dorsch:2021ubz,Dorsch:2021nje,Megevand:2009gh,Espinosa:2010hh,Leitao:2010yw,Megevand:2013hwa,Huber:2013kj,Megevand:2013yua,Leitao:2014pda,Megevand:2014yua,Megevand:2014dua}, requires an accurate modelling of the non-equilibrium properties of the plasma during the PT which are very difficult to understand.

The final speed of the wall is the result of a balance between the potential energy difference between the two phases, that drives bubble expansion and the external friction exerted by the plasma particles hitting the wall. Such friction is sourced as a back-reaction by the wall motion itself which drives the plasma out-of-equilibrium slowing down its propagation.

Depending on the size of the friction two different regimes for the wall propagation take place. In strong FOPT, characterized by a huge energy release, the wall typically never stops to accelerate and reaches ultra-relativistic velocity hence entering in the so-called runaway regime. Such situation is very interesting  from the phenomenological point of view. The huge amount of energy involved in the process allows the generation of particles that can potentially explain the observed DM abundance by further maximizing the signal of GWs emitted.

On the other hand, if the friction balances the driving force a terminal velocity is reached, and the system eventually reaches a steady state regime. This situation is particularly advantageous for the generation of the baryon asymmetry which is typically more efficient for walls moving at non ultra-relativistic speeds. 

An accurate characterization of the wall terminal velocity is hence required in order to test BSM models through the future cosmological observations. In this respect the main difficulty is ultimately represented by the out-of-equilibrium friction which crucially controls the dynamics of the bubble wall. In turn, the friction computation requires a precise characterization of the non-equilibrium properties of the plasma during the phase transition which is a highly non-trivial task.


The non-equilibrium properties of the plasma can be determined by studying the effective kinetic theory of hot gauge theories. Fermions and hard bosons, which have a momentum $p\sim T$, are successfully described by an effective Boltzmann equation, which is a master equation, that governs the statistical behaviour of a system described by a distribution function. The Boltzmann equation puts together the deviations from equilibrium caused by the external forces with the effects of multiple microscopic collisions that tend to thermalize the system. Such microscopic interactions are described by the collision operator which is an integral operator acting on the to-be-determined distribution function. From a mathematical point of view the Boltzmann equation is an integro-differential equation that is, in the vast majority of realistic cases, intractable analytically and extremely hard to solve numerically.

Various approximations and simplifications have been used in the literature to solve the Boltzmann equation. A common working hypothesis that simplifies the computation is that the system is weakly out-of-equilibrium. This situation is realized in most of the scenarios that predict a first-order EWPT and allows one to linearize the Boltzmann equation with respect to the perturbations that describe the departure from equilibrium. 
However, even in its linearized form, the Boltzmann equation remains highly challenging because of the presence of the collision operator, whose treatment is computationally very demanding.

The traditional treatment to cope with the linearized Boltzmann equation for the evolution of bubbles in first-order PT relies on an ansatz for the shape of the deviations from local equilibrium.
The seminal work of Moore and Prokopec~\cite{Moore:1995ua,Moore:1995si} contains the first computation of the friction induced by the interactions between particles and wall in terms of the out-of-equilibrium fluctuations of the different particle species in the plasma. To deal with the collision integrals, the authors introduced a formalism, that we will denote as the ``old formalism'', that employs the perfect fluid ansatz for the distribution function.
From a mathematical point of view the fluid ansatz corresponds to a truncation at the linear order of the momentum expansion of the to-be-determined distribution function around the local equilibrium distribution. Within such approach, the out-of-equilibrium properties of each particle species in the plasma are encoded in the deviations of three macroscopic quantities, namely the chemical potential, the temperature and the plasma velocity, that correspond to the coefficients in the momentum expansion.
The deviations are then computed by integrating the Boltzmann equation with a set of weights, effectively reducing the integro-differential equation to a set of ordinary differential equations. 

The fluid ansatz presents two distinctive features. The first feature is the presence of a singularity in the integrated friction, that arises from the plasma light degrees of freedom, for walls moving at the speed of sound. Such singularity, as pointed out in ref.~\cite{Moore:1995si} and recently confirmed by a more refined analysis~\cite{Dorsch:2021nje}, signalizes the break down of the naive linearization procedure for the light degrees of freedom. 
The second feature is the prediction that perturbations are suppressed in front of the bubble for supersonic walls, 
which would imply that baryogeneis is efficient only for subsonic walls.

These results have been the common lore for many years until the authors of refs.~\cite{Cline:2020jre,Laurent:2020gpg} questioned the physical nature of the singularity. They argued that the latter is an artifact of the fluid approximation and of the particular set of weights chosen to integrate the Boltzmann equation. To compute the out-of-equilibrium perturbations they proposed a new method, that they dubbed ``new formalism'', based on a different parameterization of the momentum dependence of the out-of-equilibrium perturbations, together with a factorization ansatz~\cite{Cline:2000nw} to deal with the plasma velocity perturbation. In addition they employed a different set of weights to integrate the Boltzmann equation. As a result the singularity disappears and the integrated friction has a smooth behaviour across the whole range of velocities. 
The new formalism results agree with the old formalism ones only for slowly-moving walls.


In light of refs.~\cite{Cline:2020jre,Laurent:2020gpg}, ref.~\cite{Dorsch:2021nje} revisited the problem of the singularity. In that work the authors showed that the singularity is related to a ``sonic-boom'' taking place in the plasma, where non-linear effects dominate the dynamics of the plasma light degrees of freedom. The speed of sound, where the linearization procedure breaks down, represents a critical threshold in the bubble wall evolution where the macroscopic properties of the plasma presents an abrupt transition. During such transition many non-linear effects, such as shock waves and a discontinuity in the temperature and plasma velocity, kick in and a complete characterization of the plasma behaviour using hydrodynamics must be employed. This was finally realized in ref.~\cite{Laurent:2022jrs}.

To analyze the non-equilibrium properties of the plasma the authors of ref.~\cite{Dorsch:2021nje} computed the friction acting on the bubble wall employing a generalization of the fluid ansatz, first presented in ref.~\cite{Dorsch:2021ubz} for the computation of the baryon asymmetry, where higher order terms in the momenta expansion are included. The inclusion of higher order terms determines major differences in the friction with respect to the fluid approximation. This confirms that the fluid approximation 
is not fully reliable, neither quantitatively nor qualitatively. In addition, perturbations are not 
really suppressed in front of the bubble for supersonic walls. This enlarges the region of the parameter space of many BSM models where the correct amount of baryon asymmetry can be reproduced. In addition it allows to generate the baryon asymmetry and a strong GW signal at the same time.

The approaches discussed above are clearly affected by ambiguities. They all rely on particular ansatzes for the momentum dependence of the perturbation, 
with important quantitative and qualitative consequences on the friction. In addition, the set of weights used to integrate the Boltzmann equation is to a large extent arbitrary.
Employing different sets of weights has a significant impact on the perturbations. Depending on the chosen set the integrated friction arising from heavy particle species develops a peak for velocities not necessarily related to any distinctive physical process. This is the case also for the extended fluid approximation presented in ref.~\cite{Dorsch:2021nje,Dorsch:2021ubz} where additional peaks, on top of the one present at the speed of sound, are developed for different velocities.

The presence of such peaks signalizes an intrinsic limitation of the moment method. In particular, a more refined analysis of the non-equilibrium properties of the plasma shows that the peaks in the integrated friction disappear when the system is closer to the hydrodynamic regime, namely when particle collisions are very frequent. In this situation the differences between the fluid approximation and its extension in the integrated friction reduce, signalizing that the momentum expansion is converging to the actual solution of the Boltzmann equation. However, the system is not close to the hydrodynamic regime in realistic scenarios where a FOPT takes place. In such a situation a correct characterization of the non-equilibrium properties of the plasma can be obtained only through the determination of the full solution of the linearized Boltzmann equation.

Finding a solution to the Boltzmann equation without imposing any ansatz on the perturbation constitute the main topic of this thesis. In ref.~\cite{DeCurtis:2022djw,DeCurtis:2022llw,DeCurtis:2022hlx} we presented, for the first time, the full solution to the linearized Boltzmann equation without imposing any specific ansatz on the shape of the perturbations. To account for the collision integrals we devised an iterative procedure where convergence is achieved in a small number of steps. The method we proposed avoids the ambiguities of the previous approaches allowing for a precise computation of the friction as well as for an accurate determination of the out-of-equilibrium perturbations. In particular, the full solution to the Boltzmann equation shows that the integrated friction presents a smooth behaviour across the whole velocity range, confirming that the peaks predicted by the fluid approximation and its extension are unphysical. Moreover, important quantitative and qualitative differences with both the old and new formalism in the integrated friction are present. This confirms that a precise characterization of the GW signal and the amount of baryon asymmetry generated can only be achieved by solving the full Boltzmann equation. 

Avoiding to rely on a specific ansatz for the momentum dependence of the perturbation clearly reintroduces complexity in the computation of the collision integral. This operator represents the bottleneck in the determination of the solution and its evaluation can become rapidly time-consuming if not properly manipulated. A significant improvement in the time required for its evaluation can be achieved by exploiting the operator symmetries as we showed in refs.~\cite{DeCurtis:2023hil,DeCurtis:2023aaa}.

The collision operator is endowed with two symmetries. It is rotational invariant and symmetric in the exchange between the initial and final states in collision processes. The latter symmetry allows one to interpret the operator as a Hermitian operator and allows one to spectral decompose it on the basis of its eigenfunctions. This has the main advantage to reduce a cumbersome and slow nine-dimensional integration to a fast and straightforward matrix multiplication. Furthermore, in virtue of its rotational invariance, the collision operator takes a block diagonal form when expressed on the basis of 
spherical harmonics (i.e.~Legendre polynomials), contributing to a more efficient evaluation. 
In particular, the use of the Legendre polynomial basis reveals an interesting feature of the out-of-equilibrium perturbations, which present a hierarchy in the Legendre modes. This suggests that it is possible to truncate the Legendre expansion to a finite order allowing one to simplify the equations and to improve the speed of the algorithm without affecting the accuracy of the results. 

By exploiting such symmetries a remarkable improvement in the computational performances can be achieved. The resulting algorithm allows for a fast and accurate evaluation of the out-of-equilibrium perturbations which can be computed in less than a hour on a standard desktop computer. This represents an important milestone towards a reliable and quantitative method to test the predictions of a particular BSM theory regarding the GW signal and the relics mentioned before. In particular, because the algorithm that solves the Boltzmann equation does not require a large amount of computational resources, the method presented in refs.~\cite{DeCurtis:2022djw,DeCurtis:2022hlx,DeCurtis:2022llw,DeCurtis:2023hil,DeCurtis:2023aaa} can be employed to scan the parameter space of BSM theories in search for the optimal points where the GW signal is maximized or where the correct amount of baryon asymmetry is reproduced.

An improvement in the timing performances is not the only advantage given by the exploitation of the collision operator symmetries. A deeper analysis of the hierarchy in the Legendre modes of the perturbation reveals that the latter is more pronounced close to the hydrodynamic regime. By exploiting this result it is possible to provide a semi-analytic solution to the Boltzmann that clarifies the relation between the full solution and the fluid approximation and its extensions. Such analysis confirms that the moment method is justified only when the collision processes that takes place in the plasma are very efficient in driving the system towards thermal equilibrium. When this is not realized, an accurate modelling of the non-equilibrium properties of the plasma can be given only through a full solution of the Boltzmann equation.

The characterization of the plasma dynamics through the Boltzmann equation has a series of limitations, which we will carefully analyze in this thesis to get an assessment of the reliability and uncertainty of our results. First of all the validity of the underlying assumptions needed to model the plasma as a collection of weakly interaction point particles must be checked.
This description relies on the key assumption that a large separation of scales in the plasma takes place. It is assumed that the characteristic scale of the system, which for the particular case of the EWPT is set by the domain wall (DW) width, is much larger than the microscopic scales given by the range of plasma interactions and particle wavelength. While the presence of short-range interactions in the plasma is a necessary condition to model plasma collisions as local processes, the ratio between the particle wavelength and the wall width controls the impact of quantum corrections. When the latter become relevant a description based on the Boltzmann equation is no longer suitable and different effective kinetic description, based on the Schwinger-Keldysh-Kadanoff-Baym formalism should be employed~\cite{Schwinger:1960qe,Keldysh:1964ud,Kadanoff:1962aa}.

Additional limitations come from the different approximations that are usually considered to simplify the computation of the collision integrals. Because the Boltzmann equation describes hard particles, whose momentum is typically larger than their mass, it is usually assumed that particles are massless in the computation of collision rates. The amplitude of the rates, in turn, are usually evaluated in the so called leading-log approximation where only $t$ and $u$ channels are considered since they receive a log enhancement from soft and collinear configurations.

Another possible issue arises in the presence of soft bosonic degrees of freedom. The Boltzmann equation is not a suitable effective theory to describe such excitations since their dynamics is dominated by different effects with respect to hard particles, namely the Landau damping and Debye screening, that are not captured by the Boltzmann equation. This situation is realized by the W and Z bosonic species during the EWPT since the majority of their population occupies the soft region.
A more refined theory, which correctly takes into account the damping and screening effects, was first provided by B\"odecker and Moore~\cite{Moore:2000wx,Bodeker:1998hm,Bodeker:1999ey,Bodeker:1999ud} ultimately leading to a description of the soft bosons, that can be treated as classical fields given their large occupation number, in terms of a Langevin equation.


The resulting theory nevertheless suffers of IR divergences caused by the ultra-soft modes whose wavelength is larger than the bubble wall width $L$ and that cannot be described with an effective kinetic theory. During the EWPT these divergences are naturally regularized in the broken phase by the mass that particle acquires, but is not cancelled in the symmetric phase, where particles are massless. The integrated friction generated by the W and Z bosons thus present a divergent contribution that arises from these ultra-soft gauge bosons in the symmetric phase, namely from the plasma out-of-equilibrium perturbations far from the DW where perturbations are supposed to be largely suppressed. In this case the IR divergence is simply regularized cutting off the contributions arising from all those modes that cannot be described by the effective kinetic theory. However, this result is far from being satisfactory being the friction computed from such effective description very sensible on the IR cutoff.


Due to this large sensitivity and the lack of a theory that correctly describes the behaviour of ultra-soft modes, the contribution of weak gauge bosons to the friction constitutes the main source of theoretical uncertainty on the terminal velocity. This uncertainty is more difficult to handle than the one provided by the different approximations employed to simplify the computation of the collision integrals, such as the massless and the leading-log approximation. To reduce this uncertainty it is in fact necessary to improve our description of the very IR modes which is a highly non-trivial task since the main effects that dominate their dynamics are still not completely understood. 



\bigskip

This thesis is structured in the following way. In Chapter~\ref{ch:phase_transitions} we discuss the physics of cosmological phase transitions. We begin by shortly reviewing the effective potential and the finite temperature corrections which constitute the basic ingredients for the analysis of cosmological phase transitions. We later focus on FOPTs by highlighting their main differences with respect to second-order phase transitions (SOPT) and presenting the main quantities that characterize their dynamics. We then specialize to the study of the EWPT and its realization in the SM and its scalar singlet extension. Finally we conclude the chapter reviewing GW explaining which are the relevant mechanisms that generate the signal and the quantities that control it.

In Chapter~\ref{ch:bubble_dynamics_and_plasma_hydrodynamics} we address the dynamics of the DW. We derive the equation of motion of the background fields using the WKB approximation and determine the expression of the friction in terms of the out-of-equilibrium perturbations. After a discussion regarding the main forces that determine the DW dynamics, we consider the hydrodynamic properties of the plasma and present the strategy to compute the temperature and plasma velocity without relying on any linearization procedure. 
At last we present the algorithm that we will use to compute the DW terminal velocity.

In Chapter~\ref{ch:effective_kinetic_theory} we derive the effective kinetic theory that describes hard particles assessing at the same time the validity domain of the effective description we provide. We begin the chapter by reviewing the impact of thermal corrections to the dynamics of fields in the plasma and we discuss which processes should be included in the collision integral to properly characterize the effective kinetic theory. In the second part of the chapter we discuss the Boltzmann equation. After a short review of its basic properties, we present the equation that describes the behaviour of the plasma during the EWPT. At last we present the multipole expansion in the limit of slow moving walls and discuss the hierarchy in the angular momentum modes of the perturbation.

We present in Chapter~\ref{ch:boltzmann_ansatz} the old and new formalism. 
After a review of the two strategies, we compare the main results that the two method provide assessing at the same time their main drawbacks. In particular, we show the origin of the singularity and the relation between the peaks and the set of weight used to integrate the Boltzmann equation. 

Finally in Chapter~\ref{ch:exact_solution_to_liearized_Boltzmann_equation} we present the algorithm that we developed to solve the Boltzmann equation without imposing any ansatz on the shape of the perturbation. The first part of the chapter is dedicated to the discussion of the algorithm and of its main features. We first discuss the iterative procedure and we later focus on the spectral decomposition of the collision operator. 
To assess the validity of our approach we compare our results with the old and new formalism. We then adopt our improved strategy to compute the terminal velocity of the bubble wall by including also the friction contribution arising from light particle species following the method outlined in Chapter~\ref{ch:bubble_dynamics_and_plasma_hydrodynamics} assessing the importance of the out-of-equilibrium. At last we include the gauge bosons. We show their impact on the DW dynamics and we conclude the chapter by analyzing the main theoretical limitations in the computation of the friction.

We finally include two technical appendices. 
Appendix~\ref{ap:exact_solution}
discusses the integration of the Boltzmann equation with a generic set of weights. 
In addition we provide the derivation of the analytic solution of the moment method that we discuss in Chapter~\ref{ch:boltzmann_ansatz}. In Appendix~\ref{app:evaluation_collision_integrals} we present the computation of the kernel of the collision integrals.

\chapter{Phase Transitions in the Early Universe}
\label{ch:phase_transitions}



In this chapter we review the physics of cosmological phase transitions with particular focus on transitions of first-order, which constitute one of the main topics of this thesis. As we already mentioned in the introduction, FOPT, along with many other relics like DM, primordial black holes, topological defects and  baryon asymmetry, can generate a background of GWs  detectable at future space-based interferometers.

The SM does not predict any FOPT. The EWPT, which we will discuss in-depth during this thesis is actually a weak crossover yielding poor experimental signatures. Nevertheless, as explained in the introduction, many BSM theories actually predict a first-order EWPT which provides an extremely rich phenomenology and makes such scenarios very promising for solving some of the puzzles of cosmology.

We begin this chapter by reviewing the derivation of the SM effective potential at $1-$loop by first including only the zero temperature corrections. After a short discussion regarding the basic aspects of thermal field theories, we discuss the properties of FOPT and the impact of finite temperature corrections on the SM effective potential analyzing the dynamics of the EWPT. 

To realize a first-order EWPT we consider the scalar singlet extension of the SM, where the spectrum of the theory is enriched by an additional scalar and discuss the thermal history of the universe in this particular scenario. We finally conclude the chapter by discussing the relevant mechanisms that generate the GW signal and the key quantities that characterize it.

\section{The zero-temperature Effective Potential}
\label{sec:zero_temperature_corrections}


The study of the early universe dynamics requires a careful investigation of the effective potential governing the scalar dynamics. The effective potential indeed determines whether the universe underwent a phase transition and also characterizes some of its properties. Such characterization, however, cannot rely on the tree level effective potential alone. Radiative corrections can indeed generate additional minima in the effective potential or a barrier between the symmetric and the broken phase, crucially affecting the properties of potential PT. In addition, in presence of a high temperature and high density plasma broken symmetries can be restored at sufficiently high-temperatures, as we discussed in the introduction.

An adequate characterization of the PT properties thus requires to account also for loop and thermal corrections to the effective potential of the theory which we are going to briefly review in the first part of this chapter. For this review of the effective potential we mainly follow ref.~\cite{Quiros:1999jp}
while additional details about finite-temperature quantum field theory are also discussed in~\cite{Kapusta:2006pm}. We begin our analysis by first focusing on $1-$loop corrections to the tree level effective potential.

To understand why 1-loop corrections are important let us consider a real scalar theory invariant under the reflection symmetry $\phi\leftrightarrow-\phi$, whose Lagrangian is
\begin{equation}
    {\cal L} = \frac{1}{2}\partial_\mu\phi\partial^\mu\phi - V_0\,,
\label{eq:scalar_theory}
\end{equation}
with
\begin{equation}
\label{eq:tree_level_scalar}
    V_0 = \frac{\mu^2}{2}\phi^2 - \frac{\lambda}{4}\phi^4\,,
\end{equation}
where  we require the quartic coupling $\lambda$ to be positive  to ensure the stability of the theory. Such example has the additional purpose to illustrate the main strategy adopted to compute 1-loop corrections.

The above theory already at tree level presents an interesting phenomenology controlled by the sign of the quadratic coupling $\mu^2$. Indeed, when $\mu^2 > 0$ the potential has one stable minimum at the origin where the vacuum expectation value (VEV) of the field $\phi$ is $\langle \phi\rangle$ = 0. On the other hand, if $\mu^2 < 0$ the configuration $\langle \phi \rangle = 0$ is unstable and two stable minima at $\langle \phi \rangle = \pm\sqrt{-\mu^2/\lambda}$ are present.

The case $\mu^2 < 0$ is a typical example of a theory where SSB occurs, that is when the ground state is not invariant under the symmetries of the theory. 
In this particular case, SSB happens at tree level. On the other hand
the case $\mu^2 = 0$, where the theory is also conformal invariant, is of particular interest since it provides a first example that highlights the importance of radiative corrections for the study of the effective potential.

To analyze this case in more detail, we need to compute the $1-$loop corrections to the effective potential. For that, one needs to evaluate the $1$PI action which yields the following result~\cite{Quiros:1999jp}
\begin{equation}
    V_{1} = - \sum_{n= 0}^\infty\frac{1}{n!} \phi_c^n\Gamma^{(n)}_{1\textrm{PI}}(p_i = 0) 
\end{equation}
where $\phi_c$ identifies the classical value of the field, $n$ is the number of external legs to be considered in the diagrams, while $p_i$ are the momenta in the loop. 
For the scalar theory in~(\ref{eq:scalar_theory}) the resulting  $1-$loop effective potential correction is
\begin{equation}
    V_{1} = -\int\sum_{n=1}^\infty\frac{d^4 p}{(2\pi)^4}\frac{1}{2n}\left[\frac{3\lambda\phi_c^2}{p^2-\mu^2 + i\epsilon}\right] = -\frac{i}{2}\int\frac{d^4p}{(2\pi)^4}\log\left[1-\frac{3\lambda\phi_c^2}{p^2-\mu^2+i\epsilon}\right]\,.
\end{equation}
To obtain the final result we perform a Wick rotation, $p^0 = -ip^0_E$ and discard the terms which are field independent yielding
\begin{equation}
\label{eq:zero_temperature_potential_correction}
    V_{1} = \frac{1}{2}\int\frac{d^4p_E}{(2\pi)^4}\log[p^2_E + m^2(\phi_c)]
\end{equation}
where we introduced the shifted mass 
\begin{equation}
    m^2(\phi_c) \equiv \frac{d^2 V(\phi_c)}{d\phi_c^2} = \mu^2 + 3\lambda\phi_c^2
\end{equation}

The $1-$loop potential suffers of UV divergences which must be regularized. By adopting the $\overline{\rm MS}$ scheme and imposing that the quartic coupling is $\lambda$ at the renormalization scale $\mu_R$, we obtain the well known Colemann-Weinberg (CW) potential
\begin{equation}
\label{eq:CW_correction}
    V_{\rm CW} = \frac{m^4(\phi_c)}{64\pi^2}\left[\log\left(\frac{m^2(\phi_c)}{\mu^2_R}\right) - c\right]\,,
\end{equation}
where $c = 3/2$ . The final effective potential of the theory is then obtained by summing the CW correction to the tree-level effective potential in eq.~(\ref{eq:tree_level_scalar}). 

It would appear that the loop corrections have the effect to transform the minimum in the origin to a maximum and to generate a new minimum. However, it is important to emphasize that radiative corrections, in this case, do not provide SSB. The new minimum, in fact, is generated at $\lambda \log(m^2(\phi_c)/\mu^2_R) = -16\pi^2/9$ where the logarithm takes a large negative value, namely in a region far outside the validity domain of perturbation theory. In fact higher order corrections are expected to contribute bringing higher powers of $\lambda \log(m^2(\phi_c)/\mu^2_R)$.

Nevertheless, there are theories where SSB is actually triggered by radiative corrections at $1-$loop order in some regions of their parameter space. A simple example is provided by scalar QED, as shown in ref.~\cite{Coleman:1973jx}. The crucial difference between such a theory and the simple real scalar theory that we analyzed before is the presence of the additional gauge interaction between the complex scalar and the photon. This allows to generate for such a theory a minimum in a region where the perturbative expansion is under control. 

Although to show that SSB occurs for arbitrary but small values of the quartic coupling $\lambda$ and the gauge coupling $e$ one must rely on the renormalization group equations~\cite{Coleman:1973jx}, it is possible to illustrate the mechanism of SSB in scalar QED by analyzing its effective potential in the tuned scenario $\lambda \sim e^4$. For that, we require the generalization of eq.~(\ref{eq:CW_correction}) to account for the presence of additional particles in the spectrum of the theory. The general result is
\begin{equation}
\label{eq:CW_correction_general}
    V_{\rm CW} = N_i\frac{m^4_i(\phi_c)}{64\pi^2}\left[\log\left(\frac{m^2_i(\phi_c)}{\mu^2_R}\right)-c_i\right]\,,
\end{equation}
where $N_i$ are the degrees of freedom of the corresponding field, with an extra minus if the field is a fermion, while the constant $c_i$ is $3/2$ for scalar fields and fermions and $5/6$ for gauge bosons.

By applying eq.~(\ref{eq:CW_correction_general}) to the specific case of scalar QED we can show that the effective potential of the theory at $1-$loop level is given by
\begin{equation}
    V = V_0 + V_{\rm CW} = \frac{\lambda}{4}\phi^4_c + \frac{(3e^2\phi_c^2)^2}{64\pi^2}\left[\log\left(\frac{3e^2\phi_c^2}{\mu^2_R}\right) - \frac{5}{6}\right]\,.
\end{equation}
Once again, such a theory predicts the presence of a symmetry breaking minimum. However, differently from the real scalar theory, the minimum is in a region where the logarithm is small hence keeping under control the perturbative expansion. This is a consequence of the fact that the perturbative expansion is controlled by the gauge coupling and not by the quartic coupling $\lambda$ as in the previous case making possible for the radiative corrections to be small compared to the tree-level contribution. Moreover, because the renormalization scale is arbitrary, it is possible to choose it in such a way that $\mu^2_R = 3e^2\langle\phi\rangle^2$, with $\langle\phi\rangle$ the VEV of the field in the new phase. In such a way the renormalized value of $\lambda$ is completely determined by the gauge coupling $e$. Nevertheless it is important to emphasize that the renormalized theory is still described by two parameters, namely $e$ and $\langle\phi\rangle$, having traded the dimensionless coupling $\lambda$ with the VEV of the field in the broken phase. This is the well-known phenomenon of dimensional transmutation. Hence the effective potential of scalar QED is
\begin{equation}
    V=\frac{3e^4\phi_c^4}{64\pi^2}\left[\log\left(\frac{\phi_c^2}{\langle\phi\rangle^2}\right)-\frac{1}{2}\right]\,.
\end{equation}

\begin{figure}
    \centering
    \includegraphics[width = 0.47\textwidth]{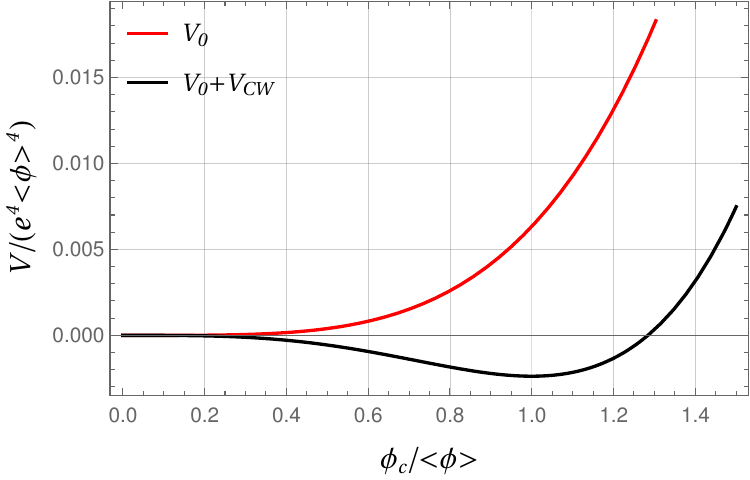}
    \caption{Effective potential of scalar QED theory at tree level (red plot) and at $1-$loop level (black plot) for the case $\mu^2 = 0$. The plot clearly shows that radiative corrections generate the SSB minimum.}
    \label{fig:CW_potential}
\end{figure}

The impact of radiative corrections is shown in Fig.~\ref{fig:CW_potential} where we plot the tree level (red plot) and $1-$loop (black plot) effective potential of scalar QED. This example, together with presenting the basic tools for the study of the effective potential, further emphasized the role of radiative corrections in the study of the SSB mechanism.

\subsection{Cut-off regularization scheme}

To study a phase transition in the SM and in BSM models the $\overline{\rm MS}$ regularization scheme that we adopted in the previous example is not the best choice. A more useful scheme for our goal is indeed provided by a cut-off regularization scheme where the UV divergences are removed by a cut-off $\Lambda$ at high energies. This setup is preferred for the study of PT since radiative corrections do not modify the position of the ground state and the masses in the true vacuum of the theory. 
By using such a scheme the $1-$loop effective potential of the SM is
\begin{equation}
    V_1 = \frac{1}{64\pi^2}\sum_{i}N_i\left[2m^2_i(h_c)\Lambda^2 + m^4(h_c)\left(\log\frac{m^2(h_c)}{\Lambda^2} - \frac{1}{2}\right)\right]\,.
\end{equation}
The counter terms are chosen in such a way that the $1-$loop corrections to the effective potential do not modify the Higgs VEV, $v = 246.22$ GeV, and its mass $m_h = 125$ GeV, namely
\begin{equation}
    \frac{d V_{eff}}{d h}\Bigr|_{h = v} = 0, \;\;\;\;\;\;\;\;\;\;\;\; \frac{d^2 V_{eff}}{dh^2}\Bigr|_{h = v} = m^2_h.
\end{equation}
By imposing the above conditions, the effective potential takes the following form
\begin{equation}
\label{eq:cutoff_regscheme_sm}
    V_{eff} = V_0(h) + \frac{1}{64\pi^2}\sum_{i}N_i\left[2m^2_i(h)m_i(v)^2 + m_i^4(h)\left(\log\frac{m_i^2(h)}{m_i(v)^2} - \frac{3}{2}\right)\right]
\end{equation}
where we dropped the subscript $c$ for the Higgs field.

The above potential cannot be used to describe the Goldstone field corrections. These suffer of IR divergences since their masses vanish in the EW symmetry breaking (EWSB) vacuum, namely when $\langle h \rangle = v $. To regularise this divergence an IR cutoff is imposed by substituting the mass $m_\chi(v)$ with $m_h(v)$ in eq.~(\ref{eq:cutoff_regscheme_sm}).

\section{Finite Temperature Field Theory}
\label{sec:finite_temperature_field_theory}

As discussed at the beginning of the previous section, the presence of a high-temperature and high-density plasma plays a crucial role in the dynamics of the early universe. In such situation the hypothesis of an empty space-time is no longer valid and we can no longer model the behaviour of the system using the tools that we described in the previous section. In particular, the thermal interactions between the system and the background environment must be considered since they provide important corrections to the effective potential.

These corrections are responsible for the restoration of broken symmetries at high temperature with important consequences on the thermal history of the universe. In fact, due to the universe expansion which cools down the plasma, restored symmetries are broken once the Universe reaches a critical temperature threshold triggering a PT. Our aim in this section is to provide a concise review of the main results of finite-temperature field theory providing the expression of the finite-temperature corrections to the effective potential. A full discussion of the whole subject is beyond our scope and we refer to~\cite{Quiros:1999jp, Kapusta:2006pm} and references therein for further reading.

\subsection{The KMS condition}

As a first step we need to characterize equilibrium states in finite-temperature quantum field theory. For that we take a preliminary step by analyzing their computation in non-relativistic quantum mechanics. For this purpose we consider a system made of a single particle species confined in a box of finite volume ${\cal V}$. Such a system, provided suitable boundary conditions on the walls of the box are imposed, is described by a Hamiltonian $H$ while its state is described by a density matrix $\rho$. The expectation value for an observable ${\cal O}$ is defined as
\begin{equation}
    \langle {\cal O} \rangle = Tr(\rho {\cal O})\,,
\end{equation}
where the trace is performed over the Hilbert space of the system.

For a system in contact with a thermal bath with temperature $T$, the equilibrium state is described by the canonical ensemble that is
\begin{equation}
    \rho = \rho_{\beta} \equiv Z^{-1} e^{-\beta H}\,,
\end{equation}
with $\beta = T^{-1}$, having set the Boltzmann constant $k_B = 1$ and with $Z$ the partition function. This setup is convenient for the description of systems with a fixed volume ${\cal V}$, particle number $N$ and temperature, where the latter is kept constant by the energy exchanges between the system and the thermal bath.

Of course different ensembles can be considered. The proper description relies on the properties of the system, in particular on which quantities are constant. The equilibrium state of an isolated system, whose energy $E$ is fixed, is described by the micro-canonical ensemble. On the other hand if the system exchanges both energy and particles with the reservoir then the equilibrium state is described by the grand-canonical ensemble.

These three different descriptions, however, are all equivalent in the thermodynamic limit, that is when we take the limit $E\rightarrow \infty,\; N\rightarrow\infty,\; {\cal V}\rightarrow\infty$ but the ratios $E/{\cal V}$ and $N/{\cal V}$ are kept finite.  
In view of this limit, which is the one we are interested in for field theories, we adopt the grand-canonical ensemble. For such a case the equilibrium state is
\begin{equation}
    \rho_{\beta\mu} = {\cal G}^{-1} e^{-\beta(H - \mu N)}
\end{equation}
where $\mu$ is the chemical potential and  ${\cal G}$ is the grand-partition function.

A generalized characterization of equilibrium states, which is also valid in the thermodynamic limit, was first provided by Kubo, Martin and Schwinger (KMS) for the canonical ensemble~\cite{Kubo:1957mj,Martin:1959jp}. The result was later generalized by Haag, Hugenwoltz and Winnick~\cite{Haag:1967sg}, that postulated a similar characterization for the grand-canonical ensemble. To present the idea we can focus on the KMS condition. The characterization of equilibrium states provided by the KMS condition has a profound impact on our discussion since it allows us to provide a suitable framework where we can describe the equilibrium states of quantum fields which interact with a thermal bath.

We begin by defining the time translate of an operator $A$,
\begin{equation}
    \alpha_t(A) = e^{iHt}Ae^{-iHt}\,,
\end{equation}
and we then observe that 
\begin{equation}
\begin{split}
        \langle \alpha_z(A)B \rangle_\beta\equiv \langle \rho_\beta\alpha_z(A)B \rangle & = Z^{-1}Tr(e^{-\beta H} e^{iHz} A e^{-iHz} B) = Z^{-1}Tr(Be^{iH(z+i\beta)}Ae^{-iHz}) \\
        & = Z^{-1} Tr(e^{-\beta H} B e^{iH(z+i\beta)}Ae^{-iH(z+i\beta)}) = \langle B\alpha_{z+i\beta}(A)\rangle_\beta\,,
\end{split}
\end{equation}
where in the above equation $B$ is a generic operator, we used the cyclic property of the trace and  we replaced the real time parameter with a complex parameter $z$. Next, for each pair of operators, we introduce two functions
\begin{equation}
\begin{split}
    & f_{AB}^{\beta}(z) = \langle \alpha_z(A)B\rangle_\beta\, \\
    & g_{AB}^{\beta}(z) = \langle B\alpha_z(A)\rangle_\beta\,.
\end{split}
\end{equation}
We find that $f_{AB}^\beta$ is analytic in the strip $-\beta <{\rm Im}(z)<0$, while the function $g_{AB}^\beta$ is analytic in $0 <{\rm Im}(z)< \beta$. The KMS condition states that, for an equilibrium state,
\label{eq:KMS_condition}
\begin{equation}
    f_{AB}^\beta(t) = g_{AB}^\beta(t+i\beta)\,,
\end{equation}
where $t$ is real.

We can now apply the KMS condition to quantum fields and discuss how the presence of a thermal bath affects the dynamics of the system. For that we begin by examining how the field propagator is modified when finite-temperature effects are considered. 
Let us first discuss the case of bosonic fields, focusing on a scalar field propagator. For a fermionic field the derivation will be similar but the final result differs due to the statistics.

We define the two-point Green function for a scalar field as 
\begin{equation}
    G_C(x-y) =\langle T_C(\hat\phi(x)\hat\phi(y)) \rangle = \theta_C(x^0-y^0)G_+(x-y) + \theta_C(y^0 - x^0) G_-(x-y)\,,
\end{equation}
where $C$ is a generic path in the complex plane (${\rm Re}(x^0)$, ${\rm Im}(x^0)$) and $T_C$ denotes that fields should be ordered accordingly to the path $C$ chosen. We further defined
\begin{equation}
    G_+(x-y) = \langle \hat\phi(x)\hat\phi(y)\rangle\;\;\;\;\;\;\;\;\;\; G_-(x-y) = \langle \hat\phi(y)\hat\phi(x)\rangle\,.
\end{equation}

An important property that we require is the analyticity of the Green function with respect to $t$. This constrains the shape of the contours $C$ in such a way that $G_C$ must be analytic in the strip $-\beta<{\rm Im}(x^0-y^0)<\beta$. The latter result easily follows by analyzing the domain where $G_+$ and $G_-$ are analytic. In addition, since we require that the system is at equilibrium, the Green functions $G_+$ and $G_-$ must satisfy the KMS condition, which implies
\begin{equation}
G_+(t, {\bf x}) = G_-(t+i\beta,{\bf x})\,.
\end{equation}

We specialize our description by choosing as a path $C$ a straight line in the imaginary direction. Such choice allows us to compute the propagators in the so-called imaginary time formalism. Of course different choices of $C$ are possible and the computation of the Green function depends on the choice of the path. Nevertheless, to illustrate the impact of finite-temperature correction, the imaginary time formalism is a particular convenient framework.

In the imaginary time formalism the KMS condition implies that the bosonic two-point Green function $G_B$ is periodic of period $\beta$, namely
\begin{equation}
    G_B(\tau + \beta) = G_B(\tau)\,
\end{equation}
where $\tau = it$. A similar result holds for the fermionic Green function $G_F$ with the only difference that $G_F$ is antiperiodic.

The main consequence of these periodicity and antiperiodicity conditions is that the energy of quantum fields is discretized and the only modes allowed are the Matsubara modes which are
\begin{equation}
\begin{split}
\label{eq:matsubara_frequencies}
    &\omega^B_n = 2n\pi T\,,\\
    &\omega^F_n = (2n +1)\pi T\,,
\end{split}
\end{equation}
for bosons and fermions respectively. Notice that for bosonic fields a mode $\omega^B_0 = 0$ is allowed. 
Such states are known as soft bosons and they play a crucial role in the determination of the plasma properties during the phase transition, as we are going to discuss further in the thesis. Modes with $n > 0$ are instead known as hard-modes and as a direct consequence of the Pauli principle, these are the only modes allowed for fermions.

By computing the propagators in the imaginary time formalism we can finally provide the Feynman rules in the presence of finite-temperature corrections. We briefly report the Feynman rules here
\begin{itemize}
    \item Boson propagator: $ \displaystyle \frac{-i}{\omega_n^2 + {\bf p}^2  + m^2}\,,$
    \item Fermion propagator: $ \displaystyle \frac{-i(\gamma \cdot p - m)}{\omega_n^2 + {\bf p}^2  + m^2}\,,$ with $ \displaystyle p^\mu = (\omega^F_n, {\bf p})$
    \item Loop: $ \displaystyle i T\sum_{n=-\infty}^{n=\infty}\int\frac{d^3{\bf p}}{(2\pi)^3}\,,$
    \item Vertex: $ \displaystyle -\frac{i}{T}\delta_{\sum_i \omega_i}(2\pi)^3\delta(\sum {\bf p}_i)$ 
\end{itemize}
Using the above set of rules we can finally compute perturbatively the finite-temperature correction to the effective potential.

\subsection{Finite-Temperature Corrections to the Effective Potential}

To compute the finite-temperature effective potential we  follow a procedure analogous to the one employed to compute the zero temperature corrections in Section \ref{sec:zero_temperature_corrections}. The only difference is in the Feynman rules adopted to evaluate the diagrams. As for the case of zero temperature radiative corrections, we first consider the corrections provided by loops involving real scalar fields and then generalize the result. Using the Feynman rules of finite-temperature field-theory eq.~(\ref{eq:zero_temperature_potential_correction}) becomes~\cite{Arnold:1992rz,Weinberg:1974hy}
\begin{equation}
    V_1^\beta = T\sum_{n=-\infty}^{\infty}\int\frac{d^3{\bf p}}{(2\pi)^3}\log(\omega_n^2 + \omega^2)\,,
\end{equation}
where $\omega_n$ are the bosonic Matsubara frequencies and where
\begin{equation}
    \omega^2 = {\bf p}^2 + m^2(\phi_c)\,.
\end{equation}
The sum over $n$ contains a field independent divergent part. To perform the sum and obtain a finite result we define a quantity $v(\omega)$ such that
\begin{equation}
    v(\omega) = \sum_{n=-\infty}^\infty\log(\omega_n^2 + \omega^2)\;\;\;\;\;\;\;\;\frac{\partial v(\omega)}{\partial \omega} = \sum_{n=-\infty}^\infty\frac{2\omega}{\omega_n^2 + \omega^2}\,.
\end{equation}
We hence perform the sum yielding 
\begin{equation}
    \frac{\partial v(\omega)}{\partial \omega} = \frac{2}{T}\left(\frac{1}{2} + \frac{e^{-\beta \omega}}{1-e^{-\beta\omega}}\right)\,,
\end{equation}
and then we integrate in $\omega$ providing the final result

\begin{equation}
\label{eq:boson_finite_temperature_corrections}
    V^\beta_1 = \int\frac{d^3{\bf p}}{(2\pi)^3}\left(\frac{\omega}{2} + T \log(1-e^{-\beta \omega})\right)\,.
\end{equation}

The first term in the sum in eq.~(\ref{eq:boson_finite_temperature_corrections}) corresponds exactly to the zero temperature corrections as one can easily show using the residue theorem. The second term, instead, describes the finite-temperature corrections. Such a term has a nice physical interpretation and indeed corresponds to the free energy of a bosonic gas. Following this interpretation it is straightforward to generalize the result to the fermionic case. It is not hard to show that in such situation the effective potential receives a correction given by

\begin{equation}\label{eq:fermion_finite_temperature_corrections}
    V_1^\beta = -\int\frac{d^3{\bf p}}{(2\pi)^3}\left(\frac{\omega}{2} + T \log(1+e^{-\beta \omega})\right)\
\end{equation}
where the temperature dependent part is indeed the free energy of a fermionic gas.

Equations~(\ref{eq:boson_finite_temperature_corrections}) and~(\ref{eq:fermion_finite_temperature_corrections}) provide the finite-temperature corrections to the effective potential in the presence of thermodynamic equilibrium. The integrals involving the temperature corrections can be further simplified and the finite-temperature corrections are usually written as
\begin{equation}
    V_T = \frac{T^4}{2\pi^2}J_{B/F}[m^2(\phi_c)/T^2]
\end{equation}
where the functions $J_{B/F}$ that account respectively for bosons and fermions loops are
\begin{equation}\label{eq:thermal_integrals}
    J_{B/F}(y) = \int_0^\infty dx x^2 \log(1\mp e^{-\sqrt{(x^2 + y^2)}})\,.
\end{equation}
The functions $J_{B/F}$ admit useful expansions for small arguments, i.e. in the high temperature limit. Studying such limit proves to be useful to understand some of the basic properties of PT. For the bosonic contribution we find
\begin{equation}\label{eq:boson_high_temperature_expansion}
    J_B[m^2/T^2]=-\frac{\pi^4}{45}+\frac{\pi^2}{12}\frac{m^2}{T^2}-\frac{\pi}{6}\frac{m^3}{T^3}-\frac{m^4}{32 T^4}\left(\log\left(\frac{m^2}{T^2}\right)-5.4076\right)+\Delta J_B(m^2/T^2)\,,
\end{equation}
with 
\begin{equation}\label{eq:fermion_high_temperature_expansion}
    \Delta J_B(x^2)=-2\pi^{7/2}\sum_{l=1}^{\infty}(-1)^l\frac{\zeta(2l+1)}{(l+1)!}\Gamma\left(l+\frac{1}{2}\right)\left(\frac{x^2}{4\pi^2}\right)^{l+2}\,,
\end{equation}
while for the fermionic case we find
\begin{equation}
    J_F[m^2/T^2]=\frac{7\pi^4}{360}-\frac{\pi^2}{24}\frac{m^2}{T^2}-\frac{m^4}{T^4}\left(\log\left(\frac{m^2}{T^2}\right)-2.6351\right)+\Delta J_F(m^2/T^2),
\end{equation}
with
\begin{equation}
    \Delta J_F(x^2)=-\frac{\pi^{7/2}}{4}\sum_{l=1}^{\infty}(-1)^l\frac{\zeta(2l+1)}{(l+1)!}(1-2^{-2l-1})\Gamma\left(l+\frac{1}{2}\right)\left(\frac{x^2}{\pi^2}\right)^{l+2}\,.
\end{equation}
\begin{figure}
\centering
    \subfloat[]{\includegraphics[width=0.23\textwidth]{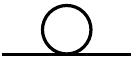}}
    \subfloat[]{\includegraphics[width=0.23\textwidth]{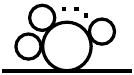}}
    \caption{Loop corrections to the propagator of a real scalar field $\phi^4$: (a) thermal correction to the mass, (b) daisy diagram with n-loops. The daisy diagrams are obtained by attaching to the diagram in panel (a) $n$ quadratically divergent loops. Figure borrowed from~\cite{Curtin:2016urg}}
\label{fig:daisy_diagrams}
\end{figure}

The inclusion of finite-temperature corrections has the important effect to restore broken symmetries in the high-temperature limit. However, at high-temperature, the perturbative expansion breaks down due to IR divergences caused by soft bosons~\cite{Weinberg:1974hy}. In particular, due to the presence of an additional scale parameter, namely the temperature $T$, the connection between the powers of the coupling and the loop order is lost. To provide a consistent perturbative theory for the study of PT one needs to provide a resummation procedure that correctly takes into account all the thermal effects. This resummation was systematically pursued in refs.~\cite{Gross:1980br,Parwani:1991gq,Arnold:1992rz} where the authors provided the so-called full dressing resummation procedure that correctly accounts for the so-called ``daisy diagrams'' depicted in Fig.~\ref{fig:daisy_diagrams}.

The full-dressing procedure does not solve all the IR issues of the effective potential. Additional strategies to address the IR behaviour of transverse gauge bosons~\cite{Linde:1980ts}, or the IR cutoff  
sensitivity to the effective potential~\cite{Croon:2020cgk} should be employed. Alternative approaches to solve such problems have been recently proposed, see for instance~\cite{Croon:2020cgk, Gould:2022ran, Ekstedt:2022zro, Ekstedt:2023anj, Ekstedt:2023oqb}. A full discussion about these issues is beyond our scope. Ultimately we will be interested in the characterization of the non-equilibrium dynamics that takes place during the EWPT which is independent from the details of the potential that triggers the PT. Since we want to compare our results with the one present in the previous literature where the daisy diagram were not resummed,  
we are not going to include the full-dressing procedure for the finite temperature corrections and we will use the cut-off renormalization scheme to regularize UV divergence in the zero temperature $1-$loop potential.

\section{Cosmological first-order and second-order phase transitions}\label{sec:cosmological_phase_transition}

Having laid down all the necessary ingredients to discuss the early universe dynamics in the last two sections, we can finally address the topic of cosmological PT. The idea that the universe underwent a PT is a very intriguing one. 
%
%
As discussed in the introduction of this thesis the dynamics of PT could provide many interesting phenomenological signatures such as a baryon asymmetry, DM, a GW signal and many other cosmological relics. Despite GW may be produced by the topological defects generated during a SOPT, the signal emitted is much stronger if the PT is first-order, namely if a barrier separating the broken and the symmetric phase in the effective potential is present allowing the coexistence of two different states. When such a barrier is absent a SOPT occurs. During a SOPT no breaking of thermal equilibrium, and hence no production mechanism of baryogenesis, take place. The minima of the effective potential in this case are smoothly connected and the broken and symmetric phase do not exist at the same time. 
In this section we review the effective potentials that characterize second-order and first-order PTs providing the reader the necessary tools to understand the properties of the EWPT we will discuss in the following section.

%
 
\subsection{Second-Order Phase Transitions}

To discuss the properties of second-order phase transitions we consider the real scalar field Lagrangian, symmetric under the $Z_2$ exchange symmetry $\phi\leftrightarrow-\phi$ in eq.~(\ref{eq:scalar_theory}). To make our discussion quantitative we consider the thermal corrections discussed in section~\ref{sec:finite_temperature_field_theory}. Using the high-temperature expansion of the thermal potential in eq.~(\ref{eq:boson_high_temperature_expansion}) and keeping only the quadratic leading term, we find the following effective potential
\begin{equation}\label{eq:second_order_pht_potential}
    V = \frac{\mu^2}{2}\phi^2 + \frac{\lambda}{8}T^2\phi^2 + \frac{\lambda}{4}\phi^4\,.
\end{equation}

\begin{figure}
    \centering
    \includegraphics[width=0.47\textwidth]{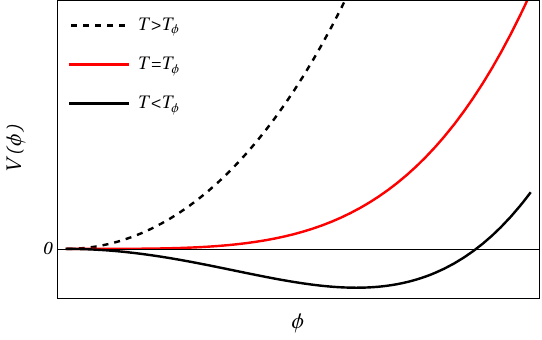}
    \hfill
    \includegraphics[width=0.47\textwidth]{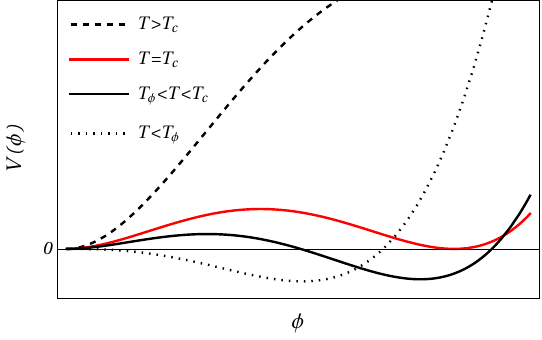}
    \caption{Plot of the potential in eq.~(\ref{eq:second_order_pht_potential}), left panel, and eq.~(\ref{eq:first_order_pht_potential}), right panel, for different values of T. Notice in the left panel that, when $T<T_\phi$, a new minimum forms for $\langle\phi\rangle\neq0$ and the stationary point at $\langle\phi\rangle=0$ becomes a maximum, and, in the right panel, the barrier between the two vacua at the critical temperature $T_c$.}
    \label{fig:first_second_order_pht}
\end{figure}

The high-temperature corrections provide a restoration of the broken symmetry $Z_2$ and modify the thermal history of the universe. The latter is described by the effective potential depicted on the left panel of figure~\ref{fig:first_second_order_pht}, where we plot, as a function of $\phi$, the potential in eq.~(\ref{eq:second_order_pht_potential}) for different choices of the temperature. The plot emphasizes the crucial impact of finite-temperature correction on the stability of the minimum in $\phi = 0$.

To study the thermal history of a SOPT, it is convenient to define a critical temperature $T_\phi$ as 
\begin{equation}
    T_\phi^2 = -\frac{4\mu^2}{\lambda}\,.
\end{equation}
At high temperature $T > T_\phi$, finite-temperature corrections overcome the importance of the tree level potential and the minimum in $\langle\phi\rangle = 0$, where the symmetry is restored, is the ground state of the theory.

The universe expansion cools down the plasma and, when the temperature $T$ reaches the critical threshold $T_\phi$, the minimum in the origin becomes extremely flat. Eventually when $T< T_\phi$ the thermal interactions are no longer sufficiently efficient to stabilize the broken phase which becomes unstable. At this stage of the evolution, the potential develops a new phase in $\langle \phi\rangle = \sqrt{-\mu^2/\lambda + T^2/4}$ and the $Z_2$ reflection symmetry is spontaneously broken. 

Because no barrier is present in the effective potential, as the left panel of Fig.~\ref{fig:first_second_order_pht} shows, the symmetric phase and the broken phase do not coexist. In fact, when the phase transition is triggered, the symmetric vacuum becomes unstable and the broken and symmetric phases are smoothly connected. Hence, the phase transition is triggered by quantum or thermal fluctuations of the field and a ``smooth'' transition between the two phases takes place.

This affects the phenomenological signatures of SOPT. Since, no energy barrier separating the two phases is present, we do not have a breaking of thermal equilibrium. The latter, as explained in the introduction of this thesis, is a fundamental ingredient for EWBG. In addition, despite a SOPT could source GW from the topological defects it potentially generates, the signal it provides is typically very weak compared to the one emitted during a FOPT. For these reasons FOPTs, that we will review in the next section, constitute a more appealing scenario from a phenomenological point of view.


\subsection{First-Order Phase Transitions}

To illustrate the properties of FOPT we consider again the real scalar model and we include the additional cubic term that the high-temperature expansion of the boson contribution in eq.~(\ref{eq:boson_high_temperature_expansion}) provides. The effective potential thus takes the following form
\begin{equation}\label{eq:first_order_pht_potential}
    V = \frac{\mu^2}{2}\phi^2+ \frac{\lambda}{8}T^2\phi^2-ET\phi^3+\frac{\lambda}{4}\phi^4\,.
\end{equation}
The additional cubic term generates a barrier between the false and the true vacuum of the theory. As a consequence, differently from a SOPT, during a FOPT the symmetry breaking and the symmetry preserving minimum coexist at the same temperature.

It is important to notice that the high-temperature expansion of thermal loops involving fermions in eq.~(\ref{eq:fermion_high_temperature_expansion}) does not provide a cubic term in the potential and, as a result, only bosons can generate a barrier at loop level. This emphasizes the importance of bosons in the universe thermal history and the necessity of an accurate modelling of the thermal corrections they provide to the effective potential.

The presence of the barrier has a significant impact on the thermal history of the universe as we can understand by plotting the potential in eq.~(\ref{eq:first_order_pht_potential}) for different choices of the temperature. The results are reported on the right hand panel of Figure~\ref{fig:first_second_order_pht}. As for the case of SOPT, at high temperature the interaction between the environment and the scalar stabilizes the symmetric phase and the broken symmetry is restored. However, differently from a SOPT, as the temperature lowers due to the universe expansion, a new symmetry breaking minimum is generated along $\phi$. Initially the new minimum is not energetically favoured. However, as the universe cools down, the new minimum gets deeper and eventually becomes degenerate with the symmetry preserving minimum in the origin at the critical temperature $T_c$. By denoting with $v_c$ the VEV of the scalar field in the newly generated minimum, namely broken phase, at the critical temperature $T_c$ the latter can be determined using
\begin{equation}
    V(0, T_c) = V(v_c, T_c)\,.
\end{equation}
For the case of the real scalar Lagrangian, the value of $T_c$ can be explicitly computed by combining the above equation with eq.~(\ref{eq:first_order_pht_potential}) providing
\begin{equation}
  T_c^2 = \frac{\lambda^2 T_\phi^2}{\lambda^2 - 8 E^2}\,.
\end{equation}
Eventually when $T < T_c$ the symmetry breaking minimum becomes the global one.
Finally, if the PT takes a long time to complete and the temperature drops below $T_\phi$, the barrier disappears and the phase transition becomes second-order.

Because of the barrier, a FOPT proceeds through tunneling. When tunneling from the symmetric to the broken phase takes place in a region of the plasma, in the corresponding region the scalar field acquires a non-trivial VEV and a bubble of true vacuum is generated. The process of bubble nucleation will be discussed in the next section but we can already summarize the main results. The main point is that small bubbles are thermodynamically disfavoured due to the positive surface energy. Hence, in order to not re-contract, bubbles must overcome a free energy barrier. Moreover, the nucleation rate must also be efficient to account for the universe expansion. For these reasons a FOPT does not start at the critical temperature $T_c$ but at a lower temperature $T_n$ called nucleation temperature. This quantity plays a major role in the dynamics of a FOPT and thus needs a careful investigation that we carry out in the next section.

Differently from a SOPT, FOPT provides a much more interesting phenomenology. In particular the presence of an expanding bubble breaks thermal equilibrium hence opening to the possibility of EWBG. In addition, when two bubbles collide, the huge energies released in the process generate a GW signal. Clearly a characterization of such experimental signatures requires to investigate the dynamics of FOPT 
that we begin to analyze studying bubble nucleation in the next section.

\subsection{Bubble nucleation}

The rate of nucleation controls the efficiency of bubble nucleation, as worked out by Linde in~\cite{Linde:1977mm, Linde:1980tt, Linde:1981zj} from the works of Coleman on vacuum decay~\cite{Callan:1977pt, Coleman:1977py}. Assuming that the mean free path of particles in the plasma is much smaller than the radius of the nucleated bubble, the nucleation rate per unit of time and volume is given by
\begin{equation}
\label{eq:nucleation_rate}
    \f{\Gamma}{\cal{V}} \sim A(T)e^\f{S_{3}(T)}{T}\,,
\end{equation}
where $A(T)\sim {\cal O}(T^4)$. Here $S_3$ is the three dimensional action of  the $O(3)$-symmetric bubble and corresponds to the free energy of a bubble configuration, namely
\begin{equation}
    S_3(T) = \int d^3{\bf x}\left[\frac{1}{2}(\nabla\phi)^2 + V(\phi, T)\right]\,,
\end{equation}
with $V(\phi, T)$ the effective potential. In order to compute $S_3$, which is also known as the bounce action, we need to solve the following equation of motion together with the boundary conditions
\begin{equation}
\label{eq:scalar_field_eom_nucleation}
    \nabla^2\phi = V'(\phi) \hspace{1cm} \phi'(r = 0) = 0\hspace{1cm} \phi(+\infty) = 0\,.
\end{equation}
In general, the free energy is computed numerically. However, in some cases, it is possible to work out an analytical solution. One particularly interesting limit is the ``thin wall'' limit, that corresponds to the case where the width of the bubble wall $L$ is much smaller compared to its radius $R$. Such an example, in addition, also clarifies why bubbles must be sufficiently large in order to be nucleated.

In the thin wall approximation, equation~(\ref{eq:scalar_field_eom_nucleation}) becomes
\begin{equation}
    \frac{d^2\phi}{dr^2} = V'(\phi,T)
\end{equation}
which is solved exactly by 
\begin{equation}
    r = \int_0^{\phi_0}\frac{d\phi}{\sqrt{2V(\phi, T)}}\,.
\end{equation}
The field profile is easily understood in the thin-wall limit: inside the bubble ($ r < R $) the field has a constant VEV $\phi_0$, with $\phi_0$ the VEV of the field in the broken phase during the phase transition\footnote{The VEV in the broken phase not always corresponds to the one computed at the nucleation temperature. As we will see in the next chapter, the energy injected by bubbles during the phase transition affects the temperature of the plasma and, as a consequence, the VEV of the field.}, while outside the bubble ($ r > R $) the scalar field VEV is still constant but equal to~$0$. The VEV rapidly changes in $r = R$, in a region of size $L \ll R$.

Inserting the thin wall solution in the free energy yields
\begin{equation}
\label{eq:bubble_eom_nucleation}
    S_3 = \left( \frac{4\pi R^3}{3}\Delta V + 4\pi{\sigma}R^2\right)\,,
\end{equation}
where $\sigma$ is the surface energy, that, in the thin wall limit is
\begin{equation}
    \sigma = \int_0^{\phi_0}d\phi\sqrt{2V(\phi, T)}\,,
\end{equation}
while $\Delta V = V(\phi_0, T) - V(0, T)$.

Equation~(\ref{eq:bubble_eom_nucleation}) has a straightforward physical interpretation. Bubble nucleation is the result of the competition of two terms: the potential energy difference which favours bubble formation when $T< T_c$ and the surface tension that hinders small bubbles. As a result, in order for bubble to be nucleated, their radius must be larger than a critical threshold $R_c$, such that they overcome the free energy barrier set by the surface tension. Extremizing eq.~(\ref{eq:bubble_eom_nucleation}) with respect to $R$ yields the critical radius 
\begin{equation}
\label{eq:critical_radius}
    R_c = -\frac{2\sigma}{\Delta V}.
\end{equation}

The nucleation temperature $T_n$ is then determined by requiring that the probability to nucleate a critical bubble in a Hubble volume is equal to~$1$. This corresponds to set
\begin{equation}
    \Gamma(T_n) \sim H^4(T_n)\,.
\end{equation}
 For the case of the EW phase transition, which is our case of interest, the above requirement roughly corresponds to~\cite{McLerran:1990zh,Dine:1991ck,Anderson:1991zb}
\begin{equation}
\label{eq:nucleation_condition}
    S_3/T_n\sim 140\,.
\end{equation}
Using the above equation it is possible to compute the nucleation temperature of the phase transition.

In this thesis we are never going to use the thin wall approximation to compute $T_n$ since it provides a poor estimate of the nucleation temperature. 
This implies that the bounce action is  computed by numerical means. In particular  we used the \texttt{Python Package CosmoTransition}~\cite{Wainwright:2011kj} which provides some useful routines to analyze the properties of FOPT.

This concludes our review of phase transitions. As a next step we are going to use the tools that we described 
in the last sections to analyze the properties the EW phase transition (EWPT).


\section{The Electroweak Phase Transition}

We finally have all the necessary tools to discuss cosmological PT in the SM. The reason we are so interested in this topic is that the SM actually predicts that the universe underwent some PTs. Among them the EWPT, that took place when the temperature reached the EW energy scale where the EW symmetry is broken by the Higgs condensate, is a very promising scenario. The different mechanisms that occurred during the EWPT could provide an explanation to the many puzzles of high-energy physics, such as the origin of DM or of the baryon asymmetry, as well as a GW signal detectable by future interferometers. Clearly, as we repeated several times during this chapter, such experimental signatures are present only if the EWPT is first-order.

In the SM the EWPT is not first order. As we will discuss in this section, lattice simulations show that the EWPT is extremely weak and it is actually a crossover. However many BSM models actually predict a first-order EWPT where a successful EWBG mechanism can take place together with the generation of a GW signal that could be used to test such SM extensions. In this section we review the EWPT in the SM and we present one of its possible extension, namely the scalar singlet extension, which provide a first-order EWPT.

\subsection{The SM case}

To discuss the properties of the EWPT in the SM, we first need to compute the SM effective potential. As we discussed in sections \ref{sec:zero_temperature_corrections} and \ref{sec:finite_temperature_field_theory}, the effective potential contains four different contributions, the tree level potential $V_0$, the 1-loop corrections $V_1$, the counter-terms that regularize the UV divergences $V_{\rm CT}$, an the thermal corrections $V_T$. The effective potential takes the following form
\begin{equation}
\label{eq:sm_effective_potential}
    V(h, T) = V_0(h) + V_1(h) + V_{\rm CT}(h,T) + V_T(h,T)\,.
\end{equation}
%

The EW is restored at high temperature due to the the finite-temperature corrections $V_T$. We recall that the latter arises from thermal loops. In sec.~\ref{sec:finite_temperature_field_theory} we analyzed, for the simple case of a real scalar theory, the finite-temperature contribution and we derived the corresponding expressions in eq.~(\ref{eq:boson_finite_temperature_corrections}) and~(\ref{eq:fermion_finite_temperature_corrections}). To apply the previous discussion to the SM case we must extend the analysis by summing all the contributions arising from thermal loops involving the different SM particles that interact with the Higgs boson.

Since only particles that interact most strongly with the Higgs provide a non-negligible correction to the effective potential, we are not going to include light degrees of freedom in our computation. We hence recognize two main contributions. A first one due to bosons which includes gauge bosons, the Higgs and the Goldstones and a second one which includes the top.
The thermal corrections are thus given by
\begin{equation}\label{eq:sm_thermal_potential}
    V_T=\frac{T^4}{2\pi^2}\left[\sum_bN_bJ_B[m^2_b(h)/T^2]+N_tJ_F[m^2_t(h)/T^2]\right]\,,
\end{equation}
where $b$ runs over the gauge bosons $W^{\pm}$, $Z$, $h$ and the Goldstone bosons $\chi$, $N_b$ is the number of degrees of freedom of the boson under consideration,  $N_t$ is the number of degrees of freedom of the top quark, while $J_B[m^2/T^2]$ and $J_F[m^2/T^2]$ are the thermal integrals defined in eq.~(\ref{eq:thermal_integrals}).

To discuss the properties of the EWPT it is sufficient to consider only the high temperature correction of the thermal potential reported in eq.~(\ref{eq:boson_high_temperature_expansion}) and~(\ref{eq:fermion_finite_temperature_corrections}). By using the high-temperature expansion the SM effective potential in eq.~(\ref{eq:sm_effective_potential}) takes the following form
\begin{equation}
    V(h,T)=\frac{c_h}{2}(T^2-T_h^2)h^2-ETh^3+\frac{\lambda_h}{4}h^4\,,
\end{equation}
with
\begin{equation}
\label{eq:sm_thermal_potential_terms}
\begin{split}
c_h & =\frac{1}{16}(3g^2+g'^2+4y_t^2+8\lambda_h),\\
    E & =\frac{1}{32\pi}({2g^3+(g^2+g'^2)^{3/2}}),\\
    T_h^2 & =\frac{\lambda_h v_0^2}{c_h}\,,
\end{split}
\end{equation}
where $v_0 = 246.22$ GeV is the Higgs VEV at $T = 0$.

The potential that we obtained describes a FOPT, as we discussed in sec.~\ref{sec:cosmological_phase_transition}. Indeed, the gauge bosons' contribution generates a barrier at loop level between the symmetry preserving and the symmetry breaking minimum. However, as we also discussed at the end of sec.~\ref{sec:finite_temperature_field_theory} we have to be careful when we deal with soft bosons. These contributions are indeed responsible for the break down of the perturbative expansion and deserve a deeper analysis. As pointed in ref.~\cite{Arnold:1994bp}, the validity of the high-temperature expansion depends on the value of the Higgs mass. In particular the parameter $m_W^2/m_h^2$, where $m_W$ and $m_h$ denotes respectively the W boson and Higgs masses, controls the perturbative expansion. Given the observed value of the Higgs mass, perturbation theory breaks down due to soft boson loops and to analyze the EWPhT we should adopt non-perturbative approaches.

\begin{figure}
    \centering
    \includegraphics[width=.6\textwidth]{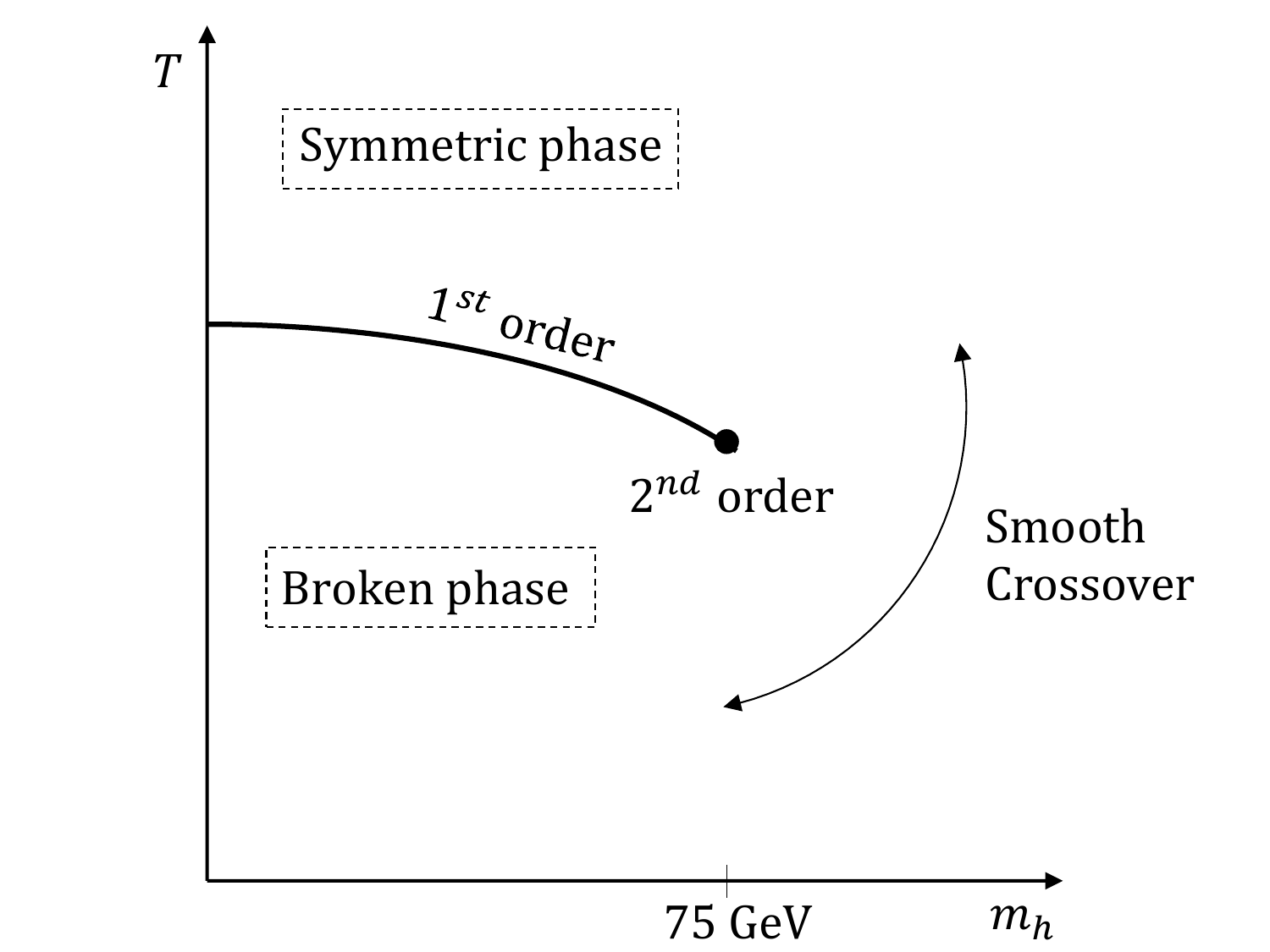}
    \caption{Schematic representation of the phase diagram of the EWPhT. The line separating the symmetric phase and the broken phase has an endpoint for $m_h=75$ GeV, where a second order phase transition occurs. For larger value of $m_h$ a smooth crossover takes place.}
    \label{fig:ewpht_phase_diagram}
\end{figure}

By applying lattice methods ref.~\cite{Kajantie:1996mn,Rummukainen:1998as} showed that the EWPT is actually a crossover. The results are schematically reported in Fig.~\ref{fig:ewpht_phase_diagram} where we show the phase diagram of the EWPT in the ($m_h$, $T$) plane. The figure clearly shows a line separating the symmetric phase and the broken phase where a first-order phase transition occurs. The line presents an endpoint at $m_h = 75$ GeV where, instead, a second-order phase transition occurs. This implies that, for larger values of the Higgs mass, the EWPhT is a smooth crossover. Therefore, given the observed value of the Higgs mass of $125$ GeV, we can conclude that the SM EWPT is a crossover.

\subsection{The SM singlet-extended case}


The SM potential lacks a barrier separating the symmetry preserving and the symmetry breaking minima. However, when we consider BSM models the additional particle content supplied by the theory modifies the potential and in some cases a barrier between the false and the true vacuum is generated.

There are two main ways to generate a barrier in the effective potential. The first one is to generate a barrier at loop level, as we discussed in section~\ref{sec:finite_temperature_field_theory}. An example of such models is the MSSM~\cite{Carena:1996wj, Delepine:1996vn} where the stop, the supersymmetric partner of the top, enhances the height of the barrier. Such models, however, require 
a large amount of additional particle content to generate a barrier and their parameter space is strongly constrained by phenomenological observations.
The second possibility, instead, is to generate the barrier already at tree level. Such possibility has been analyzed in 2-Higgs doublet models~\cite{Fromme:2006cm, Dorsch:2013wja, Dorsch:2014qja, Dorsch:2016nrg, Benincasa:2022elt}, in the singlet extension of the Higgs sector~\cite{Espinosa:2011ax,Cline:2012hg,Curtin:2014jma,Profumo:2014opa,Kakizaki:2015wua,Vaskonen:2016yiu,Kurup:2017dzf} and a variation of the latter in the context of composite Higgs models~\cite{DeCurtis:2019rxl}. 

In the present thesis we are going to focus on the $Z_2$-symmetric scalar singlet extension of the SM. In such model the SM spectrum is enriched by an additional scalar $s$, invariant under the parity symmetry $s\leftrightarrow -s$, that transforms as a singlet under the SM gauge group. The reason for our choice is twofold. First of all, such an extension allows for a first-order EWPT with a minimal additional content of new physics, namely the scalar singlet. In addition, a first-order EWPT is achieved in a large part of the parameter space of the model together with interesting experimental signatures, like gravitational waves, that can be used to probe the model~\cite{Curtin:2014jma, Friedlander:2020tnq}. 

To discuss the thermal history of such a model, we begin by studying the effective potential of the theory. The tree level effective potential is
\begin{equation}
\label{eq:tree_leve_potential_scalar_singlet_extension}
V_0(h,s) = \mu_h^2H^\dagger H + \lambda_h (H^\dagger H)^2 +\frac{\mu_s^2}{2}s^2 + \frac{\lambda_s}{4}s^4 + \frac{\lambda_{hs}}{2}(H^\dagger H)^2 s^2\,,
\end{equation}
where $H$ is the Higgs doublet
\begin{equation}
    H=\left(\chi^+, \frac{h + i\chi^3}{\sqrt{2}} \right)^T\,.
\end{equation}
$\chi$ stands of the Goldstone bosons, while $h$ is the Higgs boson. Among the three additional parameters that describe the singlet physics, namely $\mu^2_s$, the quartic coupling $\lambda_s$ and the portal coupling $\lambda_{hs}$, the latter plays a crucial role. It is in fact responsible for the generation of a barrier between the two phases already at tree level, allowing the EWPT to be first-order in a large part of the parameter space. 

As discussed in sec.~\ref{sec:zero_temperature_corrections}, the effective potential receives corrections at loop level. Using the cutoff regularization scheme as in eq.~(\ref{eq:cutoff_regscheme_sm}) to regularize the UV divergences we find
\begin{equation}
    V_1(h,s) + V_{CT}(h,s) = \frac{1}{64\pi^2}\sum_{i}N_i\left[2m^2_i(h,s)m_i(v,0)^2 + m_i^4(h,s)\left(\log\frac{m_i^2(h,s)}{m_i(v,0)^2} - \frac{3}{2}\right)\right].
\end{equation}
The sum runs over the heavy particles of the model, namely the gauge bosons $W^\pm$ and $Z$, the top quark $t$, the Higgs $h$, the singlet $s$ and the Goldstone bosons $\chi^\pm, \chi^3$. 

At last, we must include also the finite-temperature corrections which are given by
\begin{equation}
    V_T(h,s,T) = \frac{T^4}{2\pi^2}\left[\sum_b N_b J_b[m^2_b(h,s,T)/T] + \sum_f N_f J_F[m^2_t(h,T)/T^2]\right]\,,
\end{equation}
where the masses in the above equation include also the thermal mass arising from hard thermal loops and we are summing over all the degrees of freedom of the theory. As already stated in sec.~\ref{sec:finite_temperature_field_theory} we neglect correction arising from daisy diagrams.

The additional scalar singlet modifies the thermal history of the universe and provides a first-order EWPT by generating an energy barrier in the effective potential. The singlet can generate the barrier in two ways. The first one is through radiative corrections, since the singlet interacts with the Higgs through the portal coupling $\lambda_{hs}$. However, generating a sufficiently high barrier for a FOPT requires non-perturbative values for $\lambda_{hs}$. 
The second possibility, which does not require the portal coupling to be non-perturbative, is instead to generate a barrier at tree level.

\begin{figure}
    \centering
    \includegraphics[width=0.47
\textwidth]{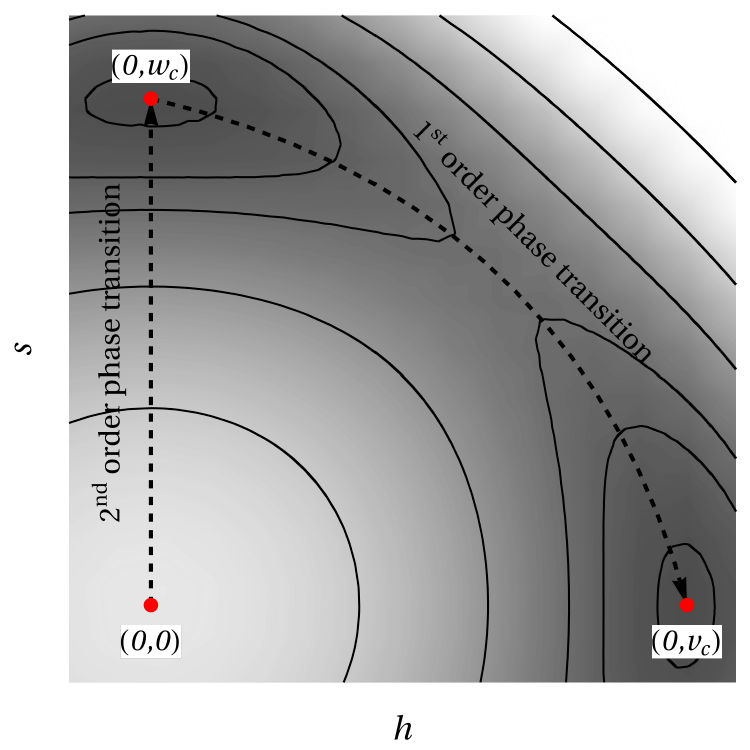}
    \caption{Schematic representation of the two step scenario described in the text. Darker colours represent a deeper potential at the critical temperature $T_c$}
    \label{fig:2step_pht_cartoon}
\end{figure}

The latter scenario requires, during the EWPT, the singlet VEV to change. Indeed, if the VEV of the singlet remained constant during the whole PT, then the effective potential would have the same shape as in the SM case and no significant barrier would be generated. Such scenario is realized if the $Z_2$ reflection symmetry of the singlet is broken before the EW one, hence providing an intermediate phase transition that precedes the EWPT. This is the so-called two-step phase transition (2SPT) scenario which is depicted in Figure~\ref{fig:2step_pht_cartoon}.

At very high temperature both the $Z_2$ and the EW symmetry are restored. Once the universe reaches a critical threshold set by a temperature $T_s$ the $Z_2$ symmetry is spontaneously broken and a first phase transition along the $s$ direction from the symmetric phase $(0, 0)$ to the broken phase $(0, w(T))$ occurs, where $w(T)$ denotes the VEV of the singlet at the temperature $T$ in the $Z_2$ broken phase. At later times, as the universe cools down, the EWSB minimum is generated along the $h$ direction. Such a minimum eventually becomes degenerate with the $Z_2$ breaking minimum at the critical temperature $T_c$, as shown in figure~\ref{fig:2step_pht_cartoon}. Eventually the EWSB minimum becomes the global minimum of the theory and once the plasma reaches the nucleation temperature $T_n$ the EWPT from the minimum $(0, w(T_n))$ to $(v(T_n), 0)$ starts.

We point out that, if the $Z_2$ symmetry is only spontaneously broken, the universe fills with an equal amount of domains where the singlet has opposite VEVs hence generating a network of DWs which could potentially affect the bubble nucleation and the dynamics of the EWPT~\cite{Blasi:2022woz}. However, in presence of a small breaking of the $Z_2$ symmetry~\cite{McDonald2}, the DW network decays before the EWPT starts. The breaking can be so small that it does not affect the dynamics of the bubble, and the $Z_2$ symmetric effective potential can be used to model the behaviour of the system. In the following we are going to assume that the $Z_2$ symmetry is broken and that, when the EWPT takes place, bubble nucleation happens on a background where the singlet has a positive VEV.

Achieving a first-order EWPT in the two-step scenario is much easier than in the single step case because of the presence of the barrier already at tree level. Moreover the two-step scenario is viable in a large part of the parameter space of the model and it is more likely to occur. Indeed, the singlet interacts with less particles than the Higgs and, as a consequence, it will lose its stability before the Higgs does if its mass is not too large. 

Clearly not the whole parameter space of the model allows for a 2SPT. Such scenario is indeed constrained by  requiring that
\begin{enumerate}
    \item the $Z_2$ symmetry is broken by the singlet VEV,
    \item the EWSB minimum eventually becomes the global one at zero temperature,
    \item the intermediate minimum, namely the $Z_2$ symmetry breaking minimum is the global minimum at intermediate temperature, namely before the EWPT occurs.
\end{enumerate}
To satisfy the first two conditions we simply need to impose the following constraints
\begin{equation}
    \mu^2_s <0 \;\;\;\;\;\;\;\;\;\;\mu^2_s+\f{\lambda_{hs}}{2}v_0^2 >0\,.
\end{equation}
The first requirement is a necessary condition for the singlet to acquire a non-vanishing VEV and trigger the intermediate phase transition, while the latter ensures that the EWSB minimum is the ground state of the theory at $T=0$. We cannot provide an analytical constraint for the third condition, because it involves also the non-trivial loop corrections to the effective potential.\ Simple formulas can be derived in the high temperature limit~\cite{DeCurtis:2019rxl}, while in the general case one must rely on numerical checks.

Additional constraints on the potential arise from experimental observations. A first bound is given by the current observed value of the Higgs invisible width that rules out the possibility for the Higgs boson to decay in two scalar singlets. This implies that the mass of the scalar singlet $m_s$ must satisfy
\begin{equation}
    m_s > \frac{m_h}{2}\,.
\end{equation}
where $m_s$ is
\begin{equation}
    m_s^2 = \mu^2_s +\frac{\lambda_{hs}}{2}v_0^2
\end{equation}

A second experimental bound is given by DM phenomenology. In the parameter space region  of interest, namely where a first-order EWPT takes place, the singlet constitutes a subleading contribution (less than $1\%$) to the total DM energy density with the main consequence that additional particles must be included in the theory to match the current experimental observations. In addition, even if the singlet abundance is very low, the current experimental observation on the spin-independent cross section of DM-nucleon~\cite{LZ:2022lsv} severely constraints the parameter space of the theory where the EWPT is first-order. In fact only  the region where $m_s > 700$ GeV is not excluded by the phenomenological bounds~\cite{Curtin:2014jma,Cline:2012hg,Cline:2013gha,Beniwal:2017eik}. One possibility to evade the stringent phenomenological bounds is to consider the presence of an additional lighter particle, coupled only to the scalar singlet, that plays the role of the DM candidate. However we are not going to delve deep into the details of such realizations as this is beyond the scope of the thesis work. In fact, our goal is to provide an accurate description of the out-of-equilibrium dynamics of a FOPT and not to analyze a specific BSM theory.

Once the effective potential of the theory is derived, we can determine the nucleation temperature by solving eq.~(\ref{eq:nucleation_condition}) and then characterize the experimental signatures provided by a FOPT.

\section{Gravitational waves}

We conclude this chapter by discussing the GW. Such experimental signatures are
one of the most compelling aspect of FOPTs since the signal generated during a first-order EWPT is potentially detectable at future space based interferometers.

In this section we aim to review our current understanding of the GWs emitted during a FOPT. In particular we are going to present which mechanisms that take place during the FOPT mainly source GWs and discuss which quantities characterize their spectrum. Good reviews on the subject can be found in ref.~\cite{Hindmarsh:2020hop,Caprini:2015zlo} while further details are provided in the references cited through the section.

\subsection{Relevant parameters for GW spectrum characterization}

The signal of gravitational waves emitted is ultimately controlled by four model-dependent parameters, namely the temperature at which GWs are emitted $T_*$, or more precisely the corresponding value of the Hubble parameter $H_*$, the inverse time duration of the phase transition $\beta$, the strength of the phase transition $\alpha$ and the bubble wall terminal velocity $v_w$. Among these parameters, the terminal velocity is the one we have less theoretical control on. Its precise computation is still an active research field and requires an accurate modelling of the out-of-equilibrium dynamics that takes place in the plasma during the EWPT. Its determination constitutes one of the main goals of this thesis and we are going to discuss its computation further in the text.

The remaining three parameters are instead much easier to compute since they are controlled by the equilibrium properties of the plasma. In the case of interest, namely EWPT, the temperature $T_*$, where bubbles collide, is different from the nucleation temperature $T_n$ at which the phase transition starts. The temperature $T_*$ approximately coincides with the percolation temperature and can be determined by the following condition~\cite{Moore:1995si,Huber:2007vva}
\begin{equation}
    \frac{S_3(T_*)}{T_*} = 131 + \log(A/T^4) - 4 \log\left(\frac{T}{100 {\rm GeV}}\right) - 4 \log\left(\frac{\beta/H}{100}\right)+ 3 \log(v_w) .
\end{equation}
where $\log(A/T^4)\simeq -14$~\cite{Carrington:1993ng}. We point out, however, that for moderately strong phase transitions the values of $H_n$ and $H_*$ at the nucleation and $T_*$ temperature are very similar and one can use $H_n$ instead of $H_*$.

The strength of the phase transition is described by the $\alpha$ parameter. This is defined as the ratio between the vacuum and radiation energy density, namely $ \alpha =\rho_{\rm vac}/\rho_{\rm rad}$. As we are going to see in the next chapter a fair good approximation to $\alpha$ is given by
\begin{equation}
    \alpha = -\frac{\Delta V}{a_+ T^4}\,,
\end{equation}
where $ \Delta V$ is the potential energy difference between the broken and the symmetric phase, while $a_+$ is the number of relativistic degrees of freedom in the symmetric phase.
Strong phase transitions are characterized by large values of $\alpha$ and large energy injections controlled by $\Delta V$. As one may expect, the GW signal is maximized for strong phase transitions.

The last relevant quantity is the inverse time duration of the phase transition which is defined as
\begin{equation}
    \beta = -\frac{dS_3}{dt}\Bigr|_{t=t_*}\simeq\frac{\dot\Gamma}{\Gamma}\,,
\end{equation}
where $t_*$ denotes the time at which GWs are produced. A key parameter is the ratio $\beta/H_*$, in particular the smaller the ratio, the stronger the phase transition is, and thus the GW signal emitted. In the limit where $T_* \sim T_n$, the ratio is approximately given by
\begin{equation}
    \frac{\beta}{H}=T\frac{dS_3}{dT}\Bigr|_{T=T_*}\,.
\end{equation}

\subsection{GW production mechanisms}

Our knowledge of GW production relies on numerical simulation. These show that during a FOPT three main mechanisms are at work for GW production namely bubble collisions~\cite{Kosowsky:1991ua,Kosowsky:1992rz,Kosowsky:1992vn,Kamionkowski:1993fg,Huber:2008hg,Caprini:2007xq}, sound waves in the plasma after collision~\cite{Hindmarsh:2013xza,Giblin:2013kea,Giblin:2014qia,Hindmarsh:2015qta, Hindmarsh:2017gnf, Hindmarsh:2019phv} and magneto-hydrodynamic turbulence effect in the plasma~\cite{Caprini:2006jb,Kahniashvili:2008pf,Kahniashvili:2008pe,Kahniashvili:2009mf,Caprini:2009yp, Cutting:2019zws,RoperPol:2019wvy, Gogoberidze:2007an,Niksa:2018ofa}. These three sources generally coexist and their contributions to the energy fraction of GWs at a first approximation linearly combine leading to
\begin{equation}
    h^2\Omega_{\rm GW} \simeq h^2 \Omega_\phi + h^2 \Omega_{\rm SW} + h^2 \Omega_{\rm MHD}\,.
\end{equation}
Each of these mechanisms takes place at a different stage of the FOPT and their impact on the GW spectrum depends on the bubble wall terminal velocity $v_w$.

The first mechanism that takes place is bubble collision. The energy injected during bubble collisions originates from the latent heat stored in the bubble wall. In fact, during the phase transition at the bubble front the VEV of the field interpolates between the symmetric and broken phase. The gradient of the scalar field induces a shear stress that is subsequently released when two bubbles collide. A reasonable approximation to deal with this term is the so called ``envelope approximation'' where it is assumed that the latent heat is concentrated on a thin shell in front of the bubble wall and is quickly dispersed when two bubbles collide. Within such approximation numerical simulations suggest that a reasonable fit of the power spectrum is given by~\cite{Konstandin:2017sat} 
\begin{equation}
    \Omega_{\phi}^{\rm env}(f) = \tilde\Omega_\phi^{\rm env} \frac{(a + b)\tilde f^b f^a}{b \tilde f^{a+b} + a f^{a+b}}\,,\;\;\;\;\;\;\;\; a = 3\,,\; b = 1.51\,,\; c = 2.18
\end{equation}
with $\tilde\Omega_\phi^{\rm env}$ and $\tilde f$ denoting the peak amplitude and frequency given by
\begin{equation}
\begin{split}
    &h^2\tilde \Omega_\phi^{\rm env} = 1.67 10^{-5}\times \frac{0.44 v_w^3}{1+8.28 v_w^3}\kappa^2\left(\frac{H}{\beta}\right)^2\left(\frac{\alpha}{\alpha + 1}\right)^2\left(\frac{100}{g_*}\right)^{1/3}\\
    &\tilde f = 2.62\times 10^{-3}\times\frac{1.96}{1-0.51 v_w + 0.88 v_w^2}\left(\frac{1}{H_*}\right)\left(\frac{T_*}{100\,{\rm GeV}}\right)\left(\frac{g_*}{100}\right)^{1/6}{\rm mHz}
\end{split}
\end{equation}
where $H_*$ denotes the Hubble parameter at the moment of bubble collision and $g_*$ the number of relativistic degrees of freedom. The parameter $\kappa$ measures the energy deposited on the wall and is defined as the ratio $\rho_\phi/\rho_{\rm tot}$ between the energy density of the wall and the total energy density. This mechanism represents the leading contribution for walls that never stop to accelerate until collision, namely in the so called runaway regime, and it is active for a short amount of time.

After bubbles have merged, the ''acoustic'' mechanism starts. On this stage sound waves generated during the bubble dynamics continue to propagate in the plasma eventually overlapping and generating additional gravitational waves. Differently from the collision mechanism, this stage has a much longer typical duration, of the order of the Hubble time and dominates the gravitational wave spectrum for a DW that eventually reaches a terminal velocity $v_w$.

The GW spectrum in this case is given by~\cite{Hindmarsh:2017gnf,Caprini:2019egz}
\begin{equation}
    \frac{d\Omega_{\rm SW}}{d\ln f} = \begin{cases}
        0.687 F_{{\rm gw},0}K^2(H_* R_*/c_s)\tilde \Omega_{\rm gw}C(f/f_{p,0}) & \tau_{\rm sh} > 1/H_* \\
        0.687 F_{{\rm gw},0}K^{3/2}(H_* R_*/\sqrt{c_s})^2\tilde \Omega_{\rm gw}C(f/f_{p,0}) & \tau_{\rm sh} > 1/H_*   
    \end{cases}
\end{equation}
where $\tau_{\rm sh}$ is the time scale of shock formation, and ${\tilde \Omega}_{\rm gw} \simeq 10^{-2}$  as determined by  numerical simulations. The kinetic energy fraction $K$ regulates the efficiency of GW production in the acoustic stage and can be expressed in term of the phase transition strength, the enthalpy $w$ and the plasma velocity $v_p$ is given by 
\begin{equation}
    K = \frac{\kappa_{\rm sh}\alpha}{1+\alpha},\;\;\;\;{\kappa_{\rm sh}} = \frac{3}{-\Delta V v_w}\int w(\xi) v_p^2 \gamma_p^2 \xi^2
\end{equation}
where $\gamma$ is the Lorentz gamma factor, while $\xi = r/t$, with $r$ the radius of the bubble and $t$ the time passed since its nucleation. The mean bubble separation is defined as $R_* = (8\pi)^{1/3}{\rm Max}(v_w,c_s)/\beta$, with $c_s$ the speed of sound, while the factor $F_{{\rm gw},0} = \Omega_{\gamma,0}\left(\frac{g_{s0}}{g_{s*}}\right)^{4/3}\left(\frac{g_*}{g_0}\right) \simeq 3.57 \times 10^{-5}\left(\frac{g_*}{100}\right)^{1/3}$ redshifts the signal emitted to the one we would observe today. Finally the spectral shape $C$ is 
\begin{equation}
    C(s) = s^3\left(\frac{7}{4+3s^2}\right)^{7/2}
\end{equation}
while the frequency peak is
\begin{equation}
    f_{p,0} \simeq 26\left(\frac{1}{H_* R_*}\right)\left(\frac{z_p}{10}\right)\left(\frac{T_*}{100\,{\rm GeV}}\right)\left(\frac{g_*}{100}\right)^{1/6}\mu {\rm Hz}\,,
\end{equation}
with $z_p\simeq 10$ determined by simulations.

The final stage of GW production is the turbulent phase where non-linearities of fluid equations become important. This stage is the one we have less control on since numerical simulations are quite challenging and the resulting models thus rely on untested assumptions. In particular it is not clear how vorticity, which eventually generates turbulence, is generated in the first place and how important it is with respect to the other part of the GW spectrum.

We conclude this section with a final remark. The research field on GWs emitted during FOPT is very active and there is still a large uncertainty on the spectra we provided through the section. There is still a lot of work that must be done to compute the power spectrum for a phase transition at any strength since simulations and models 
have a reasonable agreement in the region $0.4\lesssim v_w\lesssim 0.9$ and $\alpha_n \equiv \alpha|_{T=T_n}\lesssim 0.1$. 

Furthermore, an accurate modelling of the GW spectrum requires a precise determination of the terminal velocity $v_w$. As we already mentioned, this parameter is the one we have less theoretical control on and its determination requires an accurate description of the bubble wall dynamics during the phase transition. In the following chapter we start to address this problem by analyzing the dynamics of the scalar fields during the phase transition.


\chapter{Bubble Dynamics and Plasma Hydrodynamics}
\label{ch:bubble_dynamics_and_plasma_hydrodynamics}

We concluded the previous chapter by highlighting the important quantities that determine the spectrum of GW generated. Among them a crucial role is played by the terminal velocity at which the bubbles of true vacuum expand since it controls the gravitational wave signal as well as the amount of baryon asymmetry generated by EWBG mechanism. It is clear that an accurate determination of the terminal velocity is necessary to properly characterize the signatures of the first-order EWPT. However, such a task is highly non-trivial because it requires to study the dynamics of the bubbles nucleated during the phase transition. This is the topic of the present chapter.

The dynamics of the Higgs and the scalar singlet DWs is described by the IR condensate of the two scalar fields that we treat as background fields. The equations of motion of such condensates are determined using the WKB approximation which splits in a natural way the system in two parts: the scalar fields and a plasma made of all the SM particles. The interactions between the plasma and the scalar fields represent the biggest theoretical challenge to determine the terminal velocity since they drive out-of-equilibrium the plasma hence generating a friction that slows down the wall motion as a back-reaction.

While heavy degrees of freedom are driven out-of-equilibrium due to their large coupling with the Higgs field, light species can be modelled as a fluid in local equilibrium which set the temperature and plasma velocity profiles of the fluid.
Such profiles, which crucially control the DW dynamics, also depend on how fast the wall moves and we present in a dedicated section their determination.

We conclude the chapter by discussing the strategy that is widely adopted in the literature to compute the terminal velocity. Using a reasonable ansatz for the field profiles during the phase transition, the terminal velocity and the bubble wall widths can be determined by taking suitable momenta of the scalar field equations of motion.

\section{Equations of motion of the scalar fields}

The equations of motion of the DW were first obtained in the WKB approximation in ref.~\cite{Moore:1995si}. We review their derivation by extending the discussion to include the scalar singlet. 
Our starting point is the Lagrangian describing the Higgs and scalar singlet which we consider to be invariant under the SM gauge group and the $Z_2$ reflection symmetry $s\leftrightarrow -s$ 
\begin{equation}
\begin{split}\label{eq:scalars_lagrangian}
    {\cal L} = & (D_\mu H)^\dagger D^\mu H +\frac{1}{2}\partial_\mu s \partial^\mu s \\
    & -\mu^2_h H^\dagger H - \lambda_h (H^\dagger H)^2 -\frac{\mu^2_s}{2}s^2-\frac{\lambda_s}{4}s^4 -\frac{\lambda_{hs}}{2}H^\dagger H s^2 - \sum_i y_i(H^\dagger{\bar \psi}_R\psi_L + H{\bar \psi}_L\psi_R)
\end{split}
\end{equation}
where the sum runs over all the SM fermions and $y_i$ is their corresponding Yukawa coupling. We point out that the singlet does not interact with any fermion, which will be important when we will discuss the friction acting on the two DWs.  As for the majority of the litterature we are going to simplify our problem assuming that the $W$ and $Z$ bosons have the same mass and we will collectively denote them as W bosons. Hence
\begin{equation}
    D_\mu H = \partial_\mu H -i g [W^\mu, H]\,,
\end{equation}
with $g$ the weak gauge coupling.
The above Lagrangian provides the following equations of motion for the scalar fields
\begin{equation}
\begin{split}
    & \Box H + \mu_h^2 H + 2\lambda_h (H^\dagger H) H + \frac{\lambda_{hs}}{2} H s^2 + ig W^\mu \partial_\mu H + \f{ig}{2}\partial_\mu W^\mu H -\f{g^2}{4}W^\mu W_\mu H + \sum_i y_i {\bar \psi}_R \psi_L = 0\,, \\
    & \Box s + \mu_s^2 s + \lambda_s s^3 + \lambda_{hs} H^\dagger H s = 0\,,
\end{split}
\end{equation}
where for simplicity we suppressed group indices. 

The equations of motion of the two DWs are obtained by analyzing the dynamics of the Higgs and the scalar singlet condensates. To provide their description, we shift the Higgs doublet $H$ and the scalar $s$ with respect to their VEVs, namely $H\rightarrow H + \delta H$, with $H = (0, h/\sqrt{2})^T$ and $ s\rightarrow s + \delta s$ and then take the thermal average of the equations of motion.
%
We next make some assumptions.
First, we assume that the thermal averages of the field fluctuations vanish: $\langle \delta H\rangle = \langle\delta s \rangle = 0$ and $\langle W^\mu \rangle = 0$.
Secondly, we suppose that the scale at which the two fields VEVs change is much larger than $T^{-1}$, with $T$ the temperature of the plasma, so that it is justified to use the WKB approximation to evaluate the thermal average of the field fluctuations. Under these assumptions the equations of motion describing the two scalar condensates are
\begin{equation}
\label{eq:scalar_condensates_eom}
\begin{split}
    &\Box h + \f{\partial V|_{T=0}}{\partial h} + \sum_i N_i\frac{\partial m_i^2}{\partial h}\int \frac{d^3{\bf p}}{(2\pi)^32E}f_i({\bf p}, x) = 0 \\
    &\Box s + \f{\partial V|_{T=0}}{\partial s} + \sum_i N_i\frac{\partial m_i^2}{\partial s}\int \frac{d^3{\bf p}}{(2\pi)^32E}f_i({\bf p}, x) = 0\,.
\end{split}
\end{equation}
The sum runs over the particle species in the plasma with mass $m_i$, degrees of freedom $N_i$ and distribution function $f_i$. Finally $V|_{T=0}$ identifies the zero temperature effective potential of the theory.

Eq.~(\ref{eq:scalar_condensates_eom}) implies the conservation of the total energy-momentum tensor of the system. For this purpose, we first recall that the stress-energy tensor of a scalar field $\phi$ is
\begin{equation}
\label{eq:definition_scalar_field_stress_energy_tensor}
    T^{\mu\nu}_\phi = \partial^\mu\phi\partial^\nu\phi -\eta^{\mu\nu}\left(\frac{1}{2}\partial_\alpha\phi\partial^\alpha\phi - V|_{T=0}\right)\,.
\end{equation}
Hence, multiplying each of the equations in~(\ref{eq:scalar_condensates_eom}) for the derivative of the corresponding scalar field we get
\begin{equation}
\begin{split}
    &\partial_\mu T^{\mu\nu}_h + \sum_i N_i\frac{\partial m^2_i}{\partial h}\partial^\nu h \int\frac{d^3{\bf p}}{(2\pi)^32 E}f({\bf p},x)=0\,,\\
    &\partial_\mu T^{\mu\nu}_s + \sum_i N_i\frac{\partial m^2_i}{\partial s}\partial^\nu s \int\frac{d^3{\bf p}}{(2\pi)^32 E}f({\bf p},x)=0\,.
\end{split}
\end{equation}
Finally, we sum the above equations to get
\begin{equation}
\label{eq:system_total_energy_conservation}
    \partial_\mu T^{\mu\nu}_h + \partial_\mu T^{\mu\nu}_s + \sum_i N_i m_i \int\f{d^3{\bf p}}{(2\pi)^3 2 E}\partial^\nu m_i f_i({\bf p},x)=0\,,
\end{equation}
which simply states the conservation of the total energy-momentum tensor as we anticipated. In fact, this follows immediately by identifying the force $F^\nu = -\partial^\nu m_i$ with the one that the DW exerts on a particle with mass $m_i$. The varying VEVs of the two scalar fields during the EWPT indeed provide a non-trivial space-dependent mass for the particles which experience a force proportional to their mass change when they encounter the DW. As a back-reaction, the momentum exchange between the two DWs and the plasma generates a damping force that slows down the wall motion.

The WKB approximation therefore splits in a natural way the system in two parts. The first one being the two scalar condensates which we treat as background field while the second one is a thermal plasma made of all the different SM particle species. This identification greatly simplifies our analysis and allows us to easily understand the impact of the presence of the two DWs on the plasma.

It is important to notice that only the Higgs DW drives the plasma out-of-equilibrium. The scalar singlet, in fact, interacts only with itself and with no SM particle but the Higgs,
providing a very limited impact to the non-equilibrium dynamics of the plasma. The largest contribution to the friction arises from heavy particle species, namely the top quark and the W bosons which, due to the Higgs mechanism, also possess the strongest interaction with the Higgs field. All the remaining degrees of freedom experience a negligible force and can be described as a plasma in local thermal equilibrium. Among the light degrees of freedom we also consider the Higgs and the scalar singlet excitations. Indeed, for non-perturbative values of the portal coupling $\lambda_{hs}$ and quartic coupling $\lambda_s$ and the observed value of $\lambda_h$, their contribution to the friction is subleading with respect to the top and the W bosons since $N_{h,s} = 1 \ll N_t, N_W$.

To study these effects in more detail it is convenient to split the distribution functions $f_i$  in an equilibrium and out-of-equilibrium part, namely $f({\bf p}, x) = f_0(\beta( x)U^\mu p_\mu) + \delta f({\bf p}, x)$. The distribution $f_0(\beta(x) U^\mu(x) p_\mu)$ identifies the distribution function of a plasma in local thermal equilibrium with local temperature $\beta(x)^{-1}$, which should not be confused with the inverse time duration of the phase transition, and local plasma four velocity $U^\mu(x)$. We recall that the plasma four velocity is related to the plasma velocity $v_p$ through
\begin{equation}
\label{eq:definition_plasma_four_velocity}
    U^\mu = \gamma_p(1, {\bf v}_p)\,,
\end{equation}
where we denote with $\gamma_i = 1/\sqrt{1-{\bf v}_i^2}$. The distribution $f_0$ is the usual Bose-Einstein (Fermi-Dirac) distribution function for bosons (fermions), namely
\begin{equation}
\label{eq:equilibrium_distribution_quantum_gas}
    f_0(y) = \frac{1}{e^y\mp 1}\,
\end{equation}
where the $-$($+$) is for bosons (fermions). The fluctuation $\delta f({\bf p}, x)$, instead, identifies the perturbations around equilibrium. 

By splitting the distribution functions in an equilibrium and an out-of-equilibrium part the equations of motion of the Higgs and singlet condensates become 
\begin{equation}
\begin{split}
\label{eq:scalar_condensates_final_eom}
    &\Box h + \frac{\partial V}{\partial h} + \sum_i N_i \frac{\partial m^2_i}{\partial h}\int \frac{d^3{\bf p}}{(2\pi)^3 2 E} \delta f({\bf p},x)\, = 0,\\
    &\Box s + \frac{\partial V}{\partial s} = 0\,.
\end{split}
\end{equation}
The equilibrium part provides the thermal correction to the effective potential, while the out-of-equilibrium perturbations provide a dissipative force. Notice that since we are going to consider only the perturbations of top quark and W bosons, which interact only with the Higgs field, the singlet condensate has no dissipation arising from the fluctuations around equilibrium.

The set of equations in~(\ref{eq:scalar_condensates_final_eom}) provides the final expression for the equations of motion of the two condensates. Their solution thus fully determines the dynamics of the two DW. We next examine each term entering the equations to further clarify their role in the DW dynamics.

\subsection{Pressure acting on the bubble wall}

The dynamics of the DWs is determined by three contributions, the surface tension, which tends to shrink the bubble, the potential energy difference between the broken and symmetric phase, which drives the bubble expansion and the damping force which slows the DWs motion. Let us begin our analysis from the surface tension which we analyze for the case of the Higgs DW since an analogous analysis can be carried out for the singlet case.

In sec.~\ref{sec:cosmological_phase_transition} we discussed the process of bubble nucleation during a FOPT. In our analysis we emphasized that, in order to be nucleated, bubbles must overcome the free-energy barrier set by the surface tension. Surface tension is indeed responsible for the collapse of small bubbles. On the other hand the potential energy difference between the symmetric and the broken phase stimulates bubble expansion. As a result, in order to be nucleated bubbles must have a radius $R$ larger than the critical radius $R_c$ in~(\ref{eq:critical_radius}).

In the situation we are interested in, namely at the time of bubble collision, the radius of the bubble is so large that we can safely consider locally planar DWs. This assumption, in the case of EWPT, is also true when bubbles are nucleated. Indeed, the parameter controlling the planar limit approximation is the ratio between $L/R$, with $L$ the width of the DW. When $L/R\ll1$ we can safely assume the wall to be planar. Such condition is easily met during the EWPT also for critical bubbles with radius $R_c$. In the planar limit, the surface tension acting on the DW is negligible compared to the other forces. Thus, in our analysis we will altogether neglect it.

The assumption of a planar wall further simplifies our discussion. Within such hypothesis we can orient our reference frame in such a way that the two DWs move in the negative direction of the $z$ axis. Furthermore, the presence of the two walls breaks the Lorentz invariance of the system along the $z$ direction with the main consequence that the Higgs VEV, the singlet VEV and the plasma temperature and four velocity profile depend only on the parameters $z$ and $t$. The momentum dependence of the out-of-equilibrium perturbations, instead, is completely parameterized by the momentum $p_z$ and by the modulus of the momentum perpendicular to the direction of the DW $p_\bot = \sqrt{p^2_x + p^2_y}$. Thus in the planar limit $\delta f({\bf p},x) = \delta f(p_\bot, p_z, z, t)$.

To derive the total pressure acting on the DWs we integrate along the $z$ direction the $\nu = z$ component of eq.~(\ref{eq:system_total_energy_conservation}). Such equation, indeed, describes the total force per unit of volume acting on the scalar fields condensate. Thus the integral along $z$ yields the total pressure $P$ acting on the wall.

The pressure acting on the system is the result of the balance of two contributions. The first one is the pressure originating from the potential energy difference between the false and the true vacuum which expands the bubble. The second one is a dissipative force that arises from the interactions between the plasma particles and the bubble. Such a force is proportional to the out-of-equilibrium perturbations whose size in turn increases with the DWs velocity $v_w$. 

Regarding the size of the dissipation we recognize two different scenarios for the bubble wall motion. The first one is the case where dissipation is not strong enough to balance the energy potential difference. In such a case the wall never stops to accelerate and it reaches ultra-relativistic velocities. At such velocities the out-of-equilibrium perturbations are usually negligible and the dynamics of the wall is controlled only by the local equilibrium terms. This regime is known as the runaway regime and it has been deeply studied in the literature (see refs~\cite{Bodeker:2009qy,Bodeker:2017cim,Hoche:2020ysm,Azatov:2020ufh,Gouttenoire:2021kjv,Azatov:2021ifm,Azatov:2021irb,Azatov:2022tii,Baldes:2021vyz} for an incomplete list). It provides interesting phenomenology since the GW signal is maximized and the large energies released by the wall during its motion can generate heavy particles that constitute good DM candidates~\cite{
Azatov:2021ifm}. Baryogenesis, instead, is suppressed for fast-moving walls, but some recent results~\cite{Azatov:2021irb,Baldes:2021vyz} have shown the viability of the mechanism also in the runaway regime. 

On the other hand, if the damping forces are large enough to balance the potential energy difference, the wall eventually reaches a steady state and moves with a terminal velocity $v_w$. In such scenario, the out-of-equilibrium perturbations are not negligible and their contribution must be taken into account for a reliable description of the DW dynamics. As a consequence, understanding the dynamics of the phase transition is a much more complex task in such a case because it requires an accurate modelling of the plasma out-of-equilibrium fluctuations.

In this thesis we focus on the latter scenario. We are going to assume that the damping force balances the expansion provided by the potential energy difference and that the wall reaches a steady state with a terminal velocity $v_w$. We will address the description of the out-of-equilibrium perturbations in the following chapters. Indeed, determining the fluctuations $\delta f$ is one of the main topic of the present thesis and thus deserves a careful and detailed discussion.

The steady state regime allows us to further simplify the problem. Choosing the wall reference frame to describe the dynamics of the two DWs 
every observable becomes time independent hence allowing us to drop altogether the variable $t$. To determine the terminal velocity $v_w$ we impose that the total pressure $P$ acting on the system vanishes. This corresponds to the requirement that
\begin{equation}
\label{eq:pressure_definition}
    P \equiv \int_{-\infty}^{\infty}dz\left[ h'\Box h + s' \Box s + h'\frac{\partial V}{\partial h} + s'\frac{\partial V}{\partial s} + F(z)\right]\, = 0,
\end{equation}
where we denoted with the `` $ ' $ ''  the derivative along $z$ and we defined
\begin{equation}
\label{eq:out_of_eq_friction}
    F(z) = \sum_i \frac{N_i}{2}\frac{\partial m^2_i}{\partial z}\int\frac{d^3{\bf p}}{(2\pi)^3E}\delta f(p_\bot, p_z, z)\,.
\end{equation}
The term $F(z)$ provides the dissipative force arising from the perturbations. We will usually refer to such a term by calling it friction.

By carrying out the integration in eq.~(\ref{eq:pressure_definition}) we obtain the following expression for the total pressure
\begin{equation}
\label{eq:total_pressure}
    P = \Delta V - \int_{-\infty}^\infty dz T'\frac{\partial V}{\partial T} + \int_{-\infty}^\infty dz F(z) \,.
\end{equation}
The above equation summarizes the contributions to the pressure acting on system. While the first term corresponds to the driving force that expands the bubble, the second and the last term instead provide the dissipation on the wall motion. The last term is the integral of the friction arising from the out-of-equilibrium perturbations. The second one, instead, is a hydrodynamic back-reaction to the wall movement which is proportional to the variation of the entropy density~\cite{Balaji:2020yrx,Ai:2021kak}.

The terminal velocity of the wall $v_w$ is thus found by imposing
\begin{equation}
\label{eq:steady_state_condition}
    \Delta V = \int_{-\infty}^\infty dz T'\frac{\partial V}{\partial T} - \int_{-\infty}^\infty dz F(z)\,.
\end{equation}
The determination of the wall velocity requires to compute the effective potential of the theory $V$, the temperature profile $T$ and the out-of-equilibrium perturbation $\delta f$. The former have been already discussed in sec.~\ref{sec:zero_temperature_corrections}~and~\ref{sec:finite_temperature_field_theory} of the previous chapter, while the latter will be addressed in the later chapters. In the next section we will deal with the temperature profile $T$ which we are going to compute using hydrodynamic equations.

\section{Plasma Hydrodynamics}

The plasma surrounding the bubble is a mixture of the different SM particle species. As we discussed in the previous section, the plasma is driven out of equilibrium by its interaction with the Higgs wall and light degrees of freedom are assumed to be in local thermal equilibrium. Under this assumption, the distribution functions $f_0$ that describe the statistical properties of the plasma are the usual equilibrium distribution functions of quantum gases reported in eq.~(\ref{eq:equilibrium_distribution_quantum_gas}).

The dynamics of a perfect fluid is well described by the hydrodynamic equations which allows us to determine the evolution of the macroscopic thermodynamic properties of the plasma. In particular, such equations allow us to compute the temperature profile which is necessary for the computation of the DWs terminal velocity. Before we start in tackling this problem we first briefly review some of the main thermodynamic quantities that will enter our computation. 

For this section we base our discussion on the analysis carried out in ref.~\cite{Espinosa:2010hh}.

\subsection{Fundamentals}

Our starting point is the definition of the stress-energy tensor of a gas described by a distribution function $f$, namely
\begin{equation}
\label{eq:definition_stress_energy_tensor_microscopic}
    T^{\mu\nu} = \int\frac{d^3{\bf p}}{(2\pi)^3 2 E} p^\mu p^\nu f({\bf p}, x)\,.
\end{equation}
For a perfect fluid, the stress energy tensor can also be written in terms of the energy density of the plasma $e$ and its pressure $p$ providing the well known expression
\begin{equation}
\label{eq:perfect_fluid_stress_energy_tensor}
    T_{pl}^{\mu\nu} = e U^\mu U^\nu - p \Delta^{\mu\nu}\,,
\end{equation}
where $U^\mu$ is the plasma four velocity and we introduced the projector $\Delta^{\mu\nu}$ along the plane perpendicular to $U^\mu$, namely\footnote{We use the signature ($+$\,$-$\,$-$\,$-$) of the metric tensor.}
\begin{equation}
\label{eq:projector_perpendicular_four_velocity}
    \Delta^{\mu\nu} = \eta^{\mu\nu} - U^\mu U^\nu\,.
\end{equation}
Using the definition of the stress-energy tensor in eq.~(\ref{eq:definition_stress_energy_tensor_microscopic}), it is possible to relate the macroscopic quantities $e$ and $p$ to the statistical properties of the plasma. Indeed we find
\begin{equation}
\begin{split}
    &e = U^\mu U^\mu T_{\mu\nu} = \int\frac{d^3{\bf p}}{(2\pi)^3 2 E}U^\mu p_\mu U^\nu p_\nu f_0\,,\\
    &p = -\Delta^{\mu\nu}T_{\mu\nu} = \int\frac{d^3{\bf p}}{(2\pi)^3 2 E}\Delta^{\mu\nu}p_\mu p_\nu f_0\,.
\end{split}
\end{equation}

Another relevant thermodynamic quantity is the enthalpy density $w$ which we can express in term of the energy density and the pressure through the well known relation
\begin{equation}
\label{eq:definition_enthalpy_density}
    w = e + p\,.
\end{equation}
The thermodynamic quantities $w, e$ and $p$ in turn can be determined using the effective potential.
In sec.~\ref{sec:finite_temperature_field_theory}
we observed that the finite-temperature correction corresponds to the free energy of an ideal quantum gas. When we also include the zero-temperature effects we can still identify the effective potential of the scalar fields with the free energy of a free non interacting quantum gas ${\cal F}$ made of different particles species with masses $m_i(h,s)$, namely
\begin{equation}
    {\cal F}(h, s, T) = V(h, s, T) = V_0(h, s) + \frac{T^4}{2\pi^2}\sum_i N_i J_{B/F}(m^2_i(h,s)/T)\,.
\end{equation}
Moreover, since we are considering the special case where the chemical potential of each particle species vanishes, the free energy ${\cal F}$ also coincides with the grand potential that, for homogeneous systems is identical to the opposite of the thermal pressure $p$. Hence
\begin{equation}
    p = -V\,.
\end{equation}
To relate the  energy density $e$ and the enthalpy density $w$ to $V$ we introduce an additional thermodynamic quantity, namely the entropy density $s$. Such a quantity can be expressed using the free energy ${\cal F}$ of a system through the following equation
\begin{equation}
    s \equiv -\frac{\partial {\cal F}}{\partial T} = -\frac{\partial V}{\partial T}\,,
\end{equation}
where in the last equality we used the correspondence between the finite-temperature potential and the free energy.
Finally, to derive $e$ and $w$ we combine eq.~(\ref{eq:definition_enthalpy_density}) together with the following well-known thermodynamic relation 
\begin{equation}
    {\cal F} = e - Ts\,,
\end{equation}
yielding the following result
\begin{equation}
    e = V - T\frac{\partial V}{\partial T}\;\;\;\;\;\;\;\;\;\;\;w = -T\frac{\partial V}{\partial T}\,.
\end{equation}

The hydrodynamic equations are derived from the conservation of the total energy momentum tensor of the system. The latter is the sum of two contributions, namely the scalar field condensates and the SM particles' plasma which emerge naturally, as we discussed in the previous section, in the WKB approximation. The stress-energy tensor of the system is then  
\begin{equation}
    T^{\mu\nu} = T_h^{\mu\nu} + T_s^{\mu\nu}\, + T_{pl}^{\mu\nu},
\end{equation}
where $T_h^{\mu\nu}$ and $T_s^{\mu\nu}$ are defined in eq.~(\ref{eq:definition_scalar_field_stress_energy_tensor}). 
The conservation of the stress-energy tensor of the system thus implies
\begin{equation}
\label{eq:conservation_stress_energy_tensor_system}
    \partial_\mu T^{\mu\nu} = \partial_\mu T_h^{\mu\nu} + \partial_\mu T_s^{\mu\nu} + \partial_{\mu}T_{pl}^{\mu\nu} = 0\,.
\end{equation}

The derivative of the stress energy tensors of the two scalar fields yields the respective equations of motion. The derivative of the plasma stress energy tensor instead provides
\begin{equation}
    \partial_\mu T_{pl}^{\mu\nu} = U^\nu \partial_\mu(e U^\mu) - \Delta^{\mu\nu}\partial_\mu p - p(U^\nu\partial_\mu U^\mu + U^\mu \partial_\mu U^\nu)\,.
\end{equation}
By projecting the above equation along the plasma four velocity $U^\mu$ and the projector $ \Delta^{\mu\nu} $ we get the special relativistic generalization of the continuity and the Euler equation, namely
\begin{equation}
    \begin{split}\label{eq:relativistic_continuity_euler_equations}
        &\partial_{\mu}(e U^\mu) - p\partial_\mu U^\mu = 0\\
        & w U^\mu \partial_\mu U^\nu - \Delta^{\nu\lambda}\partial_\lambda p = 0
    \end{split}
\end{equation}

We apply the conservation of the stress-energy tensor in the case of a steady planar wall. In such a case equation~(\ref{eq:conservation_stress_energy_tensor_system}) provides
\begin{equation}
\label{eq:steady_planar_wall_stress_energy_tensor_conservation}
    \partial_z T^{zz} = \partial_z T^{z0} = 0\,.
\end{equation}
By integrating the above equations across the wall one gets the following useful relations that connect the plasma velocity and the other thermodynamic quantities in the broken and symmetric phase, namely
\begin{equation}
    w_+v^2_+\gamma^2_+ = w_-v^2_-\gamma^2_- + p_-\,, \;\;\;\;\;\;\;\;\;\; w_+v_+\gamma^2_+ = w_-v_-\gamma^2_-\,,
\end{equation}
where with the $+(-)$ subscript we identify the corresponding quantity computed in the symmetric (broken) phase. Using the above set of equations we can relate the velocities $v_\pm$ providing
\begin{equation}
\label{eq:velocity_relations_symmetric_broken_phase}
    v_+ v_- = \frac{p_+ - p_-}{e_+- e_-}\,,\;\;\;\;\;\;\;\;\; \frac{v_+}{v_-} = \frac{e_- + p_+}{e_+ + p_-}\,.
\end{equation}
The above set of equations however is not closed. We indeed lack an equation of state that relates the pressure $p$ to the energy density $e$. 

\subsection{Equations of state}

In the limit where the masses of the particles are small compared to the temperature $T$ of the plasma, it is possible to obtain the so called bag equation of state \cite{Espinosa:2010hh}. In this limit the plasma is well described by a perfect relativistic gas and in the symmetric phase we find 
\begin{equation}
    p_+ = \frac{1}{3}a_+ T_+^4 + \Delta V|_{T= 0}\;\;\;\;\;\;\;\;\;\; e_+ = a_+ T_+^4 + \Delta V|_{T=0}\,.
\end{equation}
where $\Delta V|_{T=0}$ denotes the potential energy difference between the false and the true vacuum at $T=0$, $T_{\pm}$ is the temperature measured in the symmetric (+) and broken (-) phase while $a_{+}$ are the plasma degrees of freedom in the symmetric phase. In the broken phase instead we find
\begin{equation}
    p_- = \frac{1}{3}a_- T_-^4\;\;\;\;\;\;\;\;\;\; e_- = a_- T_{-}^4\,.
\end{equation}
The physical meaning of the above relations is straightforward since they correspond to the pressure and energy density of a relativistic perfect fluid. Moreover, because of the presence of the DW, the pressure also contains the potential energy difference that drives the bubble expansion. The quantities $a_{\pm}$ are computed from the finite temperature corrections. We indeed find
\begin{equation}
    a_\pm = \f{\pi^2}{30}\sum_{light\;i}\left[N_i^b + \frac{7}{8}|N_i^f|\right]
\end{equation}
Where the sum runs over all the light species in the plasma, $N_i^b$ are the bosonic degrees of freedom while $N_i^f$ the fermionic ones. Particle species with mass $m_i \gg T$ do not enter in the sum since in such regime the functions $J_{B/F}(m_i/T)\sim \exp(-m_i/T)$. As a consequence, since in the broken phase particles acquire a mass, we expect $a_+ > a_-$.

In the case $m_i\sim T$, which is usually true in the broken phase, it is not possible to adopt the bag equation of state. However, since light degrees of freedom dominate the plasma, we expect that deviations from the bag equation states are small and we may adopt the following generalization~\cite{Espinosa:2010hh}
\begin{equation}
\label{eq:equation_of_state}
    a_{\pm} \equiv \frac{3}{4 T_{\pm}^3}\frac{\partial V}{\partial T}\Bigr|_{T=T_\pm}\;\;\;\;\;\;\;\;\;\; e_{\pm} = \left(-\frac{T_{\pm}^4}{4}\frac{\partial V}{\partial T} + V\right)\Bigr|_{T = T_\pm}\,.
\end{equation}
We can finally close the set of equations in~(\ref{eq:velocity_relations_symmetric_broken_phase}). By expressing $p$ and $e$ using eq.~(\ref{eq:equation_of_state}) we find the following expressions
\begin{equation}
\label{eq:final_velocity_relations}
    v_+v_- = \frac{1 - (1 - 3\alpha)r}{3 - 3 (1 + \alpha)r}\,, \;\;\;\;\;\;\;\;\;\; \frac{v_+}{v_-} = \frac{3 + (1- 3\alpha)r}{1 + 3 (1 + \alpha)r}\,.
\end{equation}
where we defined
\begin{equation}
    \alpha = -\frac{\Delta V}{a_+ T_+^4}\;\;\;\;\;\;\;\;\;\; r = \frac{a_+ T_+^4}{a_- T_-^4}\,.
\end{equation}
The quantity $\alpha$ is a measure of the latent heat released during the phase transition and we already encountered it when we discussed the spectrum of GW emitted during a FOPT. The parameter $r$ instead is the ratio between the energy of radiation in the broken and symmetric phase.

We finally notice that the set of hydrodynamic equations together with the expressions in~(\ref{eq:final_velocity_relations}) is still open. To close the system we need to relate the temperatures $T_+$ and $T_-$ and to fix the value of one of the two velocities. In order to accomplish this task we first need to further clarify the properties of the plasma that eq.~(\ref{eq:final_velocity_relations}) describes. This constitutes the subject of the next section.

\subsection{Hydrodynamic regimes}

The two equations in~(\ref{eq:final_velocity_relations}) can be combined in the following expression
\begin{equation}
\label{eq:vp_as_function_of_vm}
    v_+ = \frac{1}{1+\alpha}\left[\left(\frac{v_-}{2} + \frac{1}{6v_-}\right) \pm \sqrt{\left(\frac{v_-}{2}+ \frac{1}{6 v_-}\right)^2 + \alpha^2 + \frac{2}{3}\alpha -\frac{1}{3}}\right]\,,
\end{equation}
The above equation together with the Euler and the continuity equations in eq.~(\ref{eq:relativistic_continuity_euler_equations}) completely characterize the properties of the plasma. Assuming a spherical symmetric configuration of the bubble with negligible width, we can fully parameterize the macroscopic quantities with $x$ 
the distance from the center of the bubble and $t$, the time from nucleation. Furthermore, because there is no characteristic scale in the problem, the solution to eq.~(\ref{eq:relativistic_continuity_euler_equations}) is a similarity solution depending only on the ratio $\xi = x/t$. 
From our considerations it is possible to cast the hydrodynamic equations in the following form
\begin{equation}
    \begin{split}
    \label{eq:hydrodynamic_equations_symilarity}
        &2\frac{v}{\xi} = \gamma^2(1-v\xi)\left[\frac{\mu^2}{c_s^2}-1\right]\partial_\xi v\,,\\
        &\partial_\xi T = T\gamma^2\mu\partial_\xi v\,,
    \end{split}
\end{equation}
where $ \mu $ is the Lorentz transformed velocity
\begin{equation}
    \mu = \frac{\xi - v}{1 - v \xi}\,.
\end{equation}
while $c_s = \sqrt{(dp/dT)/(de/dT)}$ is the speed of sound in the plasma which, in terms of the effective potential is given by
\begin{equation}
    c^2_s = \frac{\partial V/\partial T}{T\partial^2 V/\partial T^2}\,.
\end{equation}
Notice that in the symmetric phase $c_s^+ = 1/\sqrt{3}$, which is the speed of sound of a relativistic gas. On the other hand in the broken phase, where particles acquire mass, $c_s^{-}\lesssim 1/\sqrt{3}$.

Provided the proper initial conditions, the hydrodynamic equations in eq.~(\ref{eq:hydrodynamic_equations_symilarity}) completely characterize the temperature and fluid velocity profiles. These equations together with the ones in~(\ref{eq:final_velocity_relations}) identify the different hydrodynamic regimes that take place during a FOPT.

\begin{figure}
    \centering
    \includegraphics[width=0.47\textwidth]{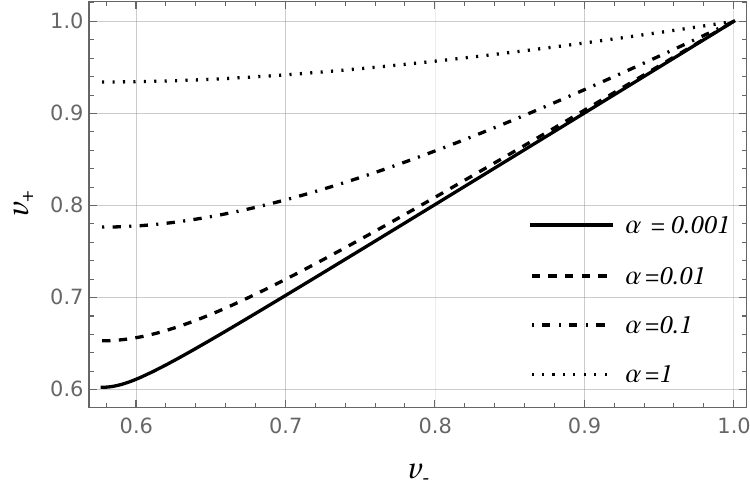}
    \hfill
    \includegraphics[width=0.47\textwidth]{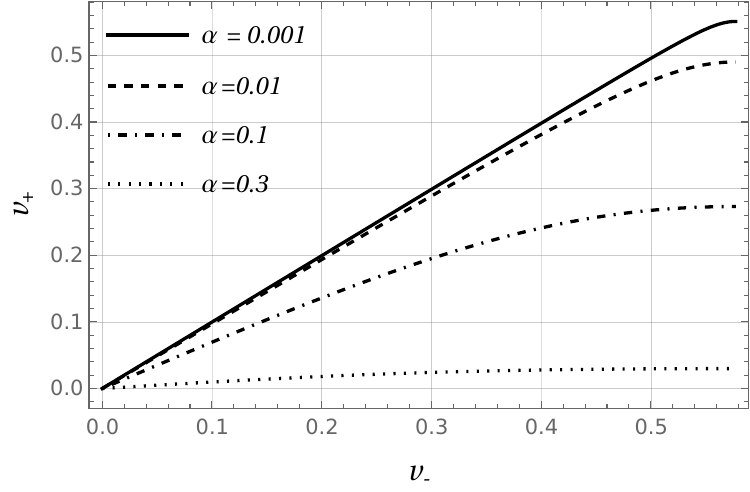}
    \caption{Plot of the velocity $v_+$ as a function of $v_-$ for detonation solutions (left plot) and deflagration solutions (right plot) for different values of the strength $\alpha$.}
    \label{fig:vpvm_detonation_deflagration}
\end{figure}

\subsubsection{Detonation}

The first regime we analyze are detonations. These correspond to the first branch of solutions in~(\ref{eq:vp_as_function_of_vm}) where $v_+>v_-$. The left plot in Fig.~\ref{fig:vpvm_detonation_deflagration} shows the velocity relation in eq.~(\ref{eq:vp_as_function_of_vm}) for the case of detonations. Consistency with the hydrodynamic equations sets a lower limit on $v_-$ given by $c_s^+$, furthermore eq.~(\ref{eq:final_velocity_relations}) sets an upper limit on $r < 1/(1+3\alpha)$. 

We identify three different types of detonations. Detonations with $v_- < c_s^+$ are called strong detonations. As pointed out in ref.~\cite{Laine:1993ey} these are not possible in cosmology. We will thus deal only with weak detonations $v_- > c_s^+$ and Jouguet detonations $v_- = c_s^+$. For such detonations, the velocity computed in the symmetric phase $v_+\equiv v_J$ using eq.~(\ref{eq:vp_as_function_of_vm}) provides the definition of the Jouguet velocity $v_J$. 
From the qualitative point of view, detonations are characterized by the presence of a rarefaction wave that trails the DW.

During a detonation, the DW moves in the unperturbed plasma. As a consequence, in the wall reference frame we can identify $T_+ = T_n$ and $v_+ = v_w$, with $v_w$ the DW velocity. Moreover, because $v_+$ has a minimum in correspondence to $v_- = c_s^+$, as shown on the left hand plot of Fig.~\ref{fig:vpvm_detonation_deflagration}, the smallest value that $v_+$ can have corresponds to the Jouguet velocity $v_J$. Thus detonation solutions take place for bubbles expanding at velocities $v_w > v_J$. As we will explore further, the Jouguet velocity has an important role in the analysis of the wall friction.

\subsubsection{Deflagration}

Solutions with $v_+ < v_-$ identify deflagrations. The right hand plot in Fig.~\ref{fig:vpvm_detonation_deflagration} shows the velocity relation for such a case. In this regime, the relations in~(\ref{eq:final_velocity_relations}) set $v_+ < c_s^-$, in addition to $v_-< c_s^-$. In contrast with detonation, in the deflagration regime the plasma is at rest inside the bubble and we thus identify $v_- = v_w$. Therefore deflagration solutions take place when $v_w < c_s^-$.

Deflagration solutions are characterized by the presence of a shock wave in front of the DW. The properties of the shock-front can be studied by using eq.~(\ref{eq:final_velocity_relations}) where $v_\pm = v_\pm^{\rm sh}$, now identify the plasma velocity in front and behind the shock wave in the reference frame of the latter. In particular, because the shock wave moves in the symmetric phase, one has $\alpha = 0$, which provides $v_+^{\rm sh}v_-^{\rm sh} = 1/3$. In the plasma reference frame this translates to
\begin{equation}
    \mu(v(\xi^{\rm sh}),\xi^{\rm sh})\xi^{\rm sh} = c_s^+\,.
\end{equation}
The above condition allows to determine the velocity of the shock wave $\xi^{\rm sh}$ measured in the bubble wall center.

Because of the shock wave, the temperature of the plasma in front of the DW is not $T_n$. Instead we find for the temperature in front of the shock wave $T_+^{\rm sh} = T_n$. 
The determination of the temperatures $T_\pm$ in this case is thus much more involved since one of the boundary conditions of the problem is defined in front of the shock wave. To determine $T_\pm$ we use the following shooting method. We first make an ansatz on the temperature $T_-$ and we compute $v_+$ and $T_+$ using eq.~(\ref{eq:final_velocity_relations}). We then evolve the temperature and velocity profiles using eq.~(\ref{eq:hydrodynamic_equations_symilarity}) from the DW at $\xi = \xi^{w}$ to the shock-wave $\xi^{\rm sh}$. Such equations are then integrated using as initial conditions $T(\xi^w) = T^+$ and $v(\xi^w) = v_+$ and their solutions determine $T_-^{\rm sh}$ and $v_-^{\rm sh}$, namely the temperature and the velocity of the plasma inside of the shock wave.
Finally we use again eq.~(\ref{eq:hydrodynamic_equations_symilarity}) to determine $T_+^{\rm sh}$ and $v_+^{\rm sh}$ and we vary $T_-$ until $T_+^{\rm sh} = T_n$.

\subsubsection{Hybrid}

The last hydrodynamic regime is the hybrid regime. Such solutions take place when $c_s^- < v_w < v_J$ and present qualitative characteristics of both detonations and deflagrations, hence the name hybrid. As in detonation a rarefaction wave trails the DW, while a shock wave is present in front of the DW as for deflagrations. The strategy to determine the boundary conditions are identical to the deflagration case with the only exception that, due to entropy conservation, we identify $v_- = c_s^-$.
We finally mention that some recent development have shown that hybrid solutions could be unstable~\cite{Krajewski:2023clt}.

\subsection{Temperature and plasma four velocity profile}
\label{sec:temperature_velocity_plasma_profile}

The analysis that we carried out so far provided all the necessary tools to determine the temperature and plasma four velocity profiles across the wall. The computation of these two quantities relies on the conservation of the stress-energy tensor of the system and in this section we review the strategy employed in ref.~\cite{Laurent:2022jrs} for the computation of the two profiles.

In view of the computation of the friction, we are going to determine the temperature and velocity profile in a perturbed plasma. We thus include in the stress-energy tensor of the system also the contribution of out-of-equilibrium fluctuations and consider the following
\begin{equation}
    T^{\mu\nu} = T_\phi^{\mu\nu} + T_{pl}^{\mu\nu} + T_{out}^{\mu\nu}\,,
\end{equation}
where $T_{out}^{\mu\nu}$ is defined as
\begin{equation}
    T_{out}^{\mu\nu} = \sum_i N_i \int\frac{d^3{\bf p}}{(2\pi)^3 2E}p^\mu p^\nu \delta f\,.
\end{equation}
By integrating eq.~(\ref{eq:steady_planar_wall_stress_energy_tensor_conservation}) across the wall we get
\begin{equation}\label{eq:tensor_conservation}
		\begin{split}
			&T^{30} = w\gamma^2v+T^{30}_{out}= c_1\,,\\
			&T^{33} = \frac{1}{2}(\partial_z\phi)^2 - V + w\gamma^2v^2+T^{33}_{out} = c_2\,,
		\end{split}
\end{equation}

We expect $T^{30}_{out}$ and $T^{33}_{out}$ to vanish at large distances of the wall. Indeed because the force between plasma and the wall is local and because we do not expect large diffusion, we expect the fluctuations $\delta f$ to vanish at $z\rightarrow \pm\infty$. As a consequence the two constants $c_1$ and $c_2$ are determined from the boundary values $v_{+(-)}, T_{+(-)}$. Hence, from the analysis of the previous section we are able to identify the correct boundary conditions for the different velocities $v_w$.

Equations~(\ref{eq:tensor_conservation}) can be recast in the following form

	\begin{equation}
 \label{eq:bg_profiles}
		\begin{split}
			&v=\frac{- w +\sqrt{4(c_1-T^{30}_{out})^2 + w^2}}{2(c_1-T^{30}_{out})}\,,\\
			\rule{0pt}{1.75em}&\frac{1}{2}(\partial_z\phi_i)^2 - V -\frac{1}{2}w + \frac{1}{2}\sqrt{4(c_1-T^{30}_{out})^2+w^2} - (c_2-T^{33}_{out}) = 0\,,
		\end{split}
	\end{equation}
and their solutions yield the velocity and temperature profiles. Because no derivative of the temperature and the plasma velocity is involved 
the former equations can be straightforwardly solved with a simple root-finding numerical algorithm.

\section{Determining the wall velocity}
\label{sec:wall_velocity_determination}

The wall terminal velocity is the result of the balance of the internal pressure and the friction as expressed by eq.~(\ref{eq:total_pressure}). In order to compute the terminal velocity we first need to determine the friction, which is proportional to the out-of-equilibrium perturbations, and the temperature and plasma velocity profiles. These quantities in turn are mostly controlled by the terminal velocity and can be determined by solving an effective kinetic theory given by the Boltzmann equation and eq.~(\ref{eq:bg_profiles}). 

We will finally determine the perturbations in Chapter~\ref{ch:exact_solution_to_liearized_Boltzmann_equation}. However we can already anticipate the strategy that we will follow to compute the DW velocity.
We begin our analysis from the scalar fields equations of motion. Under the assumption of a steady state regime it is convenient to express the equations of motion in the wall reference frame where solutions are stationary. In this frame the equations of motion read
\label{eq:higgs_singlet_eom}
\begin{equation}
    \begin{split}
         E_h &\equiv -\partial_z^2 h + \frac{\partial V(h,s,T)}{\partial h} + F(z)/h' = 0 \,, \\
        \rule{0pt}{1.75em}E_s &\equiv -\partial_z^2 s + \frac{\partial V(h,s,T)}{\partial s} = 0  \,.
    \end{split}
\end{equation}
Once solved, the above equations yield the exact profiles of the Higgs and the scalar singlet. The exact solution of such equations however is very complicated and it is not necessary for most applications. In particular, to determine the wall velocity it is possible to adopt the so-called $\tanh$ ansatz approximation which provides a good estimate for the fields profiles~\cite{Laurent:2022jrs,Friedlander:2020tnq}. Within such approximation the fields profiles are given by the solution to the equation of motion of the scalar field in the limit of thin wall. The dynamics of the system is then encapsulated into four quantities: the terminal velocity $v_w$, the widths of the Higgs and singlet DW $L_h$ and $L_s$ and finally a displacement $\delta_s$. Under this assumption the fields profiles are given by
\label{eq:field_ansatz}
\begin{equation}
\begin{split}
    h(z) &= \frac{h_-}{2}\left(1+\tanh\left(\f{z}{L_h}\right)\right)\,,\\
    \rule{0pt}{1.75em}s(z) &= \f{s_+}{2}\left(1-\tanh\left(\f{z}{L_s}+\delta_s\right)\right)\,,
\end{split}
\end{equation}
where $h_-$ and $s_+$ denote the Higgs and singlet VEVs in the broken and symmetric phase respectively.

The displacement $\delta_s$ between the Higgs and the singlet wall $\delta_s$ is introduced to take into account the possibility that a different pressure can act on the Higgs and singlet DW. Since the Higgs field interacts with many more particles with respect to the singlet this is usually the case and the collective effect originates an offset between the two walls.

To determine the Higgs VEV inside the bubble, $h_-$, and the singlet VEV in front of the wall, $s_+$, we minimize the finite-temperature effective potential leading to
\begin{equation}
    \f{\partial V(h_-,0,T_-)}{\partial h} = 0\,, \qquad \quad
    \f{\partial V(0,s_+,T_+)}{\partial s} = 0\,.
\end{equation}
The common strategy to determine the four parameters $v_w$, $L_h$, $L_s$ and $\delta_s$ consists in taking suitable moments of the equations of motion and then seek for the roots of the corresponding equations. The choice of the momenta is quite arbitrary. However a convenient choice motivated by physical considerations is~\cite{Laurent:2022jrs,Friedlander:2020tnq}
\begin{equation}
\label{eq:param_equations}
\begin{split}
    P_h & = \int dz\, E_h h' = 0\,, \qquad \quad G_h = \int dz\, E_h (2 h/h_- - 1)h' = 0\,, \\
    \rule{0pt}{1.75em}P_s & = \int dz\, E_s s' = 0\,, \qquad \quad G_s = \int dz E_s\, (2 s/s_+ -1)s' = 0\,.
\end{split}
\end{equation}

The functions $P_{h,s}$ and $G_{h,s}$ have a clear physical meaning. As we already discussed at the beginning of this chapter, $P_{h,s}$ describe the total pressure that acts on the Higgs and the scalar singlet wall respectively. Because the system reaches a terminal velocity and we are assuming that the two walls are in a steady state regime, to be consistent with our assumptions we enforce $P_{h,s}$ to vanish. We also notice that the total pressure acting on the system, given by $ P_{tot} = P_h + P_s$, corresponds to eq.~(\ref{eq:steady_state_condition}). Hence we expect that this combination mostly depends on the velocity $v_w$. On the other hand the difference $P_h - P_s$ describes the pressure difference between the two walls. Such a difference, in the steady state regime, determines the displacement between the two walls and hence we expect it to mostly depend on the displacement $\delta_s$. Our intuition is confirmed by the numerical analysis.

The last two remaining momenta, namely $G_{h,s}$ instead take into account the gradients of the pressure across the two walls. Particles impinging on the DW have the effect to stretch or shrink the wall in the $z$ direction. This effect is precisely encoded in the two momenta $G_{h,s}$. In the steady state regime these gradients must vanish and their corresponding equations determine the widths of the two walls.

As we mentioned before, to close the system of equations in~(\ref{eq:param_equations}) we also need to compute the out-of-equilibrium perturbations by solving the Boltzmann equation and eq.~(\ref{eq:bg_profiles}) to determine the temperature and plasma velocity profiles. These quantities however depend on the four observables we aim to determine which forces us to employ an iterative procedure to solve the system of equations.
\begin{enumerate}
    \item We first solve eq.~(\ref{eq:param_equations}) for the four parameters $v_w$, $\delta_s$, $L_h$, $L_s$ without including the out-of-equilibrium perturbations simultaneously with eq.~(\ref{eq:bg_profiles}). Such a task is easily performed by employing a root finding algorithm such as the Newton algorithm.

    \item We use the value of $v_w$ and $L_h$ to determine the out-of-equilibrium perturbations. The perturbations are determined by solving the corresponding Boltzmann equation. In the following chapter we are going to derive the Boltzmann equation that describes the perturbations while in Chapter~\ref{ch:exact_solution_to_liearized_Boltzmann_equation} we will discuss the algorithm that we used to determine the perturbations

    \item We insert the perturbations in eq.~(\ref{eq:higgs_singlet_eom}) and we solve again eq.~(\ref{eq:param_equations}) to determine the new values of the four quantities $v_w$, $L_h$, $L_s$ and $\delta_s$.

    \item We iterate from point 2 until convergence on the four quantities is reached.
\end{enumerate}

The last missing ingredient for the computation of the wall terminal velocity are thus the out-of-equilibrium perturbations. We stress that their computation is non-trivial and requires an accurate modelling of the plasma dynamics during the phase transition. This can be done by using an effective kinetic theory approach which leads to set of Boltzmann equations describing the fluctuations. We will discuss this aspect in the next chapter.



\chapter{Effective kinetic theory for the plasma}
\label{ch:effective_kinetic_theory}

The biggest theoretical challenge in the determination of the wall terminal velocity stems from the friction acting on the DW. As discussed in the previous chapter, an accurate description of this term requires a precise characterization of the out-of-equilibrium perturbations which in turn are determined from the effective kinetic theory that governs the plasma non-equilibrium dynamics. Solving the effective kinetic theory is a highly non-trivial task. As we are going to show in this chapter, the effective kinetic theory consists in a set of Boltzmann equations whose solution is very challenging being an integro-differential equation.


This chapter presents two parts. In the first one we present the aforementioned kinetic theory which we will use in later chapters to compute the fluctuations and hence the friction. At the same time we assess the validity of the kinetic theory we provide motivating, in particular, that such description is valid only for hard thermal particles being the latter the only excitations that can be treated as quasi-particles. 

The second part is devoted to the Boltzmann equation. In that part we review the fundamental concepts of transport equations and discuss one of the most known strategies to approximate the solution to the equation. At last, we present one of our original results, namely a multipole expansion of the Boltzmann equation.


\section{Dynamical aspects of hot gauge theories}

The presence of a hot-weakly coupled plasma during the EWPT modifies the dynamics of quantum fields as we discussed in section~\ref{sec:finite_temperature_field_theory}. The equilibrium properties of the plasma and their impact on the system are well-understood with the leading-order effect being the inclusion of the free-energy in the effective potential. Next-to-leading order effects can be studied by employing perturbation theory and provide a deeper insight on the static thermodynamic properties of hot gauge theories~\cite{Coriano:1994re,Parwani:1994zz,Parwani:1994je, Arnold:1994ps,Arnold:1994eb}. 


The dynamical properties of the plasma, such as transport phenomena, are instead much less trivial to understand. Depending on the scale of interest a large variety of effective descriptions can be used to describe the physics that takes place. In the previous chapter, for instance, we provided an effective description of the plasma light degrees of freedom using hydrodynamics and we modeled light particles as a perfect fluid in local equilibrium. Such approach clearly provides a poor description of heavy particle species because their dynamics is dominated by the DW interactions. A more accurate description of such particles is provided by an effective kinetic theory, namely a set of Boltzmann equations, that captures the relevant properties of the dynamics of such species. These in turn are determined by the relevant processes taking place in the plasma.

To understand which are the relevant processes we should account in our description, we first need to understand what are the relevant scales of our problem, namely those where the different physics effects disentangle. We thus begin our analysis by reviewing the relevant scales of hot gauge theories. Refs.~\cite{Arnold:2002zm,Arnold:2000dr,Arnold:2003zc,Arnold:1997gh} cover in deep detail this topic and we refer to those works and references therein for further discussions.

\subsection{Relevant scales}
Without loss of generality, to discuss the relevant scales of hot gauge theories we can focus on the QCD plasma at high temperature. In such regime the gauge coupling $ g(T) $ is small and the theory is weakly coupled. Such description captures the relevant features of the plasma surrounding the bubble during the EWPT and can be easily generalized to the SM plasma.

We begin our analysis by recalling that in presence of a high-temperature environment two classes of quantum states arise, hard modes, with a momentum of order $\sim T$ and soft modes with momentum $\sim g T$. In virtue of their statistics, only bosons can be soft. The dynamics of such modes is very different with respect to the hard ones and can be understood by studying how the propagators of soft excitations are modified in presence of a finite-temperature environment. This analysis is however very complicated because perturbation theory breaks down at high-temperature due to the logarithmic IR divergences caused by soft-bosons.
%
%
In particular due to these IR divergences we lose the  connection between the order of loop expansion and the power of $g$ and a resummation procedure must be pursued. A first example of this effect is represented by the daisy diagrams that we mentioned at the end of sec.~\ref{sec:finite_temperature_field_theory}, 
while an analogous effect takes place in hot gauge theories for a class of diagrams dubbed hard thermal loops (HTL) that arises from the interplay of hard and soft particles involved in loop corrections. 
The HTL resummation, discussed in ref.~\cite{Braaten:1989mz}, provides a deep insight on the relevant physical processes that characterize hot gauge theories at different scales.

There are three main effects. Thermal particles present a correction to the energy dispersion relation, soft gauge bosons are Landau damped while longitudinal soft gauge bosons are Debye screened. Each of these effects is extremely relevant to understand the effective kinetic theory that describes the plasma. Because the dynamics of soft bosons is both screened and damped their resulting effective description will be very different with respect to that of hard modes. This observation will be crucial when we will discuss the contribution arising from IR W bosons to the friction at the end of Chapter~\ref{ch:exact_solution_to_liearized_Boltzmann_equation}.




To understand how the above three effects arise, let us study how the propagator of thermal particles is modified according to the HTL resummation procedure. The HTL generate a self-energy to the particles whose complete expression is rather complicated and not necessary for the sake of our discussion. We are instead interested in the qualitative behaviour of such self-energies. For the full expression of the fermionic self energies we refer to~\cite{Klimov:1981ka, Weldon:1982aq}
while for that of the the longitudinal and transverse components of the gluon self-energy we refer to~\cite{Klimov:1982bv,Weldon:1982bn}.
%

The self energies encode the three different physical effects that we mentioned before.
Let us begin our discussion from the modification of the dispersion relation. Because of the interactions between the particles and the medium the dispersion relation is modified by an effective thermal mass. 
Such effective mass depends on the momentum of the particle but in the limit of large momenta, namely for hard particles, its dependence is well described by the temperature only.
At leading-order for a hard state of a particle species $i$ the dispersion relation is given by
\begin{equation}
    E_i=\sqrt{{\bf p}^2 + m^2_{i}(T)}
\end{equation}
where $m_{i}$ is the effective thermal mass

This modification plays an important role in the effective kinetic theory of hard particles. In particular we will use it to regularize the IR divergences of the amplitudes of the relevant processes that takes place in the plasma. For later use, we define the resulting thermal masses for gluons, quarks, weak gauge bosons, and leptons
\begin{equation}
\label{eq:thermal_masses}
\begin{split}
    &m_g^2 = 2 g_s^2 T^2\,,\;\;\;\;\;\;\;\;\;\; m_q^2 = \frac{g_s^2 T^2}{6}\,,\\
    &m_W^2 = \frac{5 g_w^2 T^2}{3}\,,\;\;\;\;\;\;\;\;\;\; m_l^2 = \frac{3g_w^2T^2}{32}\,,
\end{split}
\end{equation}
where $g_s$ and $g_W$ are the gauge couplings of strong and weak interactions respectively.

The dispersion relation of soft bosons is further modified by the Landau damping that is encoded in a logarithmic term present in the gluon self-energies. This logarithm provides a discontinuity below the light cone $|{\bf p}| > p^0 > - |{\bf p}|$ which can be physically interpreted as the absorption and emission of two particles from the thermal bath with nearly equal momenta contributing in dissipating the energy of particles. Together with the Landau damping, the dynamics of soft bosons is dominated by Debye screening which we can understand considering the qualitative behaviour of the longitudinal and transverse self-energies of gluons. The longitudinal part of the self-energy has the following behaviour
\begin{equation}
\Pi_L = 
    \begin{cases}
    4g_s^2 T^2\left(1+i\pi\displaystyle\frac{p_0}{2|{\bf p}|}\right)\,,\;\;\;\;\; &p_0/|{\bf p}|\ll 1\,,\\
    4g_s^2 T^2\,,\;\;\;\;\;  &\displaystyle p_0/|{\bf p}|\gtrsim 1\,.
    \end{cases}
\end{equation}
Which implies that longitudinal modes, namely static electric fields, are Debye screened at distances $O(1/gT)$. In other words, the hard modes cut-off the long-range Coulomb interactions for soft modes. The transverse part instead has a different behaviour. We find
\begin{equation}
    \Pi_T = 
    \begin{cases}
    4g_s^2 T^2\displaystyle\frac{p^0}{2|{\bf p}|}\left(1+i\pi\displaystyle\frac{p^2}{2|{\bf p}|^2}\right)\,,\;\;\;\;\; &p_0/|{\bf p}|\ll 1\,,\\
    4g_s^2 T^2\displaystyle\frac{p^0}{|{\bf p}|}\,,\;\;\;\;\;  &\displaystyle p_0/|{\bf p}|\gtrsim 1\,.
    \end{cases}
\end{equation}
Since $\Pi_T$ vanishes at small energies, magnetic fields are thus unscreened.

Having discussed the relevant effects that dominates the plasma dynamics at different scales we can now focus in determining which are the relevant interactions between particles that we should account in the effective kinetic theory.


\subsection{Relevant collision processes}

\begin{figure}
    \centering
    \includegraphics[width=0.20\textwidth]{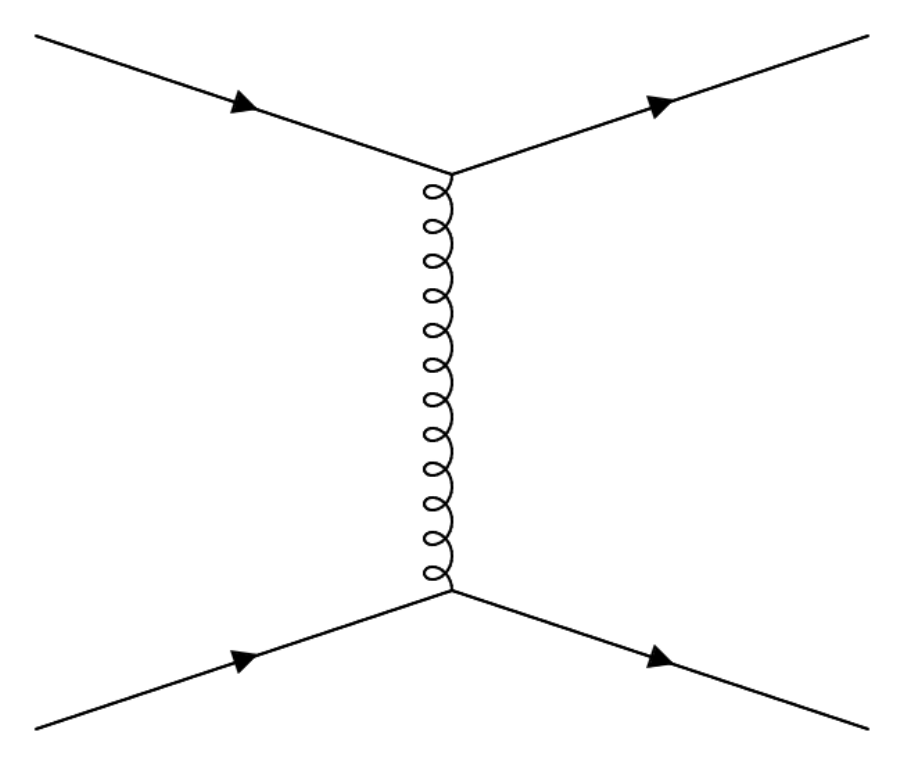}
    \includegraphics[width=0.20\textwidth]{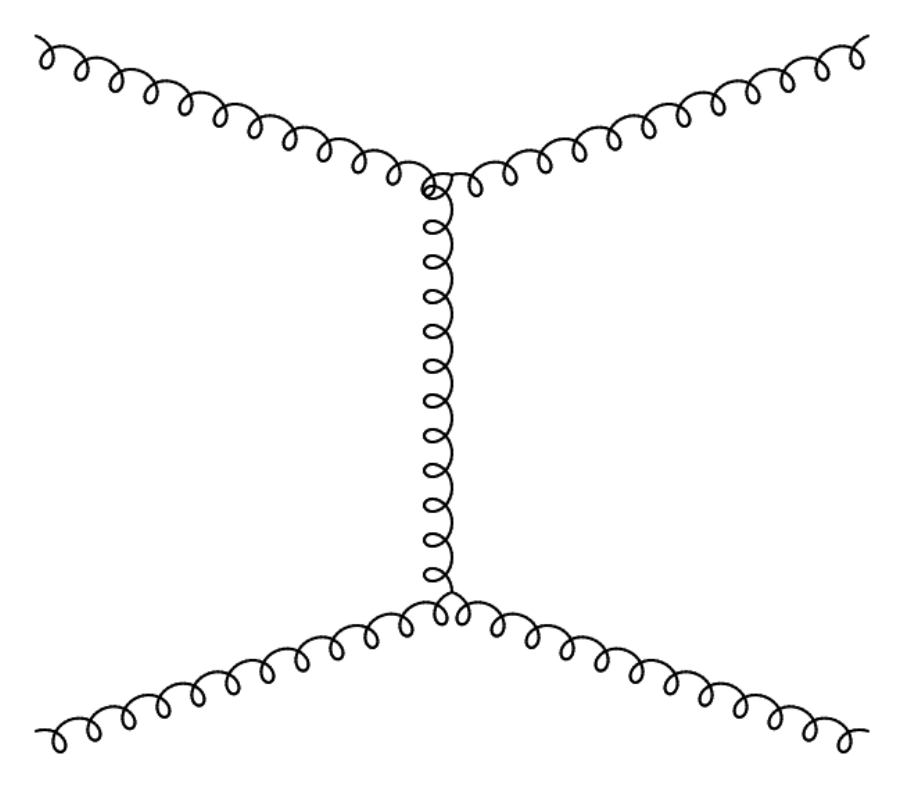}
    \includegraphics[width=0.20\textwidth]{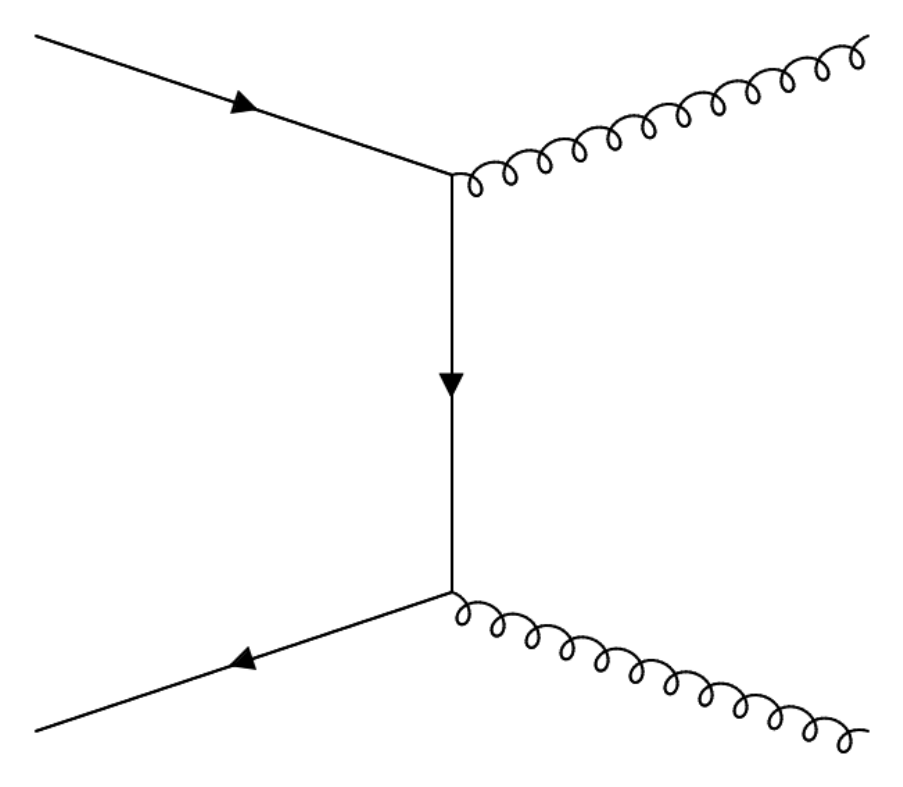}
    \caption{Compton scattering of quark (left diagram) and gluons (center diagram) and annihilation process (right diagram) in the $t$ channel for hard particles.}
    \label{fig:compton_annihilation_feynman_diagrams}
\end{figure}

To determine the relevant processes that we should consider in the effective kinetic theory we focus on the fate of the hard particles in the plasma. Compton scattering, depicted in the left and central diagram of Fig.~\ref{fig:compton_annihilation_feynman_diagrams}, is one of the most important processes. When two hard particles scatter with each other exchanging a momentum ${\bf q}$ their direction of propagation changes by an angle $\theta \sim O(|{\bf q}|/T)$. The differential scattering rate of the process hence can be estimated as
\begin{equation}
    d\Gamma \sim g^4 T^3\frac{dq}{q^3} \sim g^4 T^3\frac{d\theta}{\theta^3}\,.
\end{equation}
We recall that the above expression is valid only if the momentum exchanged is hard, namely $q\gtrsim O(gT)$. For $q \lesssim O(gT)$ Debye screening and Landau damping dominate the dynamics and provide important deviations 

The rate of hard scattering, namely the one where $q\sim O(T)$ and thus $\theta\sim O(1)$, is then $O(g^4 T)$, while for soft scattering we find $O(g^2 T)$.
The inverse of such rates defines the characteristics mean free times of the relative processes. We therefore define $\tau_g = 1/(g^2 T)$ and $\tau_* = 1/(g^4 T)$ the mean free time of small and large angle deviations respectively. Such definitions should be understood as a parametric estimation and not a quantitative definition. A proper quantitative definition, in fact, would require to compute the collision rates of the corresponding processes. 

Annihilation processes are also relevant in constructing the effective kinetic theory. An example for such a process is provided by the right diagram in Fig.~\ref{fig:compton_annihilation_feynman_diagrams} where we depict the case of $q\bar{q}\rightarrow gg$. When the virtual particle has a soft momentum, the mean free time of such a process is $O(1/g^4 T)$~\cite{Arnold:2000dr}, which corresponds to the large scattering angle mean free time.

The $s-$channel versions of the diagrams in Fig.~\ref{fig:compton_annihilation_feynman_diagrams} have a corresponding mean free time $O(1/(g^4 T))$. As a consequence also such diagrams should be considered in the effective kinetic theory. Nevertheless, as we will discuss later, $t$ and $u$ channel processes are logarithmic enhanced with respect to the $s$ channel diagrams. As a consequence, in the so called leading-log approximation it is possible to neglect altogether $s-$channel diagrams.

\begin{figure}
    \centering
    \includegraphics[width=0.75\textwidth]{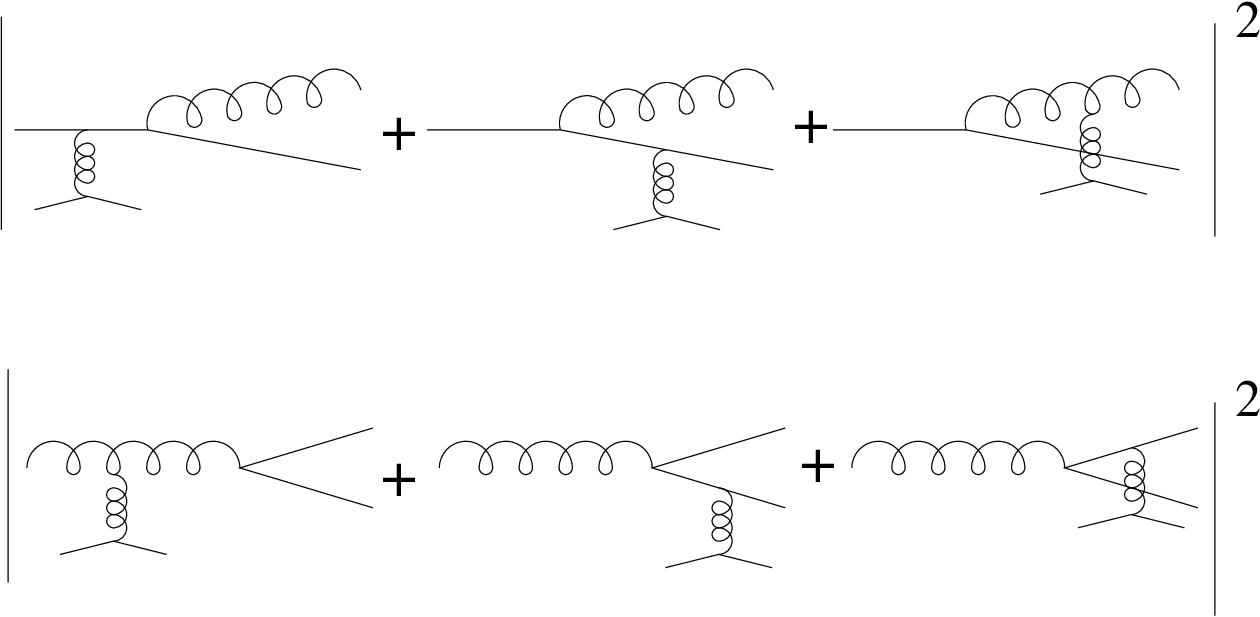}
    \caption{Leading order contributions to near collinear $1\rightarrow 2$ processes. Upper row depicts the splitting of hard quark to a nearly collinear gluon and quark after the exchange of a soft gluon with another quark. Lower row, instead, shows the splitting of a hard gluon in a couple of collinear quark and antiquark after a soft scattering with a plasma particle. The figure is borrowed from~\cite{Arnold:2002zm}.}
    \label{fig:bremmstrahlung}
\end{figure}

Together with the $2\to2$ processes, 
also the $1\to2$ processes where a hard particle splits in two hard collinear particles should be considered in an effective kinetic theory. Such processes are forbidden in the vacuum because of energy momentum conservation, but allowed in the plasma because of soft particle exchange. An example of such processes is the brehmmstrahlung process depicted in Fig.~\ref{fig:bremmstrahlung} for gluons and quarks. The upper row of the figure shows the scattering of a hard quark with another hard plasma excitation via the exchange of a soft gluon and the subsequent emission of a hard gluon. The two final states thus travel collinearly in a cone of amplitude $\theta\sim g$. The lower row of Fig.~\ref{fig:bremmstrahlung} instead shows the correspondig process for a gluon which, after a soft scattering, splits in a couple of hard collinear quark and anti-quark. The mean free time of such processes is again $O(1/(g^4 T))$ which is comparable to the one of $2\rightarrow2$ processes.

Despite the time scale of bremmstrahlung and $2\rightarrow2$ processes are the same, there is a crucial difference between the two processes. Due to the soft exchange of gluon in the processes, the momentum of the virtual particle connecting the splitting vertex and the emission vertex is off-shell by an amount $O(g^2 T^2)$. As a consequence, the formation time of the collinear process is $O(1/(g^2 T^2))$. Such a time is comparable to the small angle deviation mean time $\tau_g$ with the result that additional soft exchanges take place during the splitting process and destroy the coherence of the collinear particles.

This effect is known as the Landau-Pomeranchuck-Migdal effect~\cite{Landau:1953um,Migdal:1956tc} and takes place at high energies when collinear particles are produced during bremmstrahlung. Because the time duration of the collinear process and of the soft scattering is of  the same order, the additional soft scatterings cannot be treated as independent events. Therefore, to correctly evaluate the $1\rightarrow 2$ amplitude we should also take into account the interference between all the $N+1 \rightarrow N+2$ diagrams. Following the notation of~\cite{Arnold:2002zm} we denote these processes as ``$1\rightarrow2$'' processes where the quotes remind that in addition to the splitting of collinear particles we are also including the additional soft scatterings. Clearly also joining processes ``$2\rightarrow1$'' should be included. We collectively denote splitting and joining processes with ``$1\rightarrow2$''. The evaluation of the rate of these processes is discussed in ref.~\cite{Arnold:2001ba,Arnold:2001ms,Arnold:2002ja}.

\begin{table}
    \centering
    \begin{tabular}{l r}
    \hline
    hard particle wavelength & $ T^{-1} $\\ 
    typical particle wavelength & $ T^{-1} $ \\
    typical particle separation & $ T^{-1} $ \\
    \hline
    soft particle wavelength & $ (gT)^{-1} $ \\
    Debye length & $ (g T)^{-1} $ \\
    inverse thermal mass & $ (gT)^{-1} $ \\
    \hline
    mean free path: small-angle scattering ($\theta\sim g$, $q\sim gT$) & $ (g^2 T)^{-1} $ \\
    mean free path: very-small-angle scattering ($ \theta \sim g^2 $, $ q\sim g^2 T $) & $ (g^2 T)^{-1} $ \\
    duration (formation time) of ''$ 1\rightarrow 2 $'' collinear processes & $ (g^2 T)^{-1} $ \\
    \hline
    mean free path: large-angle scattering ($ \theta \sim 1$, $ q \sim T $ ) & $ (g^4 T)^{-1} $\\
    mean free path: hard ''$ 1\rightarrow 2 $'' collinear processes & $ (g^4 T)^{-1} $\\
    \hline
    \end{tabular}
    \caption{Relevant scales for a hot weakly coupled relativistic plasma discussed in the text.}
    \label{tab:relevant_hot_plasma_scales}
\end{table}

This exhausts the list of the processes we should include in the effective kinetic theory of hard particles and we refer to ref.~\cite{Arnold:2002zm} for motivations about this fact. Table~\ref{tab:relevant_hot_plasma_scales} provides a summary of the relevant scales of the plasma which we analyzed. 

We finally emphasize that our analysis provided just a simple estimate of the relevant scales involved in the processes. Such estimate works pretty well in the case of $g\ll 1$ but it is not very helpful when $g\sim 1$. In such a situation, the numerical factors we  ignored in our analysis are important and must be taken into account to asses the applicability of the kinetic theory. Nevertheless, it is extremely hard to compute such factors and provide a quantitative theoretical prediction of the mean free path. Such a task would require to compute the collision operator which is impracticable using only analytic methods. We can still give some estimates by numerically evaluating the collision integral, but ultimately we can only test the quality of our kinetic theory a posteriori by comparing the out-of-equilibrium perturbations and the equilibrium distributions.

\section{The effective kinetic theory}
\label{sec:effective_kinetic_theory}

The effective kinetic description we use is a set of Boltzmann equations which describe the evolution of the particle distribution functions. These equations have the following form
\begin{equation}
\label{eq:boltzmann_equation_general_covariant}
    {\cal L}[f]\equiv p^\mu\partial_\mu f + m \partial_{p_\mu}(F^\mu f)  = -{\cal C}[f]\,,
\end{equation}
where the Liouville operator $\cal{L}$ describes the evolution of the distribution function $f$ of a particle in $x$ with momentum ${\bf p}$ and mass $m$ subject to the external force $F^\mu$ while ${\cal C}$ is the collision integral that accounts for the relevant processes taking place in the plasma. Such a description cannot be applied at every scale. The Boltzmann equation is valid only for those plasma excitations which can be treated as point-like particles and whose interactions are local. This assumption is valid only in presence of a large separation between the characteristic macroscopic length scale $L$ of the system and the microscopic scales set by the interaction range $l_p$ and the particle wavelength $\lambda$. The Boltzmann equation can thus be applied only if the relevant collision processes in the plasma are short-range, namely $l_p \ll L$ and for those particles with $\lambda \ll L$. Quantum states with $\lambda \sim L$ have important wave properties that must be accounted for their accurate description and that the Boltzmann equation does not capture. A more refined analysis of such states can be carried out by using the Schwinger-Keldysh-Kadanoff-Baym equations~\cite{Schwinger:1960qe,Keldysh:1964ud,Kadanoff:1962aa} that correctly accounts for their quantum effects.

One last important scale that deserves a careful analysis, but that does not enter in the definition of the applicability domain of the kinetic theory, is the mean free path $\ell$ of particles between two successive collisions. Such a length sets the scale at which the hydrodynamic description emerges from the underlying kinetic theory. For systems where $L\gg \ell$, collisions are very frequent, thermalization occurs fast and a description based on hydrodynamics can be applied instead of the kinetic approach. On the other hand, when the mean free path is comparable or even larger than the characteristic length scale $L$ single particle collisions become important and a proper description is instead given by the effective kinetic theory.


As we will verify in the next chapters, the SM plasma light degrees of freedom during the EWPT are well described by the hydrodynamic equations we provided in the previous chapter. The large number of collisions that light particles experience determines a mean free path that is much smaller than the characteristic scale of the system set by the bubble wall width. Hence we can assume that the distribution function of light particles is the one of a perfect fluid.

The same is not true for heavy particle species. In such a case collision processes are much less efficient and the mean free path is, in some cases, comparable to the bubble wall width. Nevertheless the particle wavelength $\lambda$ of hard particles and the interaction range of the processes are much smaller than the width of the bubble and the Boltzmann equation can be applied to study the out-of-equilibrium properties of the plasma hard states.


\subsection{Effective kinetic theory at EW scale}

We now apply the analysis of the previous sections to the plasma surrounding the bubble during the EWPT. We begin by estimating the scales of our problem. Two types of interactions dominate the dynamics of the heavy particle species in the plasma: the strong and the weak interaction. The strength of such interactions is set by the temperature of the plasma, since we are interested in the dynamics of hard modes. Typically, bubbles are nucleated at a temperature of the order of the EW scale, namely $T_n\sim 100 $ GeV. Moreover, the energy injected by the wall in the plasma does not change much the temperature of the latter so $T_+\sim T_-\sim T_n$. As a result the average energy of a particle is of the order of the EW scale. At the EW scale, QCD is still perturbative but $g_s\sim 1$, while $g_w\sim 0.65$. For such values of the coupling there is  not a large separation of the scales in tab.~\ref{tab:relevant_hot_plasma_scales} for QCD and a mild separation for the weak interactions. 

Nevertheless, as we argued in the previous section, we can still provide an effective kinetic theory based on the Boltzmann equation as long as $ L\gg\lambda $, $ L\gg l_p $. Because our ultimate goal is the determination of the out-of-equilibrium friction, we focus on those scales where the background scalar fields profiles have a sizeable change. The width of the bubble wall thus sets a natural scale for our problem that we can adopt to asses the validity of our description.

Let us first focus on the top quark. The typical wavelength of such particles is $ \lambda \sim T^{-1} $. Because $ g_s\sim 1 $ also $ l_p \sim T^{-1} $. As we will see further, the typical width of the wall is $ (5-10)T^{-1} $ which satisfies the applicability condition of the Boltzmann equation.

The gauge bosons case, instead, deserves a more careful analysis. Since during the EWPT $ g_w\lesssim 1 $, we refer to momenta $ O(g T) $ as semi-hard and $ O(g^2T) $ as soft.
In contrast to fermions, bosons can have soft momenta. As we previously discussed, Debye screening and Landau damping become relevant for the determination of the particle dynamics at such scales~\cite{Arnold:1996dy,Huet:1996sh,Son:1997qj}. 

This problem has been deeply discussed in the literature~\cite{Bodeker:1998hm,Bodeker:1999ey,Bodeker:1999ud,Arnold:1999jf,Arnold:1999ux}. The main conclusion is that soft-bosons can be described using a Langevin equation. Such description results from the integration of hard and semi-hard modes and correctly takes into account the Debye screening and Landau damping. In addition such particles have a large occupation number. Indeed
\begin{equation}
    f_0(\beta p)=\frac{1}{e^{\beta p} - 1}\sim \frac{1}{\beta p}\gg 1\,,
\end{equation}
which allows us to treat soft bosons classically.

The dynamics of soft particles is extremely relevant to determine the out-of-equilibrium friction~\cite{Moore:2000wx}. Nevertheless we are not going to provide a detailed analysis of the problem yet. We refer to Chapter~\ref{ch:exact_solution_to_liearized_Boltzmann_equation} for further discussions where we also provide a deeper and critical study on the theoretical limitations in the computation of the friction.

Despite we argued that ``$1\rightarrow2$'' processes are equally important as $2\rightarrow2$ processes, we are going to neglect the former in the following. The main reason for this choice is to simplify the problem and in almost the whole literature regarding the out-of-equilibrium dynamics during EWPT the ``$1\to2$'' processes are systematically dropped. The Boltzmann equation is indeed very complex and to develop the necessary techniques for its analysis it is important to first consider a simplified version of the problem. Once a solid theoretical control has been gained, we should add one by one the neglected terms. We thus drop altogether the ``$ 1\rightarrow2 $'' collision term for this thesis but we point out that an interesting continuation of our work would be to include again the ``$1\rightarrow2$'' processes that take place in the plasma during the EWPT. 

From the above discussion we conclude that the effective kinetic theory of the hard particles is described by the set of Boltzmann equations given by eq.~(\ref{eq:boltzmann_equation_general_covariant})
where the collision operator ${\cal C}$ takes into account only $ 2 \rightarrow 2 $ processes. Assuming that collisions are uncorrelated and dubbing with $p$ and $k$ the momenta of the two incoming particles and with $p'$ and $k'$ the momenta of the outgoing particles, the collision operator is given by
\begin{equation}
\label{eq:collision_2_to_2_process}
    {\cal C}[f]=\frac{1}{4N_p}\sum_{j}\int\frac{d^3{\bf k}}{(2\pi)^3 2 E_k}\frac{d^3{\bf p}'}{(2\pi)^32E_{p'}}\frac{d^3 {\bf k}'}{(2\pi)^32E_{k'}}|{\cal M}_j|^2 (2\pi)^4\delta(p + k - p' -k'){\cal P}[f]\,,
\end{equation}
where the sum runs over all the relevant processes in the plasma, ${\cal M}_j$ is the corresponding amplitude summed over spin and internal degrees of freedom of all the four excitations entering in the process. The population factor ${\cal P}[f]$ instead is defined as
%
%
%
\begin{equation}
    {\cal P}[f] = f_pf_k(1 \pm f_{p'})(1 \pm f_{k'}) - f_{p'}f_{k'}(1 \pm f_p)(1 \pm f_k)
\end{equation}
where the $+(-)$ sign is for bosons (fermions) and $f_q$ denotes the distribution function of the species of the particle with momentum $q$.


The processes we need to consider are those that we discussed in the previous section, namely scattering and annihilation. As a preliminary step we point out that both the top quark and W bosons mostly interact with light particles. It is thus reasonable to neglect direct collisions between the top quark and the W bosons. If we include them, the resulting problem would be much more complex. Top quarks with different helicities should indeed be treated separately and the resulting system of Boltzmann equations would also couple top and W bosons' perturbations. 

The amplitudes of the relevant processes are usually evaluated in the so called leading-log approximation. This approximation arises from the observation that $t$ and $u$ channel diagrams are logarithmic IR divergent when we compute the cross-section in the vacuum. However, thermal interactions regularize such divergence and the resulting  amplitude enters in the perturbative expansion with a factor $ \alpha_s\log(\alpha_s^{-1}) $. Such logarithmic enhancement is not present for $ s $ channel processes.

The leading-log approximation consists in considering only logarithmic enhanced diagrams in the amplitude. Furthermore, interference terms between diagrams are all systematically dropped. As the majority of the literature, we use the leading-log approximation to compute the amplitudes.

To account for the thermal effects we should correct the propagators with the self-energies. Such expressions have a rather complicated dependence on the particle momenta. Nevertheless it is possible to make an approximation that makes the computation more tractable. For an exchanged top quark the dispersion relation is modified by the thermal interactions. Because fermions are hard, a good approximation of the self-energy is to replace the whole expression with the effective thermal mass of the particle.

The case of bosons is more complex. The longitudinal part of the propagator is indeed Debye screened and the transverse part is Landau damped. We still approximate this effect by correcting the propagators with the thermal masses, despite such approximation is quite rough~\cite{Moore:1995si}.

Finally we treat the outgoing and incoming particles as massless in the collision integrals. Collisions typically involve thermal particles and the log-enhancement arises from the exchange of particles with momentum between $T$ and $m$ which justifies our approximation. Under these assumptions, the collision integrals are independent from scalar fields VEV and our problem is greatly simplified.

We conclude our analysis by focusing on the gauge bosons during the EWPT. The gauge bosons have a different number of degrees of freedom in the broken and symmetric phase. By working in the unitary gauge we find that these extra degrees of freedom correspond to the Goldstone bosons. These bosons thus provide an additional contribution to the friction and we should take them into account. For that we make the following approximation. We compute the out-of-equilibrium perturbations of the transverse modes of the gauge bosons using the Boltzmann equation. Thus we consider $ N_p = 6 $ in eq.~(\ref{eq:collision_2_to_2_process}) for processes involving the weak gauge bosons. To compute the friction, instead we consider $ N_w = 9 $ in eq.~(\ref{eq:out_of_eq_friction}). In other words we include the Goldstones contributions by assuming that longitudinal and transverse components have the same distribution functions.

\begin{table}
	\centering
	\centering
    \begin{tabular}{c|c}
        process & $|{\cal M}|^2$\\
        \hline
        \rule{0pt}{1.75em}$t \bar t \to gg$ & $\displaystyle \frac{128}{3} g_s^4 \left[ \frac{ut}{(t - m_q^2)^2} +  \frac{ut}{(u- m_q^2)^2} \right ]$\\
        \rule{0pt}{1.75em}$tg \to tg$ & $\displaystyle- \frac{128}{3} g_s^4 \frac{su}{(u-m_q^2)^2} + 96 g_s^4 \frac{s^2 + u^2}{(t - m_g^2)^2}$\\
        \rule{0pt}{1.75em}$tq \to tq$ & $\displaystyle160 g_s^4 \frac{s^2 + u^2}{(t - m_g^2)^2}$\\
        \hline
        \rule{0pt}{1.75em}
        $Wq \to qg$ & $-144g_s^2 g_W^2 \displaystyle \frac{s t}{(t-m_q^2)^2}$\\
        \rule{0pt}{1.75em}
        $Wg \to \bar q q$ &
        $144g_s^2 g_W^2 \displaystyle \frac{u t}{(t-m_q^2)^2}$\\
        \rule{0pt}{1.75em}
        $WW \to \bar f f$ & $27g_W^4 \left[ \displaystyle\frac{3 ut }{(t-m_q^2)^2}+ \displaystyle\frac{ut }{(t-m_l^2)^2}\right]$\\
        \rule{0pt}{1.75em}
        $W f \to W f$ & $288g_W^4 \displaystyle\frac{s^2+u^2}{(t-m_W^2)^2} - 27g_W^4 \left[\displaystyle\frac{3 st }{(t-m_q^2)^2} + \displaystyle\frac{st}{(t-m_l^2)^2}\right]$\\
    \end{tabular}
	\caption{Amplitudes for the scattering processes relevant for the top quark and the W-bosons in the leading log approximation. In the $t q \to t q$ process we summed over all massless quarks and antiquarks.}\label{tab:amplitudes}
\end{table}

To summarize, our effective kinetic theory corresponds to an effective Boltzmann equation where only $ 2\rightarrow2 $ processes are included. The amplitudes of such processes are computed in the leading-log approximation and we assume that incoming and outgoing particles are massless inside the collision integral. The relevant amplitudes which we consider are reported in Table~\ref{tab:amplitudes}.

\section{The Boltzmann equation}

The Boltzmann equation describes the non-equilibrium properties of systems. It provides an evolution law for the particle distribution functions which in turn are used to determine the statistical properties of the system. Macroscopic quantities such as temperature, particle flow, entropy flow and many others can be computed by taking moments of the distribution function. In addition, using the same procedure it is also possible to derive the transport equations for near-equilibrium systems providing a connection between the microscopic statistical properties of the system and its macroscopic behaviour, governed by the hydrodynamics and thermodynamics equations. 

The Boltzmann equation also describes the trend to equilibrium of the system. This is the famous result of the $ {\cal H} - $theorem which describes how a system thermalizes. The key point is that out-of-equilibrium systems reach thermal equilibrium through the multiple collisions between particles. In this sense the collision operator of the Boltzmann equation is responsible for the restoration of thermodynamic equilibrium.

Our goal in this section is to provide a brief review of the Boltzmann equation. We do not provide a complete description of the subject 
and refer to the following books~\cite{Cercignani:2002aa, DeGroot:1980dk} for an in-depth discussion. We instead present some selected topics which supply the theoretical background needed to understand the strategies adopted in the literature to solve the Boltzmann equation. Moreover, at the end of this section we present some of our recent developments which may be useful for a deeper understanding of the out-of-equilibrium properties of the plasma during the EWPT.

\subsection{Basic concepts}
\label{sec:boltzmann_basic}

For the sake of simplicity we consider the special case of a monoatomic gas.
The Boltzmann equation given by eq.~(\ref{eq:boltzmann_equation_general_covariant}) is a master equation which describes the evolution of a distribution function $f$. In the case we are interested in, the external force $F^\mu$ is independent from the momentum $p^\mu$ and we can also write the Liouville operator in the following way:
\begin{equation}
    {\cal L}[f] = p^\mu d_\mu f\,,
\end{equation}
where we defined the flow derivative $d_\mu$ such that
\begin{equation}
\label{eq:flow_derivative}
    p^\mu d_\mu p^\nu = F^\nu\,.
\end{equation}
The physical interpretation of the derivative $d_\mu$ is straightforward and determines the tangent vectors of particle trajectories in the phase space in the collision-less limit. This observation will be extremely useful for the analysis we will carry out in Chapter~\ref{ch:exact_solution_to_liearized_Boltzmann_equation}.

We recall that the right hand side of the Boltzmann equation is the collision operator defined by eq.~(\ref{eq:collision_2_to_2_process}) which presents some useful symmetries. 
The amplitudes $ {\cal M} $ are symmetric in the exchange of final and initial states and in the exchange of momenta between the initial  and  the final states. As a consequence of such symmetries the collision operator presents the following property which proves very useful in studying the transport equation of a quantity $\varphi$ 
\begin{equation}
    \label{eq:collision_transport_equation}
    \begin{split}
    \int \frac{d^3 {\bf p}}{(2\pi)^3 2 E_p}\varphi(p,x){\cal C}[f] = \frac{1}{16N_p}\sum_{j}\int&\frac{d^3{\bf p}d^3{\bf k}d^3{\bf p}'d^3{\bf k}'}{(2\pi)^8 16 E_pE_kE_{p'}E_{k'}}|{\cal M}_j|^2\delta(p + k - p' -k'){\cal P}[f]\times\\
    &\times(\varphi(p,x) + \varphi(k,x) - \varphi(p',x) - \varphi(k',x)) = \cal{R}\,,
\end{split}
\end{equation}
where $\varphi$ is a generic function of the four momentum and the position. The quantity ${\cal R}$ is called production rate.

The hydrodynamic equations can be derived by studying the transport equation of the momentum $p^\nu$. To derive them we multiply the Boltzmann equation by $\varphi(p,x) = p^\nu$ and then integrate over the momentum space. From eq.~(\ref{eq:collision_2_to_2_process}) and from the conservation of momentum in particle collisions it follows that the production rate $ {\cal R}$ in such a case identically vanishes.
For the Liouville part we instead find
\begin{equation}
        \int\frac{d^3{\bf p}}{(2\pi)^3 2 E_p}p^\nu(p^\mu\partial_\mu f + m F^\mu\partial_{p^\mu}f) = 0\,.
\end{equation}
Using eq.~(\ref{eq:definition_stress_energy_tensor_microscopic}) the part involving the derivative with respect to the position yields exactly the derivative of the stress-energy tensor. Finally integrating by parts yields
\begin{equation}
    \partial_\mu T^{\mu\nu} = \int \frac{d^3{\bf p}}{(2\pi)^3 2 E_p}mF^\nu f\,.
\end{equation}
When no external force acts on the system the above equation reduces to the hydrodynamic equations. For the case of EWPT we may sum the above equation over all the particle species in the plasma to get
\begin{equation}
    \partial_\mu T^{\mu\nu}_{pl} = \sum_i N_i m_i\int\frac{d^3{\bf p}}{(2\pi)^32E_p}\partial^\nu m_i f_i\,,
\end{equation}
which, once summed with eq.~(\ref{eq:system_total_energy_conservation}), yields the conservation of the total stress energy tensor. 

Another application of the transport equation is the derivation of the conservation of the total particle flow of the system. To derive such equation we consider $\varphi = 1$ and repeat the same steps we applied to derive conservation of the stress-energy tensor. Once again from eq.~(\ref{eq:collision_transport_equation}), the production rate ${\cal R}$ identically vanishes. On the other hand the Liouville part yields
\begin{equation}
    \partial_\mu N^\mu = 0\,,
\end{equation}
where we defined
\begin{equation}
\label{eq:particle_flow}
    N^\mu = \int\frac{d^3{\bf p}}{(2\pi)^3 2 E_p}p^\mu f
\end{equation}

For the two cases we considered the production rate vanishes because
\begin{equation}
    \label{eq:summational_invariant}
    \varphi(p,x) + \varphi(k,x) - \varphi(p',x) - \varphi(k',x) = 0
\end{equation}
when $\varphi = 1$ and $\varphi = p^\mu$ in virtue of the locality of the collision operator.
A quantity $\varphi$ which satisfies eq.~(\ref{eq:summational_invariant}) is called summational invariant. We may wonder if there are other summational invariants other than $1$ and $p^\mu$. The answer is essentially no. 
Ref.~\cite{Cercignani:2002aa} provides a full characterization of summational invariants by showing that $\varphi(p,x)$ is summational invariants if and only
\begin{equation}
    \varphi(p,x) = A(x) + B^\mu(x)p_\mu\,,
\end{equation}
where $A(x)$ and $B^\mu(x)$ are respectively a scalar and a four vector depending only on the position.

There is an interesting corollary to such a theorem.
Consider the following choice for the function $\varphi$
\begin{equation}
    \varphi = -\left[\log(f) - \left(1\pm \frac{1}{f}\right)\log(1\pm f)\right]\,,
\end{equation}
where the plus is for bosons while the minus for fermions. The corresponding transport equation provides the evolution of the entropy flow. The rate of entropy production vanishes if and only if
\begin{equation}
    f_p f_k(1\pm f_{p'})(1 \pm f_{k'}) = f_{p'} f_{k'}(1\pm f_p)(1 \pm f_k)\,,
\end{equation}
Notice that the above equation corresponds to the requirement that the population factor ${\cal P}[f]$ identically vanishes. Let us take the logarithm of the above equation. The corresponding expression can then be rearranged in
\begin{equation}
\label{eq:quantum_summational_invariant_distribution}
    \log\left(\frac{f_p}{1\pm f_p}\right) + \log\left(\frac{f_k}{1\pm f_k}\right) - \log\left(\frac{f_{p'}}{1\pm f_{p'}}\right) - \log\left(\frac{f_{k'}}{1\pm f_{k'}}\right) = 0\,.
\end{equation}
As a consequence $\log[f/(1\pm f)]$ is a summational invariant. Thus from the characterization theorem of summational invariants it follows that
\begin{equation}
    \log\left(\frac{f_p}{1\pm f_p}\right) = A(x) + B^\mu(x)p_\mu\,,
\end{equation}
which eventually gives
\begin{equation}
    f_p = \frac{1}{e^{A(x)+B^\mu(x)p_\mu}\mp 1}
\end{equation}
Using the definition of the particle flow $N^\mu$ in eq.~(\ref{eq:particle_flow}) and stress-energy tensor $T^{\mu\nu}$ in eq.~(\ref{eq:definition_stress_energy_tensor_microscopic}) we can relate $A(x) = \beta(x)\mu(x)$ to the chemical potential and $ B^\mu(x) = -\beta(x)U^\mu(x) $.
Thus the characterization of summational invariants and the vanishing of entropy production rate determines the expressions of local equilibrium distribution functions.


\subsection{The Grad method}
\label{sec:chapman-enskog_grad_methodds}

For a better understanding of the methods adopted in the literature to compute the out-of-equilibrium perturbations during the EWPT we discuss one of the most famous methods to solve the Boltzmann equation, namely 
the Grad method~\cite{Grad:1949aa} or the moment method. 
%
To find a solution to the Boltzmann equation the distribution function is expanded in the series of its moments leading to an infinite set of coupled equations. To manage the problem, one truncates the series at a finite order in the expansion. There is, however, no proof of the convergence nor of the efficiency of the method although it is widely accepted that the method provides a good solution as long as non-linear phenomena, such as shock-waves, are absent.

To determine the out-of-equilibrium distribution of the particle species we maximize the entropy density $s$, which for a relativistic quantum gas is defined as
\begin{equation}
    s = U^\mu\int\frac{d^3{\bf p}}{(2\pi)^3 2E_p}p_\mu(\pm(1\pm f)\log(1\pm f) - f\log(f))\,,
\end{equation}
where the +(-) is for bosons (fermions),
under the constraint that we recover the hydrodynamic description of the system in terms of the stress energy tensor $T^{\mu\nu}$ defined in eq.~(\ref{eq:definition_stress_energy_tensor_microscopic}) and particle flow $N^\mu$ defined in eq.~(\ref{eq:particle_flow}) when we integrate the distribution function. This is expressed by imposing the following set of constraints
\begin{equation}
\label{eq:fourteen_moments}
    \begin{split}
        & N^\mu U_\mu = U^\mu \int\frac{d^3{\bf p}}{(2\pi)^32E_p}p^\mu f\,,\\
        & T^{\mu\nu}U_\nu = U_\nu \int\frac{d^3{\bf p}}{(2\pi)^32E_p}p^\mu p^\nu f\,.\\
    \end{split}
\end{equation}
%
This problem is equivalent to the maximization of the following Lagrangian
\begin{equation}
\label{eq:grad_Lagrangian}
\begin{split}
    L[f] = &U^\mu \int\frac{d^3{\bf p}}{(2\pi)^3 2E_p}p_\mu (\pm(1\pm f)\log(1\pm f) - f\log(f)\\ &- \lambda U^\mu \int\frac{d^3{\bf p}}{(2\pi)^32E_p}p^\mu f -\lambda_\mu U_\nu \int\frac{d^3{\bf p}}{(2\pi)^32E_p}p^\mu p^\nu f\\
\end{split}
\end{equation}
where $\lambda$ and $\lambda_\mu$
are Lagrange multipliers. By solving the corresponding Euler-Lagrange equation $\partial L/\partial f = 0$ we find
%
%
\begin{equation}
\label{eq:grad_solution_variational}
    f = \frac{1}{1\pm \exp(\lambda +\lambda_\mu p^\mu)}
\end{equation}

At last we need to identify the Lagrange multipliers. For that we may decompose the Lagrange multipliers $\lambda$ and $\lambda_\beta$ in an equilibrium and a non-equilibrium part, namely $\lambda = \lambda^E + \lambda^{NE}$. When the system is in equilibrium the distribution function $f$ must reduce to the Fermi-Dirac or Bose-Einstein distribution function $f_0$ defined in eq.~(\ref{eq:equilibrium_distribution_quantum_gas}). In the case where the non-equilibrium part is small we expand the distribution to find
\begin{equation}
\label{eq:grad_distribution}
    f = f_0 + f'_0(\lambda^{NE} + \lambda_\mu^{NE}p^\mu)\,.
\end{equation}
The remaining Lagrange multipliers are finally determined by taking the corresponding moments of the Boltzmann equation with the set of weights $\{1,\,p^\mu\}$ which reduces the Boltzmann equation to a set of coupled differential equations. The procedure we just outlined will be deeply analyzed in the following chapter for the case of EWPT.

We finally observe that the distribution in eq.~(\ref{eq:grad_distribution}) is equivalent to an expansion in the momenta around the equilibrium distribution $f_0$ truncated at linear order. A generalization to such solution is provided by
\begin{equation}
\label{eq:grad_small_perturbations}
    f = f_0 + f'_0[w^{(0)}+ w_\mu^{(1)} p^\mu + w_{\mu\nu}^{(2)}p^\mu p^\mu +\dots]\,.
\end{equation}
We may then apply the moment method to the general case where the distribution around which we expand does not correspond to the equilibrium distribution function but it is a generic distribution function $f^{(0)}$. The choice of  $f^{(0)}$ is crucial for the efficiency of the method. Indeed the series in the momenta can be truncated after a few terms only when the unknown distribution $f$ is close to $f^{(0)}$.

This concludes our review of the basic properties of the Boltzmann equation. We refer to the ref.~\cite{Cercignani:2002aa,DeGroot:1980dk} and references therein for further details and applications.

\subsection{The Boltzmann equation for the EWPT}
\label{sec:boltzmann_ew}

We conclude this section by discussing in more details the dynamics of the out-of-equilibrium perturbations during the EWPT. We already provided in sec.~\ref{sec:effective_kinetic_theory} the effective kinetic theory for the plasma perturbations in term of the Boltzmann equation and we discussed the relevant processes that should be included in the collision integral. As a next step we linearize the Boltzmann equation to provide an equation for the out-of-equilibrium perturbation in the limit of small departure from thermal equilibrium. We carry out 
our analysis by making the Lorentz invariance manifest.

Deviations with respect to the local equilibrium distribution are present mostly close to the
DW and are expected to vanish for $ z \rightarrow \pm \infty $. For small perturbations, the distribution function
\begin{equation}
    f(p,x) = f_0(\beta(x)U^\mu(x)p_\mu) + \delta f(p,x)\,,
\end{equation}
and the Boltzmann equation can be linearized in $\delta f$ yielding
\begin{equation}
\label{eq:lorentz_invariant_linearized_boltzmann}
     p^\mu d_\mu \delta f = -p^\mu d_\mu f_0 - {\overline{\cal C}}[\delta f]\,.
\end{equation}
In the following we omit for simplicity the variable dependence of the different quantities. In the above expression we already rearranged the terms in the Liouville operator to emphasize that out-of-equilibrium perturbations are sourced by the non-homogeneity of the equilibrium distribution and by the interactions between the plasma and the DW. Indeed by using the definition of the flow derivative in eq.~(\ref{eq:flow_derivative}) we find
\begin{equation}
\label{eq:lorentz_invariant_source}
    p^\mu d_\mu f_0 = p^\mu \partial_\mu (\beta U^\nu )p_\nu f'_0 + \beta m F^\mu U_\mu f'_0\,,
\end{equation}
where
\begin{equation}
    f'_0(y) = \frac{d}{dy}f_0(y)\,.
\end{equation}
Equation~(\ref{eq:lorentz_invariant_source}) precisely encodes the two aforementioned sources of perturbations.

The linearization of the collision operator yields instead two terms ${\cal C}[f_0]$ and ${\overline{\cal C}}[\delta f]$. We showed that the collision operator identically vanishes when it is evaluated on the equilibrium distribution function, so ${\cal C}[f_0] = 0$. The linearized collision operator ${\overline{\cal C}}[\delta f]$ is obtained by linearizing the population factor ${\cal P}[f]$ in terms of the perturbation $\delta f$. By using the conservation of the momenta and the following useful properties of the equilibrium distribution function
\begin{equation}
\begin{split}
    & f_0'(x) = -e^xf^2_0(x)\,,\\
    & (1\pm f_0(x)) = e^x f_0(x)\,,
\end{split}
\end{equation}
we find that the final expression of ${\overline{\cal C}}[f]$ is
\begin{equation}
\label{eq:linearized_collision_integral}
    {\overline{\cal C}}[\delta f] = \frac{1}{4N_p}\sum_j\int\frac{d^3{\bf k}d^3{\bf p}'d^3{\bf k}'}{(2\pi)^58 E_k E_{p'}E_{k'}}|{\cal M}_j|^2{\overline{\cal P}}[f]\delta^4(p +k -p' - k')\,,
\end{equation}
with
\begin{equation}
\label{eq:P_bar}    {\overline{\cal P}}[f]=f_0(\beta U^\mu p_\mu)f_0(\beta U^\mu k_\mu)(1\pm f_0(\beta U^\mu p_\mu'))(1\pm f_0(\beta U^\mu k_\mu'))\sum_q\mp\frac{\delta f(q)}{f'(\beta U^\mu q_\mu)}\,,
\end{equation}
where $q$ runs over the momenta involved in the process, namely $p$, $k$, $p'$ and $k'$, the $-$ sign is for incoming particles while the $+$ sign for outgoing particles. We introduced the shorthand notation $\delta f(q)$ to express the functional dependence of the perturbation.

The linearized collision operator yields two different classes of terms. The first one corresponds to the case in which the argument of the perturbation is not integrated over, so that $ \delta f $ can be brought outside the integral, while the second one corresponds to the case in which one needs to integrate over the perturbation. We then write the collision operator in the following way
\begin{equation}
\label{eq:collision_operator_structure}
    {\overline{\cal C}}[\delta f] = {\cal Q}\frac{f_0}{f_0'}\delta f +f_0(\langle\delta f\rangle_A - \langle \delta f\rangle_S)
\end{equation}
The first term, namely the one proportional to ${\cal Q}$, is easier to deal with. Because it is independent of the perturbation, we can carry out the integration over the phase space yielding a term proportional to $\delta f$. The resulting expression for ${\cal Q}$ is then provided by
\begin{equation}
\label{eq:Q_expression}
    {\cal Q} = \frac{1}{4N_p}\sum_j\int\frac{d^3{\bf k}d^3{\bf p}'d^3{\bf k}'}{(2\pi)^58 E_k E_{p'}E_{k'}}|{\cal M}_j|^2f_0(\beta U^\mu k_\mu)(1\pm f_0(\beta U^\mu p_\mu'))(1\pm f_0(\beta U^\mu k_\mu'))\delta^4(p +k -p' - k')
\end{equation}
where we point out that due to the Lorentz invariance of the above integral the quantity ${\cal Q}$ depends only on the invariant $U^\mu p_\mu$.
The second class of terms is instead much harder to compute and needs a careful analysis. It corresponds to the terms where the perturbation $\delta f$ depends on the momenta $k$, $p'$ and $k'$ and thus appears under the integral sign. We collectively denote these terms with $\langle \delta f \rangle$ and we dub them bracket terms. 

In eq.~(\ref{eq:collision_operator_structure}) we 
emphasize that there are two types of brackets which correspond to the two different types of processes that take place in the plasma. Indeed as reported in Tab.~\ref{tab:amplitudes} we recognize that particles may undergo annihilation processes, which change the number of the particles  of the species involved, 
and scattering processes, where the number of particles is left invariant. With the subscript $A$ and $S$ for the brackets we identify the bracket term computed considering respectively only the annihilation or the scattering amplitudes reported in Tab.~\ref{tab:amplitudes}.
In the current analysis, however, we are interested in studying the general structure of the Boltzmann equation. Hence, we postpone to Chapter~\ref{ch:exact_solution_to_liearized_Boltzmann_equation} further discussions regarding the bracket terms and how we can deal with them.

We finally write the linearized Boltzmann equation as
\begin{equation}
\label{eq:boltzmann_equation_at_ewpt_covariant}
    p^\mu d_\mu \delta f - {\cal Q}\frac{f_0}{f'_0}\delta f = p^\mu d_\mu f_0 + f_0(\langle \delta f\rangle_A - \langle \delta f\rangle_S)
\end{equation}
This expression emphasizes the physical meaning of the collision operator. One of the effects of collisions is to provide an exponential decay of the perturbations far from their source. Such decay length is mainly set by the ${\cal Q}$ term and provides an estimate of the mean free path of particles in the plasma. Clearly, a complete evaluation of the mean free path would require also to take into account the effects encoded in the brackets. These are impossible to deal with analytical methods and require a numerical analysis. In other words we can only determine a posteriori the mean free path of particles in the plasma.

Since we are interested in a planar wall in a steady state regime, to solve the linearized Boltzmann equation it is convenient to write the equation in the wall reference frame.
As we did in Chapter~\ref{ch:bubble_dynamics_and_plasma_hydrodynamics}, we orient the reference frame in such a way that the wall moves in the negative direction of the $z$-axis. In such a frame the Boltzmann equation in eq.~(\ref{eq:boltzmann_equation_at_ewpt_covariant}) can be written as
\begin{equation}
\label{eq:boltzmann_equation_at_ewpt}
    \left(\frac{d}{dz} - \frac{\cal Q}{p_z}\frac{f_v}{f'_v}\right)\delta f(p_\bot, p_z, z) = {\cal S} + \frac{f_v}{p_z}(\langle \delta f\rangle_A -\langle \delta f\rangle_S)\,.
\end{equation}
For later use we recall that the explicit form of the flow derivative is
\begin{equation}
\label{eq:flow_derivative_wall_frame}
    \frac{d}{dz}\delta f = \left(\partial_z - \frac{(m^2)'}{2p_z}\frac{\partial}{\partial p_z}\right)\delta f\,,
\end{equation}
and we defined
\begin{equation}
    f_v = \frac{1}{e^{\beta(z)\gamma_{p}(z)(E - v_{p}(z)p_z)}\pm 1}\,.
\end{equation}
We recall that the `` $'$ '' denotes the derivative with respect to $z$, $v_p(z)$ is the plasma velocity profile measured in the wall reference frame, while $T(z)$ the temperature profile.  

The source term ${\cal S}$ is proportional to the action of the flow derivative on $f_v$ and it is given by
\begin{equation}
\label{eq:source_term}
    \frac{d}{dz}f_v = \frac{\gamma_{p}}{T}\left(\gamma_p^2(v_p E - p_z) v_p' - (E - v_p p_z)\frac{ T'}{T}\right)f_v' + \frac{\gamma_p v_p}{T}\frac{(m^2)'}{2p_z}f'_v\,.
\end{equation}
The above expression further clarifies the two sources of the out-of-equilibrium perturbations which we already discussed. The first one corresponds to the gradients of plasma velocity and temperature. The second one instead is given by the external force acting on the particles, namely the interactions between the DW and plasma. In fact in the wall reference frame
\begin{equation}
    U^\mu F_\mu = \gamma_p v_p m'\,.
\end{equation}

Sizeable values for the source term are present only close to the DW,
where the non-trivial Higgs profile generates a non-negligible $ z $ dependence in $ m(z) $, $ T(z) $ and $ v_p(z) $. Away from
the DW, the Higgs, the temperature and velocity profiles are instead almost constant, thus giving $ (m^2)' = 0 $, $ T' = 0 $ and $ v_p' = 0 $. This behavior is in agreement with the naive expectation that deviations from
local thermal equilibrium are only present close to the DW and should decrease to zero away
from it. Hence we expect $ \delta f $ to vanish for $ z\rightarrow \pm \infty $.

As already discussed, the system relaxes to equilibrium due to the collisions that take place in the plasma. In other words, the collision term ensures that far from the DW, where the forces acting on the system
basically vanish, each particle species approaches local thermal equilibrium exponentially fast. The typical decay length of such process can be thus identified with the mean free path of particles.

\section{Multipole expansion for slow moving walls}
\label{sec:multipole_expansion}

To conclude our discussion regarding the general properties of the Boltzmann equation we discuss the multipole expansion. This analysis, that we presented in~\cite{DeCurtis:2023aaa}, yields some further insights on the structure of the Boltzmann equation and provides some useful results that can be used also to study different systems with similar settings.

As a first step we notice that the collision integral can be written as a Lorentz invariant integral operator acting on the perturbation $\delta f$. Indeed, the expression in  eq.~(\ref{eq:linearized_collision_integral}), involves, with the exception of the perturbation $\delta f$, Lorentz symmetric quantities, namely the amplitude, the integral measure, the Dirac delta and the equilibrium distribution functions. Such quantities are all rotational invariant in the plasma reference frame where the bracket term can be written as
\begin{equation}
\label{eq:collision_operator_kernel_structure}
    \langle \delta f \rangle_A - \langle \delta f \rangle_S  = \int \frac{d^3\bar{\bf k}}{(2\pi)^3 2 |\bar{\bf k}|} {\cal K}(|{\bar{\bf p}}|, |{\bar{\bf k}}|,\cos\theta_{\bar p\bar k})\frac{\delta f(k)}{f_0'(|\bar{\bf k}|)}\,,
\end{equation}
%
%
We are here  omitting the temperature dependence to streamline the notation. Furthermore, to avoid confusion between quantities computed in the wall and plasma reference frame, we introduced the notation where we identify the quantities computed in the plasma reference frame with an upper bar. Finally $\theta_{\bar p\bar k}$ denotes the angle between the driection of the momenta $\bar {\bf p}$ and $\bar{\bf k}$ measured in the local plasma reference frame.
The kernel ${\cal K}$ is the result of the integration on all the variables but the one upon which the perturbation $\delta f$ depends. As a consequence, such a kernel is universal for the process under consideration.
This result is at the core of the algorithm we will discuss in Chapter~\ref{ch:exact_solution_to_liearized_Boltzmann_equation} to solve exactly the linearized Boltzmann equation. 
For the sake of our discussion it suffices to know that the collision operator has the structure provided in~(\ref{eq:collision_operator_kernel_structure}). We refer to Chapter~\ref{ch:exact_solution_to_liearized_Boltzmann_equation} and Appendix~\ref{app:evaluation_collision_integrals} for further details.

%
%

The choice of the local plasma reference frame is very convenient to express the functional dependence of the kernel ${\cal K}$ since in such a frame, as eq.~(\ref{eq:collision_operator_kernel_structure}) shows, the kernel depends only on rotational invariant quantities. 
The rotational invariance of the kernel ${\cal K}$ has important consequences both from the numerical and theoretical point of view. First of all, the kernel results block-diagonal on the basis of the Legendre polynomials. By projecting the perturbation $\delta f$ and the kernel ${\cal K}$ on such a basis it is thus possible to reduce the number of integrations involved in the computation of the bracket to just one with a huge reduction of the time required to evaluate the brackets.  
In addition, the study of the multipole decomposition of the kernel allows to better understand the theoretical structure of the collision operator and of the Boltzmann equation in general. What we found is  that a hierarchy between the different blocks is present with blocks corresponding to higher angular momentum being mildly suppressed. 
This suggests that in the multipole expansion of the bracket term only the first terms, namely those with smaller angular momentum, are the relevant ones. Hence the series can be truncated at a finite order without affecting the solution to the equation.

Let us mention that a hierarchy is present also between the different modes of the perturbation in some particular regimes of the Boltzmann equation. The hierarchy is controlled by the collision integrals in particular by the ${\cal Q}$ term. As we are going to show in the following when such a term is large, namely when the collision processes are efficient, modes with high angular momentum are suppressed. This allows us to find a semi-analytical solution of the Boltzmann equation in this particular regime.

\subsection{Collision operator decomposition}

We begin by analyzing the multipole expansion of the Boltzmann equation starting from the collision operator. As we mentioned before, as a consequence of the rotational invariance of the kernel ${\cal K}$ we can decompose the latter by using the Legendre polynomials $P_l(\cos\theta_{{\bar p}{\bar k}})$ in the following way
	\begin{equation}\label{eq:block_diagonal_kernel}
		{\cal K}(|{\bar {\bf p}}|,|{\bar {\bf k}}|,\cos\theta_{\bar p \bar k}) = \sum_{l=0}^{\infty}\frac{2l+1}{2} {\cal G}_l(|\bar{ \bf p}|,|\bar{ \bf k}|)P_l(\cos\theta_{\bar p \bar k})\,,
	\end{equation}
	with
\begin{equation}
		{\cal G}_l(|{\bar {\bf p}}|,|{\bar {\bf k}}|) = \int_{-1}^1d\cos\theta_{\bar p\bar k} {\cal K}(|\bar{ \bf p}|,|\bar{ \bf k}|,\cos\theta_{\bar p \bar k})P_l(\cos\theta_{\bar p \bar k})\,.
	\end{equation}
We remind that the kernel ${\cal K}$ enters in the computation of the bracket as we showed in eq.~(\ref{eq:collision_operator_kernel_structure}). In order to perform such integration, it is convenient to use the spherical coordinates provided by the set $\{|{\bar{\bf k}}|, \theta_{{\bar k}},\phi\}$, where $\theta_{\bar k}$ identifies the angle between the momentum ${\bf k}$ and the $\hat z$ axis in the plasma reference frame. The main motivation for such a choice are the symmetries of the problem at hand. In fact, since the system is symmetric under rotations around the direction of the DW propagation, the unknown perturbation $\delta f$ does not depend on the azimuthal angle $\phi$. As a consequence, this integration involves only the kernel ${\cal K}$ and can be performed once. The result is 
	\begin{equation}
\label{eq:kernel_multipole_expansion}
		\int_0^{2\pi} d\phi\,{\cal K}(|{\bar{\bf p}}|,|{\bar{\bf k}}|,\cos\theta_{\bar p \bar k}) = 2\pi\sum_{l=0}^\infty\frac{2l + 1}{2}{\cal G}_l(|\bar{ \bf p}|,|\bar{ \bf k}|) P_l(\cos\theta_{\bar p})P_l(\cos\theta_{\bar k})\,,
	\end{equation}
	where we used the following property of Legendre polynomials
	\begin{equation}
		\int_0^{2\pi} d\phi\, P_l(\cos \theta_{\bar p \bar k}) = 2\pi P_l(\cos\theta_{\bar p})P_l(\cos\theta_{\bar k})\,.
	\end{equation}
Equation~(\ref{eq:kernel_multipole_expansion}) states that the kernel ${\cal K}$ is block diagonal on the basis of Legendre polynomials. As a result, it is convenient to adopt such a basis to decompose also the perturbation $\delta f$, namely
\begin{equation}\label{eq:pert_legendre}
	\delta f(p, z) = \sum_{l=0}^\infty\frac{2l + 1}{2}\psi_l(|\bar {\bf p}|,z)P_l(\cos\theta_{\bar p})\,.
\end{equation}
In such a way, the integration along $\cos\theta_{\bar k}$ in eq.~(\ref{eq:collision_operator_kernel_structure}) is performed trivially by exploiting the orthogonality of the Legendre polynomials and leads to the final result
	\begin{equation}\label{eq:bracket_term}
		\langle\delta f\rangle_A - \langle\delta f\rangle_S = 2\pi \sum_{l=0}^\infty\frac{2l+1}{2}{\cal O}_l\left[\f{\psi_l}{f'_0}\right]P_l(\cos\theta_{\bar p})\,,
	\end{equation}
	where we defined
	\begin{equation}
 \label{eq:Ol_defintion}
		{\cal O}_l[g] = \int_0^\infty \frac{|\bar {\bf k}|d|\bar {\bf k}|}{2}f_0(|\bar {\bf k}|){\cal G}_l(|\bar{ \bf p}|,|\bar{ \bf k}|) g(|\bar {\bf k}|)\,.
	\end{equation}

\begin{figure}
    \centering
    \includegraphics[width=0.47\textwidth]{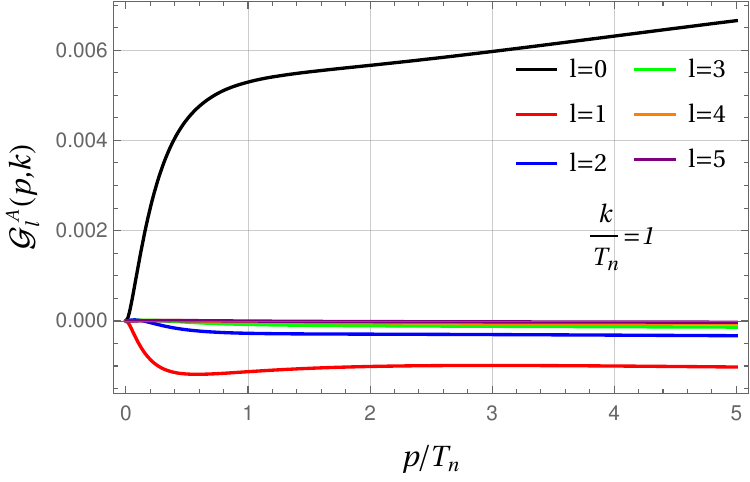}
    \hfill
    \includegraphics[width=0.47\textwidth]{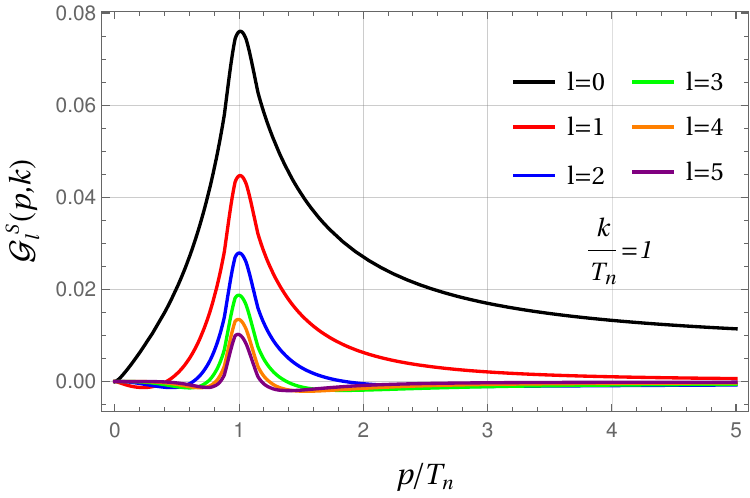}
    \caption{Comparison of the first six Legendre blocks in the multipole decomposition of the annihilation kernel (left plot) and scattering kernel (right plot). Notice the clear hierarchy in the modes for the annihilation kernel and the peak corresponding to forward scattering for the scattering processes}
    \label{fig:hierarchy_kernel}
\end{figure}

    To study how the kernel ${\cal K}$ decomposes on the Legendre polynomials, it is worth to analyze the multipole expansion of the annihilation and scattering kernel ${\cal K}_A$ and ${\cal K}_S$ separately since their structure is very different. We plot the results in Fig.~\ref{fig:hierarchy_kernel} where we specialized, for simplicity, to the case of $|{\bf k}|/T_n = 1$ and to the processes involving only the top quark. Similar results hold for the whole range of the momenta and for the W-bosons. 
    
    The left plot, which shows the blocks ${\cal G}_l^A$ of the annihilation kernel as a function of $|{\bf p}|/T_n$, highlights a strong hierarchy between the different modes of the annihilation. As a main consequence to reconstruct the whole annihilation kernel it is required to include only the first few terms, namely those with smaller angular momentum. Each block presents a smooth behaviour across the whole momentum range with a fast growth for small momenta and a slow logarithmic growth at large ones.

    On the other hand the scattering kernel, whose decomposition is shown on the right-hand panel of the same figure has a more complicated behaviour. The main feature is the presence of a peak in the forward scattering kinematic region. We notice that in this region the blocks ${\cal G}_l^S$ present a much milder hierarchy. On the contrary, away from the peak a more marked hierarchy between the different modes is present as in the case of the annihilation kernel.

    At last we analyze how the term proportional to ${\cal Q}$ decomposes along the basis of Legendre polynomials. In the plasma reference frame we easily find
    \begin{equation}\label{eq:q_term}
		{\cal Q}(E_{\bar p})\frac{f_0(E_{\bar p})}{f_0'(E_{\bar p})}\delta f(p) = {\cal Q}(E_{\bar p})\frac{f_0(E_{\bar p})}{f_0'(E_{\bar p})}\sum_{l=0}^\infty\frac{2l+1}{2}\psi_l(|{\bf {\bar p}}|,z)P_l(\cos\theta_{\bar p})\,.
	\end{equation}
    This term is responsible for the suppression of higher modes when the collision rate, namely ${\cal Q}f_0/f_0'$ is large. In particular, as we are going to show next in the large collision rate limit, the $l$-th mode is suppressed by a factor $({\cal Q}f_0/f_0')^{(l+1)}$ which originates a hierarchy between the different modes $\psi_l$.

\subsection{Liouville operator decomposition}

   As we showed in the previous section, the basis of Legendre polynomials constitutes a natural basis to decompose the collision operator due to rotational invariance. The Liouville operator, on the other hand, is not rotational invariant. Nevertheless it is interesting to study how it decomposes on the basis of Legendre polynomials.
   Differently from the collision operator, it is convenient to write the Liouville operator in the wall reference frame since the wall is in a steady state regime. However, the collision operator is block diagonal only in the plasma reference frame. Thus, to analyze the multipole decomposition of the Liouville operator we first need to boost the momenta in the plasma reference frame.
    
    We recall that the Liouville operator is
	\begin{equation}
		p_z\partial_z\delta f - \frac{(m^2)'}{2}\partial_{p_z}\delta f\,.
	\end{equation}
	We begin our analysis from the term proportional to the spatial derivative, namely
	\begin{equation}
		p_z\partial_z\delta f = \gamma_p(\bar p_z + v_p\bar E)\partial_z\delta f = \gamma_p(|{\bar{\bf p}}| P_1(\cos\theta_{\bar p}) + v_p \bar E)\partial_z\delta f
	\end{equation}
    where we boosted $p_z$ in the plasma frame.	By using eq.~(\ref{eq:pert_legendre}) to decompose the perturbation $\delta f$ it is not hard to show that\footnote{For the purpose we recall this useful multiplication property of the Legendre polynomials: $P_1(x)P_l(x) = \frac{l}{2l+1}P_{l-1}(x) + \frac{l+1}{2l+1}P_{l+1}(x)$}
	\begin{equation}
		p_z\partial_z\delta f = \frac{\gamma_p}{2}\sum_{l=0}^{\infty}P_l[|{\bar{\bf p}}|((l+1)\partial_z\psi_{l+1} + l\partial_z\psi_{l-1}) + (2l+1) v_p \bar E\partial_z\psi_l]\,.
	\end{equation}

	The above expression is valid for the whole range of velocities of the DW. To simplify our analysis we may consider the small velocity limit, namely $v_w \lesssim c_s$, where $c_s$ is the speed of sound in the plasma. As we will discuss further, this situation is realized for the two benchmark configurations analyzed in this work. In addition, this simplification is further justified by our numerical analysis which shows that the out-of-equilibrium friction increases linearly with the velocity for $v_w \lesssim c_s$. As a consequence it is sufficient to compute the perturbation $\delta f$ at the linear order in $v_w$.
 
    For slow walls the temperature and plasma velocity gradients are small and as a result we may neglect their contribution to the source term ${\cal S}$ in the Boltzmann equation and consider $T(z) = T_n$, $v_p(z) = v_w$, the terminal velocity of the DW. Hence we consider the perturbations to be sourced only by the interactions with the DW. In the slow wall limit, the source term ${\cal S}$ is then proportional to the terminal velocity of the wall $v_w$. As a consequence, to compute the perturbation at $O(v_w)$ order it is sufficient to keep $ O(1) $ terms in the Liouville operator. In such a case the term proportional to the spatial derivative simplifies and is given by
	\begin{equation}\label{eq:first_term_liouville}
		p_z\partial_z\delta f = \frac{|{\bar{\bf p}}|}{2}\sum_{l=0}^\infty P_l[(l+1)\partial_z\psi_{l+1} + l\partial_z\psi_{l-1}]\,.
	\end{equation}

    The linearization in the velocity greatly simplifies the analysis of the term proportional to the derivative with respect to $p_z$. Its decomposition on the Legendre polynomial basis is in fact given by
	\begin{equation}
		\partial_{p_z}\delta f = \frac{\bar E}{E}\left(\cos\theta_{\bar p}\frac{\partial}{\partial |{\bar{\bf p}}|} +\frac{1-\cos^2\theta_{\bar p}}{|{\bar{\bf p}}|}\frac{\partial}{\partial \cos\theta_{\bar p}}\right)\delta f\,.
	\end{equation}
	By keeping only ${\cal O}(1)$ terms in $v_w$, we can then identify $E = \bar E$ and the overall factor $ E_{\bar p} / E_p$ simplifies.
	We observe that the term involving the derivative with respect to $ |{\bar{\bf p}}| $ has the same structure of the term involving the derivative with respect to $z$ and thus can be dealt with in the same way. Hence it is immediate to show
    \begin{equation}
		\cos\theta_{\bar p}\frac{\partial}{\partial|{\bar{\bf p}}|}\delta f = \frac{|{\bar{\bf p}}|}{2}\sum_{l=0}^{\infty}P_l\left[(l+1)\frac{\partial}{\partial|{\bar{\bf p}}|}\psi_{l+1} + l\frac{\partial}{\partial|{\bar{\bf p}}|}\psi_{l-1}\right]\,.
	\end{equation}
    To deal with the term proportional to the derivative of $\cos\theta_{\bar p}$ we can use the properties of the Legendre polynomials\footnote{$(1-x^2)\frac{d}{dx} P_n(x) = \frac{n(n+1)}{2n+1}(P_{n-1}(x)-P_{n+1}(x))$}
	to show
	\begin{equation}\label{eq:second_term_liouville}
		\frac{1-\cos^2\theta_{\bar p}}{2|{\bar{\bf p}}|}\frac{\partial}{\partial \cos\theta_{\bar p}}\delta f =\frac{1}{4|{\bar{\bf p}}|}\sum_{l=0}^\infty P_l[(l+1)(l+2)\psi_{l+1}-l(l-1)\psi_{l-1}]\,.
	\end{equation}
	
	By using eq. (\ref{eq:first_term_liouville}) and (\ref{eq:second_term_liouville}) we conclude that the Liouville operator in the small velocity limit decomposes as
	\begin{equation}\label{eq:liouville_term}
		\frac{1}{2}\sum_{l=0}^\infty P_l\left[|{\bar{\bf p}}|((l+1)\bar d_z\psi_{l+1} + l \bar d_z\psi_{l-1})-\frac{(m^2)'}{2|{\bar{\bf p}}|}((l+1)(l+2)\psi_{l+1} -l(l-1)\psi_{l-1})\right]\,,
	\end{equation}
	where the flow derivative $\tilde d_z$ is defined by
	\begin{equation}
		|{\bar{\bf p}}|\bar d_z |{\bar{\bf p}}| = -\frac{(m^2)'}{2}\,.
	\end{equation}

    As a last step we need to decompose the source term ${\cal S}$. Its decomposition, however, is trivial on the basis of the Legendre polynomials and we find 
	\begin{equation}\label{eq:source_term_multipole}
		p_z {\cal S} = -v_w f_0'(\bar E)\frac{(m^2)'}{4}\sum_{l=0}^\infty\delta_{0l}P_l\,,
	\end{equation}
	Hence, by using eqs. (\ref{eq:bracket_term}), (\ref{eq:liouville_term}), (\ref{eq:source_term_multipole}) and (\ref{eq:q_term}) and by projecting along $P_l$ we finally find
	\begin{equation}\label{eq:boltzmann_multipole}
		\begin{split}
			|{\bar {\bf p}}|[(l+1)\bar d_z\psi_{l+1}+l\bar  d_z\psi_{l-1}] & -\frac{(m^2)'}{2|{\bar {\bf p}}|} [(l+1)(l+2)\psi_{l+1}-l(l-1)\psi_{l-1}]
            -\frac{{\cal Q}f_0}{f_0'}(2l+1)\psi_l = \\
			& -v_wf_0'\frac{(m^2)'}{2}\delta_{0l}+2\pi f_0(2l+1){\cal O}_l\left[\frac{\psi_l}{f'_0}\right]
		\end{split}
	\end{equation}
 %


    \begin{figure}
        \centering
        \includegraphics[width=0.32\textwidth]{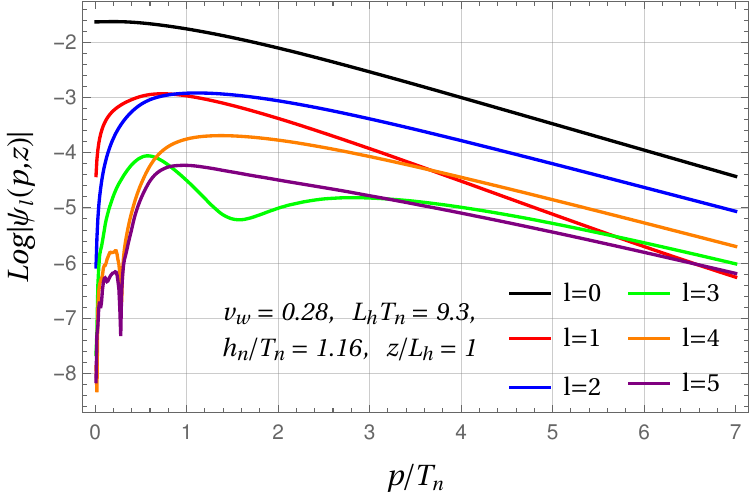}
        \hfill
        \includegraphics[width=0.32\textwidth]{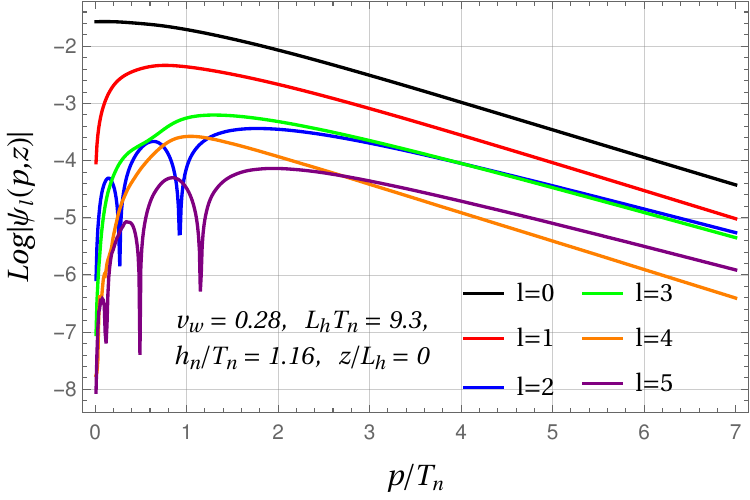}
        \hfill
        \includegraphics[width=0.32\textwidth]{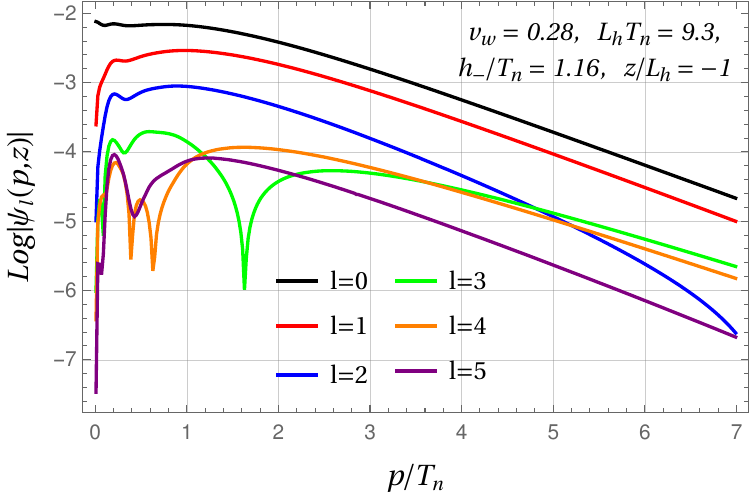}
        \\
        \includegraphics[width=0.32\textwidth]{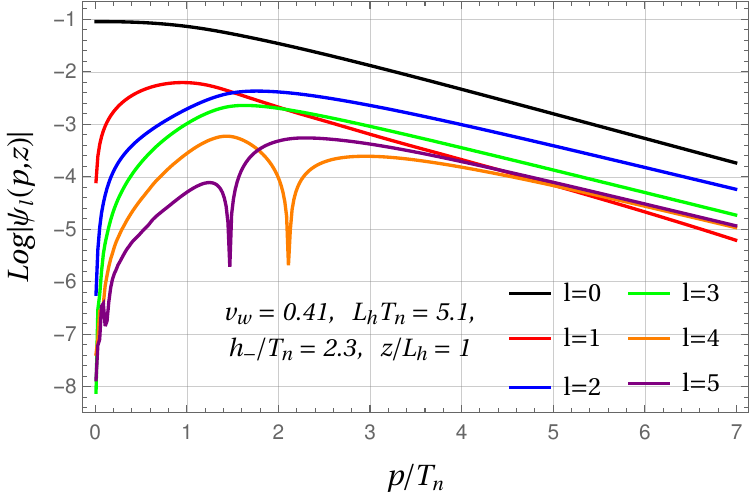}
        \hfill
        \includegraphics[width=0.32\textwidth]{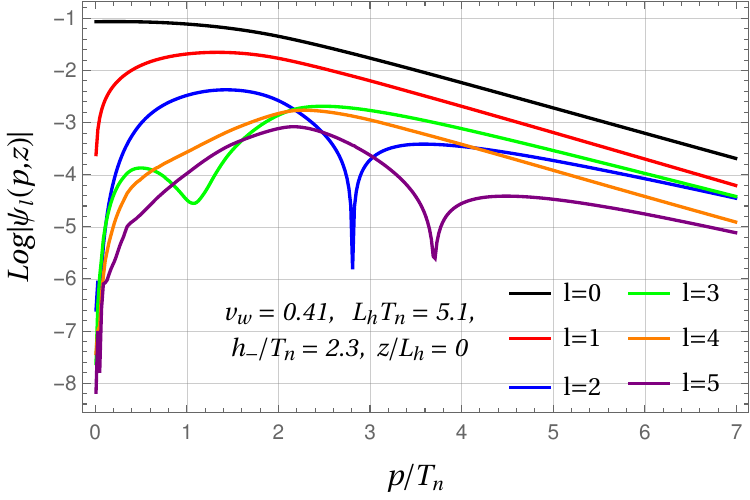}
        \hfill
        \includegraphics[width=0.32\textwidth]{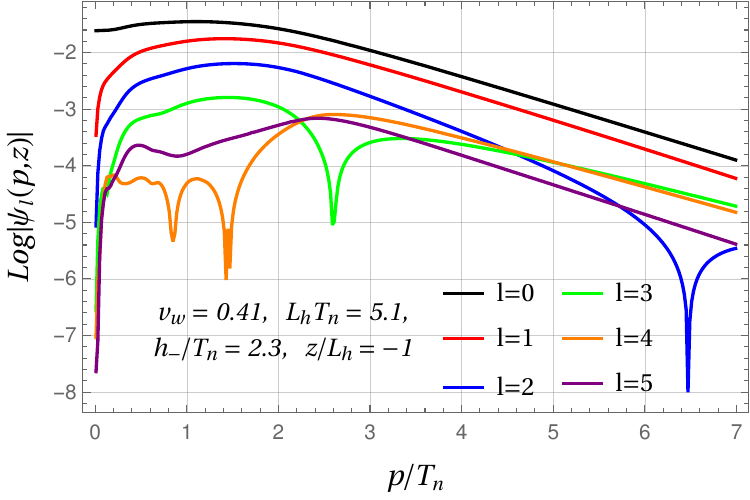}
        \caption{Comparison of the first six Legendre modes in logarithmic scale of the full solution to the linearized Boltzmann equation for the benchmark points BP1 (upper row) and BP2 (lower row) for the top quark. 
        }
        \label{fig:perturbation_legendre_decomposition}
    \end{figure}

The multipole decomposition of the perturbation $\delta f$ of the top quark is reported in Figure~\ref{fig:perturbation_legendre_decomposition}. We will discuss how to compute the perturbation in Chapter~\ref{ch:exact_solution_to_liearized_Boltzmann_equation} by solving the full Boltzmann equation but, for the sake of our discussion regarding the angular momentum decomposition, we anticipate the final result. The benchmark points at which we plot the perturbation are reported in Tab.~\ref{tab:parameters_resultsW} and are characterized by $ v_w = 0.28 $, $L_h\, T_n = 9.3$, $h_-/T_n = 1.16$ for the benchmark BP1 and by $v_w = 0.41$, $L_h\, T_n = 5.1$, $h_-/T_n = 2.3$ for BP2. As we will see in Chapter~\ref{ch:exact_solution_to_liearized_Boltzmann_equation}, such values correspond to the terminal ones obtained for a scalar potential with parameters $m_s = 103.8 \, \textrm{GeV}$, $\lambda_{hs} = 0.72$ and $\lambda_s = 1$ for BP1 and $m_s = 80.0 \, \textrm{GeV}$, $\lambda_{hs} = 0.76$ and $\lambda_s = 1$ for BP2.
    
    The plots show a hierarchy between the different modes 
    with the $ l = 0 $ mode being the largest one. This behaviour has a simple explanation. As eq.~(\ref{eq:source_term}) shows, only the zero mode is sourced, which justifies why it is the largest mode in the expansion. 

    \begin{figure}
        \centering
        \includegraphics[width=0.32\textwidth]{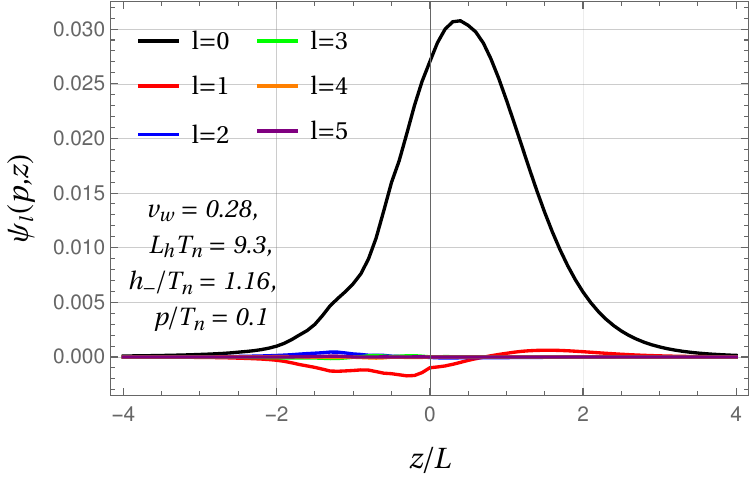}
        \hfill
        \includegraphics[width=0.32\textwidth]{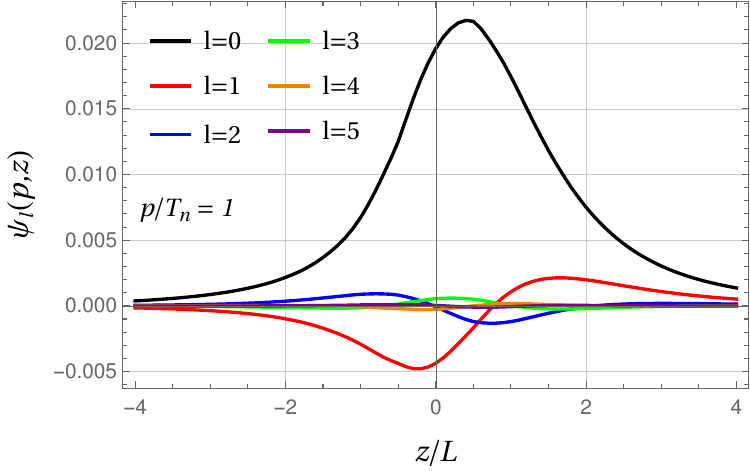}
        \hfill
        \includegraphics[width=0.32\textwidth]{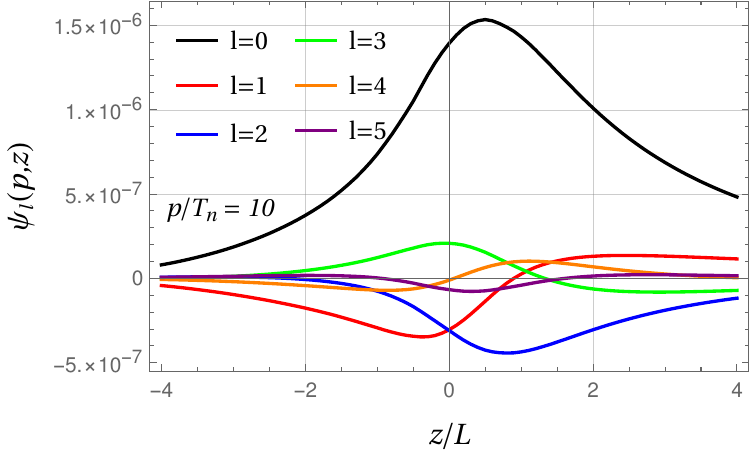}
        \\
        \includegraphics[width=0.32\textwidth]{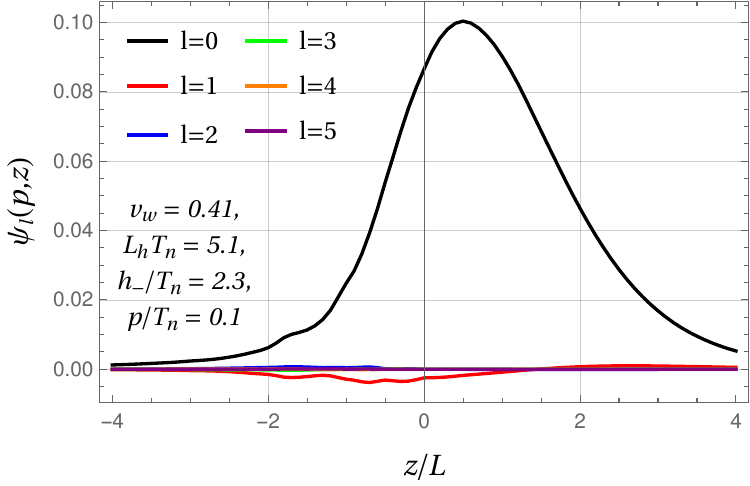}
        \hfill
        \includegraphics[width=0.32\textwidth]{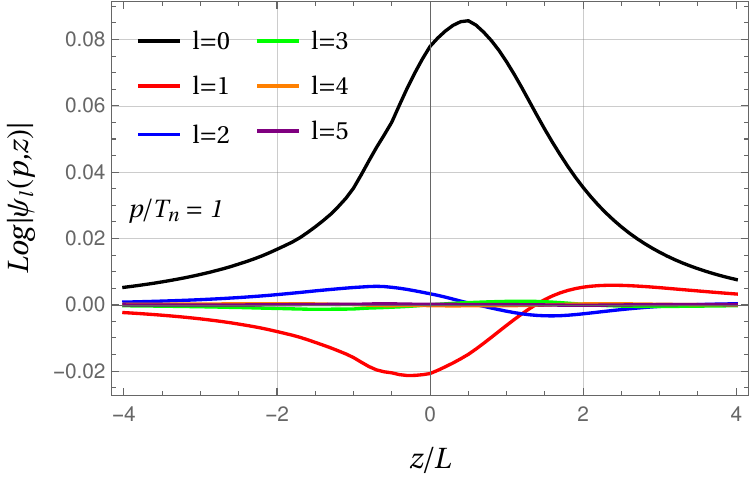}
        \hfill
        \includegraphics[width=0.32\textwidth]{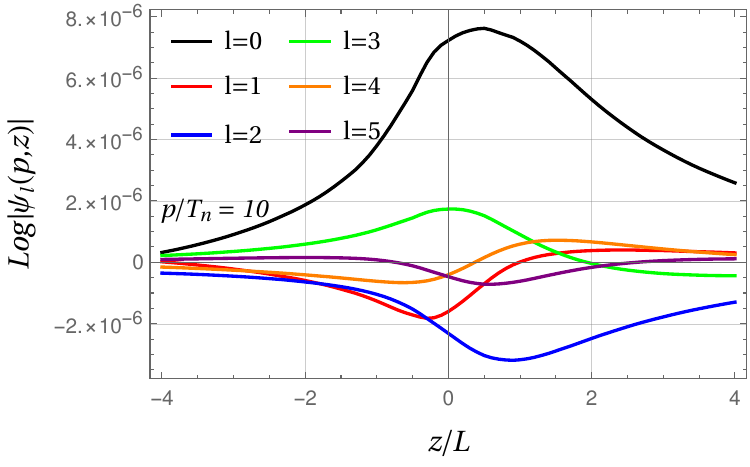}
        \caption{Plot of the first six modes of the full solution to the linearized Boltzmann equation for the benchmark points BP1 (upper row) and BP2 (lower row) for the top quark as a function of $z/L$ for different values of $p$. Notice the strong hierarchy at small values of $|\bar{\bf p}|$ and how the $l$-th mode resembles the shape of $\frac{d^{l+1}m^2}{dz^{l+1}}$.}
        \label{fig:perturbation_legendre_decomposition_zplot}
    \end{figure}
    
    One important property highlighted by the plots is that the hierarchy is more marked for small values of $ |{\bf p}|$ and much milder, instead, for large values of the momentum. This fact is more manifest in Figure~\ref{fig:perturbation_legendre_decomposition_zplot} where for the same potential benchmark points we plot the modes as  functions of $z/L$ with fixed values of the momenta. We observe that for small momenta only the $l=0$ mode is important while high angular momenta modes are essentially suppressed. The latter become more important as we consider large values for the momenta.

   One additional feature we point out is that the shape of the $l-$th mode resembles the one of $(m^2)^{(l+1)}$, where the upper script $(l+1)$ denotes the $(l+1)-$th derivative with respect to $z$. 
This resemblance improves when both the momentum and the terminal velocities are smaller. This observation provides an important hint on how the hierarchy originates. As we are going to show in the next section and as we have already anticipated, the mode hierarchy originates from the collision operator in particular  when collisions are much more efficient.

    \subsection{The large collision rate}
\label{sec:large_collision_multipole}
In order to study the large collision rate limit it is convenient to express the derivative term in eq.~(\ref{eq:boltzmann_multipole}) in terms of the adimensional variable $\xi = z/L$. Moreover, to further simplify the problem we drop the contribution of the bracket term and model all the contribution of the collision integrals through the ${\cal Q}$ term. Eq.~(\ref{eq:boltzmann_multipole}) then becomes
	\begin{equation}\label{eq:boltzmann_multipole_bgk}
		\begin{split}
			|{\bar {\bf p}}|[(l+1)\bar d_\xi\psi_{l+1}+l\bar  d_\xi\psi_{l-1}]  -\frac{( m^2)'}{2|{\bar {\bf p}}|} [(l+1)(l+2)\psi_{l+1}-\\ l(l-1)\psi_{l-1}]
            -\frac{L{\cal Q}f_0}{f_0'}(2l+1)\psi_l = 
			 -v_wf_0'\frac{(m^2)'}{2}\delta_{0l}
		\end{split}
	\end{equation}

The large collision rate limit takes place when the derivative term in the above equation can be neglected. This situation is realized when
\begin{equation}
\label{eq:large_collision_condition_modes}
    \frac{|{\bar{\bf p}}|f_0'}{L{\cal Q}f_0}\frac{l+1}{2l+1}\ll 1\,.
\end{equation}
The above condition has a straightforward physical interpretation. It corresponds to the requirement that the mean free path is much smaller than the width of the bubble wall, namely $\ell \ll L$. 
It is possible, for the particular limit we are considering, to provide a definition of the mean free path for the different modes $\psi_l$. This can be easily read from eq.~(\ref{eq:large_collision_condition_modes}). As one can expect, the mean free path is shorter when collisions are efficient and in the limit in which particles have small momenta. Thus the large collision rate limit corresponds to the situation where the plasma is close to the hydrodynamic regime and takes place when particle momenta are small or when collision processes are efficient.

It is possible to show that in the large collision rate limit a hierarchy between the modes takes place. This result is in accordance with the one that we presented in Figure~\ref{fig:perturbation_legendre_decomposition} and~\ref{fig:perturbation_legendre_decomposition_zplot} where we showed that a strong hierarchy between the modes is present at small momenta, namely in the large collision rate limit. Assuming that a hierarchy between the modes is present eq.~(\ref{eq:boltzmann_multipole_bgk}) takes the following form
\begin{equation}
\label{eq:boltzmann_multipole_hierarchy}
    |{\bar {\bf p}}|l\bar  d_\xi\psi_{l-1}+\frac{( m^2)'}{2|{\bar {\bf p}}|}l(l-1)\psi_{l-1}-\frac{L{\cal Q}f_0}{f_0'}(2l+1)\psi_l = -v_wf_0'\frac{(m^2)'}{2}\delta_{0l}
\end{equation}
which we can solve straightforwardly for the zero mode leading to
\begin{equation}
\label{eq:zero_mode}
    \psi_0(|{\bar{\bf p}}|,\xi) = v_w(f'_0)^2\frac{(m^2)'}{2L{\cal Q}f_0}\,.
\end{equation}

We can then express the higher modes in terms of the derivatives of the $\psi_0$ mode. We point out that the momentum dependence of $\psi_0$ is completely described by the energy $\bar E$. As a consequence when the flow derivative $\bar d_z$ acts on the mode $\psi_0$ it acts as a standard derivative only on the $(m^2)'$ term. The result for the $\psi_1$ mode is then
\begin{equation}
    \psi_1 = |\bar{\bf p}|\left(\frac{f'_0}{L{\cal Q}f_0}\right)^2\frac{v_w f'_0}{3}\frac{\partial_\xi^2(m^2)}{2}\,.
\end{equation}

The remaining modes can finally be found in terms of the previous ones by solving
\begin{equation}
    \frac{L{\cal Q}f_0}{f_0'}(2l+1)\psi_l = |{\bar {\bf p}}|l\bar  d_\xi\psi_{l-1} + \frac{(m^2)'}{2|{\bar {\bf p}}|} l(l-1)\psi_{l-1}\,.
\end{equation}
The above equation can be solved for a generic mode $l$ by using the properties of the flow derivative and the expression of the modes $\psi_0$ and $\psi_1$. It is not hard to show that the solution for a generic mode $\psi_l$ is given by
%
\begin{equation}
\label{eq:lmode_large_collision_rate}
    \psi_l = |2\bar{\bf p}|^l\left(\frac{f'_0}{L{\cal Q}f_0}\right)^{l+1}\frac{l!}{(2l+1)!}\frac{v_w f'_0}{2}\partial_\xi^{l+1}(m^2)\,.
\end{equation}
%
The expression of the $\psi_l$ modes clearly highlights the origin of the hierarchy that we observed in the Figures~\ref{fig:perturbation_legendre_decomposition} and~\ref{fig:perturbation_legendre_decomposition_zplot}. Close to the hydrodynamic regime, namely when the mean free path is small, the mode $\psi_l$ is suppressed by a factor $(\ell/L)^l$. In addition, in this regime the mode profiles in $z$ are given by the derivatives of $m^2$ as we pointed out in our previous analysis. 

We emphasize that eq.~(\ref{eq:lmode_large_collision_rate}) is valid only in the large collision rate limit. Away from the hydrodynamic regime the expressions we found provides a poor solution to the Boltzmann equation 
since it is no longer justified to neglect the derivative terms. In addition the hierarchy becomes milder and as a result the mixing between higher order terms becomes important.

To conclude our discussion regarding the large collision rate limit we discuss, for completeness, the impact of the bracket term that we neglected in our previous analysis. We recall that the effect of the brackets is encoded in the operator ${\cal O}_l$ whose action is defined in eq.~(\ref{eq:Ol_defintion}). To study the impact of the brackets it is convenient to introduce an integral operator ${\tilde{\cal O}}_l$ that includes also the effect of the ${\cal Q}$ term. This is easily accomplished defining ${\tilde {\cal O}}_l$ as
\begin{equation}
    {\tilde{\cal O}}_l[g] = \frac{1}{2\pi}\int |{\bf k}|d|{\bf k}| {\cal Q}\,\delta(|{\bf p}| - |{\bf k}|)\, g(|{\bf k}|) + {\cal O}_l[g]\,.
\end{equation}
In this way, in presence of a hierarchy between the modes, eq.~(\ref{eq:boltzmann_multipole}) can be written as
\begin{equation}
     |{\bar {\bf p}}|l\bar  d_\xi\psi_{l-1}+\frac{( m^2)'}{2|{\bar {\bf p}}|}l(l-1)\psi_{l-1}-(2\pi L)(2l+1)f_0{\tilde {\cal O}}_l\left[\frac{\psi_l}{f'_0}\right] = -v_wf_0'\frac{(m^2)'}{2}\delta_{0l}\,.
\end{equation}

It is possible to solve the above equation by inverting the operator ${\tilde{\cal O}}_l$ that provides the following recurrence relation for the modes $\psi_l$
\begin{equation}
\label{eq:exact_l_mode_large_rate}
    \psi_l = \begin{cases}
        \displaystyle\frac{f'_0}{2\pi L}{\cal O}_0^{-1}\left[\displaystyle\frac{v_w f'_0(m^2)'}{2}\right] & l = 0
        \\
        \\
        -\displaystyle\frac{f'_0}{2\pi L}{\cal O}_l^{-1}\left[|{\bar {\bf p}}|l\bar  d_\xi\psi_{l-1}+\displaystyle\frac{( m^2)'}{2|{\bar {\bf p}}|}l(l-1)\psi_{l-1}\right] & l\geq 1
    \end{cases}
\end{equation}
The above expression provides the exact solution to the Boltzmann equation in the large collision rate limit or in the small momentum case. Unfortunately, it does not provide a reliable solution for the perturbation away from the hydrodynamic regime but it can still be used to get further insights on the behaviour of the system when the latter is close to equilibrium.


We anticipate that to solve the linearized Boltzmann equation we adopted the multipole expansion only to deal with the bracket term. Despite the expansion proves to be useful for a theoretical understanding of the Boltzmann equation, we find that it helps our numerical analysis mainly in the computation of the collision integrals. These terms correspond to the most expensive ones from a numerical point of view and it is essential to provide an efficient method for their computation in order to have a fast algorithm that solves the Boltzmann equation numerically. 

This concludes our analysis on the effective kinetic theory and the Boltzmann equation. Before we finally discuss our method to solve the linearized Boltzmann equation we are first going to discuss the strategies that have been proposed in the literature. This will be the subject of the next chapter.

\chapter{Solving the Boltzmann equation with a specific ansatz}
\label{ch:boltzmann_ansatz}

Finding a solution to the Boltzmann equation is a highly non-trivial task. The problem of uniqueness and existence of a solution for the Boltzmann equation indeed is still an open question with important consequences for the understanding of non-equilibrium dynamics. Although we are not interested in a full solution, even solving its linearized version is still not trivial.

The main difficulty is represented by the collision operator which describes the microscopic interactions that take place in the plasma. Such operator is an integral operator which makes laborious to deal with. For this reason various authors proposed to solve the Boltzmann equation by making an ansatz on the shape of the perturbation.

In this chapter we are going to review the different methods that have been proposed to solve the Boltzmann equation. As we are going to show, these methods are essentially the Grad method, or variations of it, where the Boltzmann equation is integrated with a set of weights and reduced to a set of coupled differential equations whose solution yields the out-of-equilibrium perturbations. At the same time, we are also going to discuss the main drawbacks of such approaches which motivate the necessity of finding a full solution to the linearized Boltzmann equation.

\section{The old formalism}

The Boltzmann equation is an integro-differential equation and so far no analytic method has been found to fully solve it.  Since the main difficulties come from the collision integral, in order to deal with it and simplify the problem some authors~\cite{Moore:1995si, Moore:1995ua, Dorsch:2021nje, Dorsch:2021ubz, Cline:2000nw, Laurent:2020gpg} considered particular ansatzes to solve the equation. In this section we are going to explore the so called ``old formalism'' (OF) based on the perfect fluid ansatz first proposed in~\cite{Moore:1995si,Moore:1995ua} and extended in~\cite{Dorsch:2021nje,Dorsch:2021ubz}. A different approach based on a factorization ansatz~\cite{Cline:2000nw, Laurent:2020gpg} and dubbed ''new formalism'' (NF) by their authors will be reviewed instead in the next section.

\subsection{Fluid Ansatz}

The fluid ansatz, or fluid approximation, is a Grad method where one truncates the moment expansion, around the equilibrium distribution defined in eq.~(\ref{eq:equilibrium_distribution_quantum_gas}), at the first order in the momenta. It is assumed that the light degrees of freedom can be treated as a background fluid at local equilibrium at a common temperature $ T + T \delta \tau_{bg} (z)$ and plasma velocity $ \delta v_{bg}(z) $,  and that the shape of the unknown distribution function $f_i$ of a particle species $i$ is the one of a perfect fluid which, using eq.~(\ref{eq:grad_solution_variational}), can be expressed in the wall reference frame as
\begin{equation}
    f = f_{0,i}(\beta(\gamma_w(E-v_w p_z) + \delta_i))\,.   
\end{equation}
The quantity $\delta_i$ is defined as
\begin{equation}
    \delta_i(z) = -\{\mu_i(z)+\mu_{bg}(z) + \beta\gamma_w[(E - v_w p_z)({\delta\tau}_i(z) + {\delta\tau}_{bg}(z)) + (p_z - v_w E)({\delta v}_i(z) + {\delta v}_{bg}(z))] \}\,,
\end{equation}
and describes the non-equilibrium properties of the plasma encoded in the perturbations of the chemical potential, the temperature and the plasma velocity.

%
%

In the situation where the departure from equilibrium is small, we can linearize the equilibrium distribution function in $\delta_i$. Thus, in the fluid approximation, the shape of the perturbation $\delta f_i$ of a particle species $i$ is
\begin{equation}
\label{eq:fluid_perturbation}
    \delta f_i = (-f'_v)\{\mu_i(z)+\mu_{bg}(z) + \beta\gamma_w[(E - v_w p_z)({\delta\tau}_i(z) + {\delta\tau}_{bg}(z)) + (p_z - v_w E)({\delta v}_i(z) + {\delta v}_{bg}(z))]\}\,.
\end{equation}
Notice that the above expression corresponds to the one in eq.~(\ref{eq:grad_distribution}).

A further assumption is made in the fluid approximation to simplify the Boltzmann equation. We recall that the term proportional to $\partial_{p_z}$ in the flow derivative in eq.~(\ref{eq:flow_derivative_wall_frame}) is proportional to $(m^2)'/2p_z$. Such a factor also appears in the source of the perturbations and thus we may expect that $\delta f \sim O(m^2/T^2)$. The term involving $\partial_{p_z}$ scales as $O(m^4/T^2)$. As a consequence, since perturbations are assumed to be small such a term is subleading and dropped from the equation. Hence we identify $d_z \rightarrow \partial_z$.

The resulting Boltzmann equation for a particle species $i$, using eq.~(\ref{eq:lorentz_invariant_linearized_boltzmann})~and~(\ref{eq:source_term}) is then given by
\begin{equation}
\label{eq:linearized_boltzmann_equation_no_partialpz_term}
    p_z\partial_z\delta f = -{\bar{\cal C}}[\delta f] + \frac{(m^2)'}{2T}(-f'_v)\,,
\end{equation}
which, by inserting the fluid ansatz in eq.~(\ref{eq:fluid_perturbation}), yields
\begin{equation}
\begin{split}
    p_z\left[\partial_z\widetilde \mu_i(z) + \gamma_w\frac{(E - v_w p_z)}{T}{\delta \widetilde\tau}_i(z) + \gamma_w\frac{(p_z - v_w E)}{T}{\delta \widetilde v}_i(z)\right](-f'_v) = \\ - \bar{\cal C}[\mu, \delta \tau, \delta v ] + \frac{(m^2)'}{2T}(-f'_v)\,.
\end{split}
\end{equation}
where we introduced the shorthand notation $\widetilde\mu_i = \mu_i +\mu_{bg}$ and analogous expressions for the other macroscopic perturbations.
The expression of $\bar{\cal C}[\mu_i,\delta\tau_i,\delta v_i]$ is easily worked out from eqs.~(\ref{eq:linearized_collision_integral})~and~(\ref{eq:fluid_perturbation}) and reads
\begin{equation}
    {\bar {\cal C}}[\mu_i,\delta \tau_i, \delta v_i] =\frac{1}{4 N_p}\sum_j\int\frac{d^3{\bf k}d^3{\bf p}'d^3{\bf k}'}{(2\pi)^5 8 E_k E_{p'}E_{k'}}|{\cal M}_j|^2\bar{\cal P}[\mu_i,\delta \tau_i,\delta v_i]\delta^4(p + k - p' - k')\,,
\end{equation}
where we defined
\begin{equation}
\label{eq:population_factor_fluid_ansatz}
    \bar{\cal P}[\mu_i, \delta \tau_i, \delta v_i] = f_0(\beta U^\mu p_\mu)f_0(\beta U^\mu k_\mu)(1\pm f_0(\beta U^\mu p'_\mu))(1\pm f_0(\beta U^\mu k'_\mu))\sum\mp \delta_i\,,
\end{equation}
and where we recall that the $-\ (+)$ in front of the perturbations in the sum is for incoming (outgoing) particles.
To compute the three unknown perturbations, namely $\mu$, $\delta \tau$ and $\delta v$ we integrate the Boltzmann equation over the momentum space with a set of weights $\{\varphi_j\}$.
In such a way we reduce an integro-differential equation to a set of coupled differential equations whose solution yields the three perturbations. As we have already pointed out in sec.~\ref{sec:chapman-enskog_grad_methodds}, there is no proof of convergence nor of the efficiency of the method. We expect such a method to be effective in the case where the system dynamics is close to the hydrodynamic regime, namely when the mean free path is small compared to the macroscopic scale. However, we can only verify a posteriori whether the moment method is effective, since we cannot easily extract the value of $\ell$ from the collision integrals.

We also highlight that the choice of weights to integrate the Boltzmann equation is not unique. A natural choice, which is used in the fluid approximation, is provided by the set $\{1, E, p_z\}$ since the corresponding equations, once summed over all the plasma degrees of freedom, can be interpreted as the conservation of the number of particles, energy and momentum along the $z$-direction of the system. However, from a mathematical stand point, other choices are possible and the corresponding out-of-equilibrium friction, and thus the terminal velocity of the DW, is strongly impacted by the set of weights considered. This issue will be analyzed in a later section.
After the integration we obtain the following system of differential equations for a particle species $i$:
\begin{equation}
\label{eq:fluid_system_boltzmann}
\begin{split}
    &C_i^{1,1}\widetilde \mu_i' + \gamma_w (C_i^{0,1} - v_w C_i^{1,2}) \delta\widetilde\tau_i' + \gamma_w (C_i^{1,2} - v_w C_i^{0,1})\delta\widetilde v_i' + \int\frac{d^3{\bar{\bf p}}}{(2\pi)^3 E_{\bar p}}{\bar{\cal C}} = \gamma_w v_w C_i^{1,0}\frac{(m^2)'}{2 T}\\
    &C_i^{0,1}\widetilde \mu_i' + \gamma_w(C_i^{-1,1} - v_w C_i^{0,2})\delta\widetilde \tau_i' + \gamma_w(C_i^{0,2} - v_w C_i^{-1,1})\delta\widetilde v_i' + \int\frac{d^3{\bar{\bf p}}}{(2\pi)^3E_{\bar p}}E_p\bar{\cal C} = \gamma_w v_w C_i^{0,0}\frac{(m^2)'}{2T}\\
    &C_i^{1,2}\widetilde \mu_i' + \gamma_w(C_i^{0,2} - v_w C_i^{1,3})\delta\widetilde \tau_i' + \gamma_w(C_i^{1,3} - v_w C_i^{0,2})\delta\widetilde v_i' + \int\frac{d^3{\bar{\bf p}}}{(2\pi)^3 E_{\bar p}}p_z\bar{\cal C} = 0\,.
\end{split}
\end{equation}
We recall that barred momenta are computed in the plasma reference frame.
The coefficients $C_i^{i,j}$ are the moments of the equilibrium distribution function defined as
\begin{equation}
\label{eq:C_coefficients}
   C_i^{m,n} = T^{m-n -3}\int\frac{d^3\bar{\bf p}}{(2\pi)^3}\frac{p_z^n}{E_p^m}(-f'_0(\beta|{\bar {\bf p}}|))\,.
\end{equation}
Since they depend on the mass of the particle, the coefficients in general depend also on the position $z$. In the massless case it is possible to provide an analytical expression for all the coefficients involving the Fermi-Dirac distribution. We refer to Appendix~\ref{ap:exact_solution} for the details. 
%

The final step involves the computation of the weighted integrals of the collision operator over the momentum space. Such computation, that we recall is performed assuming massless particles and the leading-log approximation, is much easier to carry out due to the fluid ansatz and needs to be performed only once. The resulting values indeed depend only on the process under consideration whose amplitudes, for our specific case, are reported in Tab.~\ref{tab:amplitudes}. Since the top species is diluted in the plasma, the relevant processes involve the interactions of heavy species with light species only. In such a way we can treat left and right-hand top with the same distribution and eq.~(\ref{eq:fluid_system_boltzmann}) does not mix top and W bosons perturbations. 
Under these approximations the weighted integrals of the collision operator over the momentum space yield for the top quark
\begin{equation}
    \begin{split}
        & \int\frac{d^3{\bar{\bf p}}}{(2\pi)^3 E_{\bar p}}{\bar{\cal C}} = \Gamma^{(\mu_t)}_{0,0}\mu_t + \Gamma^{(\tau_t)}_{0,0}\delta\tau_t\,, \\
        & \int\frac{d^3{\bar{\bf p}}}{(2\pi)^3E_{\bar p}}E_p{\bar{\cal C}} = \Gamma^{(\mu_t)}_{-1,0}\mu_t + \Gamma^{(\tau_t)}_{-1,0}\delta \tau_t + \gamma_w v_w \Gamma_{v,t}\delta v_t\,, \\
        & \int\frac{d^3{\bar{\bf p}}}{(2\pi)^3E_{\bar p}}p_z{\bar{\cal C}} = \gamma_w v_w \Gamma_{\mu 2,t}\mu_t + \gamma_w v_w \Gamma_{\tau 2, t}\delta \tau_t + \gamma_w \Gamma_{v,t}\delta v_t\,.
    \end{split}
\end{equation}
and analogous expressions for the W bosons.
To evaluate the gamma elements it is possible to employ the rotational invariance and the Dirac delta appearing in the collision integral to perform five of the twelve integrations. The remaining seven integrals can be performed numerically by using the parameterization detailed in ref.~\cite{Arnold:1999ux,Arnold:2003zc,Moore:2001fga}.
The resulting values of the gamma elements are reported in Appendix~\ref{ap:exact_solution}.
%

The system of equations in eq.~(\ref{eq:fluid_system_boltzmann}) can be written in a compact way as
\begin{equation}
\label{eq:system_of_perturbation_matrix}
    \begin{split}
        &A_t({\vec q}'_t + {\vec q}'_{bg}) + \Gamma_t{\vec q}_t  = {\vec S}_t\,,\\
        &A_W({\vec q}_W' + {\vec q}_{bg}') + \Gamma_W{\vec q}_W +  = {\vec S}_W\,,\\
    \end{split}
\end{equation}
where we defined ${\vec q}_i = (\mu_i\,,\;\delta\tau_i\,,\;\delta v_i)^{\rm T}$, while the expressions of the matrices $A_i$ and $\Gamma_i$ and the source ${\vec S}_i$ are

\begin{equation}
\label{eq:A_Gamma_matrices}
\begin{split}
    &A_i=\left(\begin{array}{ccc}
        C_i^{1,1} & \gamma_w (C_i^{0,1} - v_w C_i^{1,2}) &\gamma_w (C_i^{1,2} - v_w C_i^{0,1})  \\
        C_i^{0,1} & \gamma_w(C_i^{-1,1} - v_w C_i^{0,2}) & \gamma_w(C_i^{0,2} - v_w C_i^{-1,1}) \\
        C_i^{1,2} & \gamma_w(C_i^{0,2} - v_w C_i^{1,3}) &  \gamma_w(C_i^{1,3} - v_w C_i^{0,2})
    \end{array}\right)\,,\\
    &\Gamma_i = \left(\begin{array}{ccc}
        \Gamma^{(\mu_i)}_{0,0} &  \Gamma^{(\tau_i)}_{0,0} & 0  \\
        \Gamma^{(\mu_i)}_{-1,0} & \Gamma^{(\tau_i)}_{-1,0} & \Gamma^{(v_i)}_{-1,0}  \\
        \Gamma^{(\mu_i)}_{0,1}& \Gamma^{(\tau_i)}_{0,1}&\gamma_w \Gamma^{(v_i)}_{0,1}
    \end{array}\right)\,,\,\,{\vec S}_i = \gamma_w v_w\frac{(m^2_i)'}{2T}\left(\begin{array}{c}
        C_i^{1,0}  \\
         C_i^{0,0} \\
        0
    \end{array}\right)\,.
\end{split}
\end{equation}
%
We point out that the matrices $A_i$ are singular for $v_w = c_s$. The value of the velocity where the determinant of $A$ vanishes depends on the choice of weights used to integrate the Boltzmann equation. However, as we are going to see further, these singularities have an important impact on the friction acting on the bubble wall since the integrated friction presents a peak in correspondence of the singular value of $v_w$. This result questions the reliability of the method of moments and thus needs a careful investigation which we are going to provide later.

We finally need to supply a set of equations for the light degrees of freedom. As a preliminary step we observe that since the background fluid has a large number of degrees of freedom the chemical potential $\mu_{bg}$ is suppressed with respect to the one of heavy species. We hence set $\mu_{bg} = 0$ and consider the background plasma to be in chemical equilibrium.
Second, we notice that $\Gamma_{bg}$ can be computed from the conservation of the total number of particle, energy and momentum along $z$ of the plasma. This is a direct consequence of the particular set of weights we chose to integrate the Boltzmann equation. Summing over all the degrees of freedom we find the following equation
\begin{equation}
\label{eq:with_background_perturbations_equation}
    A_{bg}{\vec q}_{bg}' + \Gamma_{bg,t}{\vec q}_{t} + \Gamma_{bg,W}{\vec q}_{W} = 0\,,
\end{equation}
with
\begin{equation}
    A_{bg} = N_fA_t|_{m=0} + N_b A_W|_{m=0}\,\;\;\;\;\;\Gamma_{bg,i} = -N_i\Gamma_i\,,
\end{equation}
where we recall that $N_w = 9$ and $N_t = 12$ and the $A$ matrices are evaluated at $m=0$ since light species are considered massless. The quantities $N_f$ and $N_b$ correspond instead to the total degrees of freedom of light fermions and bosons respectively. Their values are thus $N_f = 78$, $N_b = 20$.\footnote{Notice that the actual value of $N_f$ and $N_b$ depends on the spectrum of the theory under considerations. In the scalar extension of the SM the values that we provided are correct. In the SM case, instead, we find $N_b = 19$ since the singlet $s$ is absent.}

The equation in eq.~(\ref{eq:with_background_perturbations_equation}) can be inverted for the derivative of the background perturbations yielding
\begin{equation}
\label{eq:background_perturbation_equation}
    {\vec q}_{bg}' = - A_{bg23}^{-1}(\Gamma_{bg,W}\vec{q}_W + \Gamma_{bg,t}\vec{q}_t)
\end{equation}
where $A_{bg23}$ denotes the matrix where the bottom
right block of $A_{bg}$ is inverted and the rest of the
matrix elements are zero. By inserting such
result in eq.~(\ref{eq:system_of_perturbation_matrix}) we find
\begin{equation}
\label{eq:final_system_fluid_equation_perturbations}
    A\vec{q}' + \Gamma \vec{q} = \vec{S}
\end{equation}
where we defined
\begin{equation}
\begin{split}
    & A = \left(\begin{array}{cc}
        A_t & 0 \\
         0 & A_w
    \end{array}\right)\,,\;{\vec S} = \left(\begin{array}{c}
         {\vec S}_t  \\
         {\vec S}_W
    \end{array}\right)\,,\;\vec{q} =\left(\begin{array}{c}
        {\vec q}_t\\
        {\vec q}_W
    \end{array}\right)\,,\\
    & \Gamma = \left(\begin{array}{cc}
        \Gamma_t & 0 \\
        0 & \Gamma_W 
    \end{array}\right) - \left(\begin{array}{cc}
        A_t A_{bg23}^{-1}\Gamma_{bg,t} & A_t A_{bg23}^{-1}\Gamma_{bg,W}  \\
        A_W A_{bg23}^{-1}\Gamma_{bg,t} & A_W A_{bg23}^{-1}\Gamma_{bg,W} 
    \end{array}\right)\,.
\end{split}
\end{equation}

The solution to the system of equation in eq.~(\ref{eq:final_system_fluid_equation_perturbations}) provides the perturbations. These must satisfy the boundary conditions $\mu(z = \pm\infty) = \delta\tau_i(z = \pm\infty) = \delta v_i(z = \pm\infty) = 0$. In the massive case, namely when we include the mass in the computation of the $C^{m,n}_i$ coefficients defined in eq.~(\ref{eq:C_coefficients}), the perturbations can be found only by numerical algorithms. Typically one adopts a shooting algorithm, by reducing the boundary problem in eq.~(\ref{eq:final_system_fluid_equation_perturbations}) to an initial value problem. In particular one solves the system separately on the two sides of the wall, namely for $z<0$ and $z>0$, imposing two different initial conditions. Such conditions are hence varied until the two solutions match on the wall, namely in $z = 0$, to a desired precision. The shooting method provides a fast and accurate solution to the equations and we will use it in the following to compute the perturbations. Some analytical results can be found in the massless case. These are reported in Appendix~\ref{ap:exact_solution} and will be used later to discuss the impact of the different choice of weights as the presence of the mass does not affect our conclusions.

Since our aim in this chapter is to review the different ansatzes on the perturbation $\delta f$ to solve the Boltzmann equation, we can make further assumptions on the plasma to keep the discussion simpler without affecting the key results. First, we can neglect the background perturbations and consider light particles to be in global equilibrium. The background will play a crucial role only when we will analyze the singularity of the matrix $A$. Hence, we will restore its contribution only to discuss this particular aspect. For the rest of our discussion we will model the background as a thermal bath that sets the plasma temperature to $T_n$. In this limit the equations that describe top and W bosons perturbations in eq.~(\ref{eq:final_system_fluid_equation_perturbations}) decouple. In addition, in order to present the methods we will consider the top quark to be the only out-of-equilibrium species since it is the heaviest among the SM particles. Thus we are going to focus on
\begin{equation}
\label{eq:fluid_equation_top}
    A_t {\vec q}_t' + \Gamma_t{\vec q}_t = {\vec S}_t\,.
\end{equation}
These approximations do not impact our discussion since qualitative behaviour of the top and W bosons solutions are quite similar.

\begin{figure}
    \centering
    \includegraphics[width=0.47\textwidth]{fluid_approximation_solution0.2.pdf}
    \hfill
    \includegraphics[width=0.47\textwidth]{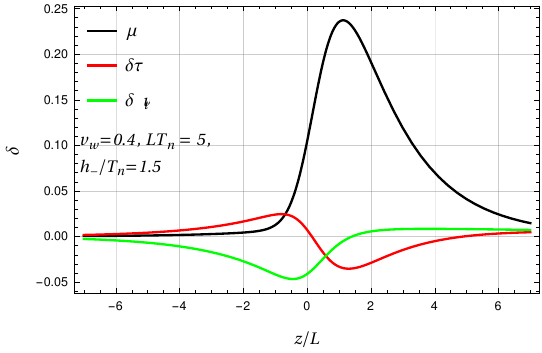}\\
    \includegraphics[width=0.47\textwidth]{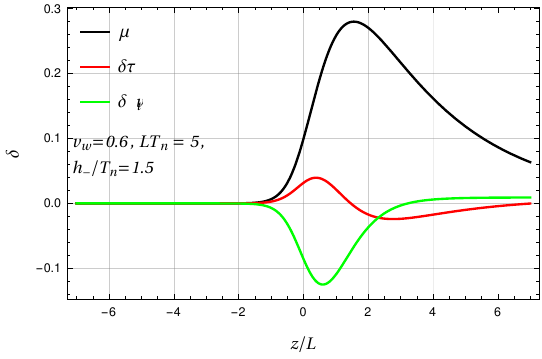}
    \hfill
    \includegraphics[width=0.47\textwidth]{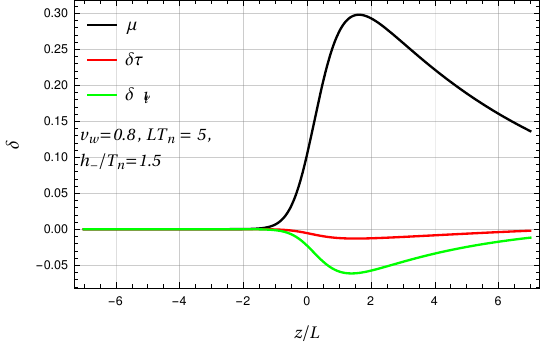}
    \caption{Plot of the three perturbations for different benchmark values of the terminal velocity. Typical values for $L$ and $h_-$ have been chosen, namely $L T_n = 5.0$, $h_-/Tn = 1.5$. }
    \label{fig:solution_fluid_approximation}
\end{figure}

In Fig.~\ref{fig:solution_fluid_approximation} we plot the three perturbations as a function of $z/L$. We fixed the values of the width of the wall and Higgs VEV in the broken phase to $L T_n = 5.0$, $h_-/T_n$, which are typical values for EWPT. The black solid line corresponds to the chemical potential, the red one to the fluctuations in temperature while the green one to the velocity. In order to have an overview of how perturbations change with the terminal velocity of the wall, we chose three benchmark values of the velocities namely, $0.2$, top left panel, $0.4$, top right panel, $0.6$, bottom left panel and $0.8$ bottom right panel.

The largest perturbation is the chemical potential $\mu$ and a mild hierarchy between the three perturbations is present. Larger perturbations are present for faster walls, as we can expect since the source term is proportional to the velocity. In addition, perturbations in front of the wall get suppressed and decay slower in the symmetric phase. For supersonic walls, as the plot on the right panel shows, all the perturbations trail the source and the plasma in front of the bubble is at equilibrium. This behaviour is a consequence of the spectrum of the matrix $A_t^{-1}\Gamma_t$ which, for supersonic walls, presents only positive eigenvalues. This has important consequences for baryogenesis since we found that the fluid approximation predicts that the baryon asymmetry is actually strongly suppressed for supersonic walls.

\begin{figure}
    \centering
    \includegraphics[width=0.47\textwidth]{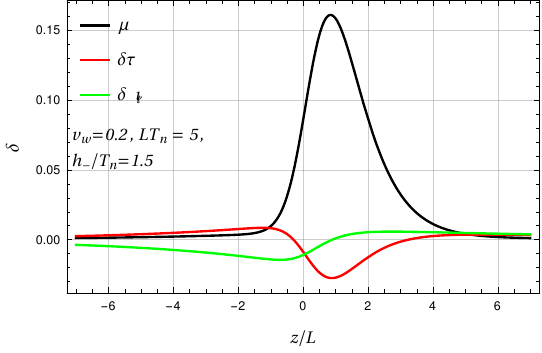}
    \hfill
    \includegraphics[width=0.47\textwidth]{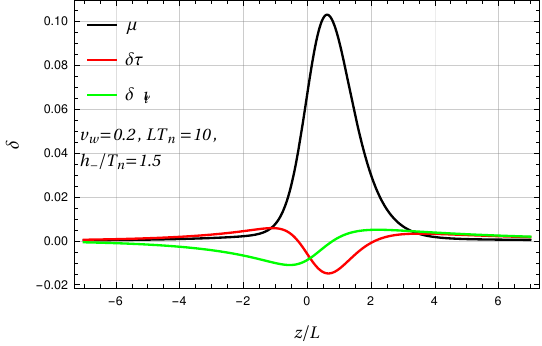}
    \\
    \includegraphics[width=0.47\textwidth]{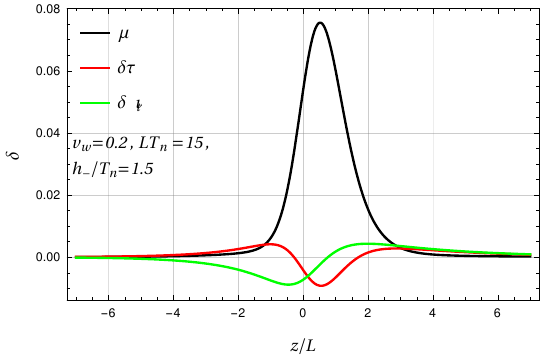}\hfill
    \includegraphics[width=0.47\textwidth]{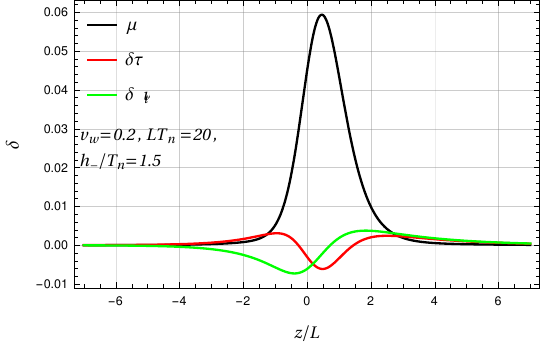}
    \caption{Plot of the three perturbations for the benchmark width $L T_n =$ $5$, $10$, $15$, $20$. For thicker walls the shapes of $\mu$ and $\delta v$ are more similar to $(m^2)'$ and $(m^2)''$ respectively.}
    \label{fig:solution_fluid_approximation_different_widths}
\end{figure}

Fig.~\ref{fig:solution_fluid_approximation_different_widths} points out another interesting property of the perturbations where we plot them by fixing the value of $v_w = 0.2$ for different benchmark values of the width of the bubble wall, namely $L T_n = 5$, $10$, $15$, $20$. A first important property is that larger walls suppress perturbations. This is not surprising. The non-equilibrium dynamics of the plasma is strongly impacted by the ratio $\ell/L$ that determines the efficiency of collision rates with respect to the force that drives the plasma out-of-equilibrium. For thick walls the collision processes are more efficient and the system gets closer to the hydrodynamic regime. As a consequence the out-of-equilibrium perturbations are smaller. A second property the figure emphasizes is that for thick walls the shapes of the chemical potential and the plasma velocity are very similar to the shapes of the first and second derivatives of the mass profile, namely $(m^2)'$ and $(m^2)''$. We encountered a similar result in sec.~\ref{sec:large_collision_multipole} when we discussed the large collision rate limit of the multipole expansion of the Boltzmann equation. In particular, in that context, we showed that the spatial dependence of $l-$th angular mode in the multipole expansion is given by $\partial_z^{l+1}(m^2)$. This provides a first connection between the fluid approximation and the full solution of the Boltzmann equation. In particular this result indicates that the fluid ansatz is a good approximation of the out-of-equilibrium perturbations when the system is close to the hydrodynamic regime, hence to local equilibrium.

\begin{figure}
    \centering
    \includegraphics[width=0.47\textwidth]{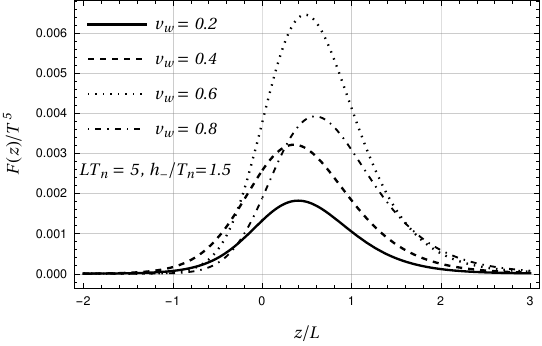}
    \caption{Plot of the friction in eq.~(\ref{eq:out_of_eq_friction}) for $v_w$ = $0.2$, $0.4$, $0.6$, $0.8$ and  with $L T_n = 5$ and $h_-/T_n = 1.5$.}
    \label{fig:friction_fluid_approximation}
\end{figure}

Fig.~\ref{fig:friction_fluid_approximation} shows the friction defined in eq.~(\ref{eq:out_of_eq_friction}) for the same four velocity benchmark points. As for perturbations, faster walls present larger friction with the largest value in correspondence to the sound barrier. Aside from a rescaling factor and a change of the position of the peak, the shape of the friction does not vary much for the four benchmark points. This is a consequence of the $(m^2)'$ term appearing in the definition of the friction in eq.~(\ref{eq:out_of_eq_friction}). Indeed one can verify that the shapes roughly agree with $d (m^2)/dz$.

\begin{figure}
    \centering
     \includegraphics[width=0.47\textwidth]{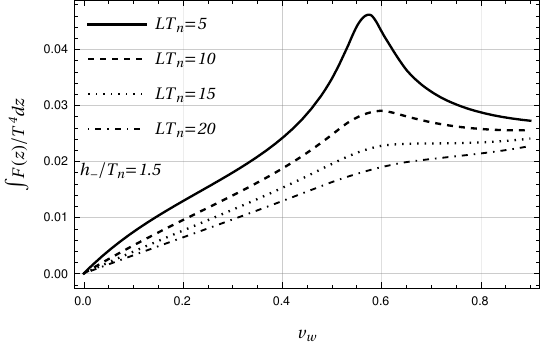}
    \caption{Plot of the integral of the friction defined in eq.~(\ref{eq:out_of_eq_friction}) as a function of the terminal velocity for $L T_n = 5$, $10$, $15$, $20$.}
    \label{fig:integral_friction_fluid_approximation}
\end{figure}

At last in Fig.~\ref{fig:integral_friction_fluid_approximation} we compare the integral of the friction in eq.~(\ref{eq:out_of_eq_friction}) for four different values of the wall width, namely $LT_n = 5$, $10$, $15$, $20$. The most relevant feature of the integrated friction is the presence of a peak corresponding to walls moving at velocity $v_w = c_s$ whose height depends on the values of the DW width. Thin walls present a sharper and higher peak with respect to thick walls. In particular for $LT_n = 20$ the integral of the friction has a smooth behaviour across the whole velocity range. Such result is not surprising. The case of thick walls actually corresponds to the situation where the system behaviour is close to the hydrodynamic regime, namely when the mean free path is small with respect to $L$. As a consequence, we expect the fluid approximation to work better in this case.
If the wall is not large enough, the fluid approximation provides a poor estimate of the friction acting on the wall and the presence of the peak signals the break down of the approximation. This result is further supported by the fact that such a peak is absent in the full solution of the Boltzmann equation, as we will show in Chapter~\ref{ch:exact_solution_to_liearized_Boltzmann_equation}, hence proving that it is unphysical and an artifact of the fluid approximation.

In our previous analysis we considered, for simplicity, only the contributions from the top quark. There are, however two relevant features that we need to comment when we include also the contribution of W bosons and the background. W bosons provide a qualitative and quantitative similar contribution to the top quark to the total friction acting on the bubble wall. This is in contrast to the expectation that W bosons contributions should be suppressed by their smaller mass and degrees of freedom. This result is a consequence of the IR divergent behaviour of W bosons which spoils the na\"ive scaling of the out-of-equilibrium friction. We refer to Chapter~\ref{ch:exact_solution_to_liearized_Boltzmann_equation} for further details.

It is also important to comment on what happens when we include the contribution of the background. As we already pointed out in eq.~(\ref{eq:total_pressure}), variations of temperature across the wall provide an additional contribution to the friction. For small gradients of the temperature this additional contribution is proportional to $\delta \tau_{bg}$, which can be computed by solving eq.~(\ref{eq:background_perturbation_equation}) providing
\begin{equation}
    \delta \tau'_{bg} =\frac{N_t(\mu_t \Gamma_{\mu2,t} + \delta \tau_t\Gamma_{\tau 2,t}) + N_W(\mu_W \Gamma_{\mu2,W} + \delta \tau_W\Gamma_{\tau 2,W})}{(N_f c_{f4} + N_b c_{b4})(1/3 - v_w^2)}
\end{equation}
The above expression diverges for $v_w = c_s$ and as result the peak in the integrated friction becomes a singularity.

This singularity signalizes the break down of the linearization procedure for the computation of the temperature and velocity fluctuations of the background for $v_w = c_s$. The analysis carried in ref.~\cite{Dorsch:2021ubz} provides an explanation of the origin of such singularity from hydrodynamic considerations which is of great help to visualize how the linearization procedure breaks down. The starting point is eq.~(\ref{eq:vp_as_function_of_vm}) which relates the velocity in the symmetric phase $v_+$ to the one in the broken phase $v_-$. The limit which we considered in this chapter corresponds to the one where the differences $v_+ - v_- = \delta v_{bg}$ and $T_+ - T_- = T_n\delta \tau_{bg}$ are small. Hence, In virtue of $\delta \tau_{bg} \ll 1$ we expand eq.~(\ref{eq:vp_as_function_of_vm}) around $\alpha = 0$ to get
\begin{equation}
    \delta v_{bg} = \frac{3v_-(1-v_-^2)}{1-3v_-^2}\alpha\,,
\end{equation}
which breaks down for $v_- = c_s$. Using eq.~(\ref{eq:velocity_relations_symmetric_broken_phase}) and inserting the above expression for $\delta v_{bg}$ yields the expression for $\delta \tau_{bg}$ which in turns diverges for $v_- = c_s$. Such situation is realized for deflagration solutions with $v_w \sim c_s$ and hybrid walls where $v_- \sim c_s$. 

The physical interpretation of such singularity is a ''sonic boom'' that takes place when the profiles solutions change from deflagration to detonation entering in the hybrid regime. To account for this regime we cannot rely on a linearization procedure and all the non-linearities must be included for a consistent description. From a mathematical point of view this result is even clearer if we consider the case $v_- = c_s$ in eq.~(\ref{eq:vp_as_function_of_vm}) which yields the expression
\begin{equation}
    v_+ = \frac{1}{1+\alpha}\left(c_s\pm\sqrt{\alpha^2 + \frac{2}{3}\alpha}\right)\,,
\end{equation}
that cannot be linearized around $\alpha = 0$. As a result, to account for the gradients of temperature and velocity of the background we must include all the non-linearities by using the procedure presented in Section~\ref{sec:temperature_velocity_plasma_profile}.

We conclude by emphasizing that the peak in Fig.~\ref{fig:integral_friction_fluid_approximation} and the singularity have different interpretations. Despite they both take place at $v_w = c_s$, the mechanisms that originate them are not the same. The peak in the integrated friction arises from the singularity in the Liouvillian term, namely on the singular value of the matrix $A_t$. As we will see in the following, different ansatzes or different choices of weights leads to the appearance of different peaks hence highlighting that they are an artifact of the fluid ansatz. The singularity, instead, as we just discussed, originates from the break down of the linearization of the temperature and plasma velocity gradients, hence it is removed by considering a procedure that computes the two profiles by taking into account all the non-linearities. 

\subsection{Extended fluid ansatz}

The main drawback of the fluid approximation is, in the limit of a plasma in global equilibrium, the appearance of a peak in the integrated out-of-equilibrium friction. Such a peak, which is smoothed out for large walls, questions the validity of the method to model the out-of-equilibrium perturbations hence requiring a different ansatz to study the non-equilibrium properties of the plasma.

One natural extension is to consider additional macroscopic fluctuations to model the out-of-equilibrium properties, rather than just the chemical potential, temperature and plasma four velocity. This corresponds to account for additional terms in the momentum expansion and consider the following ansatz for the distribution function of the particle species $i$
\begin{equation}
\label{eq:extended_fluid_ansatz}
    f_i = (-f_{i,0}')[1 + w_i^{(0)} + w_{i,\mu}^{(1)}p^\mu + w_{i,\mu\nu}^{(2)}p^\mu p^\nu + \dots]\,,
\end{equation}
where in the above expression we may identify $w_i^{(0)} = \mu_i(z)$ and $w^{(1)}_{i,\mu} = \beta(\delta \tau_i,0,0,\delta v_i)$. Once again, to simplify our discussion, we consider the light particle species to be in equilibrium at a common temperature $T_n$ and we focus only on the contributions of the top quark.\footnote{When the background is assumed to be in local equilibrium only fluctuations in temperature and fluid velocity are included just as in the fluid approximation. In fact, if additional terms in the momentum expansions were considered also for the background, the distribution function of light species would no longer be that of a local equilibrium plasma but rather an out-of-equilibrium one.}

The corresponding equations for the perturbations $ w_t^{(j)} $ are obtained by inserting the expression~(\ref{eq:extended_fluid_ansatz}) in the Boltzmann equation in~(\ref{eq:linearized_boltzmann_equation_no_partialpz_term}) and then integrating over the set of weights $\{1\,,\;E\,,\;p_z\,,\;E^2\,,\;E p_z\,,\; p_z^2\,,\; \dots\}$. The resulting set of equations still has the form of eq.~(\ref{eq:fluid_equation_top}) and 
can be solved through the same means that we discussed in the previous section.

\begin{figure}
    \centering
    \includegraphics[width=0.47\textwidth]{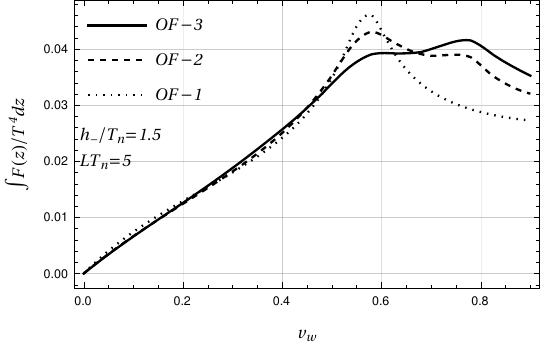}
    \caption{Integrated friction in the extended fluid approximation. The label OF-n identifies the order at which the momentum expansion in eq.~(\ref{eq:extended_fluid_ansatz}) is truncated. OF-1 corresponds to the fluid approximation. Notice the appearance of multiple peaks as additional terms in the momentum expansions are considered.}
    \label{fig:extended_fluid_approximation_integrated_friction_comparison}
\end{figure}

As we stated when we discussed the Grad method in sec.~\ref{sec:chapman-enskog_grad_methodds}, the expansion in eq.~(\ref{eq:extended_fluid_ansatz}) is expected to work better in the case where the unknown distribution function $f$ is close to $f_0$ around which we expand. For our specific case, where $f_0$ corresponds to the equilibrium distribution function, we expect the extended fluid method to work better in those situations where the plasma is close to the hydrodynamic regime. This fact is confirmed by the analysis of the macroscopic fluctuations. As for the case of the fluid approximation, for small velocities and in the large collision rate limit, the out-of-equilibrium perturbations present a hierarchy, with higher order coefficients in the momentum expansion being suppressed with respect to the lower order ones. In addition decomposing the extended fluid ansatz along the Legendre polynomial basis reveals that the shape of the corresponding modes is precisely given by the derivatives of the mass profile. These results match the expected behaviour of the full solution of the Boltzmann equation in the large collision rate limit we provided in eq.~(\ref{eq:lmode_large_collision_rate}) and~(\ref{eq:exact_l_mode_large_rate}).

The inclusion of higher order terms has an important impact on the out-of-equilibrium dynamics of the plasma.
The most relevant effect takes place on the integrated friction which we plot in Fig.~\ref{fig:extended_fluid_approximation_integrated_friction_comparison}  for the benchmark width $LT_n = 5$ and for $h_-/T_n = 1.5$. The dotted plot corresponds to the fluid approximation where the momentum expansion is truncated at linear order (OF-1), the dashed and solid line instead correspond to the integrated friction obtained by truncating the momentum expansion at second (OF-2) and third order (OF-3) respectively.
As the figure shows, the main effect of the higher order perturbation is to deliver additional peaks in the integrated friction, on top of the one already present at $v_w = c_s$. The positions of these peaks, however, are not related to any relevant physical processes hence indicating that they are an artifact of the particular momentum expansion. This is further supported by noticing that the peaks get smoother as we include additional fluctuations, 
signalizing that the momentum expansion is slowly converging to a solution of the Boltzmann equation.  Such a convergence is slower for supersonic walls, where larger fluctuations are present, while faster when the terminal velocity is small, i.e $v_w < c_s$, 

The inclusion of higher order terms is also crucial for baryogenesis. As we discussed when we analyzed the fluid approximation, the generation of the baryon asymmetry is suppressed for supersonic walls since the eigenvalues of the matrix $A_t^{-1}\Gamma_t$ are all positive and hence the perturbations trail the source. However, when we include additional perturbations the matrix $A_t^{-1}\Gamma_t$ presents eigenvalues with opposite sign also in some region of supersonic walls opening to the possibility of achieving baryogenesis also for fast moving walls. A quantitative analysis of the spectrum of the matrix $A_t^{-1}\Gamma_t$ shows that the eigenvalues become all positive for walls moving faster than the largest value of $v_c$ where the matrix $A_t$ is singular. In the case of the fluid approximation this corresponds to $v_c = c_s$, when we include second order terms in the momentum expansion instead $v_c = \sqrt{3/5}\simeq 0.77$ while at order three $v_c \simeq 0.86$. As we include additional terms the critical value $v_c$ increases. This result indicates that we can expect baryogenesis to be suppressed only for relativistic walls.


\subsection{Impact of the set of weights}

To conclude our discussion regarding the fluid approximation we discuss the impact of the set of weigths used to integrate the Boltzmann equation. We refer to Appendix~\ref{ap:exact_solution} for the expressions of the $A_t$ and $\Gamma_t$ matrices and for the vector $\vec{S}_t$ when the Boltzmann equation is integrated with weights involving a generic power of $p_z$ and $E$. Our main goal is to show that the position of the peak is determined by the zero eigenvalues of the matrix $A_t^{-1}\Gamma_t$ which in turn are determined by the roots of the determinant of $A_t$. Because the presence of the mass does not affect our conclusion, we are going to solve the Boltzmann equation in the massless approximation. In this case it is possible to provide an anlytical expression for the integrated friction given by eq.~(\ref{eq:integral_friction_fluid_approximation}) as discussed in Appendix~\ref{ap:exact_solution}.

\begin{figure}
    \centering
    \includegraphics[width=0.47\textwidth]{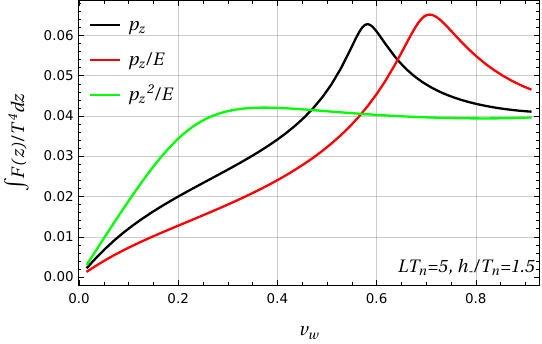}
    \hfill
    \includegraphics[width=0.47\textwidth]{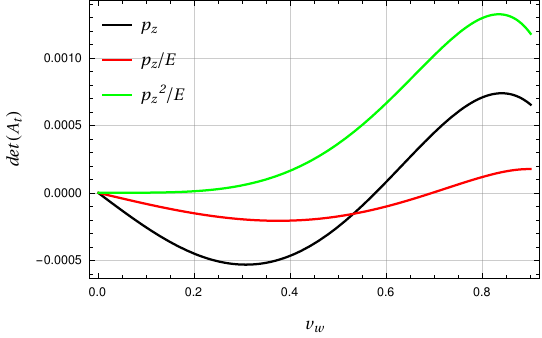}
    \caption{Left panel: comparison between the integrated friction computed using the different sets of weights discussed in the text for $LT_n=5$.
    Right panel: plot of the determinant $A_t$ matrix for different choice of weights. Notice that the position of the peak in the integrated friction is given by the root of the determinant. }\label{fig:integrated_friction_weights_comparison}
\end{figure}

We compute the integrated friction for the benchmark width $L_h T_n = 5$, $h_-/T_n = 1.5$ using the sets of weights $\{1\,,\;E\,,\;p_z\}$, $\{1\,,\;E\,,\;p_z/E\}$, $\{1\,,\;E\,,\;p_z^2/E\}$, which we label using their last weight, namely $p_z$, $p_z/E$ and $p_z^2/E$.  We report our results on the left panel of Fig.~\ref{fig:integrated_friction_weights_comparison} where we plot the integrated friction. Important quantitative and qualitative differences are present among the three cases. The set $p_z/E$ (red solid line) presents a peak in $v_w \sim 0.7$ while the set $p_z^2/E$ (green solid line) predicts a smooth behaviour across the whole velocity range and becomes of the integrated friction which reaches a constant value at $v_w\sim 0.5$. On the quantitative level large discrepancies among the three different sets are present. The sets
 $p_z$ and $p_z/E$ present a $30-50\%$ discrepancy with each other across the whole velocity range. The set $p_z^2/E$ has a factor $2-3$ difference with the other two sets at subsonic walls and then reaches similar values with the set $p_z$ for $v_w >0.8$.
 
These results provide a strong indication that the moment method is not suitable to solve the Boltzmann equation. The out-of-equilibrium dynamics is strongly affected by the set of weights adopted to integrate the Boltzmann equation with the resulting friction being controlled by this arbitrary choice. This conclusion is further confirmed by the right panel of Fig.~\ref{fig:integrated_friction_weights_comparison} which shows the determinant of the $A_t$ matrix for the three different choices of weights. The plot clearly show that the location of the peak in the integrated friction corresponds to the root of the determinant of the matrix $A_t$, further supporting that the peak is an artifact of the moment method. Notice that for the set $p_z^2/E$ the determinant never vanishes and the corresponding integrated friction has a smooth behaviour in the whole velocity range.

\section{New formalism}


To clarify the physical nature of the peaks in the integrated friction the authors of~\cite{Laurent:2020gpg,Cline:2000nw} proposed a new ansatz, which they dubbed ``new formalism'' (NF). Although the main strategy to compute the fluctuations is the same as the one we analyzed in the OF case, the NF presents some additional features that differentiate it from the OF and that we should discuss in detail.

In ref.~\cite{Cline:2020jre} the authors claimed that perturbations should not be suppressed in front of the bubble wall for supersonic values of $v_w$. They indeed argued that perturbations describe particle diffusion which is a physical process unrelated to the propagation of sound waves in the plasma. As a consequence, there is no physical reason why perturbations should be suppressed in front of the bubble wall and presented an improved set of equations to overcome this unphysical behaviour.

To avoid the suppression in particle diffusion in the NF a different ansatz on the shape of the perturbation $\delta f_i$ for a particle species $i$, which exploits a novel way to deal with the plasma velocity perturbation~\cite{Cline:2000nw, Fromme:2006cm, Fromme:2006wx}, is made. According to ref.~\cite{Cline:2000nw} the choice of a specific shape for the perturbation $\delta v$ leads to unphysical results. The author hence argued that such unphysical behaviour can be avoided by assuming no specific constraint on the shape of the velocity perturbation. On the other hand they treated the chemical potential and plasma temperature fluctuations adopting the fluid approximation. Therefore the ansatz they proposed for $\delta f$ is given by
\begin{equation}
    \delta f_i = (-f_v)'\delta\tilde X_i + \delta f_{i,u}\,,
\end{equation}
where the above expression is written in the wall reference frame and where we defined
\begin{equation}
    \delta \tilde X_i = \mu + \beta \gamma_w(E - v_wp_z)\delta\tau_i\,.
\end{equation}

The perturbation $\delta f_{i,u}$ encodes all the information regarding the plasma velocity fluctuation. In particular it is assumed that
\begin{equation}
     \delta v_i \propto \int d^3{\bf p}\,\frac{p_z}{E}\, \delta f_{i,u}\,,\;\;\;\;\;\;\;\;\;\; \int d^3{\bf p}\,\delta f_{i,u} = 0\,.
\end{equation}
Furthermore to deal with the integrals involving the perturbation $\delta f_u$ and a generic quantity $Q$ the following factorization ansatz is made
\begin{equation}
\label{eq:factorization_rules}
    \int d^3{\bf p}\,Q\delta f_{i,u} = \delta v_i\int d^3{\bf p}\,\frac{E}{p_z}\,Q\,f_v\,.
\end{equation}

To determine the three fluctuations, one still follows the strategy outlined in the previous section and solves the Boltzmann equation by taking its moments. Differently from the OF, to avoid the appearance of the unphysical peaks one should use a different set of weights to integrate the Boltzmann equation given by $\{1\,,\; E\,,\; p_z/E\}$. We point out that within such choice one can no longer identify the equation corresponding to $p_z/E$ with the conservation of the component $T^{33}$ of the stress energy tensor.

As in the OF, the resulting set of differential equations that one obtains by taking the moments of the Boltzmann equation has the form of eq.~(\ref{eq:fluid_equation_top}). Nevertheless, due to the factorization rules in eq.~(\ref{eq:factorization_rules}), the resulting expressions of the matrices $A_i$, $\Gamma_i$ and the vector source $\vec S_i$ are different. By carrying out the integration one finds
\begin{equation}
\label{eq:A_Gamma_matrices_nf}
\begin{split}
    A_i&=\left(
\begin{array}{ccc}
C_i^{1,1} & \gamma_w (C_i^{0,1} - v_w C_i^{1,2}) &D_i^{0,0}\\
        C_i^{0,1} & \gamma_w(C_i^{-1,1} - v_w C_i^{0,2}) & D_i^{-1,0} \\
        C_i^{1,2} & \gamma_w(C_i^{0,2} - v_w C_i^{1,3}) & D_i^{1,1} \\
\end{array}
\right)\,,\\
\Gamma_i & = \left(\begin{array}{ccc}
         \Gamma_{\mu,i}^{(1)} & \Gamma_{\tau,i}^{(1)} & \Gamma_{v,i}^{(1)} \\
         \Gamma_{\mu,i}^{(2)} & \Gamma_{\tau,i}^{(2)} & \Gamma_{v,i}^{(2)} \\
         \Gamma_{\mu,i}^{(3)} & \Gamma_{\tau,i}^{(3)} & \Gamma_{v,i}^{(3)}
    \end{array}\right) \;\;\;\;\;  {\vec S}_i = \gamma_w v_w\frac{(m^2_t)'}{2T}\left(\begin{array}{c}
        C_i^{1,0}  \\
         C_i^{0,0} \\
         C_i^{2,1}
    \end{array}\right)\,.
\end{split}
\end{equation}
where we defined
\begin{equation}
\label{eq:D_coefficients}
    D^{m,n}_i = T^{m-n-3}\int\frac{d^3{\bar{\bf p}}}{(2\pi)^3}f_{0,i}(\beta U^\mu p_\mu)\,.
\end{equation}
As for the OF, the expressions of the elements of the $\Gamma$ matrix are found by taking moments of the collision operator. Since in our analysis we are going to focus only on the top quark contributions we provide the expression of the matrix elements only for such species. The elements for the W bosons can be found in ref.~\cite{Laurent:2020gpg}. With the exception of $\Gamma_{v,t}^{(3)}$, the velocity dependence of such elements can be computed analytically. For the remaining term it is possible to provide a polynomial fit in $v_w$. The $\Gamma_t$ matrix entries are reported in Appendix~\ref{ap:exact_solution}.

At last one should provide the equation for light degrees of freedom. However, differently from the OF, it is not possible to use the conservation of the total energy momentum tensor in this case. In fact, as already pointed out, the equation corresponding to the weight $p_z/E$ can no longer be interpreted as the conservation of the $z$ component of the momentum. Nevertheless, the authors of ref.~\cite{Laurent:2020gpg} still assumed that eq.~(\ref{eq:with_background_perturbations_equation}) holds also in this case and they used it to describe the plasma light degrees of freedom. In addition, since $\mu_{bg}$ is not included in the perturbations, one of the three equations that describe the background is redundant. The authors of~\cite{Laurent:2020gpg} chose to keep only the equations which correspond to the weighting factors $1$ and $p_z/E$ with the main motivation that for such a choice the matrix $A_{bg}$ is invertible for all the values of $v_w$. Following this strategy one finds that the resulting equation for the background perturbations is 
\begin{equation}
\label{eq:nf_background_perturbations}
    \vec{q}_{bg}' = -\tilde A_{bg}^{-1}(\Gamma_{bg,t}{\vec q}_t + \Gamma_{bg,W}{\vec q}_W)
\end{equation}
with
\begin{equation}
    \tilde A_{bg}^{-1} = (P_2A_{bg}P_1 + P_3)^{-1} - P_3^{\rm T}
\end{equation}
and where the matrices $P_i$ are
\begin{equation}
    P_1 = \left(\begin{array}{ccc}
        0 & 0 & 0 \\
        0 & 1 & 0 \\
        0 & 0 & 1
    \end{array}\right)\,,\;\;\;\;\;
    P_2 = \left(\begin{array}{ccc}
        1 & 0 & 0 \\
        0 & 0 & 0 \\
        0 & 0 & 1
    \end{array}\right)\,,\;\;\;\;\;P_3 = \left(\begin{array}{ccc}
        0 & 0 & 0 \\
        1 & 0 & 0 \\
        0 & 0 & 0
    \end{array}\right)\,.
\end{equation}
%

\subsection{Comparison with the OF}

To compare with the OF we focus on a single particle species, namely the top quark and we consider the background to be in global equilibrium.
The perturbations can be computed by following the same strategies outlined for the OF. 
\begin{figure}
    \centering
    \includegraphics[width=0.47\textwidth]{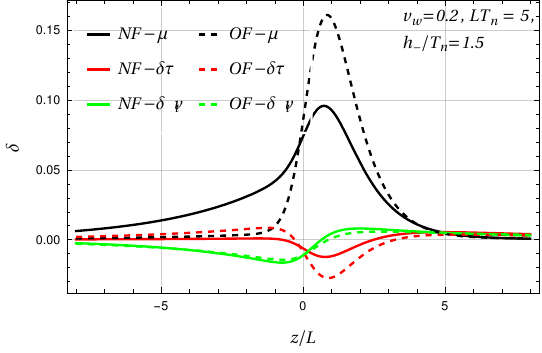}
    \hfill
    \includegraphics[width=0.47\textwidth]{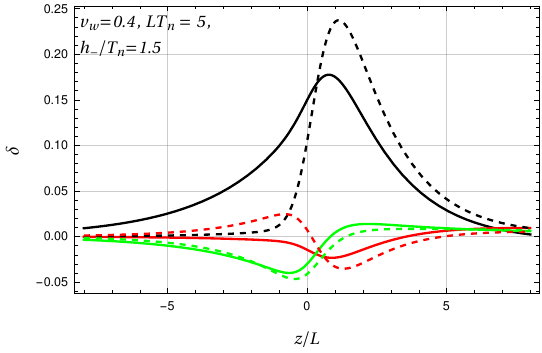}\\
    \includegraphics[width=0.47\textwidth]{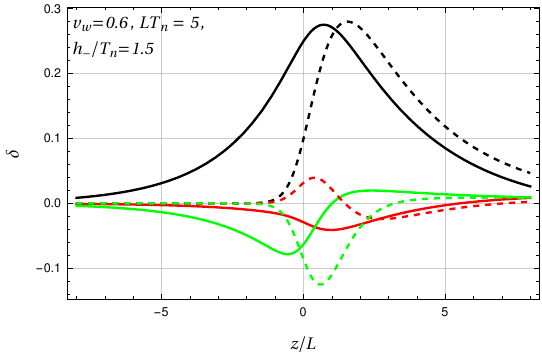}
    \hfill
    \includegraphics[width=0.47\textwidth]{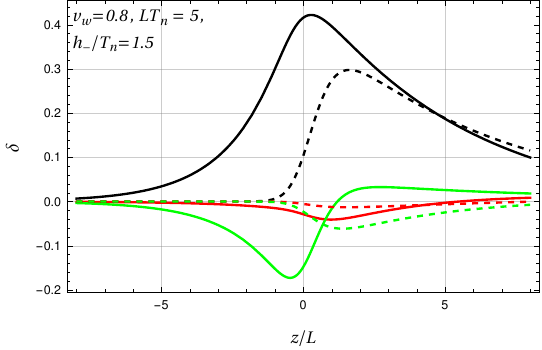}
    \caption{Comparison of the perturbations computed using the NF, solid line and the OF, dashed lines for the benchmark values of the velocity $v_w = 0.2$, $0.4$, $0.6$, $0.8$.}
    \label{fig:perturbation_nf_of_comparison}
\end{figure}
We compare the OF (dashed lines) and NF (solid lines) perturbations in Fig.~\ref{fig:perturbation_nf_of_comparison} for the benchmark velocities $v_w=$ $0.2$, $0.4$, $0.6$, $0.8$.
The black line is the $\mu$ perturbation, the red one is the $\delta \tau$ while the green one is $\delta v$. We notice that the perturbations are quite different in the two strategies. Although the overall shape is somewhat similar, large quantitative differences, are present among the three perturbations. The chemical potential predicted by the NF is almost a factor $2$ smaller for slow moving walls and presents a $20-30\%$ discrepancy for intermediate values of the velocity and for fast moving walls. The velocity perturbation roughly agrees in the two setups for subsonic walls and it is completely different in the supersonic region while the temperature perturbation presents large discrepancies for the whole velocity range. As for the case of the extended fluid ansatz, the NF predicts unsuppressed perturbations in front of the wall for supersonic velocities. This allows to generate a baryon asymmetry also for fast moving walls.

\begin{figure}
    \includegraphics[width=0.47\textwidth]{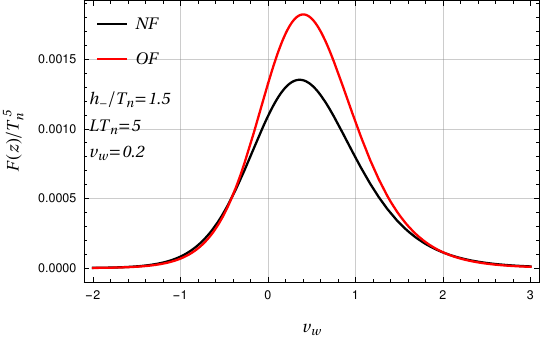}
    \hfill
    \includegraphics[width=0.47\textwidth]{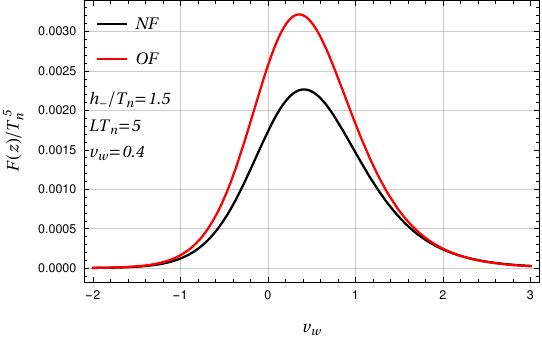}\\
    \includegraphics[width=0.47\textwidth]{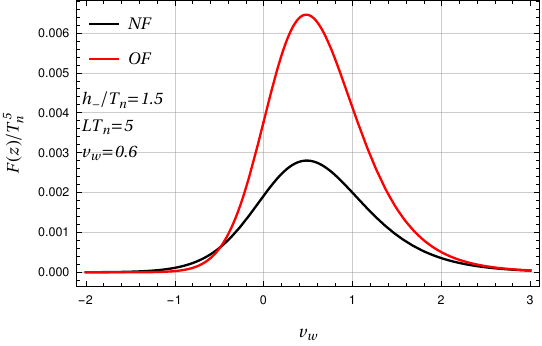}
    \hfill
    \includegraphics[width=0.47\textwidth]{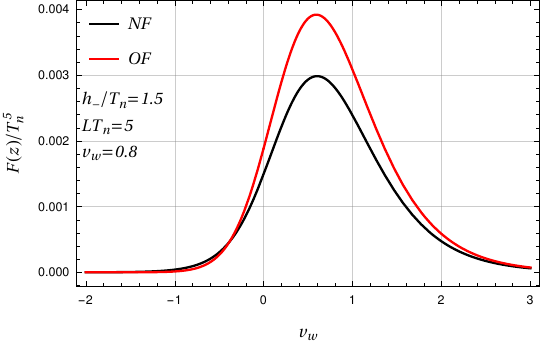}
    \caption{Comparison of the friction computed using the NF, black solid line and the OF, red solid line for the benchmark values of the velocity $v_w = 0.2$, $0.4$, $0.6$, $0.8$.}
    \label{fig:friction_nf_of_comparison}
\end{figure}

\begin{figure}
\centering
    \includegraphics[width=0.47\textwidth]{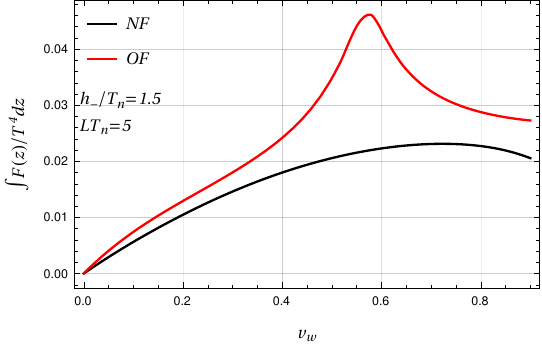}
    \caption{Comparison of the integrated friction computed using the NF, black solid line and the OF, red solid line. Notice the quantitative difference between the two appraoches and the smooth behaviour of the NF across the sound barrier.}
    \label{fig:integrated_friction_nf_of_comparison}
\end{figure}

In Fig.~\ref{fig:friction_nf_of_comparison} we compare the friction computed in the NF and and the fluid approximation for the same benchmark velocities considered for the perturbations. The plots show a large discrepancy close to the speed of sound. For $v_w = 0.6$ a factor $2$ of difference is present while a milder discrepancy of $25\%$ is present away from the sound barrier. A more refined analysis of the out-of-equilibrium dynamics predicted by the NF can be found in 
Fig.~\ref{fig:integrated_friction_nf_of_comparison} where we plot the integrated friction as a function of the velocity for the benchmark width $LT_n = 5$ and $h_-/T_n = 1.5$. Quantitative differences similar to the one we found for the friction are present. However, the most important feature highlighted by the plot is the absence of the peak in the NF which predicts a smooth behaviour of the integrated friction across the whole velocity range.


The analysis that we carried out so far highlighted the main drawbacks of the moment method. The presence of unphysical peaks smoothed out by the inclusion of higher order perturbations and whose positions depend on the set of weights used to integrate the Boltzmann equation actually indicates that the strategy is not particularly effective when the plasma is far from the hydrodynamic regime. In addition, despite the NF was proposed to solve this issue and provides an educated guess of the solution to the Boltzmann equation, it presents important discrepancies with the OF both in the subsonic and supersonic region. These limitations question the accuracy of the terminal velocity predicted by the moment method and its generalizations. The methods still provide an estimate of the magnitude of the friction acting on the DW but the theoretical uncertainty, intrinsic in the strategies that we outlined in this chapter, prevents a proper description of the non-equilibrium dynamics of the plasma. These issues are ultimately caused by the particular ansatz on the perturbation $\delta f$ and can be solved by providing a full solution to the linearized Boltzmann equation without imposing any constraints on the momentum dependence of $\delta f$. This will be the subject of the next chapter.

\chapter{Exact solution of the linearized Boltzmann equation}
\label{ch:exact_solution_to_liearized_Boltzmann_equation}

To compute the terminal velocity of the DW it is essential to provide a reliable method for solving the Boltzmann equation. As we showed in the previous chapter the moment method, both in the OF and NF case, is not particularly effective for such a task. The method itself provides an estimate of the out-of-equilibrium friction acting on the wall but its intrinsic drawbacks, which we deeply analyzed, prevent an accurate determination of the out-of-equilibrium friction.

One of the most important flaws of the OF is the appearance of peaks in the integrated friction whose position depends on the set of weights chosen to integrate the Boltzmann equation. This is a strong indication that such peaks are an artifact of the moment method. A further support to this interpretation is the behaviour of the integrated friction in the extension of the fluid approximation. As we showed in the previous chapter, when additional perturbations are included, multiple peaks appear in the integrated friction signalizing that the procedure is slowly converging to a solution of the equation.


Even the NF does not lack drawbacks. We already pointed out that due to the different choice of weights, the background fluctuations cannot be included by following the same strategy of the OF. On top of that the factorization ansatz proposed to deal with the velocity perturbations is poorly justified from a physical point of view, which inevitably leads to ambiguities in the approach.

Finally, we recall that both the OF and the NF neglect the term proportional to $\partial/\partial{p_z}$ in the Liouville term. Such a term can be neglected for weak phase transitions where $m \ll T$, with $m$ the mass of the out-of-equilibrium particle species, but should be included when we deal with strong phase transitions.

The ambiguities of both OF and NF arise from the ansatz on the momentum dependence of the perturbation $\delta f$. Solving the full linearized Boltzmann equation without considering a specific ansatz on the shape of the perturbation is the only way to clarify such ambiguities. This allows one to provide an accurate description of the non-equilibrium properties of the plasma improving, at the same time, the prediction on the terminal velocity. In ref.~\cite{DeCurtis:2022djw,DeCurtis:2022hlx,DeCurtis:2022llw,DeCurtis:2023hil} we presented, for the first time, the full solution to the linearized Boltzmann equation without imposing any ansatz on the perturbation which we are going to discuss in the present chapter.

The algorithm that we developed in~\cite{DeCurtis:2022djw,DeCurtis:2022hlx,DeCurtis:2022llw} is an iterative procedure where the solution is refined at every step of iteration by the bracket term. The latter is computed by using a spectral decomposition method, which we present in a dedicated section, and the multipole decomposition presented in Chapter~\ref{ch:effective_kinetic_theory}. To assess the validity of our approach we first compare our results with the ones obtained by the previous approaches in a simplified setup where only the top quark is included. Later we compute the terminal velocity in the singlet extended SM scenario by first including the top quark and light particles contributions and then refining our analysis including also the W bosons contributions. As our analysis will show, the largest contribution to the friction arising from such states comes from the IR region that cannot be modeled by the Boltzmann equation and is described by a different kinetic theory we discuss in the last section of this chapter. As we are going to show the theory is IR divergent providing a large theoretical uncertainty on the terminal velocity. Such uncertainty dominates over the other two sources of error we analyze in this chapter, namely the leading-log and the massless approximation in the collision integral.


\section{Flow paths and the Liouville operator}\label{sec:Liouville}

To determine the perturbation $\delta f$ without imposing any ansatz on its shape it is convenient to write the Boltzmann equation as in eq.~(\ref{eq:boltzmann_equation_at_ewpt}), namely
\begin{equation}
    \left(\frac{d}{dz} - \frac{\cal Q}{p_z}\frac{f_v}{f'_v}\right)\delta f = {\cal S} + \frac{f_v}{p_z}\langle\delta f\rangle\,,
\end{equation}
where we recall that the bracket term $\langle \delta f\rangle$ denotes the terms where the perturbation depends on the integration variables, that ${\cal Q}$ is defined in eq.~(\ref{eq:Q_expression}) while the flow derivative $d/dz$, defined in eq.~(\ref{eq:flow_derivative_wall_frame}) acts on the perturbation as
\begin{equation}
    \frac{d}{dz}\delta f = \left(\frac{\partial}{\partial z} - \frac{(m^2)'}{2p_z}\frac{\partial}{\partial p_z}\right)\delta f\,.
\end{equation}
This particular form is the well-known ``method of characteristics'' where a partial differential equation is reduced to an ordinary one introducing suitable curves known as the characteristic curves. In such a way it is possible to account for the derivative term term $\partial/\partial p_z$ which is neglected in the approaches that we discussed in the last chapter.

As discussed in sec.~\ref{sec:boltzmann_basic}, the flow derivative defines the tangent vectors of particle trajectories in the phase space in the collision-less limit. The characteristic paths of the flow derivative are determined in the wall reference frame by solving the equation
\begin{equation}
    \frac{d p_z}{dz} = -\frac{(m^2)'}{2p_z}\,,
\end{equation}
and the characteristic curves satisfy
\begin{equation}
    p^2_z + m(z)^2 = const\,.
\end{equation}
The above expression can also be inferred from the symmetries of the problem, namely time invariance and translation invariance along the DW.
It is straightforward to check that, along the flow path, the transverse momentum $p_\bot$ and the energy of the particle are conserved. Therefore the trajectories of the particles in phase space are given by
\begin{equation}
\left\{\hspace{-.25em}
\begin{array}{l}
E = \sqrt{p_\bot^2 + p_z^2 + m(z)^2} = const\\
\rule{0pt}{1.5em}p_\bot = const
\end{array}
\right.
\quad\Rightarrow\quad
p_z^2 + m(z)^2 = const\,.
\end{equation}

\begin{figure}
\centering
\includegraphics[width=.52\textwidth]{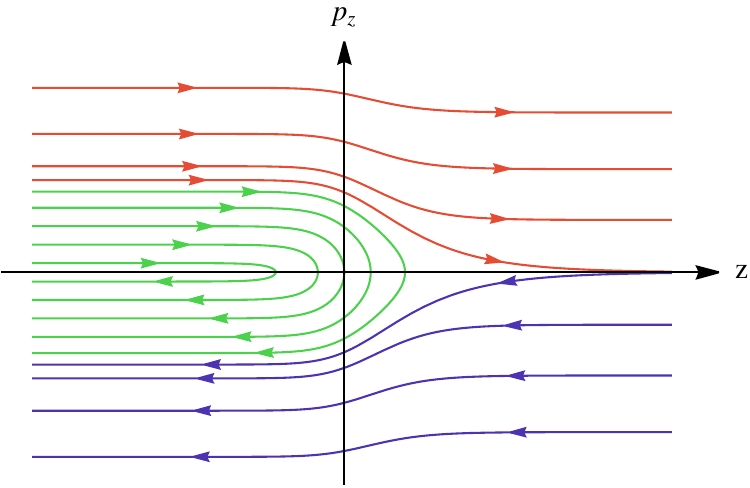}
\hfill
\raisebox{1em}{\includegraphics[width=.43\textwidth]{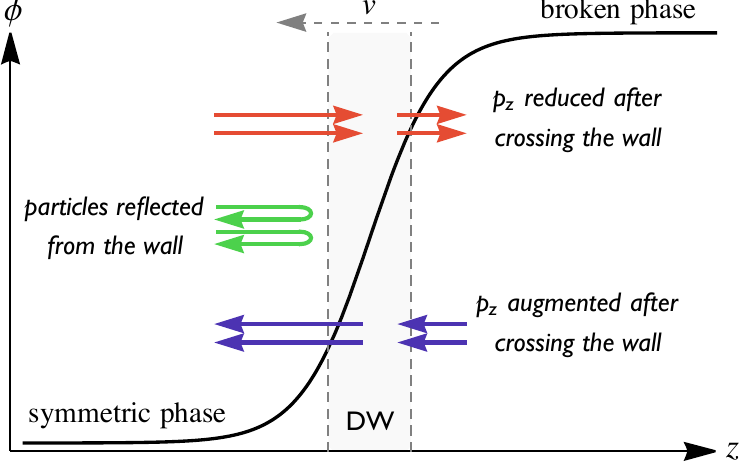}}
\caption{Left panel: Paths with fixed energy and transverse momentum in the $z - p_z$ phase space for the choice $m(z) \propto 1 + \tanh(z/L)$. The red, green and purple colors denote sets of contours with different behavior. The arrows show the flow of
a particle within the phase space. Right panel: Schematic representation of the behavior of the particles across the DW.}\label{fig:flow_paths}
\end{figure}
Due to the non-trivial profile of the Higgs VEV during the EWPT, the condition $p_z^2 + m^2(z) = const$ gives rise to different classes of \emph{flow paths}. Because the mass of the particles receives a contribution from the Higgs VEV, we expect it to smoothly increase going from the symmetric phase outside the bubble
to the symmetry-broken one inside it. In particular we are interested in the dynamics of particles whose mass comes entirely from EW symmetry
breaking (as it happens for the top quark and for the $W$ and $Z$ bosons). Hence we can assume that $m(z) \to 0$ for $z \to -\infty$,
while it approaches a constant value $m(z) \to m_0 > 0$ for $z \to +\infty$.
In this situation we recognize the presence of three different types of flow paths:
\begin{itemize}
\item[i)] for $p_z(-\infty) \geq m_0$ the path goes from $z = -\infty$ to $z = +\infty$ and has always $p_z > 0$,
\item[ii)] for $-m_0 < p_z(-\infty) < m_0$ the path goes from $z = -\infty$ to the point $\bar z$ in which $p_z(\bar z) = 0$ (i.e.~the point that solves the equation $m(\bar z) = p_z(-\infty)$) and then goes back to $z = -\infty$,
\item[iii)] for $p_z(-\infty) \leq -m_0$ the path goes from $z = +\infty$ to $z = -\infty$ and has always $p_z < 0$.
\end{itemize}
The three classes of curves are shown schematically in fig.~\ref{fig:flow_paths} for the choice $m(z) \propto 1 + \tanh(z/L)$, with $L$ denoting the wall thickness.
The paths of type i, ii and iii correspond to the red, green and purple curves respectively.\footnote{We stress that the approximation, typically used in the literature, in which the $(m^2)'/(2E) \partial_{p_z} \delta f$ term
is neglected in the Boltzmann equation could lead to an inaccurate result in our approach. Neglecting that term, in fact, modifies
the flow paths making all of them straight lines with fixed $p_z$. This completely changes the shape of the curves
in the region $|p_z| \leq m_0$, thus potentially giving a very different solution of the equation.
}

The three different classes of paths have an additional intuitive physical interpretation. Red and purple curves correspond to those particles that have enough momentum along the $z$ direction to cross the DW. Particles coming from the symmetric phase give momentum to the DW, thus after the crossing $p_z$ is reduced. On the other hand particles in the broken phase take away momentum from the DW when they cross it. Finally green paths correspond to those particles that start in the symmetric phase but do not have enough momentum to cross the DW and are thus reflected. This three different classes of particles are exactly the ones considered in ref.~\cite{Dine:1992wr} for a simplified computation of the out-of-equilibrium friction.

By exploiting the flow paths we can immediately solve any differential equation of the form
\begin{equation}
\label{eq:boltzmann_Qform}
\left(\frac{d}{dz} - \frac{\cal J}{p_z}\right) \delta f = {\cal S}\,,
\end{equation}
where ${\cal J}$ and ${\cal S}$ are generic functions of $E$, $p_z$ and $z$. This is exactly the form of the Boltzmann equation in eq.~(\ref{eq:boltzmann_equation_at_ewpt}), with ${\cal J} = {\cal Q}f_v/f'_v$ and with the bracket term encompassed in the source ${\cal S}$. The main caveat is that the bracket depends on the unknown perturbation. To appropriately address this aspect we will eventually adopt an iterative procedure where the bracket term refines the solution through successive iterations.

The general solution to eq.~(\ref{eq:boltzmann_Qform}) is
\begin{equation}\label{eq:res_S}
\delta f = \left[B(p_\bot, p_z^2 + m(z)^2) + \int_{\bar z}^z e^{-{\cal W}(z')} {\cal S}\, dz'\right]e^{{\cal W}(z)}\,,
\end{equation}
where ${\cal W}$ is given by
\begin{equation}
\label{eq:W_function}
{\cal W}(z) = \int^z \frac{\cal J}{p_z} dz'\,,
\end{equation}
and all the integrals are evaluated along the flow paths. Because the solution involves the difference between ${\cal W}$ evaluated at two different points the lower integration boundary in the definition of
${\cal W}$ can be freely chosen (for each flow path) without affecting the result in eq.~(\ref{eq:res_S}).

The arbitrary function $B(p_\bot, p_z^2 + m(z)^2)$, which is constant along the flow paths, is fixed by enforcing
the required boundary conditions. For that, let us focus separately on the three classes of flow paths.
\begin{itemize}
\item[i)] The first type of paths describes particles that travel in the positive $z$ direction with enough momentum $p_z$ to eventually enter into the bubble. For such particles it is natural
to choose the boundary conditions in such way that $\delta f$ vanishes at $z \to -\infty$, namely that we recover the equilibrium distribution in the symmetric phase far from the DW.
This can be enforced by choosing
\begin{equation}
\label{eq:bgk_approximation_solution_firstbranch}
\delta f = \left[\int_{-\infty}^z e^{-{\cal W}}\,{\cal S}\, dz'\right]e^{{\cal W}(z)}\,.
\end{equation}

\item[ii)] Also for the second type of paths, which describes particles that initially travel in the positive $z$ direction, hit the DW and are reflected because they don't have enough momentum $p_z$ to cross the wall, 
it is natural to choose the boundary conditions similarly to what we did for the previous type of paths. Differently from the previous case, however, particles at $z = -\infty$ may travel towards the DW ($p_z >0$) or away from it after being reflected ($p_z < 0$). We enforce that the distribution function of particles that have not interacted with the DW already is the equilibrium distribution. Therefore we have
\begin{equation}
\label{eq:bgk_approximation_solution_secondbranch}
\delta f = \left[\int_{-\infty_\uparrow}^z e^{-{\cal W}}\,{\cal S}\, dz'\right]e^{{\cal W}(z)}\,,
\end{equation}
where the up arrow in the lower integration boundary indicates that the integration is performed starting from $z \to -\infty$
in the half path with $p_z > 0$.

\item[iii)] The third type of paths describes particles that travel in the negative $z$ direction, and eventually exit from the bubble. In this situation it is natural to impose the boundary conditions in such way that $\delta f$ vanishes at $z \to +\infty$. This corresponds to the requirement that before particles interact with the DW their distribution function is provided by the usual equilibrium one. This can be obtained by choosing
\begin{equation}
\label{eq:bgk_approximation_solution_thirdbranch}
\delta f = - \left[\int_z^{+\infty} e^{-{\cal W}}\,{\cal S}\, dz'\right]e^{{\cal W}(z)}\,.
\end{equation}
\end{itemize}
The consistency of all these solutions requires ${\cal J} < 0$. For the
equations we are considering, this condition is always satisfied.

The form of the solution clearly shows the role of the term $({\cal J}/p_z) \delta f$ in driving the system towards local thermal
equilibrium, i.e.~in decreasing the value of $\delta f$. In fact, the exponential factor ${\cal W}$ suppresses the impact of the source term ${\cal S}$ at large distances. Furthermore the typical decay length, which is of order $\ell \sim p_z/{\cal J}$ provides an estimate of the free path of particles depending on their momenta. As expected, when the collision processes are efficient, namely when collision rates are large, the value of ${\cal J}$ is larger and the decay length is shorter. 

\section{Solution of the Boltzmann equation}

The goal of this section is to present the strategy that we adopt to solve the Boltzmann equation. As we already anticipated, the method consists in an iterative procedure where the source term in the Boltzmann equation is refined by the bracket through successive iterations. In addition, to compute the bracket we manipulate the collision integral as in ref.~\cite{DeGroot:1980dk}. Further details of the procedure are reported in Appendix~\ref{app:evaluation_collision_integrals}.

Before we present our strategy to solve the full Boltzmann equation, we analyze two simplified setups where it is possible to track an analytic behaviour of the solution, namely the collision-less limit and the modified Bhatnagar-Gross-Krook (BGK) approximation. This has a two-fold aim. On the one hand it supplements details regarding the role of the collision integral in driving the system towards equilibrium. This is particularly emphasized by the collision-less Boltzmann equation whose solution is discontinuous if we require the system to recover equilibrium far from the bubble. On the other hand, the modified BGK approximation provides additional information regarding the properties of the solution, in particular in the region where the multipole expansion is not a suitable choice for the study of the perturbation.

\subsection{Collisionless limit}

The first limit we discuss is the collisionless limit. In this regime we neglect the collision operator and the Boltzmann equation takes the following form
\begin{equation}
\label{eq:boltzmann_collisionless}
    \frac{d}{dz}f = 0\,,
\end{equation}
where we already wrote the Liouville operator as a derivative along the flow paths.
Notice that in this case it is possible to solve the full Boltzmann equation without linearizing it and a solution is given by any function that is constant along the paths. Since the quantities $p_\bot$ and $c = \sqrt{p_z^2 + m^2(z)}$ are constants along the flow paths we find that any distribution $f$ of the form
\begin{equation}
    f = f\left(p_\bot\,,\; c = \sqrt{p_z^2 + m^2(z)}\right)\,,
\end{equation}
solves eq.~(\ref{eq:boltzmann_collisionless}).

In order to determine the solution we need to impose the proper boundary conditions. These are given by the requirement that particles moving towards the DW recover the usual equilibrium distribution function far from the bubble. For particles represented by paths that start in the symmetric phase, namely those with $p_z > 0$ at $z = -\infty$ we find that the solution is thus given by
\begin{equation}
\label{eq:free_particle_solution_firstbranch}
    f = \frac{1}{e^{\beta\gamma\left(\sqrt{p_\bot^2 + p_z^2 + m^2(z)} - v\sqrt{p_z^2+m^2(z)}\right)}\pm 1}\,.
\end{equation}
Notice that the case $p_z > 0 $ at $z = -\infty$ also includes those paths that describe reflected particles. In particular it is not hard to show that particles with $p_z > - \sqrt{m^2_0- m^2(z)}$, with $m_0 = m(z = +\infty)$ are either reflected or cross the DW. This implies that eq.~(\ref{eq:free_particle_solution_firstbranch}), is valid as long as $p_z > - \sqrt{m^2_0- m^2(z)}$. On the other hand, particles with $p_z < 0$ at $z = +\infty$ 
are described by those paths where $p_z < - \sqrt{m_0^2 - m^2(z)}$, and in such a case the solution is given by
\begin{equation}
\label{eq:free_particle_solution_secondbranch}
   f = \frac{1}{e^{\beta\gamma\left(\sqrt{p_\bot^2 + p_z^2 + m^2(z)} + v\sqrt{p_z^2+m^2(z) - m_0^2}\right)}\pm 1}\,.
\end{equation}

The solution to eq.~(\ref{eq:boltzmann_collisionless}) is then given by
\begin{equation}
    f = \begin{cases}
        \frac{1}{e^{\beta\gamma\left(\sqrt{p_\bot^2 + p_z^2 + m^2(z)} - v\sqrt{p_z^2+m^2(z)}\right)}\pm 1}\;\;\;\;\;\;\; &p_z < -\sqrt{m_0^2-m^2(z)}\\
        \frac{1}{e^{\beta\gamma\left(\sqrt{p_\bot^2 + p_z^2 + m^2(z)} + v\sqrt{p_z^2+m^2(z) - m_0^2}\right)}\pm 1}\;\;\;\;\;\;\; & p_z > -\sqrt{m_0^2-m^2(z)}
    \end{cases}
\end{equation}
The above solution is clearly discontinuous on the flow path $p_z = - \sqrt{m_0^2 - m^2(z)}$ and coincides with the one found in ref.~\cite{Moore:1995si} where the solution is written in the plasma reference frame. This highlights the role of the collision operator which, as we already pointed out, is responsible for the recover of equilibrium far from the DW and provides a continuous solution of the Boltzmann equation.




\subsection{The BGK approximation}

As we showed in eq.~(\ref{eq:collision_operator_structure}) we can split the contribution of the collision operator in two terms, one where the perturbation is not integrated which is easy to manage, and another one where the perturbation is integrated, that we denoted with the brackets $\langle\delta f\rangle$ and is instead much more complicated to deal with. Although for a complete solution we must include also the bracket terms, it is very useful to discuss an approximation of the Boltzmann equation which proves to be helpful for the algorithm that solves the equation. It consists in a modified version of the BGK approximation where we write the collision operator as 
\begin{equation}
    \bar{\cal C}[\delta f] = -{\cal J}\delta f\,.
\end{equation}
While in the BGK approximation the function ${\cal J}$ is constant and is proportional to the inverse of the mean free path of particles, we relax the latter assumption by allowing ${\cal J}$ to depend on the momentum of the particle.

The Boltzmann equation in such a limit takes exactly the form we provided in eq.~(\ref{eq:boltzmann_Qform}). Since, as we previously pointed out, the equations that we consider have ${\cal J} < 0$ we can apply the solution provided by eq.~(\ref{eq:bgk_approximation_solution_firstbranch}), (\ref{eq:bgk_approximation_solution_secondbranch}) and (\ref{eq:bgk_approximation_solution_thirdbranch}). Clearly, it is impossible to provide an analytical solution to the equations for every value of the momenta and position although we can extract some useful limits for the numerical analysis.

The limit we analyze is the one where ${\cal J}$ is constant along the flow paths. This does not imply necessarily that ${\cal J}$ is a constant function, instead the following results applies also in those situations where ${\cal J}$ is a function of $p_\bot$ and the energy $E$. Although to our knowledge it is not possible to provide an analytical expression for of the perturbation $\delta f$ even in this case, in this limit it is possible to express the function ${\cal W}$ defined in eq.~(\ref{eq:W_function}) analytically. By carring out the integration we find 
\begin{equation}
    {\cal W} = \sign(p_z){\cal J}\left({\cal I}_1(p_z,z)\, +\, {\cal I}_2(p_z,z)\right)
\end{equation}
with
\begin{equation}
    {\cal I}_1(p_z,z) = -\frac{\log \left(c \left(1 + e^{-2 z}\right) \left(\left| p_z\right| +c\right)\right)}{2
   c}
\end{equation}
and
\begin{equation}
     {\cal I}_2(p_z,z) =\begin{cases} \displaystyle\frac{\log \left(\left(e^{2z}+1\right) \left( \left| p_z\right|\sqrt{c^2-m_0^2} +c^2-m_0 m(z)\right)\right)}{2 \sqrt{c^2-m_0^2}}\;\;\;\;\;\; \left|p_z\right| > \sqrt{m_0^2 - m^2(z)}\\\\
     \displaystyle\frac{\arctan\left(\frac{\left| p_z\right| \sqrt{m_0^2-c^2}}{c^2-m_0 m(z)}\right)}{2 \sqrt{m^2_0-c^2}}  \;\;\;\;\;\sqrt{m_0m(z) - m^2(z)}<\left|p_z\right| < \sqrt{m_0^2 - m^2(z)}\\\\
     \displaystyle\frac{\arctan\left(\frac{\left| p_z\right| \sqrt{m_0^2-c^2}}{c^2-m_0 m(z)}\right) + \pi}{2 \sqrt{m^2_0-c^2}}\;\;\;\;\;\;\;\;\;\;\;\;\;\;\;\;\;\;\;\; \left|p_z\right| < \sqrt{m_0m(z) - m^2(z)}
     \end{cases}\,.
\end{equation}
The analytical expression of ${\cal W}$ provides some useful insights on the solution. Focusing on paths with $\sqrt{p^2_z+m(z)^2} > m_0$ it is not hard to show that for $z\rightarrow \pm \infty$ the ${\cal W}$ function is
\begin{equation}
\label{eq:W_largez}
    {\cal W} = {\cal J}\frac{z}{p_z}\,.
\end{equation}
This asymptotic behaviour is easily justified from a physical point of view. The limit $z\rightarrow\pm\infty$ corresponds to the case where particles are far from the DW and thus the flow path is a horizontal line. Since $p_z$ does not change much in this regime the integral in eq.~(\ref{eq:W_function}) is trivial and yields the results in eq.~(\ref{eq:W_largez}).

The function ${\cal W}$ is approximately given by the expression provided by eq.~(\ref{eq:W_largez}) also in another limit. As Fig.~\ref{fig:flow_paths} shows the flow path are almost flat not only far from the DW but also in the case where $p_z$ is large, namely $|p_z| \gg \sqrt{m^2_0 - m^2(z)}$. In this situation, the derivative $\partial/\partial p_z$ in the Liouville term is negligible and the flow derivative $d/dz$ coincides with the standard derivative $\partial_z$ reducing the linearized Boltzmann equation to
\begin{equation}
\label{eq:boltzmann_largepz}
    \delta f'-\frac{{\cal J}}{p_z}\delta f = \gamma_w v_w (-f'_v)\frac{(m^2(z))'}{2Tp_z}\,,
\end{equation}
where in the above expression we made the source term explicit. Notice that the above expression has the same structure of eq.~(\ref{eq:fluid_equation_top}) and can be solved in the exact same way. It is not hard to show that the Green function of eq.~(\ref{eq:boltzmann_largepz}) is given by
\begin{equation}
    G(p_z,z-z_0) = -\sign\left(\frac{\cal J}{p_z}\right)\exp\left(\frac{\cal J}{p_z}(z-z_0)\right)\theta\left((z_0-z)\sign\left(\frac{\cal J}{p_z}\right)\right)\,,
\end{equation}
and the resulting perturbation is given by
\begin{equation}
    \delta f(p_\bot, p_z, z) = \gamma_w v_w\int_{-\infty}^{\infty} dz' (-f'_v) G(p_z, z'-z) \frac{(m^2(z))'}{2Tp_z}\,.
\end{equation}
The above equation for ${\cal J} < 0$ reduces to eq.~(\ref{eq:bgk_approximation_solution_firstbranch}) for $p_z > 0 $ and eq.~(\ref{eq:bgk_approximation_solution_thirdbranch}) for $p_z < 0$ where ${\cal W}$ is given by eq.~(\ref{eq:W_largez}). The equilibrium distribution function $f_v$ depends on the position $z'$ through the mass profile. However, in the limit we are considering, we can assume that the mass profile is essentially flat and the term can be factored out of the integral.
Therefore we can provide an analytical expression for the perturbation $\delta f$ in terms of the function $J$ defined in eq.~(\ref{eq:Hypergeometric_part}) given by
\begin{equation}
\label{eq:largepz_solution}
    \delta f = \gamma_w v_w\frac{m^2(h_T)}{4Tp_z}f'_v\,\sign\left(\frac{\cal J}{p_z}\right)J\left(\frac{\cal J}{p_z},\frac{z}{L}\right)\,.
\end{equation}

The above expression is of particular relevance in order to understand the analytical properties of the Boltzmann equation. In fact, despite we did not include the bracket terms to find the expression, the solution can still be adopted to infer the behaviour of the perturbations where flow paths are essentially flat. This situation is realized in a complementary region where the small momentum expansion we provided in eq.~(\ref{eq:lmode_large_collision_rate}) is valid. Thus, instead of the multipole expansion, we can provide some insights on the analytical behaviour of the solution by using eq.~(\ref{eq:largepz_solution}) as confirmed by our numerical analysis. 

\subsection{Bracket inclusion}\label{sec:solution}

Although the BGK approximation and the large collision rate provide a good estimate of the behaviour of the solution that can be 
used to get some theoretical insights on the properties of the out-of-equilibrium perturbations, an accurate estimate of the integrated friction, and hence of the DW velocity, requires to solve the full Boltzmann equation.
The full Boltzmann equation is not of the form of eq.~(\ref{eq:boltzmann_Qform}), we can nevertheless use the latter to implement an approximation
by steps that eventually converges to the final solution. As we showed in eq.~(\ref{eq:collision_operator_structure}), the basic idea is to split the collision integral $\overline{\cal C}[\delta f]$ in two pieces: a term analogous to $({\cal J}/p_z) \delta f$ in eq.~(\ref{eq:boltzmann_Qform}), and a second term, namely the bracket term, that is included in the source term ${\cal S}$ and is used to correct the solution through iterations.


Let us now analyze the bracket term in more detail. This term consists in a nine-dimensional integral that includes the terms in which $\delta f$ depends on one of the three integration variables, namely $k$, $p'$ and $k'$.
The numerical determination of such a term is computationally very expensive, making the solution of the Boltzmann equation impracticable if a proper simplification of the bracket is not performed.

A fast and accurate computation of the bracket term can be achieved by first performing the integration over the six variables that do not appear as argument of the unknown perturbation $\delta f$. Following the strategy outlined in Appendix~\ref{app:evaluation_collision_integrals}, it is possible to reduce the bracket term as an integral operator acting on the perturbation $\delta f$, where the kernel of such operator is the result of the first six integrations. The resulting kernel, which is determined only by the relevant processes in the plasma, is computed once and can be reused in later computations allowing to save a large amount of time.


Applying such procedure and assuming the particles to be massless inside the collision integral, the bracket term takes the following form~\cite{DeCurtis:2022hlx}
\begin{equation}\label{eq:prel_prel_bracket}
\begin{split}
    \langle \delta f\rangle = -\int\frac{d^3\bar{\bf k}}{2|\bar{\bf k}|} & \bigg[ f_0(\beta(z)|\bar{\bf k}|) \, {\cal K}_A(\beta(z)|\bar{\bf p}|, \beta(z)|\bar{\bf k}|,\theta_{\bar p \bar k}) \\
    &-  (1-f_0(\beta(z)|\bar{\bf k}|)) \, {\cal K}_S(\beta(z)|\bar{\bf p}|, \beta(z)|\bar{\bf k}|,\theta_{\bar p \bar k})\bigg]\frac{\delta f(k_\bot, k_z,z)}{f'_0(\beta(z)|\bar{\bf k}|)}\,,
\end{split}
\end{equation}
where we recall that barred momenta are computed in the local plasma reference frame, $f_0$ denotes the standard (Fermi--Dirac or Bose--Einstein) distribution functions, and
$\theta_{\bar p \bar k}$ is the relative angle in the local plasma reference frame between the particles with momentum $p$ and $k$. The momenta $k_\bot$ and $k_z$, defined in the wall reference frame, can be expressed as functions of the momenta measured in the local plasma reference frame 
through a boost along the $z$-axis, namely $k_z = \gamma_p(z)(E_{\bar k} + v_p(z){\bar k}_z)$, $k_\bot = {\bar k}_\bot$. 
The functions ${\cal K}_A$ and ${\cal K}_S$ are the annihilation and scattering kernels (we refer to Appendix~\ref{app:evaluation_collision_integrals} for the details)
\begin{equation}\label{eq:kernel_eq}
\begin{split}
    {\cal K}_A &= \frac{1}{8N_p(2\pi)^5}\int\frac{d^3{\bf k}'d^3{\bf p}'}{2E_{p'}2E_{k'}}|{\cal M}_A|^2(1\pm f_v(p'))(1\pm f_v(k'))\delta^4(p+k-p'-k')\\
    {\cal K}_S &= \frac{1}{8N_p(2\pi)^5}\int\frac{d^3{\bf k}d^3{\bf k}'}{2E_{k}2E_{k'}}|{\cal M}_S|^2f_v(k)(1\pm f_v(k'))\delta^4(p+k-p'-k') \,
\end{split}
  \end{equation}
where the matrix elements are reported in Table~\ref{tab:amplitudes}.
Using the properties of the equilibrium distribution function\footnote{$(1\pm f_0(x)) = e^x f_0(x)$ } and defining ${\cal K}(|\bar{\bf p}|,|\bar{\bf k}|,\theta_{\bar p\bar k }) = {\cal K}_A(|\bar{\bf p}|,|\bar{\bf k}|,\theta_{\bar p\bar k })- \exp(|\bar{\bf k}|){\cal K}_S(|\bar{\bf p}|,|\bar{\bf k}|,\theta_{\bar p\bar k })$ eq.~(\ref{eq:prel_prel_bracket}) written as
\begin{equation}\label{eq:prel_bracket}
    \langle\delta f\rangle = -\int\frac{d^3\bar{\bf k}}{2|\bar{\bf k}|} f_0(\beta(z)|\bar{\bf k}|) \, {\cal K}(\beta(z)|\bar{\bf p}|, \beta(z)|\bar{\bf k}|,\theta_{\bar p \bar k})\frac{\delta f(k_\bot, k_z,z)}{f'_0(\beta(z)|\bar{\bf k}|)}
\end{equation}

To take into account the effect of the bracket term we include it in the source of eq.~(\ref{eq:boltzmann_Qform}) and employ an iterative procedure that eventually converges to the solution of the equation. For that we first make an initial ansatz on the solution. A good ansatz is obtained by using the modified BGK approximation where we model the collision integral as
\begin{equation}
    \bar{\cal C}[f] = 2{\cal Q}_A\frac{f_v}{f_v'}\delta f\,,
\end{equation}
where with the subscript $A$ in ${\cal Q}$ we emphasize that the latter quantity is computed considering only the annihilation process.
The solution at the next step of the iteration is then found by solving eq.~(\ref{eq:boltzmann_Qform}) by refining the source term through evaluating the bracket using the solution at the previous step.

\section{Spectral decomposition of the collision operator}\label{sec:decomposition}

The algorithm we developed in the previous section allows us to solve the Boltzmann equation accurately with a moderate amount of time resources. From a numerical point of view the most challenging term is the bracket $\langle \delta f\rangle$ that is a three dimensional integral operator whose evaluation can be time-consuming. However, it is possible to provide a huge improvement of the timing performances by leveraging the collision operator symmetries.

In Chapter~\ref{ch:effective_kinetic_theory} we showed that that the collision operator is symmetric under rotation with the main consequence that the kernel is block diagonal on the basis of the Legendre polynomials. This was explicitly shown in eq.~(\ref{eq:block_diagonal_kernel}) for the kernel ${\cal K}$ in eq.~(\ref{eq:prel_bracket}). As a consequence of the rotation symmetry of the whole system around the $z$ axis, to perform the integration in the bracket it is convenient to use the spherical coordinates $\{|\bar{\bf k}|, \theta_{\bar k}, \phi_{\bar k}\}$, where $\theta_{\bar k}$ is the polar angle between $\bar {\bf k}$ and the $z$, direction because the perturbation $\delta f$ is independent on the azimuthal angle $\phi_{\bar k}$. Performing this last integration and decomposing the perturbation $\delta f$ as in eq.~(\ref{eq:pert_legendre}) over the basis of the Legendre polynomials allows us to write the bracket as in eq.~(\ref{eq:bracket_term}), namely
\begin{eqnarray}\label{eq:bracket_eq}
    \langle\delta f\rangle & = & \frac{2\pi}{\beta (z)^2}\sum_{l}\frac{2l+1}{2}\int\frac{|\bar{\bf k}|d|\bar{\bf k}|}{2}f_0(|\bar{\bf k}|){\cal G}_l(|\bar{\bf p}|,|\bar{\bf k}|)\frac{\psi_l(|\bar{\bf k}|/\beta(z),z)}{f'_0(|\bar{\bf k}|)} P_l(\cos\theta_{\bar p})\nonumber\\
    & \equiv &\frac{\pi}{\beta (z)^2}\sum_{l}\frac{2l+1}{2} \int {\cal D} \bar{k}\, {\cal G}_l(|\bar{\bf p}|,|\bar{\bf k}|)\,\frac{\psi_l(|\bar{\bf k}|/\beta(z),z)}{f'_0(|\bar{\bf k}|)} P_l(\cos\theta_{\bar p})\,,
\end{eqnarray}
where we recall that $\psi_l$ and ${\cal G}_l$  are the Legendre modes of the perturbation and the kernel respectively defined in eq.~(\ref{eq:pert_legendre}) and~(\ref{eq:block_diagonal_kernel}). We further introduced the notation ${\cal D} \bar{k} \equiv f_0(|\bar{\bf k}|) |\bar{\bf k}|d|\bar{\bf k}|$ and we performed the following local change of variables
\begin{equation}
    \beta(z) k\rightarrow k \;\;\;\;\;\;\;\;\;\; \beta(z) p\rightarrow p \,,
\end{equation}
to remove the position dependence of the kernel determined by the temperature profile, i.e.~ 
from the $\beta (z)$ factors.

As we pointed out in eq.~(\ref{eq:Ol_defintion}) the one dimensional integration in the bracket term that in eq.~(\ref{eq:bracket_eq}) can be reinterpreted as the action of a Hermitian operator on the perturbation defined as 
\begin{equation}
    {\cal O}_l[g] = \int{\cal D}\bar k\, {\cal G}_l(|\bar{\bf p}|,|\bar{\bf k}|)\,g(|{\bar{\bf k}}|)
\end{equation}
Due to particle exchange symmetry, the kernel function ${\cal G}_l$ is symmetric under the exchange $p \leftrightarrow k$, namely ${\cal G}_l(|\bar{\bf p}|,|\bar{\bf k}|) ={\cal G}_l(|\bar{\bf k}|,|\bar{\bf p}|)$. As a consequence, the operator ${\cal O}_l$ is Hermitian with respect to the scalar product
\begin{equation}\label{eq:scalar_product}
    (f, g) = \int {\cal D} \bar{k}\, f(|\bar{\bf k}|)\, g(|\bar{\bf k}|)\,.
\end{equation}
This conclusion is clearly also valid for other choices of the scalar product. The one we use, as we will see in the following, is motivated by the fact that it significantly simplifies the evaluation of the collision integral.

Thanks to its Hermiticity, the operator ${\cal O}_l$ can be diagonalized with an orthonormal basis of eigenfunctions
\begin{equation}
    {\cal O}_l[\zeta_i] = \int {\cal D}\bar{k}\, {\cal G}_l(|\bar{\bf p}|,|\bar{\bf k}|)\, \zeta_{l,i}(|\bar{\bf k}|) = \lambda_{l,i}\, \zeta_{l,i}(|\bar{\bf k}|)\,,
\end{equation}
and the kernel ${\cal G}_l$ can be rewritten as \cite{1953mtp..book.....M}
\begin{equation}
    {\cal G}_l(|\bar{\bf p}|,|\bar{\bf k}|) = \sum_i \lambda_{l,i}\, \zeta_{l,i}(|\bar{\bf p}|)\, \zeta_{l,i}(|\bar{\bf k}|)\,.
\end{equation}
The kernel ${\cal K}$ is then
\begin{equation}
    {\cal K}(|\bar{\bf p}|,\cos\theta_p,|\bar{\bf k}|,\cos\theta_k) = \pi\sum_l\sum_i\lambda_{l,i}\frac{2l+1}{2}\zeta_{l,i}(|{\bar{\bf p}}|)\zeta_{l,i}(|{\bar{\bf k}}|)P_l(\cos\theta_p)P_l(\cos\theta_k)
\end{equation}
Using this decomposition we can drastically simplify the computation of the bracket $\langle \delta f\rangle$. Indeed, once the eigenvectors are determined, the remaining integral involved in the computation becomes trivial due to the orthogonality property. The final expression for the bracket is then
    
\begin{equation}\label{eq:bracket_decomposition}
	\langle \delta f\rangle = -\frac{\pi}{\beta(z)^2}\sum_l\sum_i \lambda_{l,i}\frac{2l+1}{2}\, \phi_{l,i}(z)\, \zeta_{l,i}(|\bar{\bf p}|)P_l(\cos\theta_p)
\end{equation}
where $\phi_{l,i}(z)$ is the projection of the $l-$th Legendre mode of the perturbation on the eigenstate basis of the kernel ${\cal G}_l$, namely
\begin{equation}\label{eq:perturbation_decomposition}
    \phi_{l,i}(z) =  \int {\cal D}\bar{k}\, \zeta_{l,i}(|\bar{\bf k}|)\frac{\psi_l f(|{\bar{\bf k}}|/\beta(z), z)}{f_0'(|\bar{\bf k}|)} \,.
\end{equation}

As we will see in the next section, the main advantage of this decomposition method is the huge improvement in the timing performances in the computation of the bracket term.

\subsection{Numerical implementation}\label{subsec:numdec}

To implement numerically the spectral decomposition of the kernel, we need to choose a suitable basis of functions on which the Legendre modes $\psi_l$ of the perturbations can be expanded. A simple choice, which we use for our numerical analysis, is to discretize the momentum space on a regular finite lattice. The functional space is then obtained from the discretized version through a suitable interpolation.

Using a rectangular lattice with $M$ points the operators ${\cal O}_l$ are represented by a set of $M \times M$ Hermitian matrices ${\cal U}_l$ which can be diagonalized to obtain the spectral decomposition. We computed the matrix elements of ${\cal U}_l$ on an orthonormal basis of functions $\{e_{m}\}$ that vanish everywhere on the grid but on the point $|\bar{\bf p}|_m$.

Some subtleties must be taken into account in the discretization process. The measure we adopted for the scalar product, ${\cal D} \bar{k}$, contains a factor $|\bar{\bf k}|$, which vanishes for $|\bar{\bf k}| = 0$. This means that if a zeroth-order (i.e.~a piecewise constant) approximation of the basis functions and of the integration measure is used, the element $e_{m}$ corresponding to $|\bar{\bf p}|_m = 0$ becomes singular. 
This is not a significant problem, since also the kernels ${\cal G}_l(|\bar{\bf p}|, |\bar{\bf k}|)$ vanishes for $|\bar{\bf p}| = 0$ or $|\bar{\bf k}| = 0$.
Therefore the singular basis functions can be neglected in the spectral decomposition of the kernel and in the computation of the collision integral.

One of the non-trivial features of the kernel blocks, which is hard to reproduce, is the presence of a peak located at $|\bar{\bf p}| = |\bar{\bf k}|$ for each block as we showed in Fig.~\ref{fig:hierarchy_kernel}. Such peaks originate from the forward scattering of the incoming particles. However, because particle distributions are unaffected by forward scattering processes we expect the impact of such a peak on the non-equilibrium dynamics of the plasma to be limited.
To obtain a more uniform reconstruction of the peaks, whose width scales as $|\bar{\bf p}|$ we used a linearly-increasing spacing (i.e~a quadratic distribution) on the grid. The non-uniform spacing also provides a finer spacing at small momentum, where the kernel structure shows more complex features.

For our numerical analysis we considered the first $11$ blocks in the multipole expansion and we chose a grid with $M = 100$ (with the restriction $|\bar{\bf p}|/T \leq 20$). In this way each of the $11$ matrices ${\cal U}_l$ has dimension $100$. This implementation offers a good compromise to obtain a good accuracy in the reconstruction of the kernel and a fast computation of the bracket term.
\begin{figure}
\centering
\includegraphics[width=.47\textwidth]
{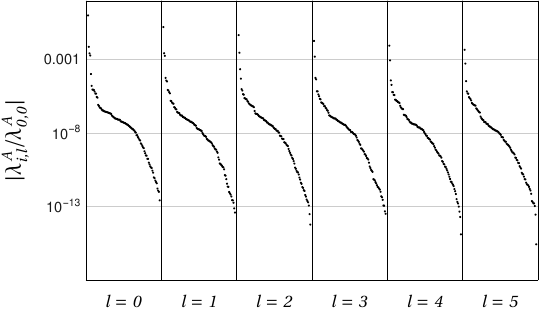}
\hfill
\includegraphics[width=.47\textwidth]
{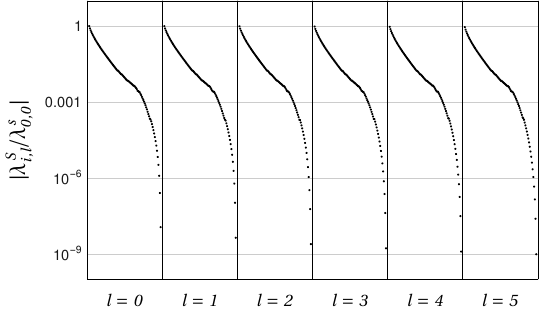}
\caption{Relative size of the eigenvalues of the kernel matrices ${\cal U}_l$ for $l = 0$, $1$, $2$, $3$, $4$, $5$. with $M = 100$ for the annihilation (left panel) and scattering (right plot) kernel. For the annihilation kernel only very few eigenvalues have a size $\gtrsim 10^{-3} |\lambda_{0,0}^A|$, while for the scattering kernel a large fraction of them (the first $\sim 90$ eigenvalues for each block) has a size $\gtrsim 10^{-3} |\lambda_{0,0}^S|$.
}\label{fig:eig_distr}
\end{figure}

We show in Fig.~\ref{fig:eig_distr} the relative size of the annihilation (left panel) and scattering (right panel) kernel eigenvalues for the first $6$ blocks. A very similar behaviour can be found for the remaining blocks in the multipole expansion. The annihilation kernel spectrum reveals a strong hierarchy between the eigenvalues. In particular, only the first $l\gtrsim 3$ blocks provide a relevant contribution in accordance to the results we showed in Fig.~\ref{fig:hierarchy_kernel}. Taking only the first $l = 3$ blocks indeed allows to reconstruct the annihilation kernel with an accuracy of order $1-2\%$ which can be easily reduced below the percent level by taking the first $6$ modes.

The eigenvalues of each block also present a very fast decrease. In the block $l = 0$ all the eigenvalues but the first four are suppressed by a $10^{-4}$ factor. The suppression is even stronger for higher terms in the mulitpole series and in the $l = 5$ block only the first two eigenvalues are not strongly suppressed. As a consequence, by taking the first $6$ blocks in the multipole expansion, in order to reconstruct the annihilation kernel with an overall accuracy of $~1\%$ it is sufficient to take just the first $2$ eigenfunctions for each block. The accuracy rapidly increases below the percent level by taking the first $5$ eigenfunctions for each block.

The scattering kernel is instead much more difficult to reproduce. As the right panel in Fig.~\ref{fig:eig_distr} shows, the blocks do not present a marked hierarchical structure as their respective eigenvalues are all of the same magnitude. For each block, differently from the annihilation case, we find that after a relatively fast decrease the size of the eigenvalues tends to decrease slowly, so that a large fraction of them as a size $\gtrsim 10^{-3}|\lambda_{0,0}^S|$. As a consequence, by taking the first $11$ blocks, in order to reconstruct the kernel with an overall accuracy of order $2-4\%$ almost all eigenvectors must be taken into account. Including in the sum the first $80$ eigenvalues from each block the typical reconstruction error is $5-10\%$ which increases above $10\%$ taking only the first $70$ eigenvalues.

We mention that significantly larger relative reconstruction errors are present in regions where the kernel is highly suppressed (such as for configurations in which $p$ is small while $k$ is large, or vice versa), or, especially for large momenta, around the peak. These regions, however have a limited impact on the computation of the collision integral, as confirmed by the numerical analysis reported in section~\ref{sec:analysis}.

When we consider the total kernel ${\cal K}$ we recover results and errors on the reconstruction similar to the scattering case, because the latter dominates over the annihilation processes. In fact we find $|\lambda^A_{0,0}| \simeq 10^{-2} |\lambda^S_{0,0}| $. The setup we provided, with $11$ blocks and a grid with $M=100$ points allows us to reconstruct the kernel with an overall $2-4\%$ accuracy. Increasing the number of blocks in the multipole expansion increases the accuracy. Taking the first $l = 21$ a $1-2\%$ accuracy can be achieved by considering all the eigenvectors of each block. However to correctly reproduce the bracket term at the $\sim 1\%$ accuracy it is sufficient to take the first $l = 11$ modes since the hierarchy in the perturbation provides a suppresion of the higher terms in the multipole expansion.

By expanding the perturbation on the Legendre polynomial basis and further using the eigenfunctions $\{\zeta_{l,i}\}$ to decompose the Legendre blocks of the perturbation $\psi_l$, the computational time of the brackets is highly reduced. Compared to the timing performance of the method presented in ref.~\cite{DeCurtis:2022hlx}, where we evaluated the bracket by performing directly the three dimensional integration, the decomposition method outlined in this section is more than two orders of magnitude faster. Such numerical improvement allows one to solve the Boltzmann equation in less than one hour on a desktop computer.

\subsection{Numerical validation}\label{sec:analysis}

To validate the method that we just described we compared it with the one we developed in ref.~\cite{DeCurtis:2022hlx}.
In particular we compared the relevant quantities that enter in the computation of the DW terminal speed, namely the out-of-equilibrium corrections to the stress-energy tensor $T^{30}_{out}$, $T^{33}_{out}$,
and the friction $F(z)$.

We choose as a benchmark scenario the $Z_2$-symmetric singlet extension of the SM we described in Chapter~\ref{ch:phase_transitions}. As we already discussed, this choice is motivated by the fact that the presence of a new scalar field $s$ affects the thermal history of the Universe and can give rise to a first-order EWPhT. We recall, from our discussion in Chapter~\ref{ch:phase_transitions} that the extra scalar, singlet under the SM gauge group, also allows for a 2SPhT, in which case the EW symmetry breaking is preceded by a $Z_2$-symmetry breaking in the extra sector. 

 Once the model parameters are chosen, the temperature and velocity profiles of the background plasma are computed by using the conservation laws of the stress-energy tensor. In addition, by solving the coupled system of the Boltzmann equation and the equations of motion of the scalar fields, one can determine the terminal speed of the DW.

\begin{figure}
    \centering
    {\includegraphics[width=.32\textwidth]{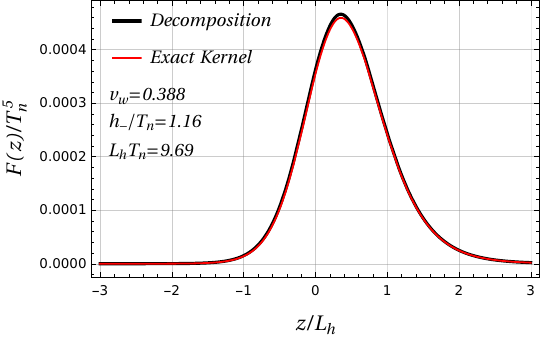}}
    \hfill
    {\includegraphics[width=.31\textwidth]{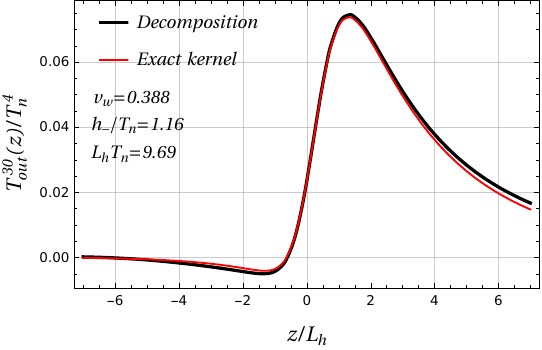}}
    \hfill
    {\includegraphics[width=.31\textwidth]{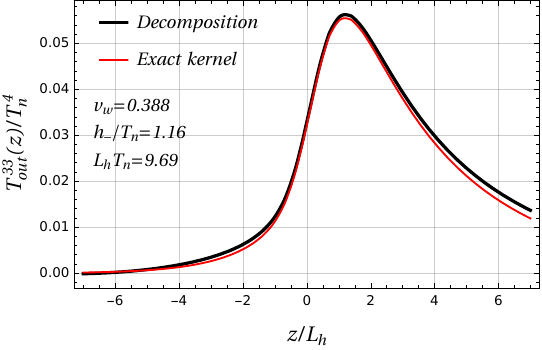}}
    \vspace{.5em}
    {\includegraphics[width=.32\textwidth]{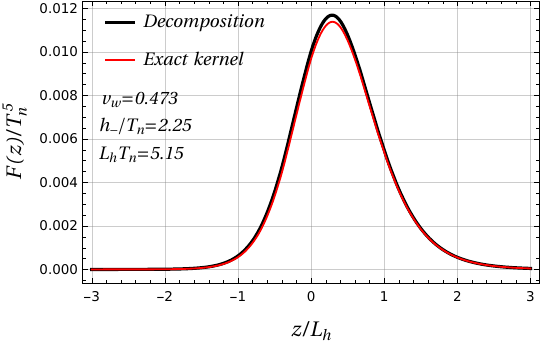}}
    \hfill
    {\includegraphics[width=.31\textwidth]{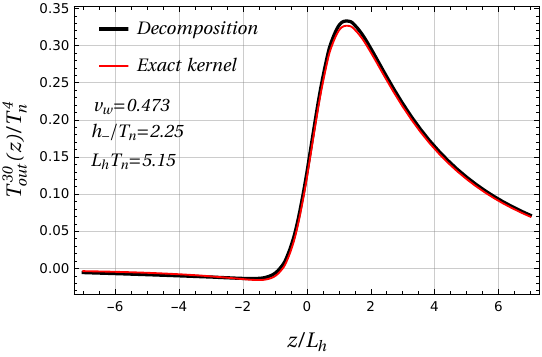}}
    \hfill
    {\includegraphics[width=.31\textwidth]{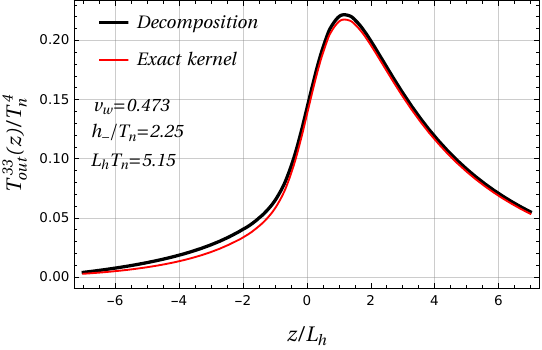}}
    \caption{Comparison of the friction (left panel), $T^{30}_{out}$ (central panel) and $T^{33}_{out}$ (right panel) as functions of $z/L_h$, computed using the procedure presented in ref.~\cite{DeCurtis:2022hlx} (red solid line) and the decomposition method (black solid line), for the benchmark point BP1 (upper row) and BP2 (lower row). The parameters defining the two benchmarks are reported in Tab.~\ref{tab:parameters_results}.}\label{fig:old_vs_new_compare}
\end{figure}

 In Fig.~\ref{fig:old_vs_new_compare} we plot $F(z)$, $T_{out}^{33}(z)$, $T_{out}^{30}(z)$ as functions of the ratio $z/L_h$, for two benchmark points
reported in Tab.~\ref{tab:parameters_results} and characterized by $v_w = 0.388$, $L_h\, T_n = 9.69$, $h_-/T_n = 1.16$ for the benchmark BP1 and by $v_w = 0.473$, $L_h\, T_n = 5.15$, $h_-/T_n = 2.25$ for BP2. As we will see in the following, such values correspond to the terminal ones obtained for a scalar potential with parameters $m_s = 103.8 \, \textrm{GeV}$, $\lambda_{hs} = 0.72$ and $\lambda_s = 1$ for BP1 and $m_s = 80.0 \, \textrm{GeV}$, $\lambda_{hs} = 0.76$ and $\lambda_s = 1$ for BP2.
The plots clearly show that the decomposition method correctly reproduces
the friction and the out-of-equilibrium corrections to the stress-energy tensor in the whole range of $z$. Although at the qualitative level an excellent agreement is found, some small quantitative differences are present. In particular, differences of order $1\%$
are present in the region corresponding to the peak ($z \sim 1$), while order $10\%$ discrepancies can be present in the tails, both inside and outside the bubble. The impact of the tails in the determination of the total friction and of the DW dynamics is however very limited, since their contribution is highly suppressed, as can be seen from the first panel in the two rows of Fig.~\ref{fig:old_vs_new_compare}.\footnote{For some choices of the model parameters, the convergence of the solution in the region $p_z<0$ and $z\ll0$ may present some numerical instabilities due to our approximation of the kernel. In this region, however, the perturbations are highly suppressed and give only a marginal contribution to the friction. Moreover possible numerical instabilities only appear after a number of iterative steps much larger than the ones needed to reach convergence in the relevant part of the perturbations, so they can be easily kept under control in the determination of the DW velocity.
}

These results thus indicate that the decomposition method is robust and can be safely used to reliably study the DW dynamics.

\section{Comparison with the OF and NF}\label{sec:numerical}


In this section we present the analysis that we carried in ref.~\cite{DeCurtis:2022hlx}. We apply the iterative method explained before to numerically solve the Boltzmann equation and we compare with the approximations discussed in Chapter~\ref{ch:boltzmann_ansatz}. To assess the validity of our method we focus for simplicity on a single species in the plasma, the top quark, which is the state with largest coupling to the Higgs
and is thus expected to provide one of the most relevant effects controlling the DW dynamics. The analysis of the top quark
distribution provides a robust framework where we can test the non-equilibrium dynamics of the plasma and determine the accuracy both qualitatively and quantitatively of the weighted approach described in Chapter~\ref{ch:boltzmann_ansatz}.
Later in the chapter we are going to include the contribution of W and Z bosons.


\subsection{Annihilation only}

It is possible to show that the non-equilibrium dynamics of the top quark is mostly affected by the annihilation process $t \bar t\to g g$. In fact, the scattering of tops on gluons and light quarks gives a smaller contribution to the collision integral. It is hence convenient, for our numerical analysis to first consider the contribution arising only from the annihilation and refine our analysis by including scattering effects afterwards.

To determine the numerical solution of the Boltzmann equation we used a dedicated C++ code, validating the results with Mathematica~\cite{Mathematica}. In a first version of our algorithm we computed the solution on a three-dimensional grid in the variables $z$, $p_\bot$ and $p_z$ restricted to the
intervals $z/L \in [-7,7]$,\footnote{The vanishing boundary conditions on the solution were imposed at the boundaries
of the considered region. We verified that this choice does not introduce a significant distortion of the solution.}
$p_\bot/T \in [0, 15]$, and $p_z/T \in [-15, +15]$. The solution was computed on a grid with $50 \times 300 \times 100$ points, which was further refined in the region $p_\bot/T < 1$ and $|p_z|\sim m_0$, where the solution showed a fast-varying behavior. Convergence of the solution (at the $\sim 0.1\%$ level) was achieved within three iterative steps for all values of the wall velocity.

We modeled the bubble wall adopting the $\tanh$ ansatz for the Higgs profile~\cite{PhysRevD.101.063525}, namely
\begin{equation}
\phi(z) = \frac{\phi_0}{2}[1 + \tanh(z/L)]\,,
\end{equation}
where we fixed $L = 5/T$ to be the thickness of the bubble wall and  $\phi_0 = 150\;{\rm GeV}$ the Higgs VEV in the broken phase. We fixed the
phase transition temperature to $T = 100\;{\rm GeV}$. This choice of parameters, as we discussed in Chapter~\ref{ch:boltzmann_ansatz}, determines the presence of friction peaks in the old formalism solution, since the system is far from the hydrodynamic regime. It is thus well suited for differentiating the various formalisms and highlighting the differences among them.

\begin{figure}
	\centering
	{\includegraphics[width=.47\textwidth]{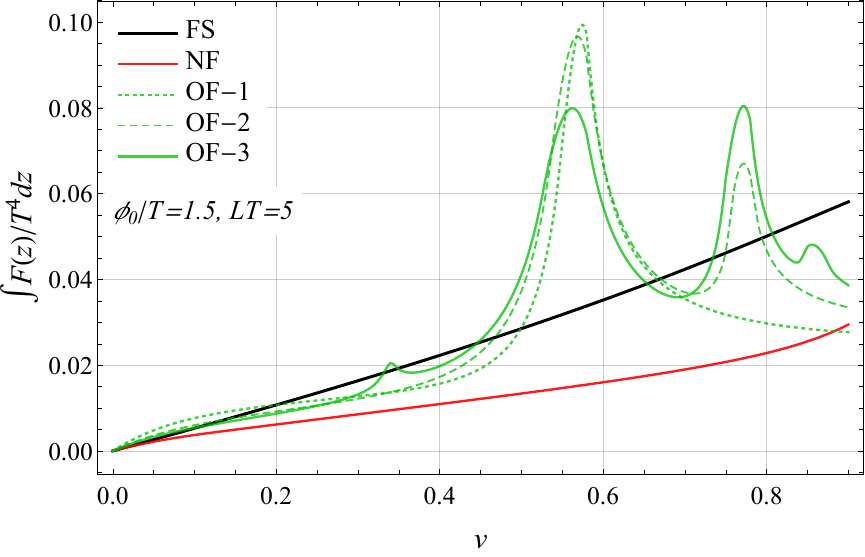}}
	\hfill
	{\includegraphics[width=.47\textwidth]{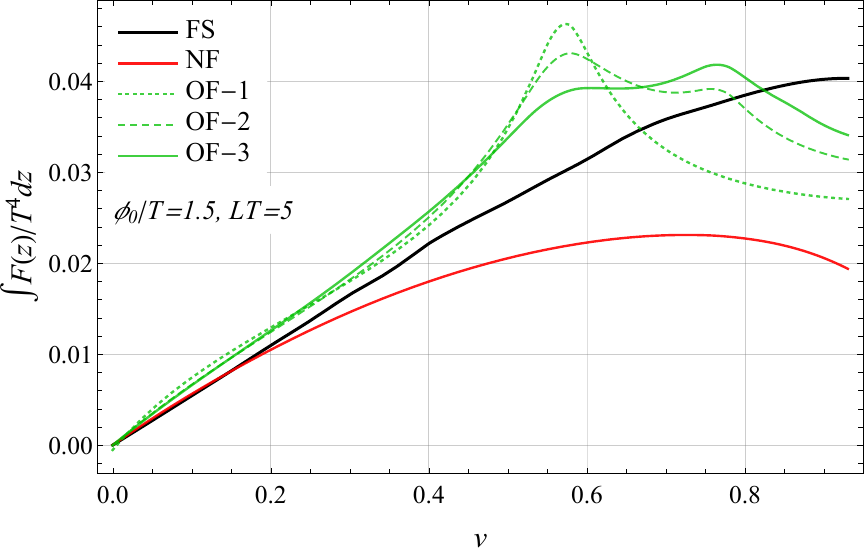}}
	\caption{Friction acting on the bubble wall as a function of the velocity. In the left plot only the top annihilation channel has been taken into account, while in the right one both annihilation and scattering are considered. The black solid line corresponds to the solution of the full Boltzmann equation (FS, our result), the dotted, dashed and solid green lines are obtained with the old formalism (OF) at order $1$, $2$ and $3$ respectively~\cite{Moore:1995si,Dorsch:2021ubz}, while the red line corresponds to the new formalism (NF)~\cite{Laurent:2020gpg}.}\label{fig:friction}
\end{figure}

To assess the impact of the non-equilibrium dynamics of the plasma on the DW we computed 
the friction given by eq.~(\ref{eq:out_of_eq_friction}) and its integral over $z$.
In the left panel of fig.~\ref{fig:friction} we show the friction integrated over $z$ as a function of the wall velocity (solid black line).
The total friction shows a smooth behavior with a (nearly) linear growth as a function of the wall velocity.

In the same plot we compare our result with the ones obtained with the weighted methods.
In particular the green lines correspond to the total friction computed in OF of ref.~\cite{Moore:1995si},
taking also into account higher-order terms in the fluid approximation~\cite{Dorsch:2021ubz}. The old formalism results
at order $1$, $2$ and $3$ are given by the dotted, dashed and solid lines respectively. The solid red line, instead, is obtained using the
NF of ref.~\cite{Laurent:2020gpg}.

To compare with our results, we included the mass dependence in the computation of the matrix $A$ of the weighted approach (see eq.~(\ref{eq:diff_matrix_of_massless}) for the definition of the matrix for a generic set of weights). 
 Our result for small and intermediate velocities, $v_w \lesssim 0.5$ is in fair numerical agreement with the old formalism ones, which
show a minor dependence on the order used for the computation. At higher velocities, instead, the old formalism develops some peaks
related to the speed of sound in the plasma and to any other zero eigenvalue of the Liouville operator. The number of peaks and their shape crucially depend on the approximation order,
denoting an intrinsic instability of the old formalism method.\footnote{Notice that the total friction shows a continuous behavior across
the sound speed thresholds, whereas in ref.~\cite{Moore:1995si} a divergence was found.
This difference was expected, since the discontinuity found in ref.~\cite{Moore:1995si} is induced by the background contributions, which are not included in our analysis.} Our results for the full solution of the Boltzmann equation show that the peaks are an artifact of
the old formalism approach and that no strong effect is present in the top contributions for velocities close to the sound speed one.

On the other hand, the new formalism correctly predicts a smooth behavior for the total friction for all domain wall velocities.
A roughly linear dependence on $v_w$ is obtained up to $v_w \simeq 0.8$, while for larger values a faster growth is found,
in contrast with the behavior of the full solution (FS) result.
The quantitative agreement with the full solution is good only for very low velocities, $v_w \lesssim 0.1$, while order $50\%$ differences
can be seen for higher velocities.

\begin{figure}
	\centering
	{\includegraphics[width=.47\textwidth]{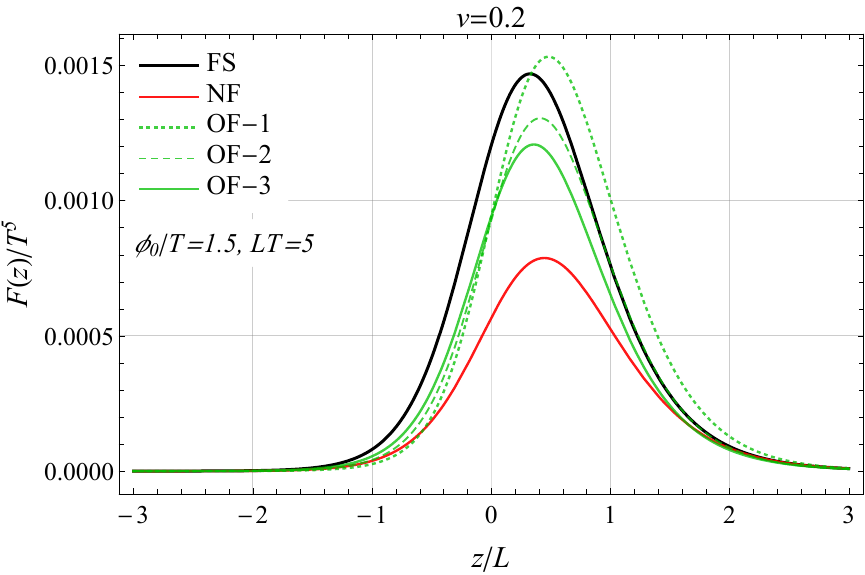}}
	\hfill
	{\includegraphics[width=.47\textwidth]{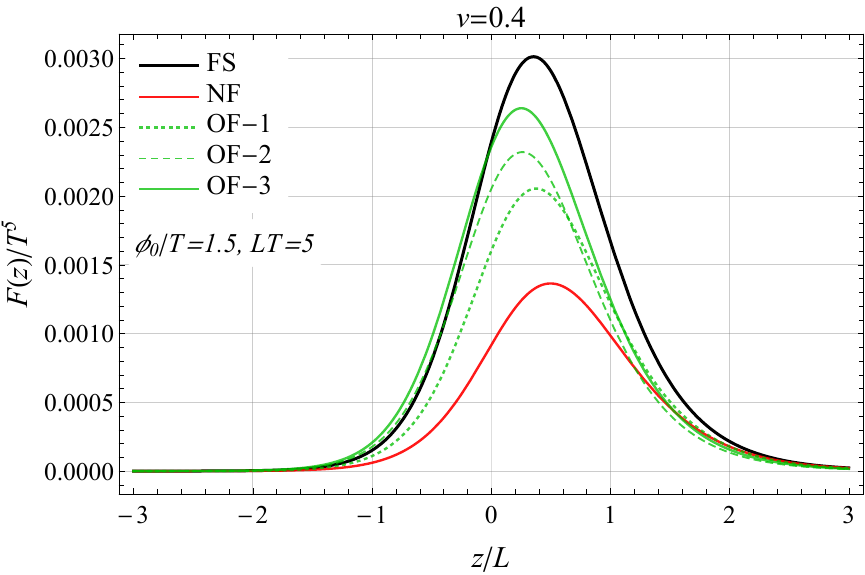}}\\
	\vspace{.5em}
	{\includegraphics[width=.47\textwidth]{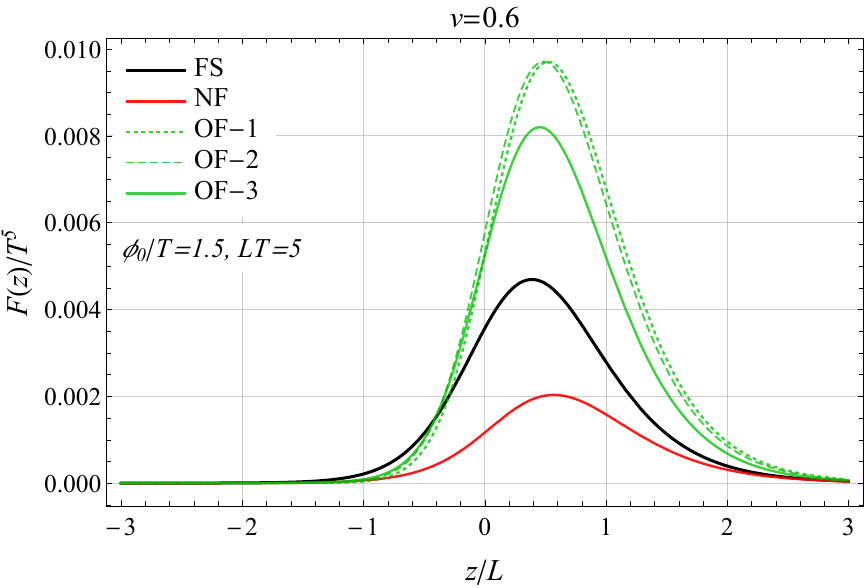}}
	\hfill
	{\includegraphics[width=.47\textwidth]{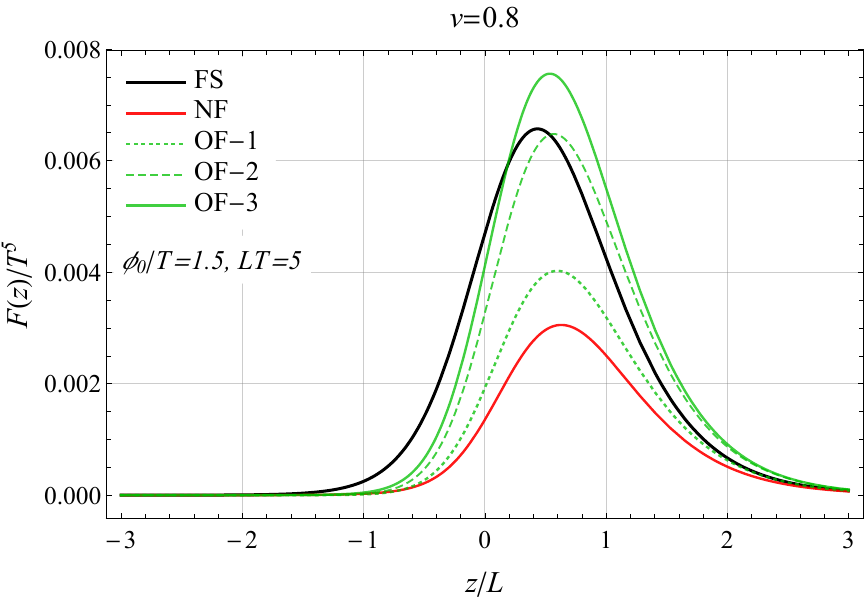}}
	\caption{Friction as a function of the position $z$ when only top annihilation processes are taken into account. The plots correspond to the wall velocities $v_w  = 0.2, 0.4, 0.6, 0.8$.}\label{fig:friction_ann}
\end{figure}

For a more refined comparison of the results we show in fig.~\ref{fig:friction_ann} the behavior of the friction $F(z)$ as a function
of the position. The plots clearly show that the overall shape of the friction is very similar in all approaches, the main difference being
the height of the peak. This property is not unexpected, since the size of the perturbation $\delta f$ is controlled by the
source term in the Boltzmann equation, whose $z$ dependence is given by $d m^2/dz$ as we already pointed out in Chapter~\ref{ch:boltzmann_ansatz}

A more detailed comparison of $\delta f$ as a function of $z$, $p_\bot$ and $p_z$ shows drastic differences among all the approaches.
Although the overall size of $\delta f$ is comparable in all formalisms (being controlled by the source term), the various solutions
significantly differ even at the qualitative level in most of the kinematic regions. We conclude from this comparison that the fluid
approximation is not reliable if we include in the Boltzmann equation only the top annihilation channel. We will see in the following that,
introducing the top scattering processes, a better agreement is found.

\subsection{Full solution}

We now consider the Boltzmann equation for the top quark distribution, including in the collision term also the main top scattering processes, namely the ones onto gluons $t g \to t g$ and onto light quarks $t q \to t q$. We found convenient to include these additional contributions treating them as source terms in the iterative steps.

To determine the numerical solution we used a grid analogous to the one described in the annihilation-only case. The convergence of the iterative procedure is somewhat slower when top scattering processes are taken into account. For $v_w \leq 0.6$ we used the
solution of the annihilation-only case as starting ansatz and we performed six iterative steps to reach a good convergence.
For higher velocities the annihilation-only solution is not a convenient choice for the first iterative step, thus we started from the
full solution determined for a lower value of $v_w$. Also in this case six iterations were sufficient to achieve convergence.

We found that the scattering processes significantly modify the solution of the Boltzmann equation, especially for large values
of the domain wall velocity ($v_w \gtrsim 0.5$). The impact on the total friction acting on the domain wall is shown in the right panel
of fig.~\ref{fig:friction}. Analogously to the annihilation-only case, an almost linear dependence on the wall velocity is present
for small and intermediate $v_w$ values, but a flattening is present at higher velocities. Quantitatively,
the scattering processes induce only minor corrections to the total friction for $v_w \lesssim 0.6$, while a decrease of order $25\%$
is found for $v_w \sim 0.8$.

The impact of the scattering processes on the solution obtained through the weighted methods is, on the contrary, much more
pronounced. The old formalism approach (green lines in fig.~\ref{fig:friction}) including only the lowest-order perturbations predicts
a strong peak for $v_w \simeq 0.57$. The peak however gets substantially reduced once higher-order perturbations are included in the
expansion, with a milder additional peak forming for $v_w \simeq 0.77$. We also point out that, differently from the annihilation case only, the peaks are much smoother when we include scattering process. This is not a suprise. As we discussed in Chapter~\ref{ch:boltzmann_ansatz} the size and sharpness of the peaks are controlled by the efficiency of the collision process which is clearly improved when the scattering processes are included. We expect that including additional higher-order perturbations
could smoothen the curve as well as considering thicker walls. This could give a qualitative behavior similar to the one we get with the full solution, with a linear behavior for $v_w \lesssim 0.6$. At the quantitative level, however, the old formalism solution differs from the one we found by
order $10 - 25\%$.

The result obtained through the new formalism (red line in fig.~\ref{fig:friction}) is also substantially modified by the scattering contributions. In particular the increase in the friction for $v_w \gtrsim 0.8$ is removed and a maximum followed by a mild decrease is now found
for $v_w \gtrsim 0.6$. The new formalism prediction is now in good quantitative agreement with our result for $v_w \lesssim 0.2$,
while differences  up to order $50\%$ are found for larger velocities.

\begin{figure}
	\centering
	{\includegraphics[width=.47\textwidth]{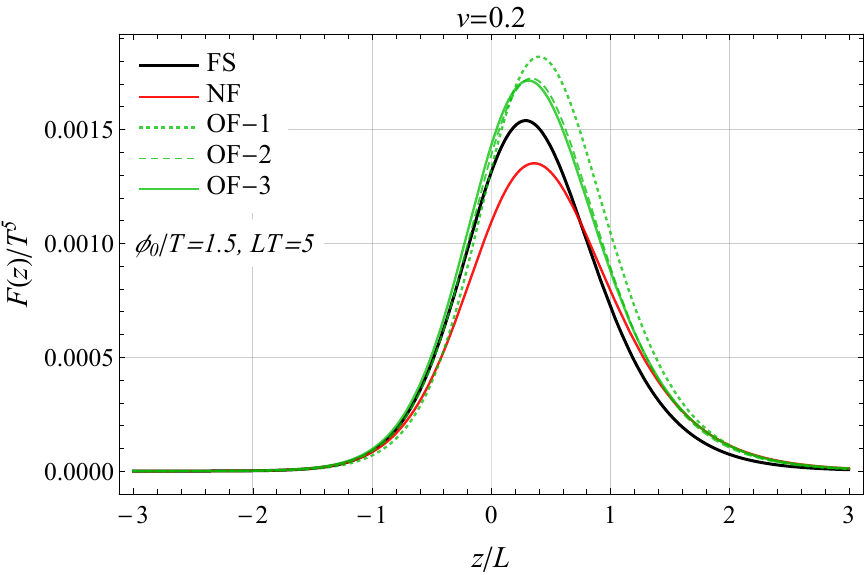}}
	\hfill
	{\includegraphics[width=.47\textwidth]{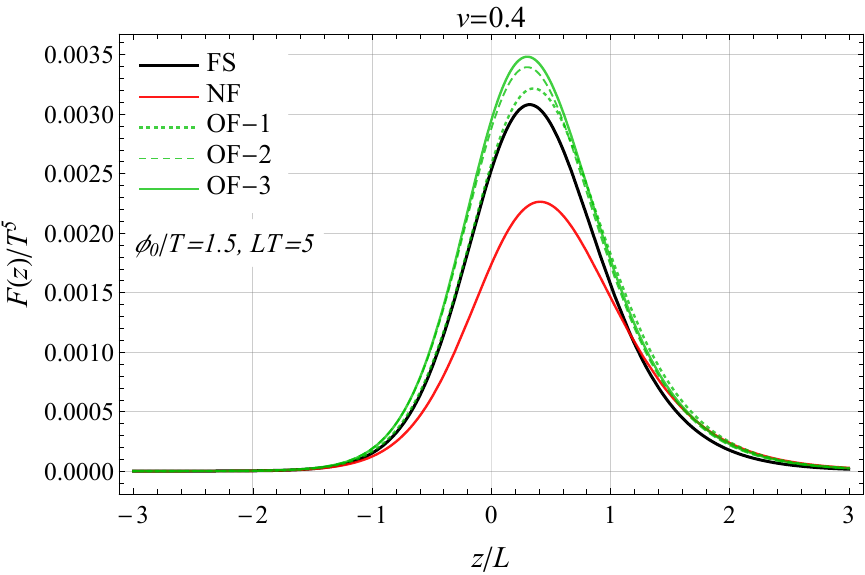}}\\
	\vspace{.5em}
	{\includegraphics[width=.47\textwidth]{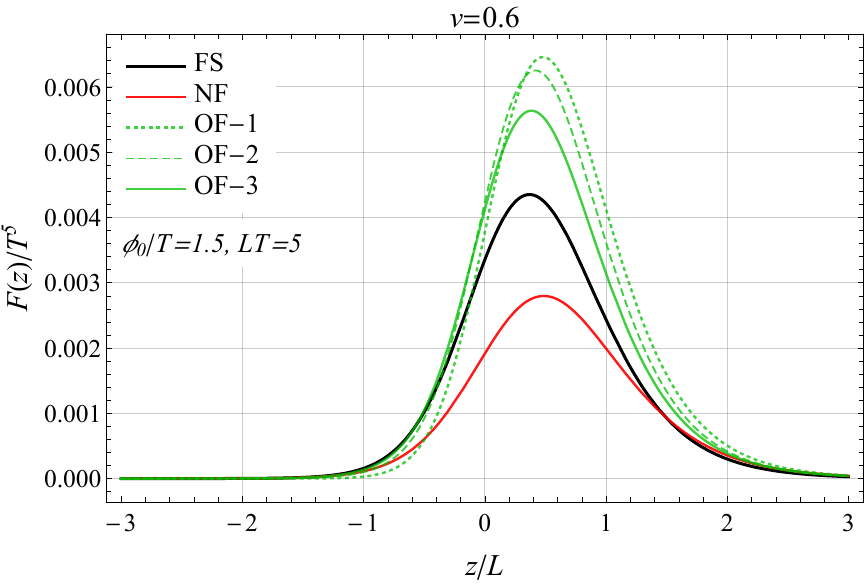}}
	\hfill
	{\includegraphics[width=.47\textwidth]{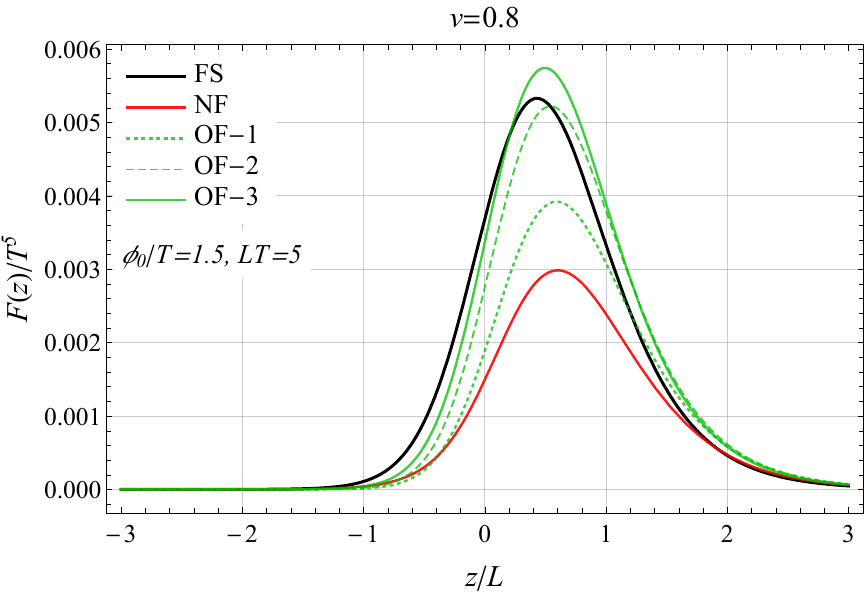}}
	\caption{Friction as a function of the position $z$ when top annihilation and scattering processes are taken into account. The plots correspond to the wall velocities $v_w  = 0.2, 0.4, 0.6, 0.8$.}\label{fig:friction_ann_scatt}
\end{figure}

The friction as a function of the position $z$ for some benchmark wall velocities is shown in fig.~\ref{fig:friction_ann_scatt}.
Analogously to what we found for the total friction, the results we obtain with our method are only mildly modified by the
scattering contributions. In particular the shape of the friction remains almost unchanged with only minor modifications in the overall
normalization. Similar considerations apply for the shape of the $z$ dependence of the friction in the old and new formalism.
In this case, however, significant changes in the overall normalization are found, as expected from the above discussion on
the total friction.

\begin{figure}
	\centering
	{\includegraphics[width=.24\textwidth]{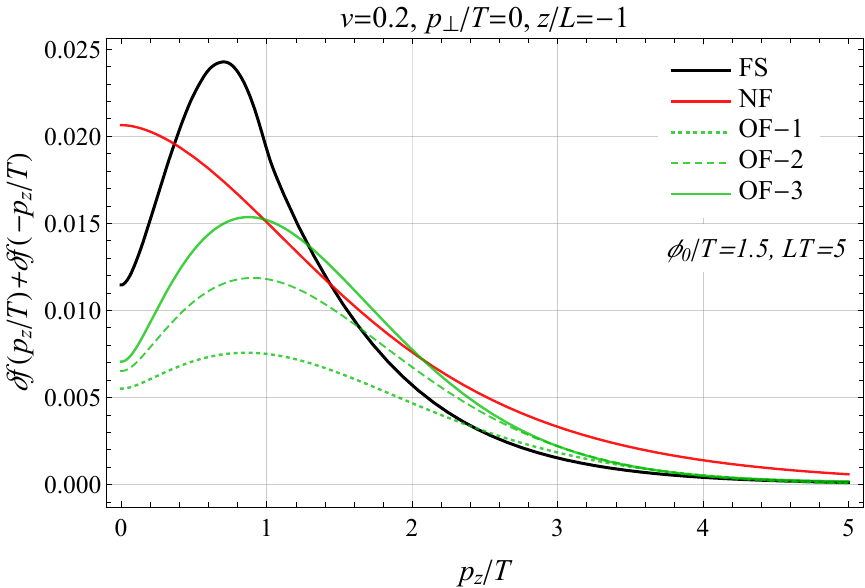}}
	\hfill
	{\includegraphics[width=.24\textwidth]{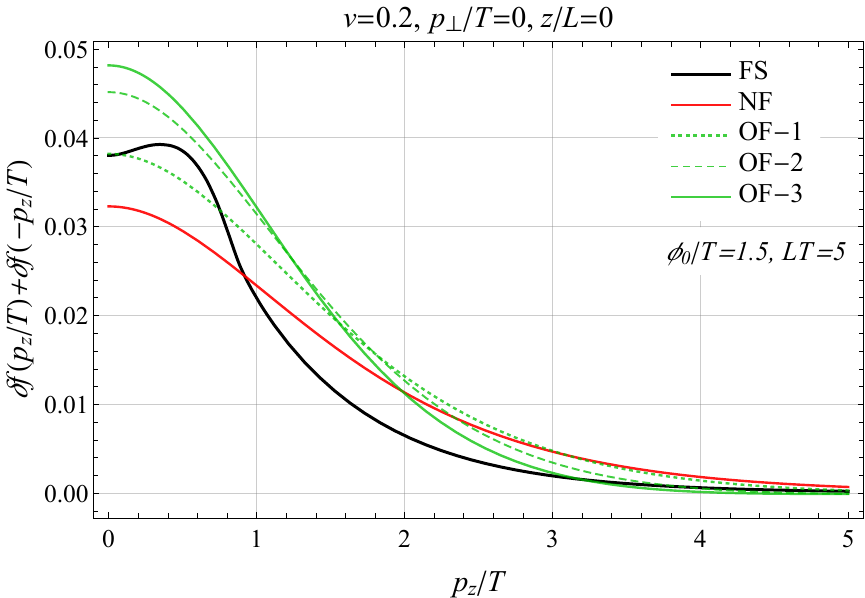}}
	\hfill
	{\includegraphics[width=.24\textwidth]{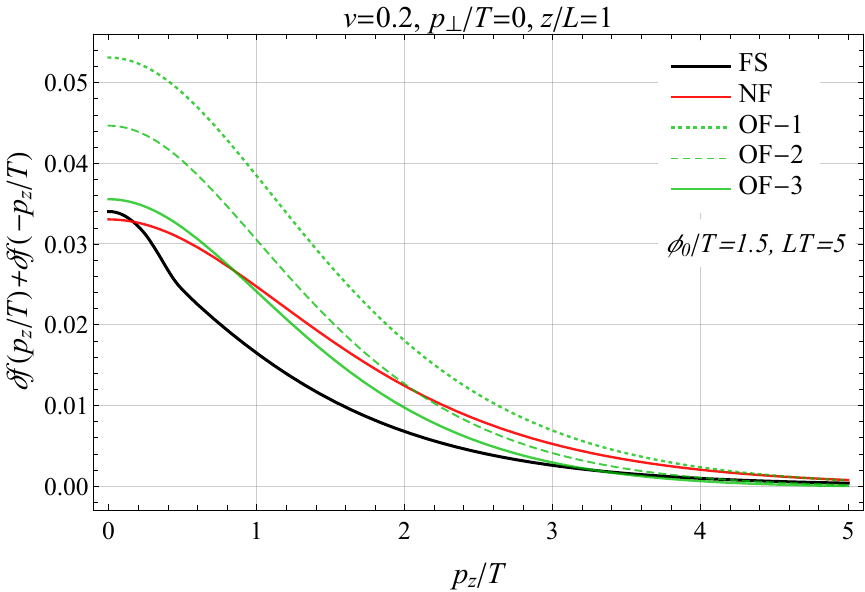}}
	\hfill
	{\includegraphics[width=.24\textwidth]{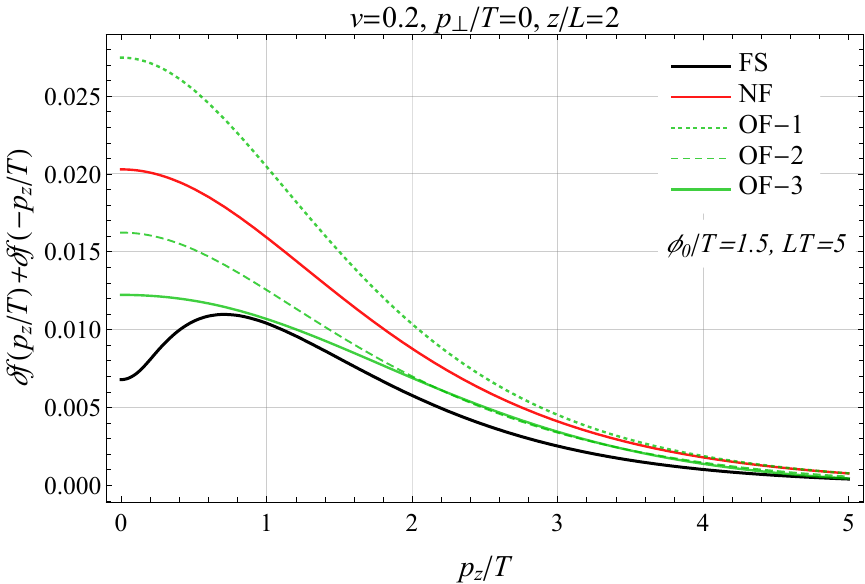}}\\
	\vspace{.5em}
	{\includegraphics[width=.24\textwidth]{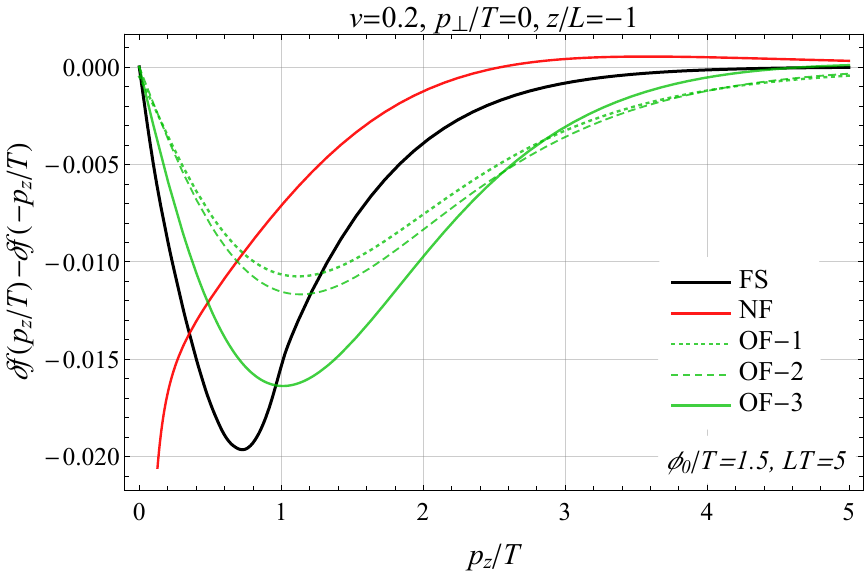}}
	\hfill
	{\includegraphics[width=.24\textwidth]{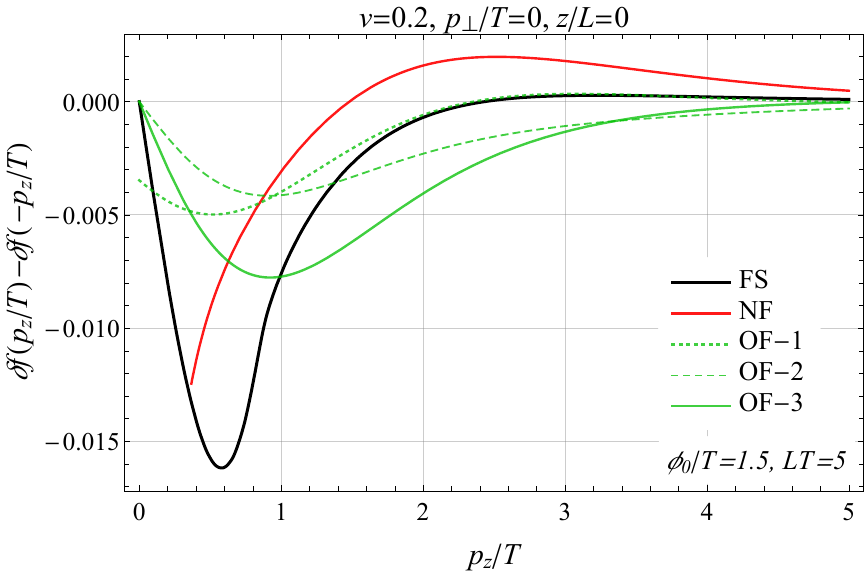}}
	\hfill
	{\includegraphics[width=.24\textwidth]{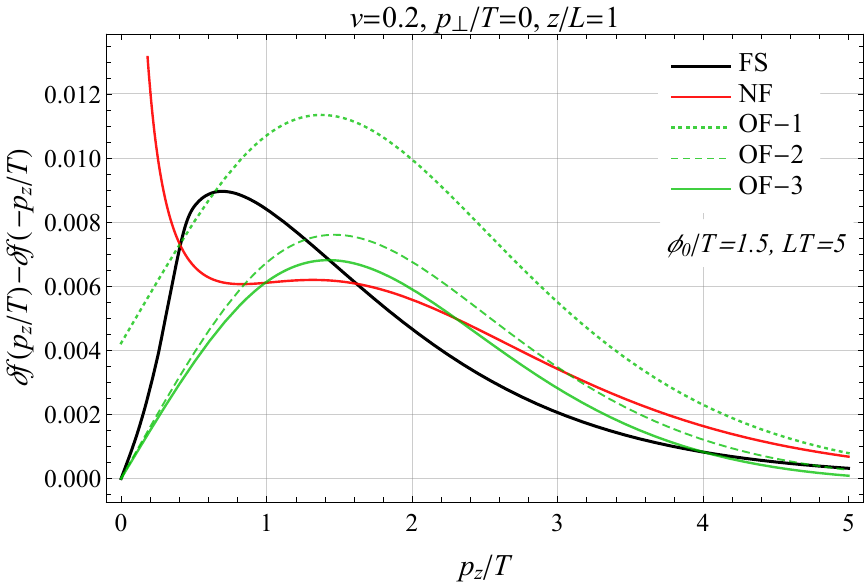}}
	\hfill
	{\includegraphics[width=.24\textwidth]{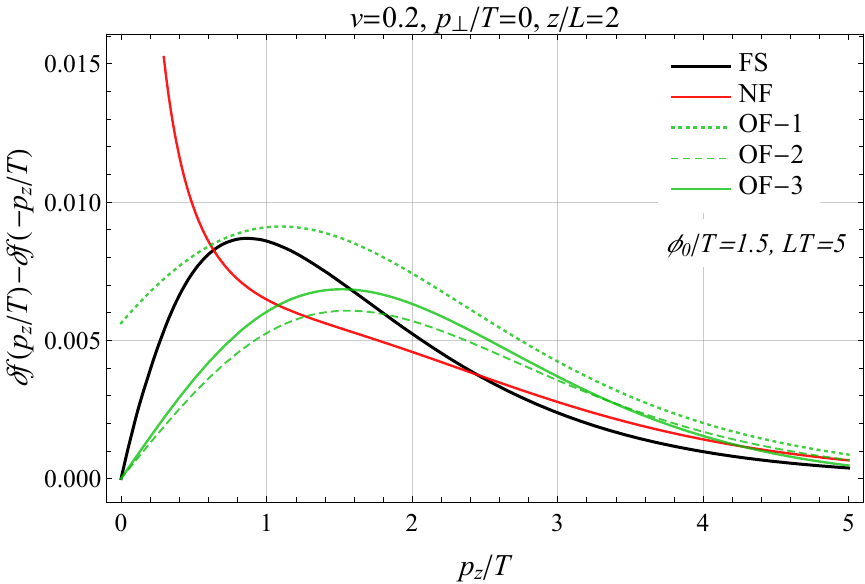}}\\
	\vspace{.5em}
	{\includegraphics[width=.24\textwidth]{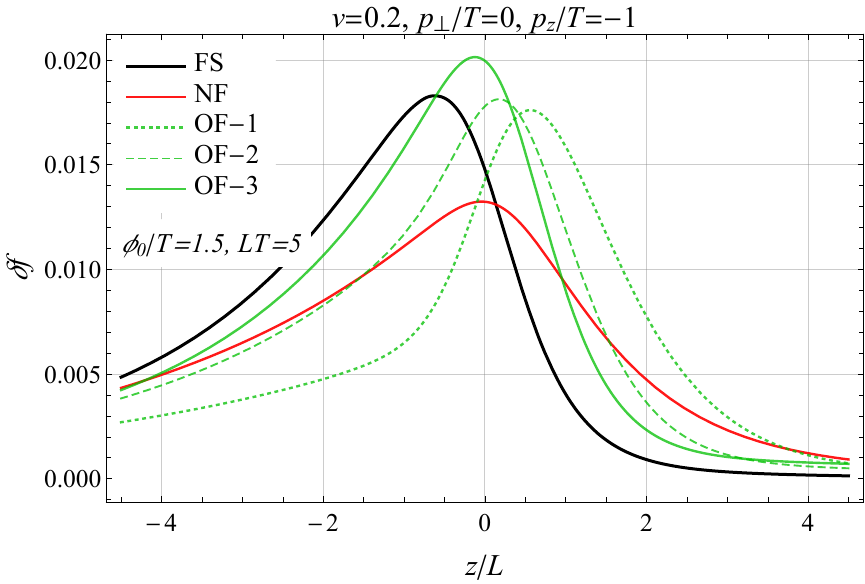}}
	\hfill
	{\includegraphics[width=.24\textwidth]{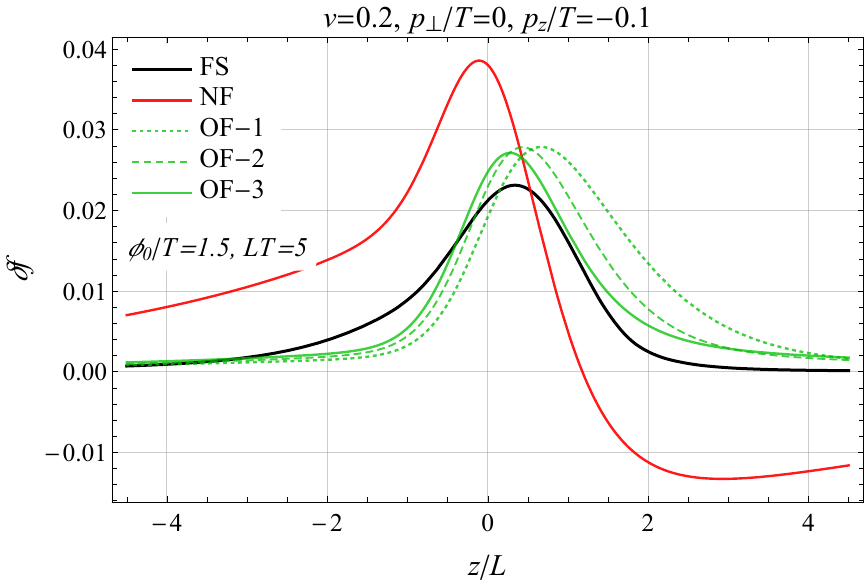}}
	\hfill
	{\includegraphics[width=.24\textwidth]{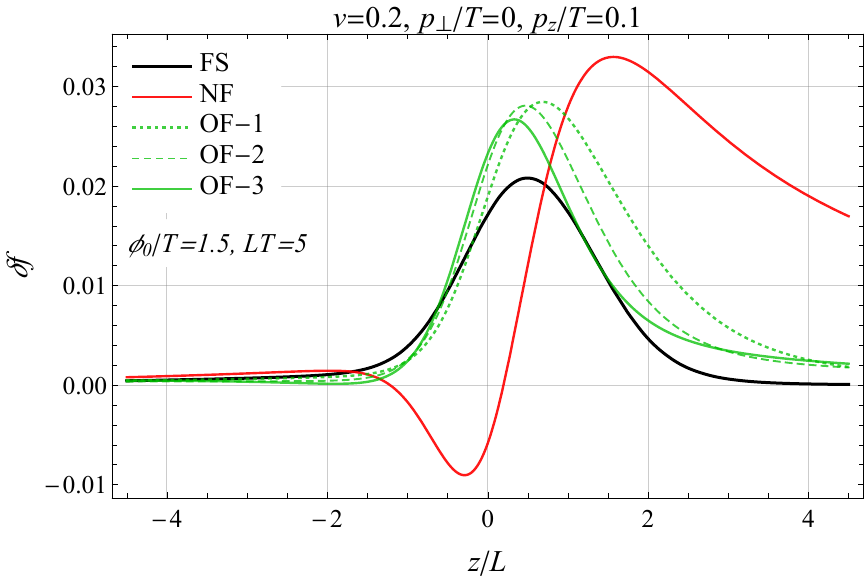}}
	\hfill
	{\includegraphics[width=.24\textwidth]{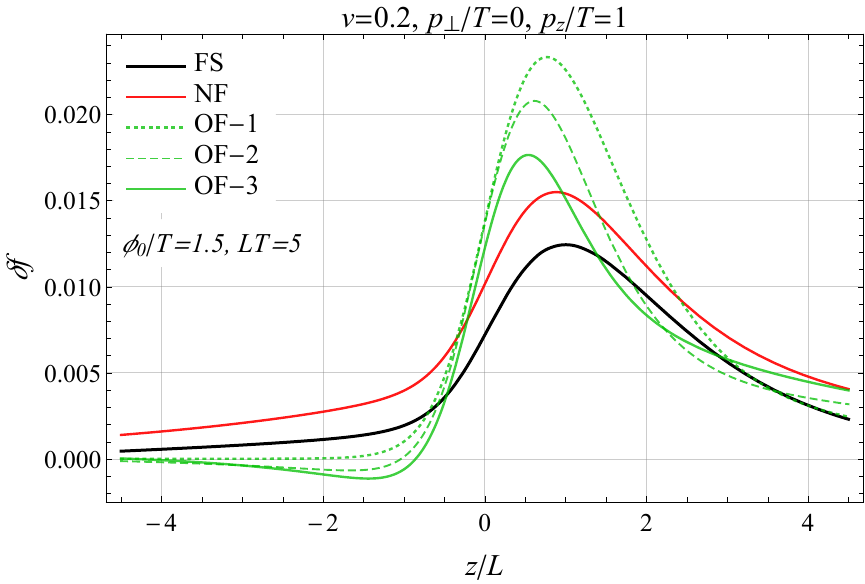}}	
	\caption{Perturbation $\delta f$ for $v_w = 0.2$ and $p_\bot = 0$. The plots on the first (second) row show the even
	(odd) part of $\delta f$ as a function of $p_z$ for $z/L = -1, 0, 1, 2$. The third row shows plots of $\delta f$ as a function of $z$ for $p_z/T = -1, -0.1, 0.1, 1$.}\label{fig:perturbation_ann_scatt}
\end{figure}

Finally we show in fig.~\ref{fig:perturbation_ann_scatt} the perturbation $\delta f$ for the benchmark velocity $v_w = 0.2$.
The results for different velocities are qualitatively analogous, the main difference being an overall rescaling with a limited change in shape. The plots show the full solution of the Boltzmann equation we got in our analysis, along with the results obtained applying
the old and new weighted approaches.
Notice that the new formalism does not fully determine the velocity perturbation, whose impact can only be computed
averaging over the momentum through a factorization ansatz~\cite{Cline:2000nw}.
To plot the solution in the new formalism we chose to identify the distribution perturbation with
\begin{equation}
\delta f = - f_v'\, [\mu (z)+\beta \gamma (E - v p_z) \delta \tau] + f_v \frac{E}{p_z} u\,,
\end{equation}
following eq.~(B5) of ref.~\cite{Laurent:2020gpg}.
This identification tends to produce a divergent behavior for small $p_z$, which however has no impact on the determination
of the friction since it is odd in $p_z$.

In the first (second) row of the figure we show the dependence of the even (odd) part of $\delta f$ on the momentum along the $z$
axis, $p_z$. The plots are obtained fixing $p_\bot = 0$, but similar results are found for $p_\bot/T \lesssim 2$
(for larger $p_\bot$ the solution is significantly suppressed and its impact on the domain wall dynamics is subleading).
The plots in the first row show that, at the qualitative level, the old and new formalisms fairly reproduce the overall shape of the even part of the solution, although a somewhat different behavior is found for small $p_z$ ($|p_z/T| \lesssim 0.5$).
This difference is most probably due to the fact that in the weighted approach the
\begin{equation}
- \frac{(m(z)^2)'}{2p_z} \partial_{p_z} \delta f
\end{equation}
term is neglected. This term, although subleading in most of the kinematic space, dominates close to $p_z = 0$, where the
$p_z \partial_z$ term vanishes. In spite of the fair qualitative agreement, large quantitative differences are present between the
full solution and the ones obtained with the weighted approaches.

The agreement of the various formalisms in the determination of the odd part of $\delta f$ proves quite poor. In particular,
marked differences are found for $z \lesssim 0$, in which case the high-$p_z$ behavior of the solution is not captured by the old formalism, even including higher-order corrections. A mildly better agreement is found for $z/L \gtrsim 1$. The new formalism tends to
reproduce the correct shape for $p_z/T \gtrsim 1$, but presents large differences for small $p_z$. It must be noticed that the
odd part of $\delta f$ does not contribute to the friction, thus the large differences found among the various solutions do not show up
in the determination of $F(z)$.

On the third row of fig.~\ref{fig:perturbation_ann_scatt} we show $\delta f$ as a function of $z$ for $p_\bot = 0$ and for the
benchmark values $p_z/T = -1, -0.1, 0.1, 1$. The old formalism reproduces the overall qualitative behavior of the solution.
Higher-order terms tend to improve the agreement, although failing to fully reproduce the full solution, especially at the quantitative level.
The new formalism  is in qualitative agreement with the full solution for $|p_z/T| \gtrsim 1$, while completely fails to
reproduce the correct shape for small $p_z$.

Our conclusion is hence that the OF and NF fail to provide an accurate and reliable solution of the Boltzmann equation. Because the perturbation controls the out-of-equilibrium friction, the weighted approach fails to achieve the necessary precision required for the computation of the terminal velocity, especially for walls moving at $v_w > 0.4$. A full solution of the Boltzmann equation is hence necessary to correctly determine the non-equilibrium properties of the plasma. 


\section{Wall velocity in the SM singlet extended model}\label{subsec:eom_scalars}

The DW terminal speed is the result of a balance between the internal pressure of the wall and the friction. 
To determine it, one has to solve the equation of motion of the two scalar fields together with the Boltzmann equation that determines the out-of-equilibrium perturbations and, through eq.~(\ref{eq:out_of_eq_friction}), the friction.

As explained in sec.~\ref{sec:wall_velocity_determination}, we model the dynamics of the scalar singlet and Higgs DW using the four parameters $v_w$, $L_h$, $L_s$, $\delta s$, where we recall that  $L_{h,s}$ and $\delta s$ identify respectively the widths and the displacement between the two DWs. Such parameters are determined by the vanishing of suitable moments of the scalar fields equation of motion that are solved simultaneously with the Boltzmann equation following the strategy outlined in sec.~\ref{sec:wall_velocity_determination}.

\begin{table}[]
    \centering
    \begin{tabular}{c|c|c|c||c|c|c|c}
     & $m_s\,$(GeV) & $\lambda_{hs}$ & $\lambda_s$ & $T_n\,$(GeV) & $T_c\,$(GeV) & $T_+\,$(GeV) & $T_-\,$(GeV)   \\
     \hline
     \rule{0pt}{1.1em}BP1 & 103.8 & 0.72 & 1 & 129.9 & 132.5 & 130.3 & 129.9 \\ 
     \rule{0pt}{1.em}BP2 & 80.0 & 0.76 & 1 & 95.5 & 102.8 & 97.5 & 95.5
    \end{tabular}
    \\
    \vspace{0.5cm}
    \begin{tabular}{c|c|c|c|c}
     & $v_w$ & $\delta_s$ & $L_h T_n$ & $L_s T_n$ \\
     \hline
     \rule{0pt}{1.1em}BP1 & 0.39\ \ (0.57) & 0.79\ \ (0.75) & 9.7\ \ (8.1) & 7.7\ \ (6.7)\\ 
     \rule{0pt}{1.em}BP2 & 0.47\ \ (0.61) & 0.81\ \ (0.81) & 5.2\ \ (4.7) & 4.3\ \ (4.1)
    \end{tabular}
    \caption{Critical and nucleation temperatures, temperatures in front and behind the DW and terminal values of the parameters $v_w$, $\delta_s$, $L_h$, $L_s$ for two benchmark points.
    The numbers in parentheses correspond to the results obtained neglecting the out-of-equilibrium perturbations.}
    \label{tab:parameters_results}
\end{table}

To numerically solve eq.~(\ref{eq:param_equations}) we implemented the Newton algorithm, while to solve the Boltzmann equation we followed the iterative procedure outlined before with the bracket term $\langle \delta f\rangle$ computed following the strategy explained in section~\ref{sec:decomposition}. 

To improve our description of the plasma dynamics, we included the effects of the light particles and modeled these species as a plasma in local thermal equilibrium. As discussed in Chapter~\ref{ch:boltzmann_ansatz}, it is not possible to linearize the temperature and plasma velocity fluctuations without affecting the dynamics of hybrid walls. To model the light degrees of freedom we applied the method described in Chapter~\ref{ch:bubble_dynamics_and_plasma_hydrodynamics} that correctly accounts the non-linearities in the analysis. Finally, we considered once again the top quark to be the only out-of-equilibrium species in the plasma and we neglected the W bosons corrections. This setup provides a more realistic scenario to study the dynamics of the DW and allows us to assess the impact of out-of-equilibrium contributions compared to the local equilibrium ones.

We report in Tab.~\ref{tab:parameters_results} our results for two benchmark points which we recall are characterized by the following choice of parameters: $m_s = 103.8 \, \textrm{GeV}$, $\lambda_{hs} = 0.72$ and $\lambda_s = 1$ for BP1 and $m_s = 80.0 \, \textrm{GeV}$, $\lambda_{hs} = 0.76$ and $\lambda_s = 1$ for BP2. The two benchmark points are characterized by different strength of the PT, with the one in BP2 stronger than the one in BP1, allowing us to assess also the impact of the perturbation when the DW releases a large amount of energy in the plasma. 
In the table we show the values of the four quantities, $v_w$, $L_h$, $L_s$, $\delta_s$ that characterize the two DWs dynamics, with and without the contributions of the out-of-equilibrium perturbations. 
The bubble speed is the parameter more strongly affected by the out-of-equilibrium contributions, followed by the width of the Higgs wall.
For the first benchmark model with $m_s = 103.8$ GeV, a difference of $\sim 30\%$ and of $\sim 20\%$ is present for the speed $v_w$ and the wall width $L_h$, respectively, with respect to the same values computed in local equilibrium. 
The width $L_s$ and the displacement $\delta_s$ show, instead, a difference of $\sim 15\%$ and a milder one of $\sim 5\%$ respectively. 
The four parameters in the second benchmark model with $m_s = 80$ GeV still present important differences with respect to the only-equilibrium case, but the impact of the out-of-equilibrium contributions is less severe. As for the previous benchmark model, the perturbations mostly impact on the speed on which they induce a change of $\sim 20\%$. 
The offset $\delta_s$, instead, is almost unaffected by the inclusion of the out-of-equilibrium corrections.

To understand why the out-of-equilibrium perturbations have a different impact on the two benchmark models, it is useful to study the total pressure acting on the system. This will also clarify why the terminal speed $v_w$ is the 
quantity more strongly affected by the out-of-equilibrium corrections. We remind that the total pressure acting on the system can be expressed using eq.~(\ref{eq:total_pressure}), where one of the fundamental contribution arises from the potential energy difference $\Delta V$ that drives the wall expansion.

\begin{figure}
    \centering
    \includegraphics[width=0.32\textwidth]{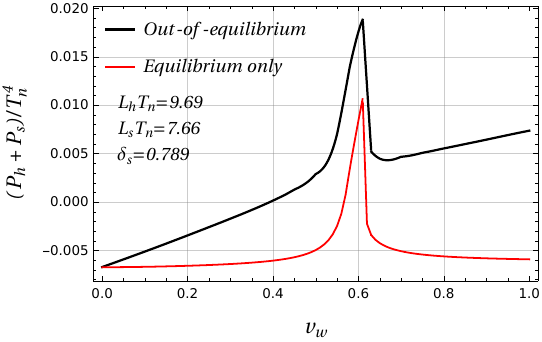}
    \hfill
    \includegraphics[width=0.32\textwidth]{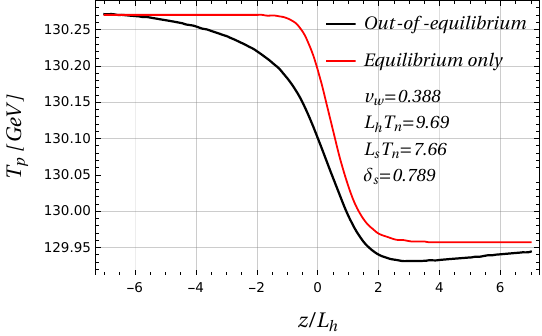}
    \hfill
    \includegraphics[width=0.32\textwidth]{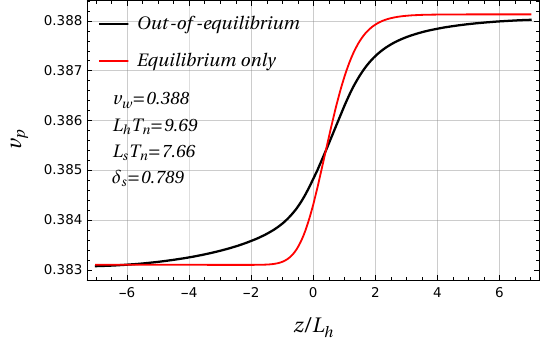}\\
    \vspace{.5em}    
    \includegraphics[width=0.32\textwidth]{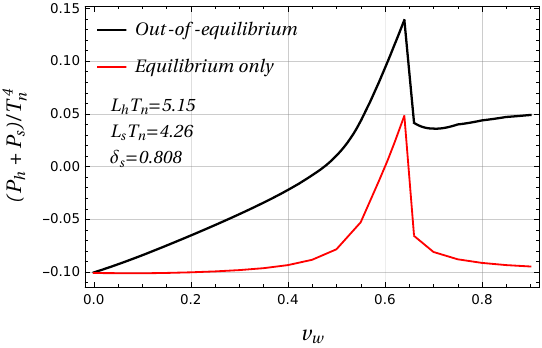}
    \hfill
    \includegraphics[width=0.31\textwidth]{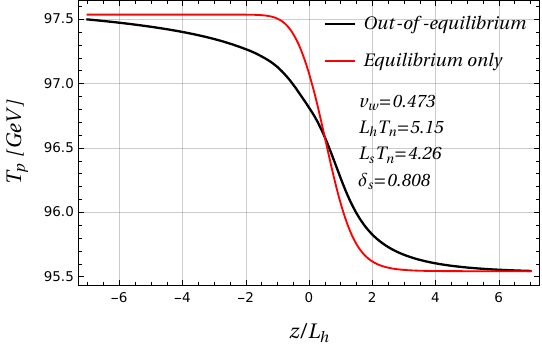}
    \hfill
    \includegraphics[width=0.32\textwidth]{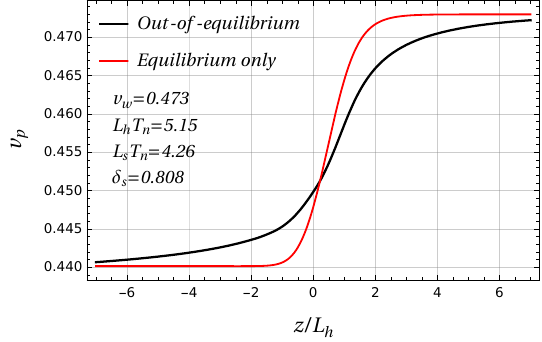}
    \caption{Total pressure as a function of the bubble speed, temperature and velocity profiles as functions of $z/L_h$ for the two benchmark models reported in Tab.~\ref{tab:parameters_results} (BP1 on the upper row, BP2 on the lower row). The red solid lines are obtained by neglecting the out-of-equilibrium perturbations, while the black solid lines correspond to the complete computation. The peak in the pressure is located at the Jouguet velocity.}
    \label{fig:pressure_and_profiles}
\end{figure}

We plot on the left panel of Fig.~\ref{fig:pressure_and_profiles} the total pressure as a function of the wall speed, with (black line) and without (red line) the inclusion of the out-of-equilibrium distributions. In both curves we can observe the presence of a peak corresponding to the Jouguet velocity. The peak originates from hydrodynamic effects that heat up the plasma, thus generating a pressure barrier that slows down the DW motion in models with small supercooling \cite{Konstandin:2010dm}. This effect is described in eq.~(\ref{eq:total_pressure}) by the term proportional to the enthalpy and its impact grows with the difference $T_+ - T_-$, so that it tends to be larger for hybrid walls. 

At small velocities, instead, the pressure at equilibrium becomes constant. In this case the temperature difference across the wall becomes negligible and the value of the total pressure settles to the potential energy difference between the false and true vacua at the nucleation temperature.

 The out-of-equilibrium perturbations provide a correction that grows linearly with the velocity. This behaviour is consistent with the one we found in Fig.~\ref{fig:friction}, where a linear dependence of the total friction was observed. From the above discussion we conclude that the perturbations have an important impact on the wall terminal speed and, therefore, an accurate modeling of the out-of-equilibrium perturbations is necessary to get a proper description of the PT dynamics.

In the central and right panel of Fig.~\ref{fig:pressure_and_profiles} we plot the temperature and velocity profiles, respectively, for the terminal values of $v_w,\,L_h,\,L_s,\,\delta_s$ of the two benchmark models reported in Tab.~\ref{tab:parameters_results}. The plots clearly show that the out-of-equilibrium perturbations impact on the shape of the profiles, mostly in the region close to the DW. Among all the three terms in eq.~(\ref{eq:total_pressure}), the modified shapes of the profiles have the largest effect on the term involving the temperature derivative.

We now give an estimation of the impact of the temperature and velocity profiles on the Boltzmann equation. They contribute to the source term through their derivatives and, thus, we expect their effect to be proportional to the relative difference between the temperature and velocity across the DW. This expectation is confirmed by our numerical analysis. We computed the integral of the friction for three benchmark velocities ($v_w=0.4, 0.6, 0.9$) and we compared with the result obtained by considering a constant temperature and plasma velocity across the system. Our analysis showed that corrections of order few $\%$ are present for $v_w = 0.4$ and $v_w = 0.9$, while we found larger corrections ($\sim 10\%$) for $v_w = 0.6$. Such result agrees with the expected behavior since the largest correction is found for velocities close to the Jouguet velocity, where the differences between the temperature and plasma velocity across the wall are larger, as we showed in Fig.~\ref{fig:vpvm_detonation_deflagration}. From these results we can conclude that, despite the profiles have to be included to compute the equilibrium part of the total pressure acting on the system, they may be typically neglected in the computation of the out-of-equilibrium friction.

Finally we comment on the comparison of our results with the previous literature. We found a fair agreement with the results given in ref.~\cite{Friedlander:2020tnq}, in which the DW terminal speed for a similar potential is computed
by solving the Boltzmann equation within the fluid approximation. This approach, as we showed previously when discussing Fig.~\ref{fig:friction}, tends to overestimate the value of the friction for subsonic walls by $\sim 10\% - 20\%$. The results reported in Tab.~\ref{tab:parameters_results} show differences with respect to the values in ref.~\cite{Friedlander:2020tnq} that are compatible up to these effects.

On the other hand, we found much larger differences with respect to the results in ref.~\cite{Laurent:2022jrs}, in which the out-of-equilibrium perturbations are found to provide only a small correction to the DW terminal velocity. To investigate the source of discrepancy, we explicitly implemented the procedure proposed in ref.~\cite{Laurent:2022jrs}, decomposing the perturbations in terms of Chebyschev polynomials. With this approach we recovered the results obtained with our method up to $\sim 20\%$ differences, which are most probably due to the different size of the grids used to discretize the perturbations. Other sources of differences could be the choice of the scalar potential (and of its renormalization procedure) and of the model parameters, which could lead to a different amount of supercooling.

\subsection{W bosons inclusion}

To complete our analysis we finally include the contribution of W bosons to the friction as we did in ref.~\cite{DeCurtis:2023aaa}. In the same work we introduced the multipole decomposition discussed in sec.~\ref{sec:multipole_expansion} and we adopted the Legendre polynomials basis to compute the bracket term. The large timing improvement provided by this strategy allowed us to refine the grid on which we interpolate the perturbation. Differently from our previous works we considered a larger grid in the $z$ direction, namely $z/L_h \in [-20, 20]$ and we imposed the boundary conditions, i.e.~the perturbation to vanish in $z/L_h = \pm 30$. We did not modify the grid in the momenta direction.

Using the above improvements to solve the Boltzmann equation, we computed the velocity, the displacement and the width of the bubble for the same two benchmark points in Tab.~\ref{tab:parameters_results} and we reported our results in Tab.~\ref{tab:parameters_resultsW}.
	\begin{table}[]
		\centering
		\begin{tabular}{c|c|c|c||c|c|c|c}
			& $m_s\,$(GeV) & $\lambda_{hs}$ & $\lambda_s$ & $T_n\,$(GeV) & $T_c\,$(GeV) & $T_+\,$(GeV) & $T_-\,$(GeV)   \\
			\hline
			\rule{0pt}{1.1em}BP1 & 103.8 & 0.72 & 1 & 129.9 & 132.5 & 130.1 & 129.9 \\ 
			\rule{0pt}{1.em}BP2 & 80.0 & 0.76 & 1 & 95.5 & 102.8 & 96.7 & 95.5
		\end{tabular}
		\\
		\vspace{0.5cm}
		\begin{tabular}{c|c|c|c|c}
			& $v_w$ & $\delta_s$ & $L_h T_n$ & $L_s T_n$ \\
			\hline
			\rule{0pt}{1.1em}BP1 & 0.28\ \ [0.39]\ \ (0.57) & 0.78\ \ [0.79]\ \ (0.75) & 9.2\ \ [9.7]\ \ (8.1) &7.4\ \ [7.7]\ \ (6.7)\\ 
			\rule{0pt}{1.em}BP2 & 0.41\ \ [0.47]\ \ (0.61) &0.81\ \ [0.81]\ \ (0.81) & 5.1\ \ [5.2]\ \ (4.7) &4.2\ \ [4.3]\ \ (4.1)
		\end{tabular}
		\caption{Critical and nucleation temperatures, temperatures in front and behind the DW and terminal values of the parameters $v_w$, $\delta_s$, $L_h$, $L_s$ for two benchmark points.
			The numbers in round parentheses correspond to the results obtained neglecting the out-of-equilibrium perturbations, the numbers in square brackets correspond to the results obtained by including only the top quark out-of-equilibrium perturbations.}
		\label{tab:parameters_resultsW}
	\end{table}
	The table clearly shows that the inclusion of the W bosons provides important corrections to the width and velocity of the DW. For the first benchmark model with $m_s = 103.8$ GeV, $\lambda_{hs} = 0.72$ and $\lambda_s = 1$ the inclusion additional contribution arising from the W bosons provides a $\sim30\%$ and $\sim4-5\%$ correction respectively to the terminal velocity $v_w$ and the two widths $L_h$ and $L_s$ computed including only the top quark. The displacement $\delta_s$ instead is mostly unaffected. Comparing still with the top quark, for the second benchmark point with $m_s = 80$ GeV, $\lambda_{hs} = 0.76$  and $\lambda_s = 1$, the W bosons provide instead a milder correction with a $\sim 12\%$ and $\sim 2\%$ difference for the velocity and the widths respectively. The displacement, also in this case is unaffected.

    The inclusion of W bosons provides a much more severe impact of the out-of-equilibrium friction on the DW dynamics. It determines a large discrepancy in the values of the four quantities that describe the plasma with respect to the  predictions of the local thermal equilibrium approximation. As Table~\ref{tab:parameters_results} shows, comparing with this latter case, we find a factor $2$ of difference in the terminal velocity for BP1 and a $\sim 30\%$ correction on the terminal velocity for BP2. These large discrepancies prove that the local thermal equilibrium approximation provides a poor estimate of the terminal velocity of the bubble, at least for some benchmark scenarios, emphasizing at the same time the necessity to provide an accurate description of the non-equilibium dynamics of the plasma during the EWPT.

    \begin{figure}
    \centering
    \includegraphics[width=0.32\textwidth]{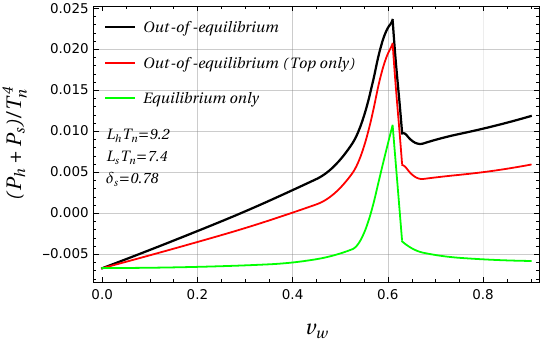}
    \hfill
    \includegraphics[width=0.32\textwidth]{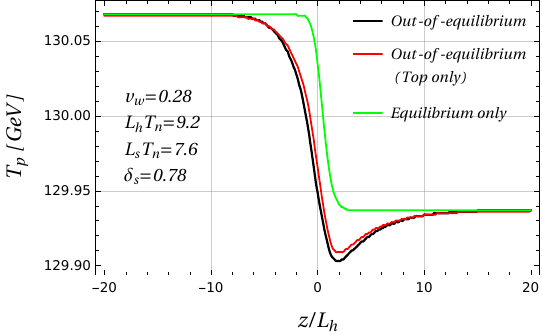}
    \hfill
    \includegraphics[width=0.32\textwidth]{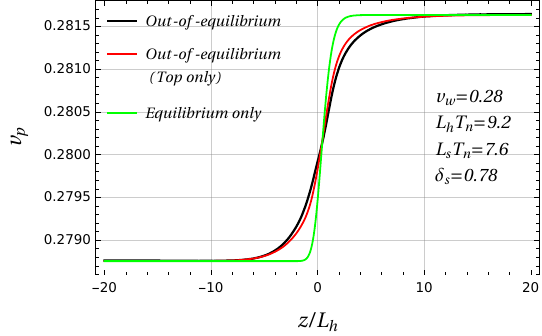}\\
    \includegraphics[width=0.32\textwidth]{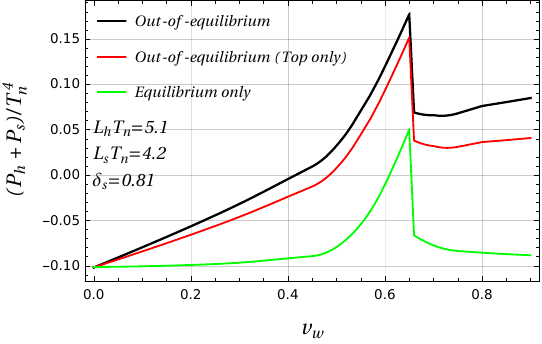}
    \hfill
    \includegraphics[width=0.32\textwidth]{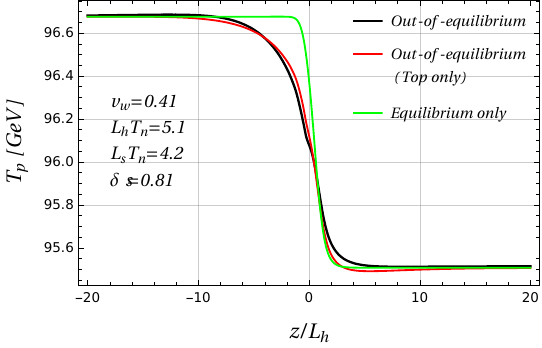}
    \hfill
    \includegraphics[width=0.32\textwidth]{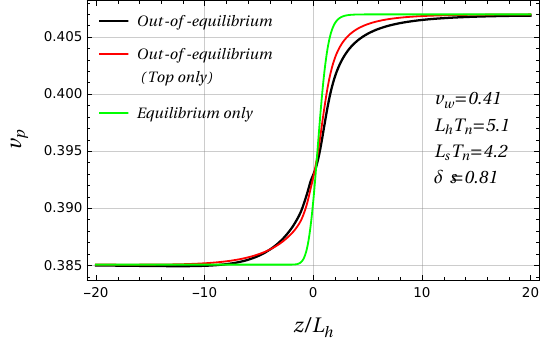}
    \caption{Total pressure as a function of the bubble speed, temperature and velocity profiles as functions of $z/L_h$
    for the two benchmark models reported in Tab.~\ref{tab:parameters_resultsW} (BP1 on the upper row, BP2 on the lower row) when also W bosons are included. The greed solid lines are obtained by neglecting the out-of-equilibrium perturbations, red solid lines are obtained by including only the top quark contributions while the black solid lines correspond to the complete computation with the inclusion of gauge bosons.}
    \label{fig:pressure_and_profilesW}
\end{figure}

In Figure~\ref{fig:pressure_and_profilesW} we plot the total pressure acting on the DW as a function of the velocity and the temperature and velocity profiles as a function of $z/L_h$ for the two benchmark points. Analogous considerations to the case where only top contributions are included hold for the out-of-equilibrium friction reported on the left panel. The total pressure acting on the DW (black solid line) grows linearly for small velocities and presents a large difference with respect to the local equilibrium case (green solid line). Comparing with the case where only the top quark is out-of-equilibrium (red solid line), the W bosons provide a contributions that grows linearly with the velocity hence rescaling by a constant factor the friction. For BP1 we find a $\sim50\%$ correction, while for the second one a $\sim 30\%$. For both benchmark points a smaller difference is found in correspondence to the peak. This is in accordance with the results we presented in Tab~\ref{tab:parameters_resultsW}. 

On the central and right panel of Figure~\ref{fig:pressure_and_profilesW} we plot, respectively, the temperature and velocity profiles of the plasma. The out-of-equilibrium corrections arising from the $W$ bosons in this case are very small with respect to the top quark correction, providing a negligible effect. The most relevant feature of the profiles is the presence of a valley for the temperature inside the DW. Such an effect was also present in Figure~\ref{fig:pressure_and_profiles} 
and arises from the energy and momentum exchange between the equilibrium and out-of-equilibrium components of the plasma. In fact as the plots of the $T^{30}$ and $T^{33}$ components of the out-of-equilibrium stress energy tensor show in Figure~\ref{fig:old_vs_new_compare}
show, the components present a peak exactly where the valley is located.

We finally point out that the impact of W bosons is a lot larger than the expected contribution justified by the na\"ive scaling of the friction. Indeed, because the perturbations are mainly sourced by $(m^2)'$, using eq.~(\ref{eq:out_of_eq_friction}) we find that for a particle species $i$ the friction $F_i$ should scale as $F_i \sim N_i m^4_i$. The contribution of W bosons should thus be $\sim 30$ times smaller with respect to top quark and hence negligible. This is in contrast with the results we presented in this section. 
    
The origin of such a discrepancy is related to the IR behaviour of the W bosons. The presence of a zero Matsubara mode in the energy spectrum of bosons allows for the presence of states with soft momentum in the plasma. Differently from fermions, the majority of the boson population in the plasma is soft with important consequences on the non-equilibrium properties of the system. In particular, this spoils the scaling of the friction we just provided which is instead valid in the hard region.

Because the friction arising from the W bosons is dominated by the soft region, the Boltzmann equation is no longer the appropriate effective kinetic theory to describe the out-of-equilibrium perturbations. Of course it can still be adopted to properly model the hard particles but since the friction arising from such modes is subleading with respect to the soft modes eventually we must rely on a different description. As we will see in the next section, the contribution of IR W bosons to the friction is the most important source of theoretical uncertainty on the DW terminal velocity.

\section{Theoretical limitations}

Our computation of the terminal velocity is affected by systematic uncertainties caused by the many approximations that we adopted through this thesis work. In particular, when we discussed the computation of the collision integrals in Chapter~\ref{ch:effective_kinetic_theory} we stressed that in order to simplify our probem we used a set of approximations, namely we neglected ``$1\to 2$'' processes, we assumed the particles to be massless inside the collision integrals, we evaluated the amplitudes at the leading-log order and we encapsulated all the thermal effects through the thermal masses in the propagator. We also saw in the previous section that the leading contribution to the out-of-equilibrium friction from W bosons arises from the soft region which cannot be properly modeled  using the Boltzmann equation.

To assess the uncertainty on the determination of the DW terminal velocity we discuss the impact of some of the approximations we discussed through the thesis. This analysis provides an estimate of the different approximation often employed in the literature emphasizing, at the same time, which aspect of the computation of the out-of-equilibrium perturbation should be improved. As a first step we are going to relax the leading-log and the massless assumption and estimate the terminal velocity in these more realistic scenarios. As we are going to see, such approximations provide a small change in the terminal velocity, justifying their use. At last we discuss the impact of IR gauge bosons and discuss the theory that can be used to describe soft bosons. Such a theory, as we will see, is still IR divergent and sensible to the IR cutoff. Our ignorance regarding the behaviour of these soft mode constitutes the main source of theoretical uncertainty on the non-equilibrium dynamics of the plasma and, in turn, on the DW terminal velocity. 


\subsection{Massive collision integral}

The first approximation we discuss is the massless approximation. In Chapter~\ref{ch:effective_kinetic_theory} we argued that such an approximation is justified for hard particles that can be considered as ultra-relativistic particles. This approximation simplified a lot the computation of the collision integral. On the one hand this operator does not depend on the VEV of the Higgs and its position dependence in the local plasma reference frame is trivial. On the other hand the kinematic space over which we perform the integrations becomes particularly simple. To assess the impact of the massless approximation we focus on the top quark.

The presence of a finite mass can have important consequences on the out-of-equilibrium friction. In fact, the presence of the mass affects the distributions of particles which control the efficiency of collision processes. Our aim in this section is to provide a simple estimate of this effect. This can be achieved by considering a simplified setup where the mass is included only in the distribution function of the out-of-equilibrium species without affecting the kinematic space. In practice in the computation of the term ${\cal Q}$ in eq.~(\ref{eq:Q_expression}) and the bracket term in eq.~(\ref{eq:prel_bracket}) we restore the mass dependence in the top equilibrium distribution, which simply corresponds to the substitution
\begin{equation}
    f_0(|\bar{\bf k}|)\rightarrow f_0(E_{\bar k})\,.
\end{equation}

The restoration of the mass has a different impact depending on the process under consideration, which we can easily understand with the help of eq.~(\ref{eq:prel_prel_bracket}). Annihilation processes becomes less efficient. Because $f(E_{\bar k}) < f(|{\bar{\bf k}}|)$ the population of the top is smaller and less particles participate in the reactions hence damping its efficiency. On the other hand the collision rate of scattering processes increases. A smaller population of top quark in fact increases the size of the factor $(1-f_0)$ multiplying the scattering kernel. The physical justification is that a smaller ``repulsion'', caused by the Pauli's principle, is present hence favouring the top quark scattering\footnote{In the case of the W bosons the efficiency of the scattering processes is reduced as well as for the annihilation ones. In fact for bosons the population factor multiplying the scattering kernel is $(1+f_0)$}.

The competition of these two effect induces a small impact on the non-equilibrium plasma dynamics. Comparing the ${\cal Q}$ term with the one we obtained in the mass-less approximation shows that the mass impact is very limited. The ${\cal Q}$ term computed by including the mass is very similar to the mass-less one and it is only $\sim 2\%$ larger in the region where $p \sim m$. To refine our analysis we also computed the integrated friction by including the massive effects. Our results showed that in the massive case the integrated friction is $\sim2\%$ smaller confirming our results for the ${\cal Q}$ term. Because the impact is very limited we did not computed the terminal velocity as we should recover results very similar to the ones presented in the previous section.

In the W bosons case the impact of the mass could be larger. However, such an impact is important in the region $p \lesssim m_W$, where the Boltzmann equation can no longer be applied. In the hard region instead, where W bosons can still be described using the Boltzmann equation, the inclusion of the mass is expected to be negligible and W bosons can still be treated as massless particles. We conclude that the massless approximation provides a good framework for the study of the non-equilibrium dynamics of the plasma during the EWPT.

\subsection{Beyond the leading-log approximation}

We next turn our attention to the leading-log approximation. We recall its main aspects before we analyze its impact on the DW dynamics and refer to Chapter~\ref{ch:effective_kinetic_theory} for a more in-depth discussion. The $t$ and $u$ channels of the scattering and annihilation amplitudes present an IR divergence which is regularized by the thermal self-energies. Such channels, differently from the $s$-channel, are enhanced by a factor $\alpha_s\log(\alpha_s^{-1})$. The leading-log approximation consists in considering only these log-enhanced contributions systematically dropping all $s$ channels and interference in diagrams. 

\begin{table}
		\centering
		\begin{tabular}{c|c}
			process & $|{\cal M}|^2$\\
			\hline
			\rule{0pt}{1.75em}$t \bar t \to gg$ & $\displaystyle \frac{128}{3} g_s^4 \left[ \frac{ut}{(t - m_q^2)^2} +  \frac{ut}{(u- m_q^2)^2} \right ]-\uline{\uline{ 96\frac{t^2+u^2}{(s+2m_q^2)^2}}}$\\
            \rule{0pt}{1.75em}$\uline{\uline{ t\bar t \to q \bar q}}$ & $\displaystyle\uline{\uline{ 80 g_s^4 \frac{t^2 + u^2}{(s + 2m_q^2)^2}}}$\\
			\rule{0pt}{1.75em}$tg \to tg$ & $\displaystyle- \frac{128}{3} g_s^4 \left(\frac{su}{(u-m_q^2)^2}+\uline{\uline{\frac{su}{(s+m_q^2+m_g^2)^2}}}\right) + 96 g_s^4 \frac{s^2 + u^2}{(t - m_g^2)^2}$\\
			\rule{0pt}{1.75em}$tq \to tq$ & $\displaystyle160 g_s^4 \frac{s^2 + u^2}{(t - m_g^2)^2}$\\			
		\end{tabular}
		\caption{Tree-level amplitudes for the scattering processes relevant for the top quark including next-to-leading-log terms; the latter are underlined. In the $t q \to t q$ process we summed over all massless quarks and antiquarks, and in the $t\bar t \to q \bar q$ process we summed over the 5 pairs of massless quark-antiquark. As explained in the text, all the new terms correspond to $s$-channel processes.}\label{tab:amplitudesNLL}
	\end{table}

   \begin{table}[]
		\centering
		\begin{tabular}{c|c|c|c||c|c|c|c}
			& $m_s\,$(GeV) & $\lambda_{hs}$ & $\lambda_s$ & $T_n\,$(GeV) & $T_c\,$(GeV) & $T_+\,$(GeV) & $T_-\,$(GeV)   \\
			\hline
			\rule{0pt}{1.1em}BP1 & 103.8 & 0.72 & 1 & 129.9 & 132.5 & 130.3 & 129.9 \\ 
			\rule{0pt}{1.em}BP2 & 80.0 & 0.76 & 1 & 95.5 & 102.8 & 97.5 & 95.5
		\end{tabular}
		\\
		\vspace{0.5cm}
		\begin{tabular}{c|c|c|c|c}
			& $v_w$ & $\delta_s$ & $L_h T_n$ & $L_s T_n$ \\
			\hline
			\rule{0pt}{1.1em}BP1 & \ 0.41\ \ [0.39] & 0.79\ \ [0.79] & 9.8\ \ [9.7] &7.7\ \ [7.7]\\ 
			\rule{0pt}{1.em}BP2 & 0.49\ \ [0.47] &0.81\ \ [0.81] & 5.2\ \ [5.2] &4.3\ \ [4.3]
		\end{tabular}
		\caption{Critical and nucleation temperatures, temperatures in front and behind the DW and terminal values of the parameters $v_w$, $\delta_s$, $L_h$, $L_s$ for two benchmark points, considering only the top quark out-of-equilibrium perturbations and including next to leading log terms. The numbers in square brackets correspond to the results obtained including only leading log terms.}
        \label{tab:parameters_resultsNLL}
	\end{table}

To assess the impact of such approximation we restored the neglected contribution and evaluated the collision operator with the full amplitudes, whose expressions are reported in Tab.~\ref{tab:amplitudesNLL}. For simplicity we focused on the top quark contribution. In Tab.~\ref{tab:parameters_resultsNLL} we report the terminal value of the velocity, width and the displacement of the two walls for the same two benchmark points 
considered earlier in this chapter. As our results show the only quantity that receives a relevant correction is the terminal velocity of the DW. This quantity presents a $\sim 5\%$ correction in both benchmark scenarios. These results are in agreement with the conclusions of ref.~\cite{Arnold:2003zc}, where the authors also found that the correction provided by the next-to-leading-log contributions is small\footnote{In evaluating the leading-log approximation the authors of ref.~\cite{Arnold:2003zc} also included the full momentum dependence of the thermal self-energies in the propagators of the fields.}.



\begin{figure}
    \centering
    \includegraphics[width=0.47\textwidth]{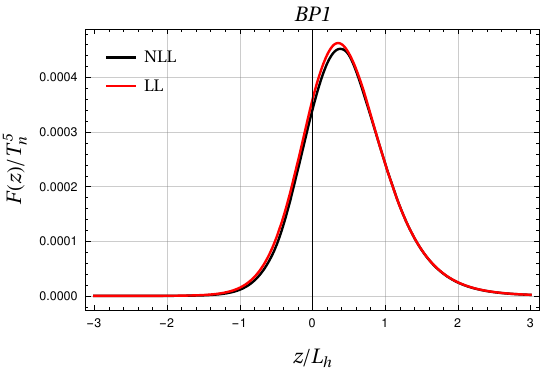}
    \hfill
    \includegraphics[width=0.47\textwidth]{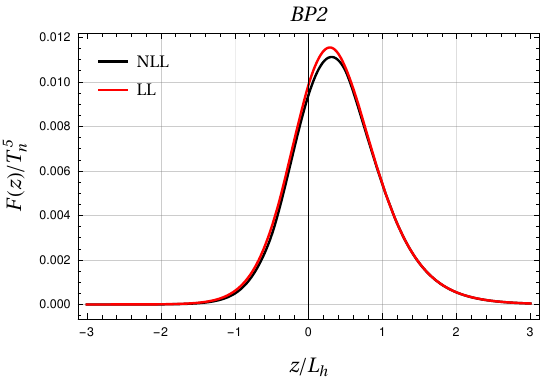}
    \caption{Comparison of the friction for the two benchmark points analyzed in the text including the next-to-leading-log contributions (black plot) and only the leading-log contributions (red plot).}
    \label{fig:frictionNLL}
\end{figure}

We provide a more refined analysis by plotting the friction acting on the bubble wall in Fig.~\ref{fig:frictionNLL}. We plot with the black solid line the friction where the next-to-leading-log corrections are included and where the terminal velocity and width correspond to the unbracketed values in Tab.~\ref{tab:parameters_resultsNLL}. The red solid line, instead, represent the friction in the leading-log approximation where the DW terminal velocity and width are given by the bracketed values in Tab.~\ref{tab:parameters_resultsNLL}. The plot confirms that the impact of the next-to-leading-log contributions is very limited since they affect the height of the peak just by a $2-5\%$ in accordance with the modifications we observed in Tab.~\ref{tab:parameters_resultsNLL}. Our analysis hence confirms that the leading-log approximation is reasonable for the description of the DW dynamics during the EWPT.

\subsection{IR gauge bosons}

Finally we focus on the IR gauge bosons and their contribution to the friction. As already pointed out in ref.~\cite{Moore:1995si} and mentioned in Chapter~\ref{ch:effective_kinetic_theory}, the Boltzmann equation is not a good effective kinetic theory to describe the dynamics of semi-hard and soft gauge bosons in the plasma, with $p\lesssim g T$, but provides a suitable characterization of the hard modes with $p \gtrsim T$. 
The main reason is, as we discussed in Chapter~\ref{ch:effective_kinetic_theory}, that the dynamics of soft modes is dominated by screening and damping effects which we did not include in our effective description. 
Because these soft modes dominate the friction arising from W bosons, the Boltzmann equation is no longer a suitable kinetic theory for such a species during the EWPT and a different description must be employed.

    \begin{figure}
        \centering
        \includegraphics[width=0.47\textwidth]{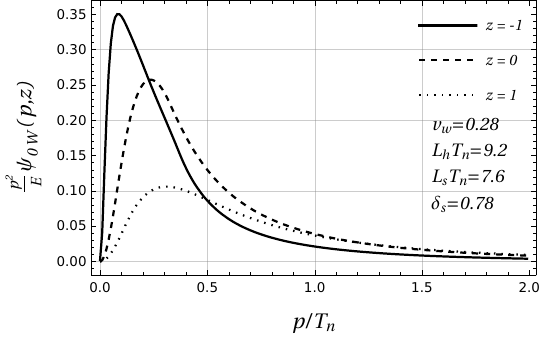}
        \hfill
        \includegraphics[width=0.47\textwidth]{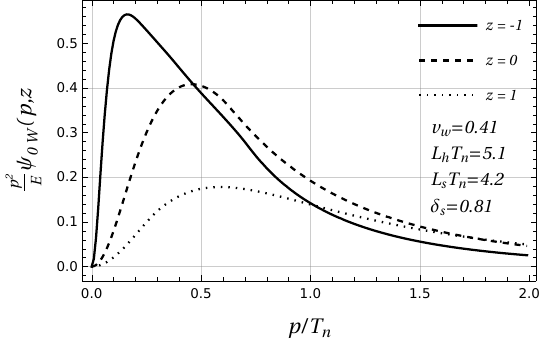}
        \caption{Plot of the zero mode of $\delta f$ times the integral measure that appear in the definition of the friction as a function of $p/T$ at different values of $z$ for the different two different benchmark points. We notice that the main contribution originates from the region $p ~ m$.}
        \label{fig:ir_friction_w_analysis}
    \end{figure}

Before we discuss the alternative theory let us first analyze the impact of the IR region on the friction. For that we focus on the contribution of the W bosons to the friction in eq.~(\ref{eq:out_of_eq_friction}) and expand their corresponding perturbation $\delta f_W$ on the spherical harmonic basis
\begin{equation}
    F(z) = \frac{N_W}{2}\frac{\partial m^2_W(z)}{\partial z}\int \frac{|\bar{\bf p}|^2d|{\bar{\bf p}}|}{(2\pi)^2E_p}\psi_{0W}(|{\bar{\bf p}}|,z)\,,
\end{equation}
where $\psi_l$ are the Legendre modes defined in eq.~(\ref{eq:pert_legendre}). The integrand in the above equation quantifies the impact of the different regions of the kinematic space to the friction. In Fig.~\ref{fig:ir_friction_w_analysis} we show the plot of the integrand for different choices of the position $z$ and for the two benchmark points BP1 (left plot) and BP2 (right plot) defined in Tab.~\ref{tab:parameters_resultsW}. A clear feature that we infer from fig.~\ref{fig:ir_friction_w_analysis} is that the main contribution to the friction arises from the IR region ($p\lesssim gT$). In addition, the friction gets larger in the symmetric phase where the majority of the soft particles are, as we can understand observing the increasing size of the peak as we move towards negative positions. This confirms that the Boltzmann equation cannot be used to describe the dynamics of W bosons and a different effective kinetic theory must be employed to assess the impact of such a species.


The dynamics of soft gauge bosons has been analyzed in \cite{Arnold:1996dy, Huet:1996sh, Son:1997qj} to determine baryon number violation, further discussed in \cite{Bodeker:1998hm,Bodeker:1999ey,Bodeker:1999ud,Arnold:1999jf,Arnold:1999ux} and applied in \cite{Moore:2000wx} to compute the friction acting on DW arising from IR. The key point is that the IR particles' dynamics is dominated by Landau damping and screening, as we already pointed out in Chapter~\ref{ch:effective_kinetic_theory}, that provide an over-damped time evolution for the gauge bosons. The main consequence of this result is that the time evolution of the distribution function $f$ is a Langevin equation, namely
\begin{equation}
\label{eq:langevin_equation_for_soft_gauge_bosons}
    \frac{\pi m^2_D}{8 p} \frac{d f}{dt} = -(p^2 + m_W^2(h)) f +\xi(t)\,,
\end{equation}
where $m_D$ is the Debye mass of the gauge bosons, which in the SM is $m_D^2 = 11g_W^2 T^2/6 $, while $m_W$ is the mass of the W bosons $m_W = g h/2$. The term $\xi(t)$, instead is a white-Gaussian noise that originates from the interaction between soft and hard modes. 

The Langevin equation in eq.~(\ref{eq:langevin_equation_for_soft_gauge_bosons}) correctly models the soft gauge bosons but it is still affected by some limitations that we will discuss later. 
We can use eq.~(\ref{eq:langevin_equation_for_soft_gauge_bosons}) to compute the friction arising from the soft modes following the strategy as outlined in~\cite{Moore:2000wx}. As a fist step we split the distribution function in an equilibrium and an out-of-equilibrium part. Next we average over the noise to get
\begin{equation}
    \frac{\pi m^2_D}{8 p}\f{df_0}{dt} = E^2 \delta f\,.
\end{equation}
where we also neglected the derivative of the perturbation as in ref.~\cite{Moore:2000wx}.
The above equation yields the expression for the perturbation $\delta f$ 
\begin{equation}
    \delta f = \f{\pi m^2_D \gamma_w v_w}{16 p E^3T}f_0(1+f_0)\f{dh}{dt}\f{dm_W^2}{dh}\,,
\end{equation}
that we can use to compute the integrated friction $P_{out}$ given by\footnote{Differently from ref.~\cite{Moore:2000wx} we set $N_W = 9$}
\begin{equation}
\label{eq:soft_bosons_oeq_pressure}
    P_{out} = \gamma_wv_w\f{9 m_D^2 T}{32\pi L_h}\int_0^1\f{1-x}{x}dx\,,
\end{equation}
where the integration variable is $x = h/h_-$. Notice that for $x = 0$ correspond to $z = -\infty$ while $ x = 1$ is $z = \infty$.

We notice that the above expression is IR divergent. This IR divergence stems from the ultra-soft particles ($p\lesssim g^2 T$) in the symmetric phase with vanishing mass and signalizes the breakdown of the effective kinetic theory we used. This is the limitation that we mentioned at the beginning of our analysis. In the derivation of the Langevin equation it is assumed that the W bosons can be treated as classical fields given the large occupation number of the soft region. When quantum effects become relevant, as for the ultra-soft particles, the effective kinetic theory breaks down since, as we discussed in Chapter~\ref{ch:effective_kinetic_theory}, the particle wavelength is comparable with the width of the wall $\lambda\sim L$. To remove the IR divergence we set a cutoff $x_{IR}$ in the integration in eq.~(\ref{eq:soft_bosons_oeq_pressure}) as in ref.~\cite{Moore:2000wx} by ignoring the contribution from ultra-soft particles in the symmetric phase. Since we can express the condition $\lambda \ll L$ as
\begin{equation}
    m_W(h) L_h \gg 1\,,
\end{equation}
the IR cutoff $x_{IR}$ is
\begin{equation}
\label{eq:IR_cutoff}
    x_{IR} = \f{1}{m_{W-} L_h}\,,
\end{equation}
where with $m_{W-}$ we denoted the mass of gauge bosons in the broken phase namely $m_{W-} = m_W(z = +\infty)$. Setting the cutoff $x_{IR}$ in eq.~(\ref{eq:soft_bosons_oeq_pressure}) cures the IR divergence.

The integrated friction computed using eq.~(\ref{eq:soft_bosons_oeq_pressure}) is smaller than the one obtained from the solution of the Boltzmann equation.
Comparing with this case, the Langevin equation predicts that the W bosons provide a friction $2$ times smaller for the benchmark point BP1 and $6$ times smaller for BP2. In addition, using eq.~(\ref{eq:soft_bosons_oeq_pressure}) we determine that W bosons contribute to the $20\%$ and $5\%$ of the total friction for BP1 and BP2 respectively. These results however are affected by an intrinsic uncertainty. The pressure in eq.~(\ref{eq:soft_bosons_oeq_pressure}) is very sensitive to the exact value of $x_{IR}$ in eq.~(\ref{eq:IR_cutoff}) since halving the value of the cutoff doubles the integrated friction. This sets a large theoretical absolute error on the friction. Assuming a $100\%$ relative uncertainty on the exact value of the integrated friction arising from W bosons we can then estimate a relative error on the real value of the out-of-equilibrium friction as $20-30\%$ for the first benchmark point and a $5-10\%$ for the second one. This translates in a large uncertainty on the final value of the terminal velocity that can be refined only by an improved characterization of the IR plasma dynamics.

Providing such an improved theory is a difficult challenge. On the one hand it would require to improve our description of the IR dynamics of hot gauge theories which is still not completely understood. On the other hand, for the case of the EWPT, such an improved theory should also account for the quantum corrections using the Schwinger-Keldysh-Kadanoff-Baym formalism which is far more complicated than the Boltzmann equation.




\chapter*{Conclusions and outlook}
\addcontentsline{toc}{chapter}{Conclusions and outlook}

In this thesis we studied the non-equilibrium dynamics of the EWPT in new-physics models in which it becomes a first-order transition. The study is of particular interest for the determination of the DW terminal velocity which is one of the main quantities that determine the experimental signatures of FOPTs. Such a quantity crucially impacts the GW signal emitted during the bubble dynamics that characterizes the EWPT, as well as many other interesting experimental signatures such as the amount of baryon asymmetry generated by the mechanism of EWBG. Despite its relevance, the terminal velocity is the observable we have less control on.

The theoretical challenge in the determination of the DW terminal velocity is represented by the friction arising from the out-of-equilibrium perturbations in the plasma. In the past many methods have been proposed to determine such perturbations by solving the Boltzmann equation that governs them through a suitable ansatz. 
All the methods proposed rely on the computation of suitable moments and are affected by many ambiguities, such as the choice of the set of weights to integrate the Boltzmann equation. This generates important quantitative and qualitative differences in their predictions, as we discussed in Chapter~\ref{ch:boltzmann_ansatz}.

We found the description of the non-equilibrium dynamics provided by the moment method unsatisfactory since it is unable to yield an accurate computation of the DW terminal velocity. For this reason we provided, for the first time, a full quantitative solution to the Boltzmann equation. Contrary to the previous approaches, we solved the Boltzmann equation without imposing any constraint on the shape of the perturbations. In such a way we were able to eliminate the ambiguities of the previous methods clarifying the non-equilibrium behaviour of the plasma, such as the absence of peaks in the integrated friction as predicted by the OF. In addition, our approach allows one to obtain a reliable quantitative solution to the Boltzmann equation which provides an accurate characterization of the non-equilibrium properties of the plasma. This is an important milestone in the computation of the terminal velocity of the DW and hence of the characterization of the GW signal, the amount of baryogenesis produced by EWBG and the other cosmological relics.

We solved the complete linearized Boltzmann equation 
by integrating it along the characteristic curves of the Liouville operator.
In addition we devised a new method to deal with the collision operator, to cope with the complexity that we reintroduced in its computation by not assuming any specific ansatz on the shape of the perturbations. We reinterpreted the collision operator as a Hermitian operator acting on the perturbation and we effectively decomposed it on the basis of its eigenfunctions. In such a way a complex nine-dimensional integration is reduced to a much faster matrix multiplication. Moreover, as a consequence of the rotational symmetries of the problem, the collision integral is block diagonal on the basis of spherical harmonics.

We showed that this basis is very useful to study not only the collision integral but also the properties of the whole Boltzmann equation and its solution. Our analysis of the multipole decomposition of the solution revealed a hierarchy in the Legendre modes as well as in the collision operator. This allowed us to truncate the angular momentum expansion at a finite order providing an error $\lesssim 1\%$ in the computation of the collision operator by taking into account the first $11$ modes in the expansion. In such a way we were able to solve the Boltzmann equation in less than a hour on a desktop computer.

The use of the spherical harmonics basis provided further insights on the structure of the Boltzmann equation itself. In the case of slow walls we showed that, by decomposing the Liouville operator on the angular momentum basis, the Boltzmann equation couples only adjacent Legendre modes. This result allowed us to solve semi-analytically the Boltzmann equation when the system is close to the hydrodynamic regime. In this situation we recovered shapes of the perturbations similar to the ones of the OF, confirming that the moment method is a valid description when the mean free path of particles is small compared to the width of the wall.
%
%

To assess the validity of our approach we first focused on a simplified setup where we considered the top quark to be the only out-of-equilibrium species. We compared the perturbation, the friction and the integrated friction computed with our method with the ones obtained following the OF and NF prescriptions. Despite small qualitative differences in the friction are found, being the shape of the latter quantity mostly dictated by the gradients of the mass, large quantitative differences between the three different approaches are found by inspecting the integrated friction. The comparison of this quantity shows a good agreement between the NF and the full solution for velocities $v_w\lesssim 0.2$ while a $\sim 10\%$ discrepancy between the old formalism and its extensions and the full solution is present in the same region. The agreement, however, worsens in the the supersonic region. The old formalism predicts the presence of peaks whose position is determined by the zero eigenvalues of the Liouville operator. Such peaks are unphysical and absent in the full solution of the Boltzmann equation that predicts a smooth behaviour for the integrated friction across the whole velocity range. On the other hand the NF correctly predicts the absence of peaks but the friction that it provides strongly disagrees with the one provided by the full solution at large velocities.
Finally a more refined analysis on the shape of the perturbation was carried out. An overall qualitative agreement between the different methods was found in some regions of the momentum space of the solution. However, the analysis highlighted the the new and old formalism poorly reproduce at the quantitative level the solution to the Boltzmann equation.

This analysis showed the main drawbacks of the previous approaches employed to solve the Boltzmann equation. In particular it confirms that the presence of peaks in the old formalism is an artifact of the moment method. 
An accurate determination of the friction, and hence of the wall terminal velocity, thus requires a full solution of the Boltzmann equation. In this regard, our results represent an important step towards a theoretical understanding of the terminal velocity. In particular, our procedure can be embedded in a much larger algorithm to accurately characterize the bubble dynamics that takes place during the EWPT. 

Having assessed the validity of our method, as a second step we computed the terminal velocity of the DW in the singlet extension of the SM for two benchmark points in the parameter space of the theory. To properly determine the velocity, we included the contribution of the light degrees of freedom which we modeled as a plasma in local thermal equilibrium as well as of the W bosons. Our analysis showed the out-of-equilibrium perturbation crucially impacts the DW dynamics. The terminal velocity presents a $\sim 30-50\%$ discrepancy with repect to the value computed ignoring the out-of-equilibrium perturbations. This emphasizes the important role played by the non-equilibrium properties of the plasma in the characterization of the cosmological relics of a first-order EWPT, further confirming the importance of an accurate modeling of the out-of-equilibrium perturbations.

In addition, we found that the impact of W bosons is comparable with the one of the top quark and reduces the value of the terminal velocity by a $10-30\%$ factor with respect to the top only case. This is in contrast to the na\"ive expectations based on the scaling of the friction which predicts a suppressed contribution for the W-bosons. The origin of such a discrepancy is in the IR behaviour of W bosons.

A careful analysis of the friction provided by the W bosons revealed that the latter receives its largest contribution from a soft bosons condensate, which, however, cannot be described by an effective Boltzmann equation. The latter is an effective kinetic theory that describes the dynamics of hard modes and cannot be applied to the study of long wave particles. A proper description of soft modes, which correctly includes Debye screening and Landau damping that dominate the dynamics of such excitations, is provided by an effective Langevin equation. We used the latter to compute the friction of soft gauge bosons and compared it with our previous result. We found that the correct approach provides a friction $2$ and $6$ times smaller that the one predicted by solving the full Boltzmann equation for the benchmark points we analyzed. However, the integrated friction we obtained is IR divergent. Such IR divergence is caused by ultra-soft modes in the symmetric phase and is regularized by setting a cutoff. The results, however, are very sensible to the IR cutoff scale. Indeed, we found a $50\%$ change of the friction by halving the value of the cutoff.

This effect is the source of the largest theoretical uncertainty among the three effects we studied in this thesis, namely the IR gauge bosons, the leading-log approximation and the massless approximation in the collision integrals. The latter two in fact provide a $5\%$ and a $2\%$ uncertainty on the terminal velocity respectively. However, a correct characterization of the IR bosons is a much more difficult challenge with respect to the other sources of uncertainties in the computation. Assessing the impact of other approximations, such as the full dependence of the thermal propagator from the self-energy or the inclusion of ``$1\to 2$'' processes in the collision integrals is straightforward, as we already know how to include these effects in the Boltzmann equation. On the other hand, a correct characterization of the properties of soft gauge bosons can only be established by the correct effective kinetic theory of such modes. Contrary to the results found in the literature, we expect such a theory to predict the correct IR behaviour of the friction hence showing that particles in the symmetric phase have a negligible impact on the DW dynamics and not a divergent contribution. Unfortunately, we still do not have an effective description of gauge bosons valid at soft and ultra-soft scale and since an accurate characterization of the DW dynamics requires such an effective theory, we believe that determining such description is one of the most important open problems of this research field.

Additional improvements in the description of the non-equilibrium properties of the plasma developments are also provided by the inclusion of ``$1\to 2$'' processes in the collision integrals and of the thermal self-energy of particles in their respective propagators. As today, their impact on the non-equilibrium properties of the plasma during EWPT has always been neglected. 


\chapter*{Acknowledgements}

First of all, I thank my supervisors, Giuliano and Stefania, for their continuous support, their invaluable help, even in the most frustrating moments, and for the patience they showed despite my mistakes.

Second, I thank Luigi for his precious advice and help and my friend and colleague \'Angel with whom I shared many frustrations and who helped me to bear the most difficult moments of these years.

I am very grateful to my group of friends at the university, Andrea, Guido, Veronica, Francisco, Flavio, Chiara, Costanza, Alessia, Luigi, Ettore, Tommaso, Matteo and Michele and to my friends of a lifetime, Chiara, Ilaria, Andrea and Lorenzo who all supported me.

A special thanks to my parents, who have always encouraged me since when I was a little kid, and my wife, Elisa, who never stopped believing in me, even in my lowest moments. Her words and help guided me during the thoughest times of my years at the university.
\appendix

\chapter{Analytical results in the fluid approximation}
\label{ap:exact_solution}

In this appendix we provide the analytical expressions of the Boltzmann equation solution and the integrated friction for the fluid approximation when we consider a generic set of the form
\begin{equation}
    \label{eq:generic_set}
    \{p^{n_1}/E^{m_1},\,p^{n_2}/E^{m_2},\,p^{n_3}/E^{m_3}\}\,.
\end{equation}
Both expressions are computed assuming that particles are massless when we integrate over the momentum space. Our results are straightforwardly generalized to extended version of the fluid approximation.

\section{Analytic solution to the Boltzmann equation}

To determine the analytic solution we need to first compute the expression of the $A$ and $\Gamma$ matrices, resulting from the integration of the Liouville and the collision operator, as well as for the source term $\vec{S}$.

\subsection{Matrix $A$ for a generic set of weights}

We begin our analysis from the matrix $A$. The latter results from the integration of the derivative term of the Boltzmann equation which we remind is given by
\begin{equation}
    \frac{p_z}{E}\partial_z(\mu_i+\beta\gamma(E-vp_z)\delta \tau_i+\beta\gamma(p_z-v E)\delta v_i)\,.
\end{equation}
Integrating the above expression with the set of weights in eq.~(\ref{eq:generic_set}) yields
\begin{equation}
   A_i=\left(\begin{array}{c c c}
   C_i^{m_1+1,n_1+1} & \gamma_w(C_i^{m_1,n_1+1}-v_wC_i^{m_1+1,n_1+2}) & \gamma_w(C_i^{m_1+1,n_1+2}-v_wC_i^{m_1,n_1+1})\\
   C_i^{m_2+1,n_2+1} & \gamma_w(C_i^{m_2,n_2+1}-v_wC_i^{m_2+1,n_2+2}) & \gamma_w(C_i^{m_2+1,n_2+2}-v_wC_i^{m_2,n_2+1})\\
    C_i^{m_3+1,n_3+1} & \gamma_w(C_i^{m_3,n_3+1}-v_wC_i^{m_3+1,n_3+2}) & \gamma_w(C_i^{m_3+1,n_3+2}-v_wC_i^{m_3,n_3+1})\,.
    \end{array}\right)
\end{equation}
modulus the factors $\beta^{m_1 - n_1 -3}$, $\beta^{m_2 - n_2 -3}$, $\beta^{m_3 - n_3 -3}$ multiplying the first second and third rows respectively and where the coefficients $C_i^{m,n}$ are defined in eq.~(\ref{eq:C_coefficients}). The coefficients present a non-trivial dependence on the terminal velocity. Such dependence, however, can be factored out by expressing $C_i^{m,n}$ as the product of a function of the velocity and a constant. For this purpose it is convenient to write the integral in eq.~(\ref{eq:C_coefficients}) using plasma variables, namely
\begin{equation}
\begin{split}
    C_i^{m,n}&=T^{m-n-3}\gamma^{n-m+1}\int\frac{d^3|\bar {\bf p}|}{|\bar {\bf p}|(2\pi)^3}\frac{(\bar p_z+v|\bar {\bf p}|)^n}{(|\bar {\bf p}|+v\bar p_z)^{m-1}}(-f'_0(\beta |\bar {\bf p}|))\\
    &=\frac{\gamma_w^{n-m+1}}{(2\pi)^2}\int_0^{\infty}dx x^{n-m+2}(-f'_{0,i}(x))\int_{-1}^{+1}d\cos\theta\frac{(\cos\theta+v_w)^n}{(1+v_w\cos\theta)^{m-1}}.
\end{split}
\end{equation}
and to define
\begin{equation}
\begin{split}
c_i^l&=\frac{1}{2\pi^2}\int_0^{+\infty}dxx^l(-f'_{0,i}(x))\,,\\
 a_{m,n}(v_w)&=\frac{\gamma_w^{n-m+1}}{2}\int_{-1}^{1}d\cos\theta\frac{(\cos\theta+v_w)^{n+1}}{(1+v_w\cos\theta)^m}\,.
\label{eq:Ca_definition}
\end{split}
\end{equation}
In such a way we can write the coefficients $C_i^{m,n}$ as
\begin{equation}
    C_i^{m,n}= a_{m-1,n-1}(v_w)c_i^{n-m+2}\,,
\end{equation}
encoding the whole velocity dependence in the functions $a_{m,n}$.

Both the coefficients $c_i^l$ and the functions $a_{m,n}$ can be computed analytically. The latter can be expressed in term of the hypergeometric functions and is well defined for $ n \neq -2k - 1$, with $k>0$ a natural number. The cases $ n = -2k $ can be handled by taking the principal value of the integral in the definition of $a_{m,n}$ in eq.~(\ref{eq:Ca_definition}).
The computation of the $c_i^l$ is instead straightforward. In the fermionic case they are well defined for $l\geq 0$ yielding
\begin{equation}
    c_f^l = \begin{cases}
        \displaystyle\frac{1}{2\pi^2}(1-2^{1-l})\Gamma(l+1)\zeta(l) &l\neq 1 \land l \geq 0\,,\\
        \\
        \displaystyle\frac{\log(2)}{2\pi^2} &l = 1\,.
    \end{cases}
\end{equation}
In the bosonic case the coefficient $c_{b}^1$ diverges. Such divergence is regularized by the mass of the bosons $m$ and at leading order in $m/T$ we find 
\begin{equation}
    c_b^l = \begin{cases}
        \displaystyle\frac{1}{2\pi^2}\Gamma(l+1)\zeta(l) &l >1\,,\\
        \\
        \displaystyle\frac{\log(2T/m)}{2\pi^2} &l = 1\,.
    \end{cases}
\end{equation}

We can further simplify the $A$ matrix by introducing an additional function
$b_{m,n}$ of the terminal velocity defined as
\begin{equation}
\label{eq:b_functions}
     b_{m,n}(v_w) =\frac{\gamma_w^{n-m+1}}{2}\int_{-1}^{1}d\cos\theta\frac{(\cos\theta+v_w)^{n+1}}{(1+v_w\cos\theta)^m}\cos\theta\,.
\end{equation}
that can be computed similarly to $a_{m,n}$. Using
\begin{equation}
\label{eq:a_properties}
\begin{split}
    &\gamma_w(a_{m-1,n}(v_w) - v_w a_{m,n+1}(v_w)) = a_{m,n}(v_w)\,,\\
    &\gamma_w(a_{m,n+1}(v_w) - v_w a_{m-1,n}(v_w)) = b_{m,n}(v_w)\,.
\end{split}
\end{equation}
we can express the $A$ matrix as
\begin{equation}
    A_i=\left(\begin{array}{c c c}
   a_{m_1,n_1}(v_w)c_i^{n_1-m_1+2} & a_{m_1,n_1}(v_w)c_i^{n_1-m_1+3} & b_{m_1,n_1}(v_w)c_i^{n_1-m_1+3}\\
    a_{m_2,n_2}(v_w)c_i^{n_2-m_2+2} & a_{m_2,n_2}(v_w)c_i^{n_2-m_2+3} & b_{m_2,n_2}(v_w)c_i^{n_2-m_2+3}\\
    a_{m_3,n_3}(v_w)c_i^{n_3-m_3+2} & a_{m_3,n_3}(v_w)c_i^{n_3-m_3+3} & b_{m_3,n_3}(v_w)c_i^{n_3-m_3+3}
      \end{array}\right)
      \label{eq:diff_matrix_of_massless}
\end{equation}
which is its final expression.

\subsection{Matrix $\Gamma$ for a generic set of weights}

Just as we did for the matrix $A$ in the previous section, it is also possible to compute the matrix $\Gamma$ for a generic set of weights and to factor out the velocity dependence of its elements. This is very convenient since it simplifies a lot the numerical computation of such terms.

We begin our analysis from the expression
\begin{equation}
    \int \frac{d^3{\bf p}}{(2\pi)^3 E_p}\frac{p_z^n}{E_p^m}\bar{\cal C}[\mu_i,\tau_i,\delta v_i] = \frac{1}{2N_p}\sum_j\int \tilde d(\bar p, \bar k, \bar p', \bar k')\frac{p_z^n}{E_p^m}|{\cal M}_j|^2\bar {\cal P}[\mu_i, \delta\tau_i, \delta v_i]\delta^4(p + k - p' - k')\,.
\end{equation}
where the sum runs over the relevant processes and where we defined
\begin{equation}
    \tilde d(p, k, p', k') = \frac{d^3|{\bf p}|d^3|{\bf k}|d^3|{\bf p}'|d^3|{\bf k}'|}{(2\pi^8)16E_{p}E_{k}E_{p'}E_{k'}}.
\end{equation}
We formally write
\begin{equation}
 \int \frac{d^3{\bf p}}{(2\pi)^3 E_p}\frac{p_z^n}{E_p^m}\bar{\cal C}[\mu_i,\tau_i,\delta v_i] = \Gamma^{(\mu_i)}_{m,n}\mu_i+\Gamma^{(\tau)}_{m,n}\delta\tau_i+\Gamma^{(v_i)}_{m,n}\delta v_i,
\end{equation}
and focus on the case of the top quark. The case of the W bosons can be trivially obtained by following the same technique. Also, for simplicity, we omit the subscript $t$ since the top quark is the only particle species that we consider in the plasma. It is convenient to write the expressions of $\Gamma_{m,n}^{(\mu)}$, $\Gamma_{m,n}^{(\tau)}$, $\Gamma_{m,n}^{(v)}$ in the plasma reference frame
\begin{equation}
    \begin{split}
        \Gamma_{m,n}^{(\mu)}&=\frac{\gamma_w^{n-m}}{N_p}\int \tilde d(p,k,p',k')|{\cal M}_1|^2\delta^4(p+k-p'-k')\bar{\cal P}_0\frac{(\bar p_z+v_wE_ {\bar p})^n}{(E_{\bar p}+v_w\bar p_z)^m},\\
    \Gamma_{m,n}^{(\tau)}&=\frac{\beta\gamma_w^{n-m}}{2N_p}\int\tilde d(p,k,p',k')[(E_{\bar p}+E_{\bar k})|{\cal M}_1|^2+(|{\cal M}_2|^2+|{\cal M}_3|^2)(E_{\bar p}-E_{\bar p'})]\delta^4 \bar{\cal P}_0\frac{(\bar p_z+v_wE_{\bar p})^n}{(E_{\bar p}+v_w\bar p_z)^m},\\
        \Gamma_{m,n}^{(u)}&=\frac{\beta\gamma_w^{n-m}}{2N_p}\int\tilde d(p,k,p',k')[(\bar p_z+\bar k_z)|{\cal M}_1|^2+(|{\cal M}_2|^2+|{\cal M}_3|^2)(\bar p_z-\bar p'_z)]\delta^4\bar{\cal P}_0\frac{(\bar p_z+v_wE_{\bar p})^n}{(E_{\bar p}+v_w\bar p_z)^m},
    \end{split}
    \label{eq:gamma_matrices_entries}
\end{equation}
where we identified with ${\cal M}_1$, ${\cal M}_2$, ${\cal M}_3$ the amplitudes of the annihilation process $tt\rightarrow gg$ and of the scattering processes $tg\rightarrow tg$, $tq\rightarrow tq$ respectively. Furthermore $\delta^4$ is a shorthand notation for the Dirac delta, while we defined
\begin{equation}
    \bar{\cal P}_0 = f_0(\beta E_{\bar p})f_0(\beta E_{\bar k})(1\pm f_0(\beta E_{\bar p'}))(1\pm f_0(\beta E_{\bar k'}))\,.
\end{equation}
 We recall that we treat particles in the collision to be massless. Within such approximation we may replace $E_{\bar p} \rightarrow |{\bar {\bf p}}|$. At last we factorize the temperature dependence by performing the following transformation for each momenta of the particles involved in the process $$\beta p\rightarrow p\,.$$
It is not hard to show that
\begin{equation}
    \Gamma^{(a)}_{m,n}\rightarrow\beta^{m-n-4}\Gamma^{(a)}_{m,n},
\end{equation}
where $a = \mu, \tau, v$.

To factor out the velocity dependence of the elements $\Gamma_{m,n}^{(a)}$ we first integrate over one of the four momenta using the Dirac delta. The choice of the momentum depends on the Mandelstram variable that appears at the denominator of $|{\cal M}|^2$ as explained in Appendix~\ref{app:evaluation_collision_integrals} and ref.~\cite{Arnold:2003zc}. Because the strategy that we are going to outline does not depend on this choice, let us integrate over the phase space $k'$. The structure of $\Gamma^{(a)}_{m,n}$ is



\begin{equation}
    \Gamma^{(a)}_{m,n}=\int_0^{+\infty}|\bar{\bf p}|^2d|\bar{\bf p}||\bar{\bf k}|^2d|\bar{\bf k}||\bar{\bf p}'|^{2}d|\bar{\bf p}|'\int d\Omega_p d\Omega_k d\Omega_{p'} f^{(a)}( {\bar {\bf p}}, {\bar {\bf k}}, {\bar {\bf p}}')\frac{(\bar p_z+v |{\bar {\bf p}}|)^n}{( |{\bar {\bf p}}|+v\bar p_z)^m}\,,
\end{equation}
where $f^{(a)}$ can be easily read from equation (\ref{eq:gamma_matrices_entries}) and $d^3{\bf p} = |{\bf p}|^2d|{\bf p}|d\Omega_p$ and where $\Omega_p$ is the solid angle. 

As a second step we orient the plasma reference frame in such a way that the momentum ${\bf p}$ lies along the $z$ direction of the new frame. In such a way we can express the momenta and the versors $\hat{x}, \hat{y}$ and $\hat{z}$ of the plasma reference frame as
\begin{equation}
\label{eq:new_frame_variables}
\begin{split}
    &{\bar {\bf p}}=|{\bar {\bf p}}|\left(\begin{array}{c}
        0\\0\\1
    \end{array}\right),\;\;\;\;\;{\bar {\bf k}}=|{\bar {\bf k}}|\left(\begin{array}{c}\sin\theta_{\bar p \bar k}\cos\tilde\phi_{\bar k}\\\sin\theta_{\bar p \bar k}\sin\tilde\phi_{\bar k}\\\cos\theta_{\bar p \bar k}\end{array}\right),\;\;\;\;\;{\bar {\bf p}}'=|{\bar {\bf p}}'|\left(\begin{array}{c}\sin\theta_{\bar p \bar p'}\cos\tilde\phi_{\bar p'}\\\sin\theta_{\bar p \bar p'}\sin\tilde\phi_{\bar p'}\\\cos\theta_{\bar p \bar p'}
    \end{array}\right).\\
    &\hat{x}=\left(\begin{array}{c}\cos\theta_{\bar p}\cos\phi_{\bar p}\\-\sin\phi_{\bar p}\\\sin\theta_{\bar p}\cos\phi_{\bar p}
    \end{array}\right),\;\;\;\;\;\hat{y}=\left(\begin{array}{c}\cos\theta_{\bar p}\sin\phi_{\bar p}\\\cos\phi_{\bar p}\\\sin\theta_{\bar p}\sin\phi_{\bar p}\end{array}\right),\;\;\;\;\;\hat{z}=\left(\begin{array}{c}-\sin\theta_{\bar p}\\0\\\cos\theta_{\bar p}\end{array}\right).\;\;\;\;\;
\end{split}
\end{equation}
where $\tilde\phi_i$ is the azimuthal angle measured in the new reference frame. We then perform the following transformation over the integration measure

\begin{equation}
    \int_{-1}^{1}d\cos\theta_{\bar p}d\cos\theta_{\bar k}d\cos\theta_{\bar p'}\int_0^{2\pi}d\phi_{\bar p}d\phi_{\bar k}d\phi_{\bar p'}\rightarrow\int_{-1}^{1}d\cos\theta_{\bar p}d\cos\theta_{\bar p\bar k}d\cos\theta_{\bar p\bar p'}\int_0^{2\pi}d\phi_{\bar p}d\tilde\phi_{\bar k}d\tilde\phi_{\bar p'},
\end{equation}
and we define
\begin{equation}
    g^{(a)}_{m,n}({\bar {\bf p}_i},\theta_{\bar p \bar k},\theta_{\bar p \bar p'})=\frac{1}{2(2\pi)^3}\int_{-1}^1d\cos\theta_{\bar p}\int_0^{2\pi}d\phi_{\bar p}d\tilde\phi_{\bar k}d\tilde\phi_{\bar p'}f^{(a)}({\bar {\bf p}},{\bar {\bf k}},{\bar {\bf p}}')\frac{(\cos\theta_{\bar p}+v_w)^n}{(1+v_w\cos\theta_{\bar p})^m}\,.
\end{equation}
%

We can now factor out the velocity dependence from the $\Gamma_{m,n}^{(a)}$ elements.
The functions $f^{(\mu)}$ and $f^{(\tau)}$ are already written in terms of rotational invariants thus using eq.~(\ref{eq:Ca_definition}) we find
\begin{equation}
\begin{split}
    g^{(\mu,\tau)}_{m,n} & =\frac{1}{2}\int_{-1}^1d\cos\theta_{\bar p}\frac{(\cos\theta_{\bar p}+v_w)^n}{(1+v_w\cos\theta_{\bar p})^m}f^{(\mu,\tau)}(|{\bar{\bf p}}_i|,\theta_{\bar p \bar k},\theta_{\bar p \bar p'})\\ &=\frac{1}{\gamma_w^{n-m}}a_{m,n-1}(v_w)f^{(\mu,\tau)}(|{\bar{\bf p}}_i|,\theta_{\bar p \bar k},\theta_{\bar p \bar p'})\,.
\end{split}
\end{equation}
For the function $f^{(v)}$ we can express $k_z$ and $p'_z$ using eq.~(\ref{eq:new_frame_variables}). Carrying out the integration over the azimuthal angles and using eq.~(\ref{eq:b_functions}) we find
\begin{equation}
\begin{split}
    g^{(v)}_{m,n} & =\frac{1}{2}\int_{-1}^1d\cos\theta_{\bar p}\frac{(\cos\theta_{\bar p}+v_w)^n}{(1+v_w\cos\theta_{\bar p})^m}\cos\theta_{\bar p}f^{(v)}(|{\bar{\bf p}}_i|,\theta_{\bar p \bar k},\theta_{\bar p \bar p'})\\ &=\frac{1}{\gamma_w^{n-m}}b_{m,n-1}(v_w)f^{(v)}(|{\bar{\bf p}}_i|,\theta_{\bar p \bar k},\theta_{\bar p \bar p'})\,.
\end{split}
\end{equation}
We conclude that the expression of the $\Gamma^{(\mu,\tau,v)}_{m,n}$ is given by
\begin{equation}
  \begin{split}
        \Gamma_{m,n}^{(\mu)}&=a_{m,n-1}(v_w)\beta^{m-n-4}\tilde\Gamma_{m,n}^{(\mu)}\\
        \Gamma_{m,n}^{(\tau)}&=a_{m,n-1}(v_w)\beta^{m-n-4}\tilde\Gamma_{m,n}^{(\tau)}\\
        \Gamma_{m,n}^{(u)}&=b_{m,n-1}(v_w)\beta^{m-n-4}\tilde\Gamma_{m,n}^{(v)},
    \end{split}
    \label{eq:gamma_matrices_entries_vfact}
\end{equation}

with

\begin{equation}
  \begin{split}
        \tilde\Gamma_{m,n}^{(\mu)}&=\frac{1}{N_p}\int \tilde d(p,k,p',k')|{\cal M}_1|^2\delta^4(p+k-p'-k'){\bar{\cal P}}_0p^{n-m},\\
        \tilde\Gamma_{m,n}^{(\tau)}&=\frac{1}{2N_p}\int\tilde d(p,k,p',k')[(E_{\bar p}+E_{\bar k})|{\cal M}_1|^2+(|{\cal M}_2|^2+|{\cal M}_3|^2)(E_{\bar p}-E_{\bar p'})]\delta^4{\bar{\cal P}}_0|{\bar{\bf p}}|^{n-m},\\
        \tilde\Gamma_{m,n}^{(v)}&=\frac{1}{2N_p}\int\tilde d(p,k,p',k')[(|{\bar{\bf p}}|+|{\bar{\bf k}}|\cos\theta_{\bar p\bar k})|{\cal M}_1|^2+(|{\cal M}_2|^2+|{\cal M}_3|^2)(|{\bar{\bf p}}|-|{\bar{\bf p}}|'\cos\theta_{\bar p\bar p'})]\delta^4{\bar{\cal P}}_0|{\bar{\bf p}}|^{n-m}.    \end{split}
\end{equation}
The elements $\tilde \Gamma^{(a)}_{m,n}$ can be computed following ref.~\cite{Arnold:2003zc}.

In conclusion for a given set of weights $\{p_z^{n_1}/E^{m_1}$, $p_z^{n_2}/E^{m_2}$,$p_z^{n_3}/E^{m_3}\}$ the matrix $\Gamma$ is
\begin{equation}
    \Gamma=\left(\begin{array}{l l l}
    a_{m_1,n_1-1}(v_w)\tilde\Gamma^{(\mu)}_{m_1,n_1} & a_{m_1,n_1-1}(v_w)\tilde\Gamma^{(\tau)}_{m_1,n_1} & b_{m_1,n_1-1}(v_w)\tilde\Gamma^{(v)}_{m_1,n_1}\\
    a_{m_2,n_2-1}(v_w)\tilde\Gamma^{(\mu)}_{m_2,n_2} & a_{m_2,n_2-1}(v_w)\tilde\Gamma^{(\tau)}_{m_2,n_2} & b_{m_2,n_2-1}(v_w)\tilde\Gamma^{(v)}_{m_2,n_2}\\
    a_{m_3,n_3-1}(v_w)\tilde\Gamma^{(\mu)}_{m_3,n_3} & a_{m_3,n_3-1}(v_w)\tilde\Gamma^{(\tau)}_{m_3,n_3} & b_{m_3,n_3-1}(v_w)\tilde\Gamma^{(v)}_{m_3,n_3}
    \end{array}\right)\,,
\end{equation}
modulo a factor $\beta^{m_1-n_1-4}$, $\beta^{m_2-n_2-4}$, $\beta^{m_3-n_3-4}$ respectively for the first, second and third row. The values of the $\tilde\Gamma$ elements in the fluid approximations for the top quark are
\begin{equation}
\begin{split}
    & \tilde\Gamma^{(\mu_t)}_{0,0} = 0.00128\,,\; \tilde\Gamma^{(\tau_t)}_{0,0} = \tilde \Gamma^{(\mu_t)}_{-1,0} = \tilde \Gamma^{(\mu_t)}_{0,1} = 0.00303\,,\\
    &\tilde\Gamma^{(\tau_t)}_{-1,0} = \tilde \Gamma^{(\tau_t)}_{0,1} = 0.0141\,,\;\tilde\Gamma^{(v_t)}_{-1,0} = \tilde \Gamma^{(v_t)}_{0,1} = 0.00992\,.
\end{split}
\end{equation}
%

In the NF we can factor out the velocity dependence of every matrix elements of $\Gamma$ in eq.~(\ref{eq:A_Gamma_matrices_nf}) but $\Gamma_{v,t}^{(3)}$. We can fit the velocity dependence of the latter using a quartic polynomial. We find
\begin{equation}
\begin{split}
    \Gamma_{\mu,t}^{(1)} & = \Gamma^{(\mu_t)}_{0,0}\,, \;\;\;\;\; \Gamma_{\mu,t}^{(2)} = \Gamma^{(\mu_t)}_{-1,0}\,, \;\;\;\;\; \Gamma_{\tau,t}^{(1)} = \Gamma^{(\tau_t)}_{0,0}\,, \;\;\;\;\; \Gamma_{\tau,t}^{(2)} = \Gamma^{(\tau_t)}_{-1,0}\,,\\
    \Gamma_{\mu,t}^{(3)} & = \Gamma^{(\mu_t)}_{1,1} = 0.00128 \left(-\frac{1}{\gamma_w^2 v_w} \tanh ^{-1}(v_w)+\frac{1}{v_w}\right)\,,\\
    \Gamma_{\tau,t}^{(3)} & = \Gamma^{(\tau_t)}_{1,1} = 0.00303 \left(-\frac{1}{\gamma_w^2 v_w} \tanh ^{-1}(v_w)+\frac{1}{v_w}\right)\,,\\
    \Gamma_{v,t}^{(1)} &= -0.000758 \left(\frac{1}{\gamma_w^2} \log \left(\frac{1-v_w}{v_w+1}\right)-2v_w\right)\,,\\
    \Gamma_{v,t}^{(2)} &= -0.000139 \f{v_w}{\gamma_w} \left(v_w^2-1\right) \left(v_w \log \left(\frac{2}{v_w+1}-1\right)+2\right)\\
    &+0.000830 \gamma_w\left(-2 v_w^3-\frac{1}{\gamma_w^4}
   \log \left(\frac{2}{v_w+1}-1\right)+4 v_w\right)\\
   &-0.000898\gamma_w \left(\frac{1}{\gamma_w^2} \log \left(\frac{1-v_w}{v_w+1}\right)-2v_w\right)\,,\\
   \Gamma_{v,t}^{(3)} &= 0.000634 + 0.00102 v_w^2 + 0.000835 v_w^3 - 0.000964 v_w^4\,.
\end{split}
\end{equation}

\subsection{Source term}

At last we need the expression for $\vec{S}$ for a generic set of weights. We recall that the source term is given by$$\vec{S}_t=\gamma_w v_w\beta\frac{(m_t^2)'}{2E_p}.$$
By integrating the above expression with the set of weights in eq.~(\ref{eq:generic_set}) and using the definition of the functions $a_{m,n}$ in eq.~(\ref{eq:Ca_definition}) we find

\begin{equation}
    \vec{S}=\gamma_w v_w\frac{(m_t^2)'}{2}\left(\begin{array}{c} \beta^{m_1-n_1-1}a_{m_1,n_1-1}(v_w)c_f^{n_1-m_1+1}\\\beta^{m_2-n_2-1}a_{m_2,n_2-1}(v_w)c_f^{n_2-m_2+1} \\\beta^{m_3-n_3-1}a_{m_3,n_3-1}(v_w)c_f^{n_3-m_3+1} \end{array}\right)\,.
\end{equation}

\subsection{Analytical solution of the perfect fluid}

We now discuss how to determine the analytic solution of eq.~(\ref{eq:fluid_equation_top}). We first focus on a delta function response, namely the limit where the width of the wall is $L\rightarrow 0$, that is
\begin{equation}
\label{eq:delta_function_response}
    A_t \partial_zG(z-z_0) + \Gamma_t G(z-z_0) = \mathbb{I}\delta(z - z_0)\,,
\end{equation}
where $G(z-z_0)$ is the Green function of the operator $A_t\partial_z + \Gamma_t$. To determine the Green function 
it is convenient to write eq.~(\ref{eq:delta_function_response}) on the basis where the matrix $A^{-1}_t\Gamma_t$ is diagonal. By identifying with $\chi_t$ the matrix of eigenvectors, and with $\tilde G = \chi^{-1}_tG\chi$ the Green function in the basis of the eigenvectors we find
\begin{equation}
    \partial_z \tilde G_{ij}(z - z_0) + \lambda_i \tilde G_{ij}(z-z_0) = (\chi_t^{-1}A_t^{-1}\chi)_{ij}\delta(z-z_0)\,.
\end{equation}
The Green function $\tilde G$ is then a linear combination of exponential tails with decay lengths set by the eigenvalues $\lambda_i$ of $A^{-1}_t\Gamma_t$. The values of the coefficients entering in the linear combination are then set
by the boundary conditions. We require the system to recover equilibrium far from the bubble wall, which corresponds to the requirement that the vector ${\vec q}$ vanishes at $z = \pm \infty$. Thus, to ensure an exponential decay, we should keep only the positive eigenvalues for $z>z_0$ and the negative ones for $z < z_0$. The final expression for $G$ is found by returning to the original basis and is given by
\begin{equation}
    G_{ij}(z-z_0) = \sum_k\chi_{ik}(A_t^{-1}\Gamma_t)_{kj}\textrm{sign}(\textrm{Re}(\lambda_k))e^{-\lambda_j(z-z_0)}\theta((z-z_0)\textrm{sign}(\textrm{Re}(\lambda_k)))
\end{equation}

The Green function provides some interesting information on the structure of the perturbations. These indeed are linear combinations of exponential tails. The decay lengths given by the inverse of the eigenvalues $\lambda_i$ also provide an estimate of the typical mean free path of the particles in the plasma. In fact, such length sets the typical scale where the plasma recovers equilibrium.

It is important to notice that for supersonic walls, namely $v_w > c_s $, in the fluid approximation, particles don't spread in front of the wall. For such values of the velocity, the three eigenvalues $\lambda_i$ are all positive. As a consequence the perturbations trails the source since the Green function vanishes in front of the wall and the resulting fluctuations in the symmetric phase are largely suppressed providing no net baryon asymmetry as discussed in Chapter~\ref{ch:boltzmann_ansatz}.

We can determine the solution ${\vec q}_t$ for a generic wall width by exploiting the Green function. We find
\begin{equation}
	{\vec q}_t(z) = \gamma_w v_w\int_{-\infty}^\infty dx G(z - x) {\vec F}_t \frac{m_t^2(h_-)}{4T}\frac{d}{dx}\left(1+\tanh\left(\frac{x}{L}\right)\right)^2
\end{equation}
where $m_t = y_t h_-/\sqrt{2}$ where $h_-$ denotes the Higgs VEV in the broken phase $y_t$ is the top quark Yukawa coupling, $L$ is the width of the DW while ${\vec F}_i = (c_{i1}\,,\;\gamma_wc_{i2}\,,\;0) $. The above integral, can be solved exactly by using
\begin{equation}
\begin{split}{}_2F_1(a,b;c;x)=\frac{\Gamma(c)}{\Gamma(b)\Gamma(c-b)}\int_0^1y^{b-1}(1-y)^{c-b-1}(1-xy)^{-a}dy, \\ \;\;\;\;\textrm{with}\;\;\;{\rm Re}(c)>{\rm Re}(b)\land|\arg(1-x)|<\pi
\end{split}
\end{equation}
which provides
\begin{equation}
\label{eq:fluid_solution}
    q_i(z)=\gamma_w v_w\frac{m_t^2(h_-)}{4T}\sum_k\chi_{ik}(\chi^{-1}A^{-1}\vec{F})_k\sign(\lambda_k)J\left(L\lambda_k,\frac{z}{L}\right)
\end{equation}
where we defined
\begin{equation}
\label{eq:Hypergeometric_part}
    \begin{split}
    & J(a,x) = \sign(\Re(a))(1+a+\tanh(x))(1+\tanh(x))\\
     &2(2+a)\left[\frac{-1-\sign(\Re(a))}{2}+{}_2F_1\left(1,\frac{\sign(\Re(a))a}{2};\frac{\sign(Re(a))a}{2}+1;-e^{2x\sign(\Re(a))}\right)\right]\,.
    \end{split}
\end{equation}

\section{Analytical expression for the integrated friction}

The friction in eq.~(\ref{eq:out_of_eq_friction}) can be written using the definition of the coefficients in eq.~(\ref{eq:C_coefficients}) and~(\ref{eq:D_coefficients}) as
\begin{equation}
\label{eq:friction_top_fluid_approximation}
	F(z) = \frac{dm^2_t}{dz}\frac{N_t}{2}{\vec C}_t\cdot{\vec q}_t
\end{equation}
where for the case of the OF $\vec{C}_t = (C_t^{1,0}, C_t^{0,0},0)^{\rm T}$ while in the NF $\vec{C}_t = (C_t^{1,0}, C_t^{0,0},D_t^{0,-1})^{\rm T}$.

The integral of the friction plays a crucial role in the determination of the terminal velocity $v_w$, as we explained in Chapter~\ref{ch:bubble_dynamics_and_plasma_hydrodynamics} and can be determined analytically in the massless case as we are now going to show. As shown in ref.~\cite{Moore:1995si}, the integral of the friction can be evaluated using the Fourier transform. The transformed perturbations are easily computed from eq~(\ref{eq:fluid_equation_top}) and the final result is given by
\begin{equation}
\label{eq:integral_friction_fluid_approximation}
    \int_{-\infty}^{\infty} dzF(z) = N_t\frac{m_t^4(h_T)}{16}C_{ti}\sum_{i,j}\chi_{ij}(\chi^{-1}A^{-1}F)_jI(L\lambda_j)\,,
\end{equation}
where
\begin{equation}
    I(a) = 4a\left[\left(1-\frac{a^2}{4}\right)I_1\left(\frac{a\pi}{2}\right)+\frac{1}{3}\right]\,,
\end{equation}
with the function $I_1$ defined as
\begin{equation}
    I_1(a) = - 2 + \frac{a\pi}{\sin(a)^2}\sign({\rm Re}(a)) - \sum_{n=1}^{\infty}\frac{n(2\pi a)^2}{((n\pi)^2 - a^2)^2}\,.
\end{equation}
The authors in ref.~\cite{Moore:1995si} provide some useful asymptotic behaviours of the series. However, we can find an analytic result in term of the trigamma function $\psi_1$.\footnote{$\psi_1$ must not be confused with the $l = 1$ Legendre mode of the perturbation in eq.~(\ref{eq:perturbation_decomposition}).}
Using the following useful identities for the trigamma function
\begin{equation}
    \psi_1(z) = \sum_{n=0}^{\infty}\frac{1}{(n+z)^2}\,,\;\;\;\;\; \psi_1(1-z)+\psi(z)=\frac{\pi^2}{\sin^2(\pi z)}\,,\;\;\;\;\psi_1(1+z)=\psi_1(z)-\frac{1}{z^2}\,,
\end{equation}
we find that 
\begin{equation}
        \sum_{n=1}^{\infty}\frac{n(2\pi a)^2}{((n\pi)^2-a^2)^2}= \sign({\rm Re}(a))\left(\frac{\pi}{a\sin(a)^2}-\frac{\pi^2}{a^2}-2\psi_1(1+\sign({\rm Re}(a))\frac{a}{\pi})\right)\,.
\end{equation}
The above expression can then be used to provide the final expression for the function $I(a)$ that is given by
\begin{equation}
     I(a)=4a\left[\left(1-\frac{a^2}{4}\right)\left(-2+\sign(\Re(a))\left(\frac{2}{a}+a\psi_1\left(1+\sign(\Re(a))\frac{a}{2}\right)\right)\right)+\frac{1}{3}\right]\,.
\end{equation}
\chapter{Evaluation of the collision integrals}\label{app:evaluation_collision_integrals}

\section{The term proportional to $\delta f(p)$}\label{sec:C_deltafp}

We focus, at first, on the term of the collisional integral proportional
to $\delta f(p)$, which, for a single matrix element, reads
\begin{equation}\label{eq:C_bar_int1}
{\cal J}[\delta f]=\frac{-\delta f(p)}{4 N_p }\frac{f_{v}(p)}{f_{v}'(p)}\int\!\!\frac{d^{3}{\bf k}\,d^{3}{\bf p'}\,d^{3}{\bf k'}}{(2\pi)^{5}2E_{k}\,2E_{p'}\,2E_{k'}}|{\cal M}|^{2}\delta^{4}(p+k-p'-k'){\cal \,}f_{v}(k)(1\pm f_{v}(p'))(1\pm f_{v}(k'))\,.
\end{equation}
To evaluate the integral it is convenient to change variables through a boost, going to the
plasma frame, in which the Boltzmann distribution is the standard
equilibrium one $f_{v}$. We denote by a bar the momenta in the plasma frame, namely
\begin{equation}
\bar{p}_{0}=\gamma(E_p-vp_{z})\,,\quad\bar{p}_{z}=\gamma(p_{z}-v E_p)\,,\quad\bar{p}_{\bot}=p_{\bot}\,,
\end{equation}
and analogously for $k$, $p'$ and $k'$. We thus get
(notice that the integration measure $d^{3}{\bf p}/E_p$ is invariant
under boost)
\begin{eqnarray}
\bar{\cal J}[\delta f]&=&\frac{-\delta f(p(\bar{p}))}{4 N_p}\frac{f_{0}(\bar{p})}{f_{0}'(\bar{p})}\label{eq:C_bar_int2}\\
&&\times\int\!\!\frac{d^{3}{{\bf \bar k}}\,d^{3}{{\bf \bar p'}}\,d^{3}{{\bf \bar k'}}}{(2\pi)^{5}2E_{\bar{k}}\,2E_{\bar{p}'}\,2E_{\bar{k}'}}|{{\cal M}}|^{2}\delta^{4}(\bar{p}+\bar{k}-\bar{p}'-\bar{k}'){\cal \,}f_{0}(\bar{k})(1\pm f_{0}(\bar{p}'))(1\pm f_{0}(\bar{k}'))\,.\nonumber
\end{eqnarray}

In order to evaluate the integrals, we follow the approach of ref.~\cite{Moore:1995si}, including only leading-log contributions.
In this approximation we can also neglect the masses of the particles involved in the scattering. This approximation significantly
simplifies the numerical evaluation, since it removes any explicit dependence on the $z$ coordinate in the integrals.
Closer inspection of the integral appearing in eq.~(\ref{eq:C_bar_int2}) shows that it is invariant under rotation of the
three-momentum components of $\bar p$, thus it is just a function of $E_{\bar p}$.\footnote{Rotation invariance is an immediate
consequence of the fact that the Boltzmann distribution $f_0$ depends only on the energy of the particle, while $|{\cal M}|^2$
is a function of the kinematic invariants (i.e.~the Mandelstam variables).}

The evaluation of the integral can be simplified by exploiting the delta function and the symmetries of the integrand.
In this way one can perform analytically five of the nine integrals. An efficient parametrization for performing the integration
is presented in ref.~\cite{Arnold:2003zc}.

In the leading log approximation, only t-channel and u-channel scattering amplitudes are relevant (see Table~\ref{tab:amplitudes}).
So we can focus on these two types of contributions and neglect s-channel processes (and interference terms).

\subsection*{t-channel parametrization}

We start by considering amplitudes coming from t-channel diagrams.
The integration over $d^{3}\bar{{\bf k'}}$ can be easily performed exploiting
the $\delta$-function. The remaining integrals can be handled through a change of variables.
As in ref.~\cite{Arnold:2003zc}, we introduce
the three-momentum $\mathbf{q}\equiv\mathbf{\bar{p}}'-\mathbf{\bar{p}}=\bar{\mathbf{k}}-\mathbf{\bar{k}}'$.
Rotational invariance allows us to trivially integrate on the orientation
of $\mathbf{q}$. Fixing $\mathbf{q}$ to be along a $z'$ axis, we
can express the orientation of the $\mathbf{\bar{p}}$ and $\mathbf{\bar{k}}$
momenta in terms of the polar angles $\theta_{\bar{p}q}$ and $\theta_{\bar{k}q}$ and
the azimuthal angle $\phi$ between the $\mathbf{\bar{p}}$-$\mathbf{q}$
and the $\mathbf{\bar{k}}$-$\mathbf{q}$ plane.

The remaining delta function can be handled by introducing an additional variable $\omega$
linked to the $t$ Mandelstam variable as $t \equiv \omega^2 - q^2$, where $q \equiv |{\bf q}|$. In this way the
integrations on the angles $\theta_{\bar{k}q}$ and $\theta_{\bar{p}q}$ can be performed analytically
and one is left with the final expression for the integral on the second line of eq.~(\ref{eq:C_bar_int2}):
\begin{equation}
\bar{\cal Q}=\frac{1}{8(2\pi)^{4}E_{\bar{p}}}\int\limits _{-E_{\bar{p}}}^{+\infty}d\omega\int\limits _{|\omega|}^{\omega+2E_{\bar{p}}}dq\int\limits _{\frac{q+\omega}{2}}^{+\infty}d E_{\bar{k}}\int\limits _{0}^{2\pi}d\phi\,|{{\cal M}}|^{2}f_{0}(\bar{k})(1\pm f_{0}(\bar{p}'))(1\pm f_{0}(\bar{k}'))\,.
\end{equation}

As alternative parametrization, which can help in the numerical integration and in studying the behavior of the integral,
one can define
\begin{equation}
\chi_{\pm}\equiv q\pm\omega\,,
\end{equation}
in terms of which
\begin{equation}
\int\limits _{-E_{\bar{p}}}^{+\infty}d\omega\int\limits _{|\omega|}^{\omega+2E_{\bar{p}}}dq\int\limits _{\frac{q+\omega}{2}}^{+\infty}d E_{\bar{k}}\quad\rightarrow\quad\frac{1}{2}\int\limits _{0}^{+\infty}d\chi_{+}\int\limits _{0}^{2 E_{\bar{p}}}d\chi_{-}\int\limits _{\chi_{+}/2}^{\infty}d E_{\bar{k}}\,.
\end{equation}
The $\chi_{\pm}$ parametrization can be also useful to leave as last integration the one on $E_{\bar{k}}$:
\begin{equation}
\frac{1}{2}\int\limits _{0}^{+\infty}d\chi_{+}\int\limits _{0}^{2 E_{\bar{p}}}d\chi_{-}\int\limits _{\chi_{+}/2}^{\infty}d E_{\bar{k}}\quad\rightarrow\quad\frac{1}{2}\int\limits _{0}^{\infty}d E_{\bar{k}}\int\limits _{0}^{2 E_{\bar{k}}}d\chi_{+}\int\limits _{0}^{2 E_{\bar{p}}}d\chi_{-}\,.
\end{equation}
This choice of integration order clearly shows the symmetric role of $E_{\bar{p}}$ and $E_{\bar{k}}$ in the collisional integral.

The expressions for the $s$ and $u$ Mandelstam variables as a function of $\omega$, $q$ and $E_{\bar k}$ are given by
\begin{eqnarray}
s & = & -\frac{t}{2q^{2}}\left\{ \left[(2 E_{\bar{p}}+\omega)(2 E_{\bar{k}}-\omega)+q^{2}\right]-\cos\phi\sqrt{(4 E_{\bar{p}}(E_{\bar{p}}+\omega)+t)(4 E_{\bar{k}}(E_{\bar{k}}-\omega)+t)}\right\}\,,\hspace{1.5em}\\
t & = & \omega^{2}-q^{2}\,,\\
u & = & -t-s\,,
\end{eqnarray}
while the relative angles between the three-momenta are given in ref.~\cite{Arnold:2003zc} (see Appendix A.2, eqs.~(A21a)-(A21e)),
among which
\begin{equation}
\cos\theta_{pq}=\frac{\omega}{q}+\frac{t}{2 E_{\bar{p}}q}\,,\qquad\cos\theta_{kq}=\frac{\omega}{q}-\frac{t}{2E_{\bar{k}}q}\,.
\end{equation}

\subsection*{u-channel parametrization}

Analogous formulae can be found for the u-channel parametrization,
by exchanging $\bar{\mathbf{p}}'$ and $\bar{\mathbf{k}'}$ in the
t-channel parametrization. In this way the integral becomes
\begin{equation}
{\cal Q}=\frac{1}{8(2\pi)^{4} E_{\bar{p}}}\int\limits _{- E_{\bar{p}}}^{+\infty}d\omega\int\limits _{|\omega|}^{\omega+2 E_{\bar{p}}}dq\int\limits _{\frac{q+\omega}{2}}^{+\infty}d E_{\bar{k}}\int\limits _{0}^{2\pi}d\phi\,|{{\cal M}}|^{2}f_{0}(\bar{k})(1\pm f_{0}(\bar{p}'))(1\pm f_{0}(\bar{k}'))\,,
\end{equation}
with $\mathbf{q}\equiv\mathbf{\bar{k}}'-\mathbf{\bar{p}}=\mathbf{\bar{k}}-\mathbf{\bar{p}}'$
and
\begin{equation}
\omega=E_{\bar{k}'}- E_{\bar{p}}=E_{\bar{k}}- E_{\bar{p}'}\,.
\end{equation}
 The expressions for the $s$ and $u$ Mandelstam variables are given
by
\begin{eqnarray}
s & = & -\frac{u}{2q^{2}}\left\{ \left[(2 E_{\bar{p}}+\omega)(2 E_{\bar{k}}-\omega)+q^{2}\right]-\cos\phi\sqrt{(4 E_{\bar{p}}(E_{\bar{p}}+\omega)+u)(4E_{\bar{k}}(E_{\bar{k}}-\omega)+u)}\right\}\,,\hspace{1.5em}\\
u & = & \omega^{2}-q^{2}\,,\\
t & = & -u-s\,,
\end{eqnarray}
 while
\begin{equation}
\cos\theta_{pq}=\frac{\omega}{q}+\frac{u}{2E_{\bar{p}}q}\,,\qquad\cos\theta_{kq}=\frac{\omega}{q}-\frac{u}{2 E_{\bar{k}}q}\,.
\end{equation}

\subsection*{Structure of the contribution}

From the above formulae we can easily infer the global structure of
the collisional term proportional to $\delta f(p)$. The quantity
(we consider the $t$-channel parametrization for definiteness)
\begin{equation}
\bar{\cal Q}=\frac{1}{8(2\pi)^{4} E_{\bar{p}}}\int\limits _{-E_{\bar{p}}}^{+\infty}d\omega\int\limits _{|\omega|}^{\omega+2 E_{\bar{p}}}dq\int\limits _{\frac{q+\omega}{2}}^{+\infty}d E_{\bar{k}}\int\limits _{0}^{2\pi}d\phi\,|{{\cal M}}|^{2}f_{0}(\bar{k})(1\pm f_{0}(\bar{p}'))(1\pm f_{0}(\bar{k}'))\,,
\end{equation}
only depends on $E_{\bar{p}}$, as we already anticipated.
Therefore we get
\begin{equation}
\overline{\cal J}[\delta f]=- \frac{\delta f(p(\bar{p}))}{4N_p}\frac{f_{0}(\bar{p})}{f_{0}'(\bar{p})}{\cal K}[E_{\bar{p}}]\,.
\end{equation}

We can now go back to the wall frame, obtaining
\begin{equation}
{\cal J}[\delta f]=-\frac{\delta f(p(\bar{p}))}{4N_p}\frac{f_{v}(p)}{f_{v}'(p)}{\cal Q}[\gamma_w(E_p-v_wp_{z})]
= \frac{1}{4N_p} \frac{\delta f(p)}{E_p}\left(1\pm e^{-\beta\gamma_w(E_p-v_wp_{z})}\right){\cal Q}[\gamma_w(E_p-v_wp_{z})]\,.
\end{equation}
Notice that the massless-limit approximation introduced a small `mismatch' in this expression, since we chose $f_{v}(p)$
in the prefactors to have the full mass dependence (from the definition of $E_p$). In the approach to the solution via the use
of weights, instead, $f_{v}$ is treated in the massless limit for all the factors in the collisional integrals.
This problem could be solved by also considering the massive form for all the $f_{v}$ factors inside the collisional integral.
This however is computationally more demanding, since it introduces an explicit $z$ dependence in the integrand, so that
the kernel should be evaluated also as a function of $z$.\footnote{Notice that a full treatment would also need a redefinition
of the matrix element $|{\cal M}|^{2}$ and of the integration boundaries.}

The numerical analysis shows a behavior
\begin{equation}
\bar{\cal Q}(E_{\bar{p}}) \sim \log E_{\bar{p}}+const
\end{equation}
which, as expected, has a logarithmic divergence for mass and thermal mass going to zero. We can thus infer the rough
behavior (at least for small $v$)
\begin{equation}
\bar{\cal J}[\delta f] \sim \delta f(p)(\log E_{\bar{p}}+const)\,.\label{eq:C_behavior}
\end{equation}

\section{The terms $\langle \delta f\rangle$}\label{sec:bk_deltafp}

The second ingredient we need in order to compute the collision integrals is the determination of the terms $\langle\delta f\rangle$
in which the perturbation appears under the integral sign. The generic structure of the term that depends on $\delta f(k)$ is
\begin{equation}\label{eq:CI_dfk}
\langle\delta f(k)\rangle = \frac{-1}{4N_p E_p}\int\!\!\frac{d^3{\bf k}\,d^3{\bf p'}\,d^3{\bf k'}}{(2\pi)^5\,2E_k\,2E_{p'}\,2E_{k'}}|{\cal M}_A|^2\delta^4(p+k-p'-k')f_v(k)(1\pm f_v(p'))(1\pm f_v(k'))\frac{\delta f(k)}{f'_v(k)}\,,
\end{equation}
and analogous expressions are valid for the $\delta f(p')$ and $\delta f(k')$ contributions.

The above integral can in principle be evaluated using the same manipulations we described in section~\ref{sec:C_deltafp}.
However, the integrand, due to the $\delta f(k)$ factor, is not rotationally invariant, and an additional integration over the
direction of $\bf k$ with respect to the $z$ axis remains.
The final result is (also in this section we treat all the particles as massless)
\begin{equation}
\langle\delta f(k)\rangle = -\frac{1}{32N_p}\frac{f_0(p)}{(2\pi)^5 E_p} {\cal I}[p_\bot, p_z, z]\,,
\end{equation}
where ${\cal I}$, written as a function of the $p$ momentum in the plasma frame, reads
\begin{equation}
{\cal I} = \int_{-\bar E_p}^{+\infty}d\omega\int_{|\omega|}^{\omega+2 E_{\bar p}}dq\int_{\frac{q+\omega}{2}}^{+\infty}d E_{\bar k}\int_0^{2\pi} d\phi\int_0^{2\pi}d\phi_k|{\cal M}|^2f_0(\bar k)(1\pm f_0(\bar p'))(1\pm f_0(\bar k'))\frac{\delta f(\bar k)}{f'_0(\bar k)}\,,
\end{equation}
in which $\phi_k$ denotes the angle between the vector ${\bf k}$ and the plane where ${\bf p}$ and $\hat z$, the direction along which the wall moves, lie.
The integral ${\cal I}$ depends on the three variables $p_\bot$, $p_z$ and $z$, and requires five numerical integrations.
Therefore its evaluation on a fine grid, as required in our numerical approach, is quite cumbersome.

An alternative procedure to manipulate the integral can be used to reduce the number of numerical integrations.
This can be done by performing the integration over ${\bf p}'$ and ${\bf k}'$ in eq.~(\ref{eq:CI_dfk}) and leaving the integral over
${\bf k}$ as a last step. In this way the expression for $\langle \delta f(k)\rangle$ can be brought to the form
\begin{equation}
\langle\delta f(k)\rangle = -\int\frac{d^3{\bf k}}{2 E_k}{\cal K}_A\, f_v(k)\frac{\delta f(k)}{f'_v(k)}\,,
\end{equation}
where we recall
\begin{equation}
{\cal K}_A = \frac{1}{8N_p(2\pi)^5}\int\frac{d^3{\bf k}'d^3{\bf p}'}{2E_{p'}2E_{k'}}|{\cal M}_A|^2(1\pm f_v(p'))(1\pm f_v(k'))\delta^4(p+k-p'-k')\,.
\end{equation}
Since ${\cal K}_A$ is a Lorentz scalar, it will be a function of the only Lorentz scalars that can be obtained from the four vectors $p^\mu$, $k^\mu$ and the plasma velocity $U^\mu$, namely, $U^\mu p_\mu$, $U^\mu k_\mu$ and $p^\mu k_\mu$.
These quantities are related, respectively, to the energies $E_{\bar p}$ and $E_{\bar k}$ of the incoming particles in the plasma reference frame
and to the angle $\theta_{\bar p \bar k}$ between the momenta $\bar p$ and $\bar k$. Putting everything together we find
\begin{equation}
\langle\delta f(k)\rangle = -\int \frac{d^3{\mathbf{\bar{k}}}}{2 E_{\bar{k}}}{\cal K}_A(E_{\bar{p}}, E_{\bar{k}}, \theta_{\bar{p}\bar{k}})\,f_0(\bar{k})\frac{\delta f(k_\perp,\gamma_w(\bar{k}_z+v_w E_{\bar{k}}),z)}{(-f'_0(\bar{k}))}\,.
\end{equation}
which can be rewritten as
\begin{equation}
\langle\delta f(k)\rangle= -\frac{1}{2}  \int_0^{\infty}\!\!\! E_{\bar{k}}\, d E_{\bar{k}}\int_{-1}^1\!\!\!d\cos\theta_{\bar{p}\bar{k}}\,{\cal K}_A(E_{\bar{p}}, E_{\bar{k}} ,\theta_{\bar{p}\bar{k}})\int_0^{2\pi}\!\!\!\!d\phi_{\bar{k}}\,f_0(\bar{k})\frac{\delta f(k_\perp,\gamma_w(\bar{k}_z+v_w E_{\bar{k}}),z)}{(-f'_0(\bar{k}))}\,.
\end{equation}

The collision integral for the scattering processes includes an additional set contributions in which $\delta f(p')$
or $\delta f(k')$ appears. In analogy to the previous case, for the $\delta f(p')$ terms, we can first perform the integrals over ${\bf k}$ and ${\bf k'}$,
obtaining the following expression
\begin{equation}
\langle\delta f(p')\rangle= \int \frac{d^3{\mathbf{\bar{p}}}'}{2 E_{\bar{p}'}}{\cal K}_A(E_{\bar{p}}, E_{\bar{p}'}, \theta_{\bar{p}\bar{p}'})\,f_0(\bar{p}')\frac{\delta f(p_\perp',\gamma_w(\bar{p}_z'+v_w E_{\bar{p}}),z)}{(-f'_0(\bar{p}'))}\,,
\end{equation}
which can also be rewritten as
\begin{eqnarray}
\langle\delta f(p')\rangle &=& \frac{1}{2}\int_0^\infty E_{\bar{p}'}\,d E_{\bar{p}'}\int_{-1}^1 d\cos\theta_{\bar{p}\bar{p}'}{\cal K}_S(E_{\bar{p}}, E_{\bar{p}'},\theta_{\bar{p}\bar{p}'})\times\nonumber\\
&& \hspace{5em}\times\int_0^{2\pi} d\phi_{\bar{p}'}(1 \pm f_0(\bar{p}'))\frac{\delta f(p'_\perp,\gamma_w(\bar{p}'_z+v_w E_{\bar{p}'}),z)}{(-f'_0(\bar{p}'))}\,.
\end{eqnarray}
The contributions from  $\delta f(k')$ can be treated in an analogous way.

\subsection{Evaluation of the ${\cal K}_A$ kernel}

The evaluation of the kernel ${\cal K}_A$ can be performed as in ref.~\cite{DeGroot:1980dk}. As a first step we perform the integration over ${\bf k}'$ exploiting the Dirac delta:
\begin{equation}
{\cal K}_A=\frac{1}{8N_p(2\pi)^5}\int \frac{d^3{\bf p'}}{2 E_{p'}}\frac{1}{2 E_{k'}}|{\cal M}|^2(1\pm f_0(U^\mu p'_\mu))(1\pm f_0(U^\mu k'_\mu))\delta(E_p+E_k-E_{p'}-E_{k'}).
\end{equation}
Notice that in the above expression we expressed the energies $E_{p'}$ and $E_{k'}$ in the Lorentz-invariant form
$U^\mu p'_\mu$ and $U^\mu k'_\mu$. As we will see, this is useful to keep track of the changes of reference frame.

As a second step, we rewrite the Dirac delta (in the center-of-mass (COM) frame) as
\begin{equation}
\delta(E_p+E_k-E_{p'}-E_{k'})=\delta(\sqrt{s}-2E_{p'}) = \frac{1}{2}\delta\left(\frac{1}{2}\sqrt{s}-E_{p'}\right)\,,
\end{equation}
with $s = (p + k)^2$ the usual Mandelstam variable.
The integration over ${\bf p}'$ can be performed by rewriting $d^3{\bf p}' = E_{p'}^2\, d E_{p'}\, d\cos\theta\, d\phi$,
where $\theta$ is the angle between ${\bf p}'$ and ${\bf p}$ in the COM frame of the scattering process:
\begin{equation}
{\cal K}_1=\frac{1}{(2\pi)^5}\frac{1}{64N_p}\int_{-1}^1d\cos\theta\int_0^{2\pi}d\phi|{\cal M}|^2(1\pm f_0(U^\mu p'_\mu))(1\pm f_0(U^\mu k'_\mu))\,.
\end{equation}

As a last step we need to compute $U^\mu p'_\mu$ and $U^\mu k'_\mu$ in the COM frame.
We conveniently choose the orientation of the COM frame axes such that $U^y=0$ leading to
\begin{equation}
U^\mu p'_\mu = U^0\frac{\sqrt{s}}{2} - U^x\frac{\sqrt{s}}{2}\sin\theta\cos\phi - U^z\frac{\sqrt{s}}{2}\cos\theta 
\end{equation}
We then introduce the four-vectors $P^\mu$ and $Q^\mu$ defined as
\begin{equation}
P^\mu = p^\mu+k^\mu\,,\qquad\quad
Q^\mu =p^\mu-k^\mu\,.
\end{equation}
In the COM frame we find that
\begin{equation}
P^\mu = \left(\begin{array}{c}\sqrt{s}\\0\end{array}\right)\,,
\qquad\quad
Q^\mu = \left(\begin{array}{c}0\\ \mathbf{Q}\end{array}\right)\,.
\end{equation}
We can get a further simplification by choosing the frame such that ${\bf Q}$ lies along the $z$ axis.
Since, in the massless case, $| {\bf P} |=|{\bf Q}|=\sqrt{s}$, the vectors $P^\mu/\sqrt{s}$ and $Q^\mu/\sqrt{s}$ coincide with the versors
along the first and fourth Minkowski directions.

The $U^0$ and $U^z$ components can be easily computed in terms of the momenta of the particles in the plasma frame
:
\begin{equation}
U^0=\frac{U^\mu P_\mu}{\sqrt{s}}=\frac{E_{\bar{p}}+ E_{\bar{k}}}{\sqrt{s}}\,\qquad \quad
U^z =-\frac{U^\mu Q_\mu}{\sqrt{s}}=-\frac{(E_{\bar{p}}- E_{\bar{k}})}{\sqrt{s}}\,.
\end{equation}
The $U^x$ component can be determined from the condition $U^\mu U_\mu = 1$:
\begin{eqnarray}
U^x&=&\sqrt{U_0^2-U_z^2-1}=\frac{1}{\sqrt{s}}\sqrt{(E_{\bar{p}}+E_{\bar{k}})^2-(E_{\bar{p}}-E_{\bar{k}})^2-s}\nonumber\\
&=&\frac{1}{\sqrt{s}}\sqrt{4 E_{\bar{p}} E_{\bar{k}}-s}=\frac{1}{\sqrt{s}}\sqrt{2 E_{\bar{p}} E_{\bar{k}} (1+\cos\theta_{\bar{p}\bar{k}})}\,.
\end{eqnarray}

Putting everything together we find that $u^\mu p'_\mu$ is given by
\begin{equation}
\begin{split}
U^\mu p'_\mu & =U^0 \frac{\sqrt{s}}{2} - U^x\frac{\sqrt{s}}{2}\sin\theta\cos\phi - U^z\frac{\sqrt{s}}{2}\cos\theta\\
    & = \frac{E_{\bar{p}}+E_{\bar{k}}}{2}-\frac{1}{2}\sqrt{2 E_{\bar{p}} E_{\bar{k}}(1+\cos\theta_{\bar{p}\bar{k}})}\sin\theta\cos\phi + \frac{(E_{\bar{p}} - E_{\bar{k}})}{2}\cos\theta\\
    & = \frac{1}{2}\left(E_{\bar{p}} (1+\cos\theta) + E_{\bar{k}} (1-\cos\theta) - \sqrt{2 E_{\bar{p}} E_{\bar{k}}(1+\cos\theta_{\bar{p}\bar{k}})}\sin\theta\cos\phi\right)\,.
\end{split}
\end{equation}
Similarly we find
\begin{equation}
U^\mu k'_\mu = \frac{1}{2}\left(E_{\bar{p}} (1-\cos\theta) + E_{\bar{k}}(1+\cos\theta) + \sqrt{2E_{\bar{p}} E_{\bar{k}} (1+\cos\theta_{\bar{p}\bar{k}})}\sin\theta\cos\phi\right)\,.
\end{equation}
Finally, the Mandelstam variables are given by
\begin{equation}
t = -\frac{s}{2}(1-\cos\theta)\,\qquad\quad
s = 2E_{\bar{p}} E_{\bar{k}} (1-\cos\theta_{\bar{p}\bar{k}})\,.
\end{equation}

\subsection{Evaluation of the ${\cal K}_S$ kernel}

We now discuss the evaluation of the ${\cal K}_S$ kernel:
\begin{equation}
    {\cal K}_S=\frac{1}{8N_p(2\pi)^5}\int\frac{d^3{\bf k}\,d^3{\bf k'}}{2E_k\,2E_{k'}}|{\cal M}|^2f_0(U^\mu k_\mu)(1\pm f_0(U^\mu k'_\mu))\delta^4(p+k-p'-k')
\end{equation}
Also in this case we follow ref.~\cite{DeGroot:1980dk}. We introduce the four-vectors
\begin{equation}
\begin{split}
   &K^\mu = k^\mu+k'^\mu\\
   &P^\mu=p^\mu+p'^\mu\\
   &Q'^\mu=k^\mu-k'^\mu\\
   &Q^\mu=p^\mu-p'^\mu
\end{split}\,.
\end{equation}
Recalling that
\begin{equation}
\frac{d^3{\bf k}\,d^3{\bf k'}}{2E_k\, 2E_{k'}}=d^4k\,d^4k'\,\theta(E_k)\theta(E_{k'})\delta(k^2)\delta(k'^2)\,,
\end{equation}
we can use as integration variables $K$ and $Q'$ finding
\begin{equation}
\frac{d^3{\bf k}\,d^3{\bf k'}}{2E_k\,2E_{k'}}=\frac{1}{4}d^4K\,d^4Q'\,\theta(K_0)\theta(K^2)\delta(K^2+Q'^2)\delta(K^\mu Q'_\mu)\,.
\end{equation}
Since $\delta^4(p+k-p'-k')=\delta^4(Q+Q')$, we can integrate over $Q'$ obtaining
\begin{equation}
{\cal K}_S=\frac{1}{8N_p(2\pi)^5}\int \frac{1}{4}d^4K\theta(K_0)\delta(K^2+Q^2)\delta(K^\mu Q_\mu)|{\cal M}|^2 f_0(U^\mu k_\mu)(1\pm f_0(U^\mu k'_\mu))
\end{equation}
In the massless case
\begin{equation}
Q^2=-P^2=t\,,
\end{equation}
hence
\begin{equation}
\delta(K^2+Q^2)=\delta(K^2+t)\,.
\end{equation}
Using the identity
\begin{equation}
d^4K\,\theta(K^0)\delta(K^2+t)=\frac{d^3{\bf K}}{2\sqrt{{\bf K}^2-t}}\,,
\end{equation}
we can rewrite ${\cal K}_2$ as
\begin{equation}
{\cal K}_S=\frac{1}{64N_p(2\pi)^5}\int\frac{d^3{\bf K}}{\sqrt{{\bf K}^2-t}}\delta(K^\mu Q_\mu)|{\cal M}|^2f_0(U^\mu k_\mu)(1\pm f_0(U^\mu k'_\mu))\,.
\end{equation}

We can now rewrite this formula in the COM frame, in which $P^\mu=(\sqrt{-t},0,0,0)$ and $Q^\mu=(0,0,0,\sqrt{-t})$.
Introducing polar coordinates for ${\bf K}$, 
with polar angles $\theta$ and $\phi$, one gets
\begin{equation}
\delta(K^\mu Q'_\mu)=\delta(|{\bf K}|\sqrt{-t}\cos\theta)=\frac{1}{|{\bf K}|\sqrt{-t}}\delta(\cos\theta)\,,
\end{equation}
which allows to trivially perform the integration over $\cos \theta$, leading to
\begin{equation}
{\cal K}_S=\frac{1}{64N_p(2\pi)^5}\int\frac{|{\bf K}|\,d|{\bf K}|\,d\phi}{\sqrt{{\bf K}^2-t}\sqrt{-t}}|{\cal M}|^2 f_0(U^\mu k_\mu)(1\pm f_0(U^\mu k'_\mu))\,.
\end{equation}

As a last step, we need to determine the expressions for the $U^\mu$  components.
Focusing on $U^\mu k_\mu$ we find
\begin{equation}
U^\mu k_\mu =\frac{U^\mu}{2} (K_\mu+Q'_\mu)=\frac{U^\mu}{2}(K_\mu-Q_\mu)=\frac{1}{2}\left(U^0 \sqrt{{\bf K}^2-t}-U^x|{\bf K}|\cos\phi+U^z\sqrt{-t}\right)\,.
\end{equation}
Where we chose the orientation of the COM frame in such way that $u^y=0$. In an analogous way we find
\begin{equation}
U^\mu k'_\mu=\frac{U^\mu}{2} (K_\mu-Q'_\mu)=\frac{U^\mu}{2}(K_\mu+Q_\mu)=\frac{1}{2}\left(U^0 \sqrt{{\bf K}^2-t}-U^x|{\bf K}|\cos\phi-U^z\sqrt{-t}\right)\,.
\end{equation}
Exploiting the fact that $P^\mu/\sqrt{-t}$ and $Q^\mu/\sqrt{-t}$ coincide with the versors in the time and $z$ directions, we can write
\begin{equation}
\begin{split}
U^0&= \frac{U^\mu P_\mu}{\sqrt{-t}}=\frac{ E_{\bar{p}}+ E_{\bar{p}'}}{\sqrt{-t}}\,,\\
U^z&=-\frac{U^\mu Q_\mu}{\sqrt{-t}}=-\frac{E_{\bar{p}}- E_{\bar{p}'}}{\sqrt{-t}}\,.
\end{split}
\end{equation}
Finally, from $U^\mu U_\mu=1$, one gets
\begin{equation}
U^x=\sqrt{\frac{(E_{\bar{p}}+ E_{\bar{p}'})^2}{-t}-\frac{(E_{\bar{p}}- E_{\bar{p}'})^2}{-t}-1}=\frac{1}{\sqrt{-t}}\sqrt{2 E_{\bar{p}} E_{\bar{p}'}(1+\cos\theta_{\bar{p}\bar{p}'})}\,.
\end{equation}

In order to make the numerical evaluation of the kernel more stable, we used the following coordinate change $|{\bf K|}=\sqrt{-t}\tan\theta$, and then we defined $1/\cos\theta=x$. The expression for ${\cal K}_S$ becomes
\begin{equation}
{\cal K}_S =\frac{1}{8(2\pi)^5}\int_1^\infty \int_0^{2\pi}dx\,d\phi\,|{\cal M}|^2f_0(U^\mu k_\mu)(1\pm f_0(U^\mu k'_\mu))
\end{equation}
with
\begin{equation}
\begin{split}
U^\mu k_\mu &=\frac{1}{2}\left((E_{\bar{p}}+E_{\bar{p}'})x-\sqrt{x^2-1}\sqrt{2E_{\bar{p}} E_{\bar{p}'}(1+\cos\theta_{\bar{p}\bar{p}'})}\cos\phi+(E_{\bar{p}}-E_{\bar{p}'})\right)\,,\\
U^\mu k_\mu &=\frac{1}{2}\left((E_{\bar{p}}+E_{\bar{p}'})x-\sqrt{x^2-1}\sqrt{2E_{\bar{p}} E_{\bar{p}'}(1+\cos\theta_{\bar{p}\bar{p}'})}\cos\phi-(E_{\bar{p}}-E_{\bar{p}'})\right)\,,\\
s &=\frac{-t}{2}(x+1)\,,\\
u &=\frac{t}{2}(x-1)\,.
\end{split}
\end{equation}

\clearpage

\providecommand{\href}[2]{#2}\begingroup\raggedright\endgroup
\end{document}